\documentclass[11pt,a4paper]{amsart}
\usepackage{amssymb}
\usepackage{mathptmx}
\usepackage{fullpage}
\usepackage{longtable}
\usepackage{aliascnt}
\usepackage[hidelinks]{hyperref}
\usepackage[nameinlink,capitalize]{cleveref}
\allowdisplaybreaks
\renewcommand{\Pr}{\operatorname*{\mathbf{Pr}}}
\newtheorem{theorem}{Theorem}[section]
\newaliascnt{proposition}{theorem}\newtheorem{proposition}[proposition]{Proposition}\aliascntresetthe{proposition}
\newaliascnt{lemma}{theorem}\newtheorem{lemma}[lemma]{Lemma}\aliascntresetthe{lemma}
\newaliascnt{corollary}{theorem}\newtheorem{corollary}[corollary]{Corollary}\aliascntresetthe{corollary}
\theoremstyle{definition}
\newaliascnt{definition}{theorem}\newtheorem{definition}[definition]{Definition}\aliascntresetthe{definition}
\newaliascnt{remark}{theorem}\aliascntresetthe{remark}
\newtheorem{openproblem}{Open Problem}
\crefname{openproblem}{open problem}{open problems}
\Crefname{openproblem}{Open Problem}{Open Problems}
\title{\textbf{The Complexity of ReLU Network Synthesis from a Parameter Multiset}}
\author[Dimitrios Myrisiotis]{Dimitrios Myrisiotis\\
Great Bay University\\
China}
\date{}
\hypersetup{pdftitle={The Complexity of ReLU Network Synthesis from a Parameter Multiset},pdfauthor={Dimitrios Myrisiotis}}
\begin{document}
\begin{abstract}
We ask how hard it is to build a ReLU network when its parameter values are given in advance.
An instance supplies an appropriate encoding of a target behavior, a multiset of rational numbers, and a bound on the number of neurons (output included).
The question is whether one can choose the graph of a ReLU network within the bound, and place the numbers on its edges and biases so that it meets the target.
This separates the numbers from the architecture.
Matching two labeled examples exactly is already \(\mathsf{NP}\)-complete, so we ask what changes when the target must hold on every input or on average.
Agreeing with a Boolean circuit on every Boolean input is \(\Sigma_2^{\mathsf P}\)-complete, already for four neurons.
Agreeing with a given ReLU network on every real input lies in \(\mathsf{DP}\) for every fixed neuron bound and is \(\mathsf{DP}\)-complete for three neurons.
Approximating a ReLU network on the unit cube within absolute error one everywhere is \(\Sigma_2^{\mathsf P}\)-complete, even with no hidden neurons and all numbers equal and positive.
Deciding whether some network has expected squared loss at most a threshold is \(\mathsf{NP}^{\mathsf{PP}}\)-complete for uniform Boolean inputs.
For uniform inputs on the unit cube and for standard Gaussian inputs with a positive threshold, the same holds for two neurons when a positive risk gap, given in binary and possibly exponentially small, separates yes from no instances.
Without a gap, the cube version is complete for guessing a polynomial-length string, then comparing a sum of exponentially many rationals with a threshold, all computable in polynomial time from the input and the guess.
\end{abstract}
\maketitle

\setcounter{tocdepth}{2}

\begingroup
\makeatletter
\def\@pnumwidth{2.5em}
\def\@tocrmarg{3.5em}
\def\@dotsep{4.5}
\def\l@section#1#2{%
\@dottedtocline{1}{0pt}{2.8em}{\bfseries\numberline{}#1}{#2}}
\def\l@subsection#1#2{%
\@dottedtocline{2}{2.8em}{3.3em}{\numberline{}#1}{#2}}
\renewcommand{\tocsection}[3]{%
\@ifnotempty{#2}{\llap{\hb@xt@\@tempdima{#2.\hfil}}}#3}
\let\tocsubsection\tocsection
\makeatother
\tableofcontents
\endgroup

\section{Introduction}

Knowing every parameter value of a neural network does not yet determine the network.
One must still decide which values become weights and which become biases, which neurons each weight connects, and which computed values later neurons share.
This paper asks how hard these decisions are when the required behavior of the network is given by a compact specification.

An instance consists of a multiset \(W\) of rational numbers, called the inventory, a cap on the number of computational neurons, and a target.
A repeated value supplies separate occurrences: three copies of \(1\) permit three uses.
A candidate is a network built from the inventory: each of its edges consumes one occurrence, and so does each neuron's bias, which may also be absent.
We study two rules for the inventory: at-most-once use, which may leave occurrences unused, and exactly-once use, which must consume every occurrence; every problem below has a version for each rule.

Within the cap, the graph may have any depth, width, skip connections and fan-out, and its scalar output is affine.
The cap counts the output neuron, but it does not bound the input dimension or the size of the target.
When the target is itself a network, its parameters need not belong to \(W\) and do not consume occurrences.

The model separates which numbers are available from how the network is built.
Related formulations constrain the structure more.
In permutation training, Cai, Chen, and Qiao \cite{CaiChenQiao2025} rearrange the output coefficients over a prescribed univariate hinge basis.
Arithmetic expression construction \cite{AlcockEtAl2020} assigns supplied numbers to the positions of an expression tree.
Here one inventory supplies every weight and bias, and computed values may be shared.
The simplest specification, a list of labeled examples, is already hard: two explicit samples give an \(\mathsf{NP}\)-complete feasibility problem with positive integer parameters (\Cref{thm:hardness}).
Our main question is how specifications that constrain the behavior on all inputs, or on average over a probability distribution, change this baseline.

The specifications we study ask for different kinds of agreement.
Exact reproduction of a Boolean circuit requires agreement on every Boolean input, and exact reproduction of a supplied ReLU network requires agreement on every real input.
Uniform approximation requires agreement within a tolerance on a box.
A population-risk specification instead compares the mean squared error with a rational threshold, under the uniform distribution on the Boolean cube, the uniform distribution on the unit cube, or the standard Gaussian distribution.
In each case the candidate's graph is part of the witness.
Hardness for a prescribed graph therefore does not give hardness here, because another legal placement of the same numbers may meet the same specification.
Excluding every such alternative graph and placement is the central difficulty in our reductions.

Even an elementary change of realization can violate the inventory constraint.
Write \(\rho(t)=\max\{0,t\}\).
The identity \(\rho(x)=\tfrac12\rho(2x)\) rescales one hidden neuron without changing its function.
The left side uses two occurrences of \(1\), one on the input edge and one on the output edge.
The right side uses \(2\) and \(1/2\) instead, which the inventory \(W=(1,1)\) does not supply.
A functional identity therefore need not preserve a legal assignment of occurrences.
Our reductions must control every realization that the actual inventory permits, including graphs different from the one used to encode a source witness.

The existential-universal verification pattern behind our second-level upper bounds is established.
Existential choice followed by universal verification already arises in succinct circuit minimization \cite{AllenderHoldenKabanets2017}, and it underlies the second-level completeness results for formula minimization \cite{BuchfuhrerUmans2011,Jaakkola2026}.
Equivalence of ReLU networks has a standard \(\mathsf{coNP}\) verification procedure based on activation patterns and linear constraints \cite{Wurm2024}.
Our contribution is to determine how a consumable numerical inventory interacts with these verification requirements, and with the counting that population risk involves.
The hardness results distinguish these sources of complexity at small caps: every synthesis hardness result holds with at most four computational neurons, except the exactly-once versions of Boolean risk, whose cap is the size of the inventory.

Two findings organize the classifications below.
The first concerns exact agreement with a ReLU network on all real inputs.
For a fixed cap, a polynomial-size set of test inputs separates all candidates that compute different functions, which gives a \(\mathsf{DP}\) upper bound, tight at cap three, instead of only the general existential-universal bound.
Uniform approximation on a compact box behaves differently: even one affine candidate with identical positive inventory values gives second-level completeness.
In both comparisons the input dimension and the target size are unrestricted.
The second finding concerns population risk, where the main issue is to preserve a numerical choice through an average.
Exact continuous synthesis must retain the sign of a rational weighted sum even when that sum is zero.
The construction achieves this by canceling a source-dependent second moment, without assuming a positive accuracy gap.

\subsection{Our Results}

\Cref{tab:results} summarizes the classifications, with all numbers encoded in binary and running times measured in the binary input length.
\begin{longtable}{p{0.31\textwidth}p{0.57\textwidth}}
\caption{Synthesis from a parameter multiset. Caps count the affine output. Both inventory rules are included unless stated otherwise.}\label{tab:results}\\
\hline
Specification & Guarantee \\
\hline
\endfirsthead
\hline
Specification & Guarantee \\
\hline
\endhead
\hline
\endfoot
Two explicit samples & \(\mathsf{NP}\)-complete with positive integer inventory and cap one (\Cref{thm:hardness}).\\
Boolean-circuit equality & \(\Sigma_2^{\mathsf P}\)-complete with cap four, even for monotone threshold targets (\Cref{thm:boolean-complete}).\\
Full-space ReLU equality & In \(\mathsf{DP}\) for every fixed cap; cap three is \(\mathsf{DP}\)-complete; cap two is \(\mathsf{coNP}\)-complete under at-most-once use (\Cref{thm:fixed-cap-dp,thm:three-vertex-dp,thm:two-vertex}).\\
Exact Boolean squared risk & \(\mathsf{NP}^{\mathsf{PP}}\)-complete: integer inventory, cap two for at-most-once use, cap \(|W|\) for exactly-once use (\Cref{thm:population-risk}).\\
Uniform approximation on a box & \(\Sigma_2^{\mathsf P}\)-complete with cap one and identical positive inventory values (\Cref{thm:compact-relu-hardness}).\\
Exact continuous uniform squared risk & Complete for the discrete existential closure of exact rational weighted counting; integer-inventory cap-two hardness under both inventory rules (\Cref{thm:continuous-exact-synthesis}).\\
Continuous uniform or standard Gaussian squared risk with a binary gap & \(\mathsf{NP}^{\mathsf{PP}}\)-complete with integer inventory and cap two (\Cref{thm:continuous-gap-complete,thm:gaussian-choice}).\\
Standard Gaussian squared risk with fixed absolute gap \(1/4\) & \(\mathsf{NP}^{\mathsf{PP}}\)-complete with integer inventory and cap two; unbounded loss and a variable threshold (\Cref{cor:gaussian-fixed-gap}).\\
\end{longtable}

The first row is the numerical baseline.
Its two-sample construction excludes every hidden graph, even when a larger cap is permitted.
Our main results concern the implicit specifications in the remaining rows.
We first treat the universal requirements and then ask how averaging changes their verification.

\paragraph{\emph{Universal specifications.}}
The Boolean target in the next theorem is a monotone threshold function with positive coefficients, which is simple to realize when real weights may be chosen freely.
Its difficulty arises from the prescribed occurrences and their possible placements.
\begin{theorem}[Informal; restated as \Cref{thm:boolean-complete}]
\label{ithm:boolean-complete}
Boolean-circuit exact synthesis is \(\Sigma_2^{\mathsf P}\)-complete under either inventory rule.
Hardness holds with four computational neurons, including the affine output, even when the target is a monotone threshold with positive coefficients.
\end{theorem}

A target defined on all real inputs changes the classification at a fixed cap.
In the next theorem the candidate cap is constant, while the rational ReLU target and the input dimension remain part of the input.
\begin{theorem}[Informal; restated as \Cref{thm:fixed-cap-dp}; see also \Cref{thm:three-vertex-dp,thm:two-vertex}]
\label{ithm:fixed-cap}
Full-space ReLU-target synthesis lies in \(\mathsf{DP}\) for every fixed computational cap under either inventory rule.
Cap three is \(\mathsf{DP}\)-complete under both rules.
Under at-most-once use, cap two is \(\mathsf{coNP}\)-complete.
\end{theorem}
A problem belongs to \(\mathsf{DP}\) when it is the conjunction of an \(\mathsf{NP}\) condition and a \(\mathsf{coNP}\) condition.
The upper bound comes from finitely many separating evaluations, not from listing all candidate functions in polynomial time.

The two conditions have separate sources in the hardness proof.
An explicit affine target already gives \(\mathsf{NP}\)-completeness at cap three (\Cref{thm:three-vertex-affine}), and requiring equality with an arbitrary ReLU target adds the universal condition.
When the cap grows with the input, the full-space problem lies between the \(\mathsf{DP}\) lower bound and the second-level upper bound, and its exact classification is open.
The Boolean theorem does not resolve it, because agreement at the vertices of the cube need not extend to agreement on all real inputs.

\paragraph{\emph{Uniform approximation on a compact domain.}}
Changing the domain and the error requirement yields a different fixed-cap classification.
\begin{theorem}[Informal; restated as \Cref{thm:compact-relu-hardness}]
\label{ithm:compact-relu}
Uniform approximation of a supplied rational ReLU target on the unit box is \(\Sigma_2^{\mathsf P}\)-complete, under either inventory rule, already at tolerance one with one affine computational neuron and an inventory of identical positive rational occurrences.
\end{theorem}
The construction first establishes completeness for explicitly listed multilinear polynomial targets of degree at most five and affine candidates, and a Lov\'asz extension then turns these targets into ReLU targets.
The domain and the error requirement therefore matter even when the candidate has no hidden neuron.

\paragraph{\emph{Population-risk specifications.}}
The risk results keep the same occurrence constraint, but replace universal agreement by a threshold on an average.
For Boolean inputs, verifying a candidate means summing over a finite cube.
For continuous inputs, the risk becomes a weighted count whose terms may have different rational denominators.
Exact rational weighted counting, the counting model used for continuous inputs, sums a rational weight over all binary strings of a polynomially bounded length.
Each weight is a short rational computed exactly in polynomial time, and different weights may have different denominators.
The associated decision class asks whether this sum is at least a rational threshold, also computed exactly in polynomial time.
Its discrete existential closure allows one polynomial-length binary choice before a single such comparison; it does not allow arbitrary adaptive oracle access.
The formal definitions appear in \Cref{sec:continuous-risk,sec:continuous-exact-synthesis}.
\begin{theorem}[Informal; restated as \Cref{thm:population-risk}; see also \Cref{thm:continuous-risk,thm:continuous-exact-synthesis}]
\label{ithm:population-risk}\label{ithm:continuous-risk}\label{ithm:continuous-exact-synthesis}
Under uniform Boolean inputs, exact squared-risk synthesis is \(\mathsf{NP}^{\mathsf{PP}}\)-complete under either inventory rule.
Hardness uses an integer inventory and target labels in \(\{-1,+1\}\), with cap two for at-most-once use and cap \(|W|\) for exactly-once use.
Under continuous uniform inputs on the unit cube, comparing the squared risk between two supplied networks with a threshold is complete for exact rational weighted counting, even in one dimension when one of the two networks is zero.
Synthesis is complete for its discrete existential closure, with integer-inventory cap-two hardness under either inventory rule.
\end{theorem}
For Boolean inputs, target labels in \(\{0,1\}\) also give \(\mathsf{NP}^{\mathsf{PP}}\)-completeness when the inventory consists of integers and exactly one occurrence of \(1/2\) (\Cref{cor:binary-risk}).

Exact continuous risk also has an obstruction of a different kind, concerning the size of the risk value itself.
Some bounded one-dimensional dyadic targets have descriptions of length \(O(n\log n)\), yet their exact risk denominators require \(\Omega(2^n)\) bits (\Cref{thm:risk-denominator}).
This shows only that the exact value can be too long to print; it does not separate classes of threshold decision problems.
Neither exact-risk classification asserts hardness at a fixed bounded-loss accuracy.

\paragraph{\emph{Accuracy gaps and Gaussian inputs.}}
The remaining risk results relax the exact threshold by a gap.
In the binary-gap version, in YES instances some legal candidate has risk at most \(\tau\), and in NO instances every legal candidate has risk at least \(\tau+\delta\), where \(\delta>0\) is supplied in binary and may be exponentially small.
For such a promise problem, membership means that some problem in the stated class, a total separator, accepts every YES instance and rejects every NO instance.
Hardness is under deterministic polynomial-time many-one promise reductions, which map YES and NO instances of the source to YES and NO instances, respectively.
\begin{theorem}[Informal; restated as \Cref{thm:continuous-gap-complete}; see also \Cref{thm:gaussian-choice}]
\label{ithm:continuous-gap-complete}\label{ithm:gaussian-choice}
For continuous uniform inputs and for standard Gaussian inputs, binary-gap squared-risk synthesis is \(\mathsf{NP}^{\mathsf{PP}}\)-complete under either inventory rule, with integer inventory and cap two.
\end{theorem}
For Gaussian inputs, the exact positive-threshold problem has the corresponding lower bound, but \Cref{thm:gaussian-choice} supplies no upper bound for it.

Scaling the inventory and the target keeps the Gaussian node cap and enlarges the absolute gap to the fixed value \(1/4\), while retaining \(\mathsf{NP}^{\mathsf{PP}}\)-completeness under both rules (\Cref{cor:gaussian-fixed-gap}).
This consequence uses unbounded squared loss and a supplied positive threshold that may be numerically large.
It gives no classification for a fixed threshold, a relative gap, or a clipped loss.

\subsection{Our Techniques}

Because the graph is chosen together with the placement of the numbers, each reduction must control every network that the inventory can build, not only the intended one.
The paragraphs below describe the device each reduction uses for this purpose.

\paragraph{\emph{Controlling every occurrence placement.}}
The finite-sample reduction adds a large common positive offset to every source number.
An all-ones sample then excludes every path that multiplies two parameters, so all occurrences must lie in a single affine output.
A second sample, through its zero and one coordinates, selects a subset of prescribed cardinality.
This establishes the numerical baseline without prescribing the graph.

The Boolean reduction needs a stronger converse.
Its source problem asks for a small set of forbidden occurrences that intersects every subset with a specified sum.
We prove this restricted blocker problem complete by a quantified-CNF construction, together with congruence and nonpower promises that the network proof uses.
Two shifted ReLUs implement a threshold ramp, an established representation \cite{Williams2018}.
With threshold \(P\), width \(w>0\), and shared affine score \(S\), the intended output is
\[
\frac{\rho(S-(P-w))-\rho(S-P)}{w}.
\]
It is zero when \(S\le P-w\) and one when \(S\ge P\).
Small changes to the score encode forbidden source occurrences.
The challenge is to force every successful candidate with four computational neurons into this common-score form, with the two copies of each coefficient paid for by distinct supplied occurrences.
An alternative placement of a gain as a bias, or an extra path through a shared vertex, would otherwise invalidate extraction of the source choice.

The target's inputs include several query blocks, groups of coordinates that are repeated so that some block stays free of unintended parameters.
The proof begins with an obstruction visible on one two-coordinate face of the target restricted to a query block.
On this face the target is an OR: its four values are \(0,1,1,1\), so the mixed difference is \(1-1-1+0=-1\).
Helmbold and Long's supermodularity property \cite[Lemma 12]{HelmboldLong2018} implies that a ReLU network with nonnegative edge weights has nonnegative mixed differences on Boolean inputs, even with arbitrary biases (\Cref{lem:boolean-supermodular}).
It therefore cannot realize this face.

This obstruction alone does not identify the candidate, and two separated numerical scales give stronger restrictions: Boolean output equalities become formal identities under each witness's actual activation masks.
Replication leaves an uncontaminated query block, on which these identities exclude all unintended gain and negative-parameter positions.
The full target then identifies the shared score, and anchor inputs recover a source blocker.
\Cref{app:boolean-placements} gives a structural map and the complete case analysis.

\paragraph{\emph{Separating finitely many candidate behaviors.}}
For a fixed cap, a disagreement between two candidates has a witness supported on a constant number of input coordinates.
The bound counts their hidden preactivations and one output difference.
Enumerating local graphs on those subspaces yields a polynomial set of separating inputs.
All legal candidates consistent with the target on the separating inputs compute the same function on the full space.
Synthesis therefore requires a consistent candidate and the absence of any consistent candidate that disagrees with the target elsewhere.
Existence of a consistent candidate is one \(\mathsf{NP}\) condition; existence of a consistent disagreeing candidate is a second \(\mathsf{NP}\) condition.
Taking the first condition and the complement of the second gives the \(\mathsf{DP}\) upper bound.

For the matching lower bound at cap three, shifted numerical matching forces the two output gains to be unit occurrences.
The remaining required positions exhaust the inventory and exclude an internal hidden edge.
A separate target slice encodes unsatisfiability, supplying the second conjunct of \(\mathsf{DP}\) hardness.

\paragraph{\emph{Uniform error through covering.}}
For compact approximation, a one-node candidate chooses feature coordinates for a succinct set cover.
Two selector coordinates calibrate all possible bias and nonfeature placements away.
Local circuit constraints form a multilinear penalty, whose consistent Boolean vertices require the chosen sets to cover every input.
Multilinearity extends the inequalities to the entire box.
The standard Lov\'asz extension \cite{GrabischMarichalRoubens2000} produces a rational ReLU target with the same vertex values and the same uniform-error test against each affine candidate.

\paragraph{\emph{Preserving choice in an average.}}
For Boolean risk, parity blocks hide source random bits from every candidate reading only the number of coordinates its inventory permits.
We use the parity device of Koriche, Lagniez, and Tran \cite{KoricheLagniezTran2026} with existential-majority satisfiability \cite{LittmanGoldsmithMundhenk1998}.
An integer-valued pointwise quadratic inequality excludes all unintended candidates.

For the upper bound on continuous risk, activation masks and complete flags of polytope faces write the risk as a sum of simplex integrals, each a short rational summand of a counting problem.
For the lower bound, tent folding \cite{Telgarsky2016} decodes words into dyadic intervals, and narrow triangular payloads realize reciprocal denominators and numerator multiplicities.
Balanced occurrence pairs restrict low-risk candidates with two computational neurons to corner functions encoding existential strings.
A continuous modular-sharing construction, based on the uniform sharing of Chor and Kushilevitz \cite[Section 4]{ChorKushilevitz1993}, keeps a virtual uniform coordinate independent of every affordable raw-coordinate observation.
The exact construction must do more than the binary-gap reduction.
Smoothing an encoded bit perturbs the risk, and such a perturbation can reverse a comparison in which the source sum equals zero.
Writing \(T\) for the target and \(N\) for a legal candidate, expand the loss as
\[
\mathbb E[(T-N)^2]=\mathbb E[T^2]+\mathbb E[N^2]-2\mathbb E[NT].
\]
The aim is to keep the source sum in the cross term while the target energy \(\mathbb E[T^2]\) stays independent of that sum.
For a fixed nonnegative envelope height \(U\), the payload \(B\), integrated over its hidden uniform coordinate, satisfies
\[
3\int B^2=2U\int B.
\]
Two positive copies on zero background and one negative copy on background \(U\), each occupying an interval of length \(\lambda\), change the energy by
\(\lambda(3\int B^2-2U\int B)=0\).
Their first-moment change is still \(\lambda\int B\).
Conditional independence makes this retained mean available against every permitted candidate; the numerical separation argument then excludes all candidates except those encoding source choices.
Thus the threshold comparison preserves the sign of the source sum, including equality, rather than replacing it by a promised gap (\Cref{thm:continuous-exact-synthesis}).

For Gaussian inputs, a sign decoder and an independent homogeneous amplitude give the forced-zero counting lower bound.
The use of Boolean-to-Gaussian lifting follows earlier learning-hardness constructions \cite{DanielyVardi2021,ChenGollakotaKlivansMeka2022}.
To encode existential choice, opposite unit occurrences on paired coordinates define rare hinges.
Their small cross moments preserve the source signal, while a second Hermite coefficient excludes every competing cap-two placement.
Uniform growth, tail and Lipschitz bounds turn continuous risk with a binary gap into a polynomial-time fair-bit counting experiment.
Finally, scaling all inventory occurrences and the target enlarges the Gaussian absolute gap; the proof separately controls the candidates' affine skip terms.

\subsection{Other Related Work}

We compare our results with three further lines of work: circuit minimization and neural verification, numerical resource constraints and training, and integration and counting.

\paragraph{\emph{Minimization and neural verification.}}
Circuit minimization from truth tables \cite{MurrayWilliams2017} differs from a succinctly supplied target; the latter already leads to existential-universal verification \cite{AllenderHoldenKabanets2017}.
Buchfuhrer and Umans \cite{BuchfuhrerUmans2011} and Jaakkola \cite{Jaakkola2026} establish second-level completeness for Boolean formula minimization.
Their formula-tree occurrence constraints do not identify the possible sharing and numerical placements of a ReLU graph.
Williams's exact Boolean representation results \cite{Williams2018} concern freely chosen real coefficients and therefore do not impose the supplied numerical resource constraint.

Wurm's neural verification framework \cite{Wurm2024} includes equivalence via activation patterns and linear constraints.
His finite-alphabet universal-trigger problem is \(\Sigma_2^{\mathsf P}\)-complete \cite[Theorem 5.2]{Wurm2026}, so second-level neural decision problems precede this work.
Those choices modify inputs of a fixed network; ours consume weight and bias occurrences while choosing a graph with a constant computational cap.
Lee \cite{Lee2026} studies minimizing the hidden-neuron count under positive-tolerance \(L^p\) approximation with freely chosen real parameters and two hidden layers.
Its approximation guarantee and numerical freedom differ from full-space exact synthesis with a fixed inventory.

Parameter symmetries give a different tractable equivalence regime.
Gegenfurtner, Grillo, and Mont\'u\-far \cite[Proposition 14]{GegenfurtnerGrilloMontufar2026} give a polynomial-time equivalence test for two generic parameter choices in the same three-layer bottleneck architecture.
In their architecture, the first hidden layer is no wider than the input.
Our fixed-cap synthesis theorem searches over occurrence placements and arbitrary permitted graphs, allows degenerate parameters, and leaves the target size unrestricted.

\paragraph{\emph{Numerical resources and training.}}
Permutation training \cite{CaiChenQiao2025} assigns output coefficients over a prescribed hinge basis, whereas our single inventory governs every parameter position.
Arithmetic expression construction \cite{AlcockEtAl2020} uses expression trees and supplied occurrences, rather than shared ReLU graphs.
Doron-Arad and Mossel \cite{DoronAradMossel2026} prove \(\mathsf{NP}\)-completeness for finite-sample ReLU training with polynomial-bit rational parameters on a supplied architecture.
Our finite-sample theorem restricts all parameters to an input multiset and permits alternative architectures; the higher classifications concern implicit specifications.
Subset-sum interdiction fits the general blocking framework of Gr\"une and Wulf \cite{GrueneWulf2026}; our source lemma proves the additional arithmetic promises directly.
The distinct-number numerical matching premise for cap-three hardness is due to Hulett, Will, and Woeginger \cite[Appendix, Corollary 8]{HulettWillWoeginger2008}.

\paragraph{\emph{Integration and counting.}}
We use the exact polynomial-time rational-weight model of de Campos, Stamoulis, and Weyland \cite{DeCamposStamoulisWeyland2019}, and distinguish it from their broader approximation-defined model.
The general arithmetic upper bound we quote is Meiburg's recent preprint \cite{Meiburg2026}.
Feischl, Gr\"a\ss le, and Zehetgruber already use folding to establish fixed-dimensional ReLU quadrature hardness \cite{FeischlGrassleZehetgruber2026}.
Morettin, Passerini, and Sebastiani formulate probabilistic neural verification through weighted model integration \cite{MorettinPasseriniSebastiani2024}.
Our exact continuous classification gives reciprocal rational-weight realization and then preserves existential numerical choice in synthesis.
Lawrence's long polytope-volume denominators \cite{Lawrence1989} precede the output-size phenomenon; our example uses one real coordinate and dyadic network parameters.

The Gaussian counting reduction concerns threshold comparison, whereas Rannen-Triki et al.\ \linebreak\cite{RannenTrikiEtAl2019} study whether a supplied ReLU realization has zero norm.
Gaussian agnostic learning of a freely chosen ReLU \cite{GoelKarmalkarKlivans2019} has sample access and a different output requirement.
Neither that learning problem nor smooth perturbations of Boolean data replace the single standard Gaussian probability distribution used here.

\subsection{Discussion}

The same finite-inventory model accommodates four different requirements: universal Boolean agreement, agreement as piecewise-affine functions on all real inputs, uniform approximation on a compact box, and an average squared loss.
In each case the proof must combine the verification mechanism with a numerical choice that survives every alternative architecture.
The resulting classes are not ordered by neuron count alone.
The input domain, the target representation and the promise guarantee change the class to which our results assign the problem, and the inventory rule changes the caps at which our hardness proofs apply.
For example, the Boolean-risk hardness uses cap two only under at-most-once use, and cap \(|W|\) under exactly-once use.
Two-neuron hardness under both inventory rules holds for the continuous uniform and Gaussian results.

The exact continuous classification and the binary-gap classifications answer different questions.
A binary gap may be exponentially small, and the fixed absolute Gaussian gap allows both unbounded squared loss and growing thresholds.
Neither result establishes hardness of bounded-loss approximation at a fixed accuracy.
Nor can one decide whether a Gaussian risk equals a rational threshold by enumerating candidates and approximating each risk, since approximations do not certify equality.

\subsection{Open Problems}

We close with four open questions.
The first concerns exact synthesis on the full space when the cap grows with the input; the other three concern risk synthesis and arise from the distinctions just drawn between the exact and the gap classifications.

\paragraph{\emph{Unrestricted full-space exact synthesis.}}
Fixed-cap membership in \(\mathsf{DP}\) and cap-three hardness leave open how the complexity changes when the cap grows with the input.
With the cap part of the input, the problem is \(\mathsf{DP}\)-hard by \Cref{thm:three-vertex-dp} and belongs to \(\Sigma_2^{\mathsf P}\) by \Cref{thm:relu-target}; the paper does not determine where it lies between these bounds.
\begin{openproblem}
\label{op:relu-hardness}
Determine the exact deterministic polynomial-time many-one reduction complexity of full-space synthesis for a supplied rational ReLU target, a rational inventory used at most once, arbitrary permitted graphs, and an affine output.
In particular, determine whether it is \(\Sigma_2^{\mathsf P}\)-complete.
\end{openproblem}

\paragraph{\emph{Clipped risk with an inverse-polynomial gap.}}
For uniform Boolean inputs, define the clipped risk by
\[
\overline R_T(N)=\mathbb E[\min\{1,(N(X)-T(X))^2\}].
\]
The input supplies \(0<\delta\le1\), \(0\le\tau\le1-\delta\), and a unary string of length \(\lceil1/\delta\rceil\), so the gap is at least inverse-polynomial in the input length.
In YES instances some legal candidate has clipped risk at most \(\tau\); in NO instances every candidate has clipped risk at least \(\tau+\delta\).
\begin{openproblem}
\label{op:gap-risk}
Determine the deterministic polynomial-time many-one promise-reduction complexity of this clipped-risk synthesis problem for a rational inventory used at most once, an independent rational ReLU target and an affine output.
In particular, determine whether it is complete for \(\mathsf{MA}\).
\end{openproblem}
Here \(\mathsf{MA}\) permits a polynomial certificate followed by a randomized polynomial-time check.
The exact-risk reductions may distinguish only an exponentially rare assignment fiber; clipping does not amplify its probability.

\paragraph{\emph{Exact standard-Gaussian comparison.}}
The binary-gap theorem gives no algorithm for deciding exact equality with a positive threshold.
\begin{openproblem}
\label{op:gaussian-risk}
Determine whether standard-Gaussian squared-risk synthesis is decidable for an independent rational ReLU target, a rational inventory used at most once, a node cap, an explicitly listed dimension and a positive rational threshold.
If it is, determine its deterministic polynomial-time many-one reduction complexity.
\end{openproblem}
\Cref{thm:gaussian-choice} already gives an \(\mathsf{NP}^{\mathsf{PP}}\) lower bound with integer inventory and cap two.

\paragraph{\emph{Fixed Gaussian thresholds.}}
Scaling to a fixed absolute gap changes the threshold; it does not settle the following normalized question.
\begin{openproblem}
\label{op:relative-gaussian-risk}
Determine the exact deterministic polynomial-time many-one promise-reduction complexity of standard-Gaussian squared-risk synthesis, with YES risk at most one and NO risk at least \(5/4\), an independent rational ReLU target, a rational inventory, an explicitly listed dimension and an affine output.
Treat both inventory rules.
In particular, determine whether it is \(\Sigma_2^{\mathsf P}\)-complete.
\end{openproblem}

\phantomsection
\addcontentsline{toc}{subsection}{AI Disclosure}
\begingroup
\makeatletter
\let\addcontentsline\@gobblethree
\makeatother
\subsection*{AI Disclosure}

This work was developed with substantial assistance from Codex and Claude Code.
These tools were used to discover almost all of the proofs in this work, find relevant literature, and write parts of the prose; the author wrote the rest.
The author posed the problem, guided the development, and is responsible for any errors.
\endgroup

\subsection{Paper Organization}

\Cref{sec:model} defines the network and inventory conventions, and \Cref{sec:complexity} proves the finite-sample baseline.
\Cref{sec:circuit} proves the Boolean exact classification, with its complete placement converse in \Cref{app:boolean-placements}.
\Cref{sec:relu-target} treats exact ReLU targets on all real inputs and the fixed-cap classifications.
\Cref{sec:compact} treats uniform approximation on a compact box.
\Cref{sec:population-risk,sec:continuous-risk,sec:gaussian-risk} treat population squared risk under the three input probability distributions.

\section{Preliminaries}
\label{sec:model}

We write \(\mathbb R,\mathbb Q,\mathbb Z\) for the real numbers, rational numbers and integers.
Unless another encoding is explicitly specified, integer inputs are encoded in binary, and a rational input is encoded by a signed integer numerator and a positive integer denominator.
The binary input length is the total number of bits in these encodings and the explicitly supplied graphs and coordinate lists.
For a nonnegative integer \(r\), write \([r]=\{1,\ldots,r\}\), with \([0]=\varnothing\).
The symbols \(\lfloor t\rfloor,\lceil t\rceil\) denote the greatest integer at most \(t\) and the least integer at least \(t\), respectively.
For an integer \(r\ge0\), \(r!=\prod_{i=1}^r i\), with \(0!=1\); empty sums and products have values zero and one, respectively.
An expression \(\operatorname{poly}(L)\) denotes a fixed polynomial bound in the input length \(L\).
For nonnegative functions of an integer variable, \(f=O(g)\) means \(f\leq Cg\) eventually for a constant \(C>0\), while \(f=\Omega(g)\) means \(f\geq cg\) eventually for a constant \(c>0\).
Logarithms are natural unless a base is displayed.

Let \(d\geq1\) be an integer, the input dimension, and write \(x=(x_1,\ldots,x_d)\in\mathbb R^d\) for an input.
For vectors \(a,x\in\mathbb R^d\), their dot product is \(a\cdot x=\sum_{j=1}^d a_jx_j\).
The vector \(e_j\) has coordinate one at \(j\) and zero elsewhere; \(\mathbf1\) denotes the all-ones vector in the stated dimension.
The support of a vector is the set of coordinates where it is nonzero.
Write \(\|x\|_2=(\sum_jx_j^2)^{1/2}\) for its Euclidean length and \(\|x\|_\infty=\max_j|x_j|\) for its infinity norm.

An affine form is \(b+a\cdot x\), where \(a\in\mathbb R^d\) and \(b\in\mathbb R\); it is linear when \(b=0\).
An affine hyperplane is \(\{x:a\cdot x=b\}\) with \(a\ne0\).
The vector \(a\) is a normal vector to this hyperplane.
The corresponding closed and open affine halfspaces are \(\{x:a\cdot x\leq b\}\) and \(\{x:a\cdot x<b\}\).
An open ball with center \(x\) and radius \(\epsilon>0\) is \(\{y:\|y-x\|_2<\epsilon\}\).
A set is open if each of its points has an open ball contained in the set.
The interior of any set consists of all points with such a ball contained in the set.

Write \(\rho(t)=\max\{0,t\}\) for the ReLU function.
A function \(f:\mathbb R\to\mathbb R\) is convex if
\[
f(\lambda u+(1-\lambda)v)\leq\lambda f(u)+(1-\lambda)f(v)
\quad(u,v\in\mathbb R,0\leq\lambda\leq1).
\]
For \(f=\rho\), the right side is both nonnegative and at least \(\lambda u+(1-\lambda)v\), so it is at least their maximum \(\rho(\lambda u+(1-\lambda)v)\), the left side; hence \(\rho\) is convex.

Write \(W=(w_1,\ldots,w_m)\) for the inventory, an explicitly listed multiset of \(m\geq1\) rational numbers, where \(m\) is an integer, and let \(n_{\max}\geq1\) be an integer, the cap on the number of computational vertices (also called the computational-node budget).
A multiset permits repeated values, and its indices distinguish their occurrences.
The operation \(\uplus\) joins multisets while retaining every occurrence.

A directed graph consists of vertices and edges \((u,v)\) pointing from \(u\) to \(v\).
A directed path follows a sequence of such edges; a directed cycle is a nonempty path returning to its starting vertex.
A graph is acyclic when it has no directed cycle.
A topological order lists its vertices so every edge points forward.
The empty graph has the empty order.
Every nonempty finite acyclic graph has a vertex with no incoming edge, since otherwise repeatedly following incoming edges would revisit a vertex and give a cycle.
Removing that vertex and repeating lists the vertices of the graph in a topological order.
A predecessor of \(v\) is a vertex \(u\) with an edge \((u,v)\).

\begin{definition}[Permitted network]
\label{def:network}
A permitted network is a finite directed acyclic graph with \(d\) input vertices and between one and \(n_{\max}\) computational vertices.
Input vertices have no incoming edges; every edge enters a computational vertex.
One computational vertex is the scalar output.
The other computational vertices are called hidden vertices.
Every computational vertex lies on a directed path to the output and has at least one incoming edge or a present bias.
For each ordered pair of vertices, there is at most one edge.
A hidden vertex \(v\) computes
\[
z_v=\max\left\{0, b_v+\sum_{u:(u,v)\in E}a_{uv}z_u\right\}.
\]
Here \(E\) is the edge set, \(a_{uv}\) is the weight of edge \((u,v)\), and \(b_v\) is the bias at \(v\), with \(b_v=0\) when that bias is absent.
Input vertex \(j\) has value \(z_j=x_j\).
The output computes an affine form in its predecessors' values.
The sample-based results and the bounds of \Cref{thm:circuit} also hold if the output applies ReLU to that form.
The completeness result, \Cref{thm:boolean-complete}, uses the affine-output convention.

Each edge and each present bias is assigned a distinct index from \(\{1,\ldots,m\}\) and takes the corresponding value \(w_i\).
Unassigned occurrences are unused.
No other numerical weights or present biases are allowed.
In the exactly-once variant, every index must be assigned.
\end{definition}
Throughout, a network or an assignment of inventory occurrences is called \emph{legal} when it is permitted in the sense of \Cref{def:network}, under whichever inventory rule is in force.
A computational vertex is called legal when it satisfies the incoming-edge-or-present-bias and path-to-output conditions of \Cref{def:network}.
In \Cref{sec:circuit} and \Cref{app:boolean-placements}, legal refers to the at-most-once rule unless the exactly-once rule is stated explicitly; for the exactly-once case, the proof of \Cref{thm:boolean-complete} enlarges the blocker in the forward direction and, in the converse, uses that every exactly-once witness is an at-most-once witness.

The zero implicit in an absent bias is not an extra supplied parameter.
A present bias equal to zero does consume a zero occurrence.
Hidden activations and summation are operations; they do not provide free weighted edges.
A vertex may have several outgoing edges, which is called fan-out, but each outgoing edge consumes its own weight occurrence.
A direct input-to-output edge is called a skip edge.
The convention on vertices excludes disconnected storage of unused parameters.
It still permits an assigned edge whose contribution vanishes at a particular sample.

If input vertices are included in the requested total node limit, replace \(n_{\max}\) here by that limit minus \(d\).
A budget below one computational vertex is infeasible in this model.
An optional output ReLU has no separate numerical parameter.
All classifications below use the affine output unless explicitly stated otherwise.

The affine expression supplied to a hidden ReLU is its preactivation.
At a fixed input, the actual activation indicator \(\eta_v\in\{0,1\}\) is one when the preactivation of hidden vertex \(v\) is positive, and zero otherwise.
The list of these indicators over all hidden vertices is the activation pattern at that input.
Freezing a pattern replaces each active ReLU by the identity and each inactive ReLU by zero.
For fixed numerical parameters, the vertex values of the frozen network are affine forms in the input, by induction in topological order.
The closed activation region of a pattern is the set of inputs at which every vertex labeled active has nonnegative preactivation and every vertex labeled inactive has nonpositive preactivation.
When a verification argument guesses a pattern and imposes only these weak inequalities, as in the proof of \Cref{thm:relu-target}, a zero preactivation may receive either label: both substitutions give the correct value zero.

Let \(q\geq1\) be an integer, the sample count, and let
\[
D_q=((x_1,y_1),\ldots,(x_q,y_q))
\]
be an explicitly listed sample, with \(x_i\in\mathbb{Q}^d\) and \(y_i\in\mathbb{Q}\).
For a network \(N\), its total squared empirical loss is
\[
L_{D_q}(N)=\sum_{i=1}^{q}(N(x_i)-y_i)^2.
\]
The decision problem asks whether a permitted \(N\) satisfies \(L_{D_q}(N)\leq\tau\), for a rational threshold \(\tau\geq0\).
Dividing by \(q\) gives mean squared error and only rescales the threshold.

We next bound the search space.
For a fixed inventory and budget, let \(\mathcal H_{W,n_{\max}}\) be the class of functions computed by permitted networks.
Distinct graphs can compute the same function; a count of graph descriptions is therefore an upper bound on the class size.
Put
\[
n=\min\{m,n_{\max}\},\qquad K=n(d+n+1),\qquad
N_{\mathcal H}=(n+1)^2\min\{(K+1)^m,(m+1)^K\}.
\]
Here \(K\) bounds the possible edge and bias positions after labeling the computational vertices.

\begin{proposition}[Finite class size]
\label{prop:count}
Every permitted network has at most \(m\) computational vertices, and
\[
|\mathcal H_{W,n_{\max}}|\leq N_{\mathcal H}.
\]
The same bound holds for the exactly-once variant.
\end{proposition}
\begin{proof}
For each computational vertex choose one incoming edge or its present bias.
The positions chosen for different vertices are distinct because they have different target vertices.
Each consumes a different occurrence of \(W\).
Thus the number of computational vertices is at most \(m\), even if \(n_{\max}\) is encoded in binary.

Label at most \(n\) computational vertices.
There are at most \(n(d+n)\) directed edge positions and \(n\) bias positions, hence at most \(K\) positions.
This count includes self-loops and other forbidden edges; retaining them only increases the upper bound.
There are at most \(n\) choices of the number of computational vertices and at most \(n\) choices of the output label.

One encoding sends each of the \(m\) occurrence indices to a position or declares it unused.
There are at most \((K+1)^m\) such assignments.
A second encoding assigns to each position either an occurrence index or the declaration that the position is absent.
There are at most \((m+1)^K\) such assignments.
In each encoding, reject assignments with repeated use, collisions, cycles, invalid vertices, or the wrong number of vertices.
Every permitted labeled network has an encoding of both kinds.
Multiplying the smaller bound by \(n^2\leq(n+1)^2\) proves the assertion.
Repeated numerical values and functionally identical descriptions decrease the actual class size.
Requiring every occurrence to be assigned restricts the class further.
\end{proof}

For the finite-sample problem, both encodings also give exhaustive search algorithms: enumerate assignments, discard invalid ones, and evaluate the survivors.
If \(L\) denotes the total binary input length, the running time is at most
\[
N_{\mathcal H}\operatorname{poly}(L).
\]
The counting bound may be loose.
Taking logarithms in the definition of \(N_{\mathcal H}\) gives
\[
\log N_{\mathcal H}=O\left(\min\{m\log(K+1),K\log(m+1)\}+\log(n+1)\right).
\]
For fixed \(d\) and \(n_{\max}\), \(K\) is constant and the search is polynomial in the input length.
This does not conflict with hardness below, where \(d\) grows.

The linear span of finitely many vectors consists of their linear combinations, with arbitrary real coefficients.
They are linearly independent when a linear combination equal to zero has all coefficients zero.
A basis is a linearly independent spanning family; the dimension of the span is the size of a basis.
The rank of a finite vector family is the dimension of its span, and the rank of a matrix is the rank of its columns.
The affine span consists of combinations whose coefficients sum to one.
A convex combination has nonnegative coefficients summing to one.
For a nonzero vector \(a\), write \(\ker a=\{z:a\cdot z=0\}\).
Two vectors are proportional when one is a scalar multiple of the other.

For an integer \(r\ge0\), a permutation of \([r]\) is a rearrangement in which each element appears once.
For an \(r\times r\) matrix \(B\), its determinant is
\[
\det B=\sum_{\pi}(-1)^{\operatorname{inv}(\pi)}\prod_{i=1}^r B_{i,\pi(i)},
\]
where the sum ranges over permutations \(\pi\) of \([r]\), and \(\operatorname{inv}(\pi)\) counts pairs \(i<j\) with \(\pi(i)>\pi(j)\).
The empty determinant is one.
A square matrix is called nonsingular, or invertible, when its determinant is nonzero.
The formula is linear in each row and column and changes sign when two rows or two columns are exchanged; pairing the permuted terms proves these properties.
It therefore vanishes when two rows or two columns coincide.
Subtracting a multiple of one row from another leaves the determinant unchanged.

Elimination chooses a nonzero entry, called a pivot, in the column currently being processed, among the rows not yet used.
Exchange its row into the next diagonal position, then subtract suitable multiples of that row from the rows below it to make their entries in that column zero.
These row operations are invertible, meaning that they can be reversed, and therefore preserve every linear dependence or independence relation among the columns.

Elimination shows that an \(r\times r\) matrix \(B\) with linearly independent columns has \(\det B\ne0\).
If \(r=0\), the determinant is already one.
For \(r\ge1\) and independent columns, the first column is nonzero, so a first pivot exists.
After this step the lower-right square block also has independent columns: a dependence there would give a nontrivial combination of the remaining full columns supported only in the first row, which could be canceled using the first column.
That would contradict the independence of the full columns.
Repeating the argument on each remaining block gives a triangular matrix, with every entry below the diagonal zero and every diagonal pivot nonzero.
Only the identity permutation can contribute a nonzero term to its determinant, so that determinant is the product of these pivots.
Row exchanges change only the determinant's sign and the other elimination operations leave it unchanged, proving \(\det B\ne0\).

For such a matrix \(B\), if \(Bz=y\), replace column \(j\) of \(B\) by \(y=\sum_i z_i B_i\), where \(B_i\) is column \(i\), and denote the resulting matrix by \(B^{(j)}\).
Column linearity and vanishing for repeated columns give \(\det B^{(j)}=z_j\det B\), so
\[
z_j=\frac{\det B^{(j)}}{\det B}.
\]
This determinant-ratio formula is Cramer's rule.

The elimination argument also shows that a rectangular matrix with \(k\) independent columns contains a nonsingular \(k\times k\) submatrix formed by selecting \(k\) original rows.
At each stage the residual columns remain independent by the same lifting argument used for the square block above, so the next column has a nonzero entry among the unused rows.
Choose that row as the next pivot row.
After \(k\) stages, these selected rows form a square triangular system with nonzero diagonal pivots.
Each transformed selected row is its original row minus a linear combination of earlier selected rows.
Thus those same row operations, applied just to the selected original square submatrix, give the nonsingular triangular matrix.
Their invertibility proves that the original submatrix is nonsingular as well.
For \(k=0\), choose the empty submatrix, whose determinant is one.

Later verification and integration arguments use rational linear programming.
A rational linear program consists of finitely many non-strict affine inequalities and a linear objective, all with binary-encoded rational coefficients.
A point is feasible if it satisfies every inequality; maximizing the objective asks for its largest value on the feasible points.
The following standard result does not require the feasible region to have an interior or to be bounded.
\begin{theorem}[Rational linear programming {\cite{GacsLovasz1979,GrotschelLovaszSchrijver1981}}]
\label{thm:rational-lp-import}
Feasibility of a rational linear program is decidable in deterministic polynomial time in its binary input length.
If the program is feasible and its objective has a finite optimum, an optimal rational solution and the optimum value can be computed exactly in polynomial time.
\end{theorem}

The integer-coefficient formulation applies to rational coefficients by clearing denominators in each row and in the objective.
Unrestricted variables can be written as differences of nonnegative variables.
These transformations increase the binary encoding length by at most a polynomial factor.

\section{Computational Complexity}
\label{sec:complexity}

A decision problem, also called a language, is a set of finite binary strings, its yes instances.
Its complement contains precisely the binary strings outside that set.
A polynomial-time many-one reduction maps inputs of one decision problem to inputs of another, preserving the yes or no answer.
The class \(\mathsf P\) consists of problems decided by deterministic algorithms in time polynomial in the input length.
The class \(\mathsf{PSPACE}\) consists of problems decided by deterministic algorithms using polynomially many working-memory bits, with no prescribed time bound.
For a binary string \(a\), write \(|a|\) for its length.
A decision problem \(A\) is in \(\mathsf{NP}\) when there are a nonnegative integer-valued polynomial \(p\) and a polynomial-time predicate \(V\) such that, for every binary input \(a\),
\[
a\in A\quad\leftrightarrow\quad
\exists z\in\{0,1\}^{p(|a|)}\, V(a,z)=1.
\]
The string \(z\) is a certificate, padded to the stated length if necessary.
For a class \(\mathcal C\) of decision problems, a decision problem \(A\) is \(\mathcal C\)-hard if every problem in \(\mathcal C\) has a polynomial-time many-one reduction to \(A\).
The problem \(A\) is \(\mathcal C\)-complete if it is both \(\mathcal C\)-hard and in \(\mathcal C\).

An integer \(c\) divides an integer \(b\) when \(b=ct\) for some integer \(t\).
For integers \(a,b\) and a positive integer \(q\), the congruence \(a\equiv b\pmod q\) means that \(q\) divides \(a-b\).
The residue of an integer \(a\) modulo \(q>0\) is its remainder in \(\{0,\ldots,q-1\}\) upon division by \(q\).
Two integers are coprime when their only common positive divisor is one.
A fraction with positive denominator is reduced when its numerator and denominator are coprime; that denominator is called its reduced denominator.

The exact linear-algebra computations used below also have polynomial bit complexity.
To test whether a rational vector is in the span of supplied rational columns, solve for their linear-combination coefficients by rational linear-program feasibility, writing each equality as two inequalities.
Process the columns in order and retain precisely those that are not in the span of the columns already retained; the retained columns form an independent spanning family.
This computes rank and a column basis with polynomially many calls to \Cref{thm:rational-lp-import}; the same procedure applies to rows viewed as vectors.
A matrix with no rows or no columns has rank zero.

For a square matrix, rank below its size gives determinant zero.
Otherwise clear denominators row by row and apply elimination to the resulting integer matrix.
A minor is the determinant of a square submatrix obtained by selecting equally many rows and columns.
After each pivot stage, every remaining entry is a signed ratio of two minors of that integer matrix.
Indeed, take an invertible pivot block \(P\), a column \(u\), a row \(v\), a scalar \(a\), and the vector \(z\) with \(Pz=u\).
Clearing the lower-left row by the first rows gives
\[
\det\begin{pmatrix}P&u\\v&a\end{pmatrix}
=\det(P)(a-vz).
\]
Thus the residual entry is the bordered minor divided by the pivot minor, with a possible sign from restoring the original row order.

If the integer entries have absolute value at most \(2^B\), every minor of size \(h\) has absolute value at most \(h!2^{hB}\), by its determinant expansion.
Its binary length, and hence the lengths of the residual numerators and denominators after cancellation, are polynomial in the original matrix encoding length.
Elimination uses polynomially many rational operations; its final diagonal product, the row-swap sign, and the known row-denominator factors recover the original determinant.
For an empty square matrix return one.

Common factors can be canceled by the Euclidean algorithm: on nonnegative integers not both zero, replace a pair \((u,v)\), with \(v>0\), by \((v,u-\lfloor u/v\rfloor v)\) until the second entry is zero.
The common divisors are unchanged at every step, so the last nonzero entry is the greatest common divisor, meaning the largest common positive divisor.
The positive remainders fall by at least a factor of two within two divisions: if a remainder exceeds half of the remainder before it, the next remainder is their difference and is smaller than that half.
The number of binary divisions is therefore polynomial in the bit length.
This makes the preceding exact rational elimination a polynomial-time bit algorithm.

We use the following imported starting point, with positive integers and binary encoding.
\begin{theorem}[Subset sum; Karp \cite{Karp1972}]
\label{thm:subset}
Given positive integers \(c_1,\ldots,c_r\) and a positive target \(B\), deciding whether some subset sums to \(B\) is \(\mathsf{NP}\)-complete.
\end{theorem}

The following elementary conversion makes the number of selected entries explicit.
It is convenient that \(W\) is a multiset: equal padding entries present no difficulty.
\begin{lemma}[Prescribed cardinality]
\label{lem:cardinality}
Given nonnegative integers \(a_1,\ldots,a_m\), an integer \(k\) with \(1\leq k\leq m\), and a target \(B\) with \(0\leq B\leq\sum_j a_j\), deciding whether exactly \(k\) entries sum to \(B\) is \(\mathsf{NP}\)-complete.
\end{lemma}
\begin{proof}
Membership in \(\mathsf{NP}\) follows by listing the selected indices and checking their count and sum.
For hardness start from \Cref{thm:subset}.
Instances with target larger than the sum are immediate no instances and can be mapped to a fixed no instance of the present problem, for example \(a=(2,2)\), \(k=1\), \(B=1\).

For all other instances append \(r\) zero entries to \(c_1,\ldots,c_r\), set \(m=2r\), and require \(k=r\) selected entries.
A subset of the original entries of size \(t\leq r\) and sum \(B\) can be supplemented by \(r-t\) of the new zeros.
Conversely, removing all selected zeros leaves an original subset with the same sum.
The construction is polynomial and satisfies the stated ranges.
\end{proof}

We now give a reduction in which deeper networks cannot circumvent the intended numerical choice.
\begin{theorem}[Finite samples]
\label{thm:hardness}
Squared-error feasibility for permitted networks is \(\mathsf{NP}\)-complete.
The \(\mathsf{NP}\)-hardness holds with zero threshold, two samples, cap \(n_{\max}=1\), positive integer inventory, and input coordinates in \(\{0,1\}\).
It holds both for at-most-once use and for exactly-once use.
Moreover, the same reduction remains valid for any larger node budget.
\end{theorem}
\begin{proof}
First consider membership.
A certificate describes at most \(m\) computational vertices, at most \(m\) parameter positions, the output, and the assigned occurrence indices.
It has polynomial length.
The graph and inventory rules are checkable in polynomial time.

Evaluation uses exact rational arithmetic of polynomial bit length.
To justify this point, fix a sample.
Inactive ReLUs have value zero; replace active ReLUs by the identity.
Each vertex value then expands as a sum over directed paths beginning at an input or a bias.
Each term is a product of an input coordinate, if any, and distinct parameter occurrences.
No path repeats an edge or bias because the graph is acyclic.

The product of all denominators in the inventory and this sample's coordinates is a common denominator for all terms.
Its bit length is at most the input length.
A term's numerator after scaling to this common denominator also has polynomial bit length, since it uses a subset of the supplied numerators and denominators.
There are at most exponentially many directed paths in a graph with at most \(m\) edges, so the logarithm of the number of summands is polynomial.
All vertex values thus have polynomial bit length.
Evaluation in topological order determines the active ReLUs without enumerating paths, and exact comparison of the squared loss with \(\tau\) is polynomial.

For hardness take an instance from \Cref{lem:cardinality}, and put
\[
A=\sum_{j=1}^{m}a_j,\qquad M=A+B+m+3,\qquad
w_j=M+a_j.
\]
Supply the multiset \(W=(w_1,\ldots,w_m)\).
Define the two requested labels by
\[
T=kM+B,\qquad U=mM+A.
\]
Use dimension \(d=2m\).
The first input \(u\) has its first \(m\) coordinates equal to one and its last \(m\) equal to zero.
The second input \(v\) has all \(2m\) coordinates equal to one.
The samples are \((u,T)\) and \((v,U)\).
All these integers have polynomial bit length.

The inequalities needed to control hidden paths are
\[
A<M,\qquad B<M,\qquad 0<T\leq U<M^2.
\]
The first two follow from the definition of \(M\).
Since \(k\leq m\) and \(B\leq A\), we have \(T\leq U\).
Finally, \(M^2-U=M(M-m)-A>0\), because \(M-m=A+B+3\).

Suppose a permitted network fits both samples.
Every assigned parameter is positive.
On \(v\), every computational vertex has a positive value: an induction in topological order applies because it has either a positive bias or an incoming edge from a positive predecessor.
Thus hidden ReLUs are identities on \(v\).
A directed input-to-output path containing at least two edges contributes at least \(M^2\) to the output.
A contribution from a hidden bias followed by an edge likewise contributes at least \(M^2\).
All contributions are nonnegative.
Either contribution would make the output at least \(M^2\), contradicting the required output \(U<M^2\).

Every hidden vertex has a path to the output and ultimately has an input predecessor or a present bias.
Such a vertex would create one of these forbidden contributions.
Consequently there are no hidden vertices in any fitting network, even when a larger node budget is permitted.

On \(v\), each assigned parameter now contributes exactly its value, either through an input edge or through the output bias.
For the at-most-once problem, omitting any occurrence gives a sum strictly smaller than \(U=\sum_jw_j\).
Hence all occurrences must be used.
In the exactly-once problem this is already required.

On \(u\), the output is the sum of a subset \(I\) of the occurrences: those assigned to edges from coordinates equal to one, together with the output bias if present.
Writing \(s=|I|\) and \(C=\sum_{j\in I}a_j\), exact fitting gives
\[
sM+C=kM+B.
\]
Both \(C\) and \(B\) lie in \(\{0,\ldots,M-1\}\).
Reduction modulo \(M\) gives \(C=B\), and then \(s=k\).
A fitting network therefore supplies a yes answer to the cardinality instance.

Conversely, given a subset \(I\) of size \(k\) and sum \(B\), use one affine output neuron without a bias.
Assign the \(k\) selected occurrences to distinct edges from the first \(m\) input vertices, and the other \(m-k\) occurrences to distinct edges from the last \(m\) input vertices.
There are enough vertices in both groups.
This uses every occurrence once and gives outputs \(T\) and \(U\).
It respects \(n_{\max}=1\), and an output ReLU would leave both positive labels unchanged.
Every assigned parameter contributes positively at \(v\).

This proves both reductions and \(\mathsf{NP}\)-completeness.
The proof also shows that a mandatory output bias would not change hardness: put one of the \(k\geq1\) selected occurrences in that bias and assign the others to edges.
\end{proof}

Optimization asks for a permitted network of minimum loss, or a report that no permitted network exists.
When some permitted network exists, a minimum is attained because the class of functions computed by permitted networks is finite by \Cref{prop:count}.
The construction yields a quantitative gap as well as a feasibility result.
\begin{corollary}[Optimization]
\label{cor:optimization}
Finding a global minimum of total squared empirical loss is \(\mathsf{NP}\)-hard.
On the integer instances in \Cref{thm:hardness}, computing a permitted network with loss at most the optimum plus any fixed \(0\le\eta<1\) is \(\mathsf{NP}\)-hard.
\end{corollary}
\begin{proof}
The outputs of every permitted network on integer samples are integers, by induction over additions, multiplications by integer parameters, and ReLUs.
Therefore its total squared loss is a nonnegative integer.
A yes instance has optimum zero, whereas a no instance has optimum at least one.
A feasible output whose loss is less than one must have loss zero, so either optimization guarantee decides the \(\mathsf{NP}\)-complete problem.
For mean squared loss on the two samples the corresponding additive gap is \(1/2\).
\end{proof}

\section{A Boolean Target Supplied by a Circuit}
\label{sec:circuit}

In this section the target is supplied by a Boolean circuit rather than by an explicit sample list.
A Boolean bit is an element of \(\{0,1\}\), with one meaning true and zero meaning false.
AND outputs one precisely when both inputs are one; OR does so when at least one is one, and NOT exchanges zero and one.
A Boolean formula combines variables with these operations; a literal is a variable or its negation.
A clause is an OR of literals, and a formula in conjunctive normal form (CNF) is an AND of clauses.
The width of a clause is its number of literals.
An assignment satisfies a formula when the formula takes the value one at that assignment.
A tautological clause contains both a variable and its negation and is true on every assignment.

Let \(C\) be an explicitly encoded Boolean circuit with \(d\) input bits and one output bit.
Its gates are AND, OR, and NOT, with the meanings just given.
Such a circuit is a directed acyclic graph computing \(C:\{0,1\}^d\to\{0,1\}\); evaluation on one input takes polynomial time.
The synthesis input is \((C,W,n_{\max})\), and the requirement is
\[
N(x)=C(x)\qquad\text{for every }x\in\{0,1\}^d.
\]
Equality means that the numerical output is exactly zero or one.
There is no free rounding or threshold gate.
A thresholded-output formulation is a different convention.
The network convention remains \Cref{def:network}.
The preliminary bounds below also hold with a ReLU output; the completeness theorem uses an affine output.

A problem belongs to \(\mathsf{coNP}\) if its complement belongs to \(\mathsf{NP}\).
The class \(\mathsf{DP}\) consists of languages expressible as the intersection of a language in \(\mathsf{NP}\) and a language in \(\mathsf{coNP}\).
The class \(\Sigma_2^{\mathsf P}\) consists of problems expressible by an existentially quantified polynomial-length string followed by a universally quantified polynomial-length string, with a polynomial-time predicate.
We recall the satisfiability theorem before using it as a hardness source.
\begin{theorem}[Satisfiability; Cook {\cite{Cook1971}}]
\label{thm:satisfiability-import}
Deciding whether a Boolean formula in conjunctive normal form has a satisfying assignment is \(\mathsf{NP}\)-complete under polynomial-time many-one reductions.
\end{theorem}

The \(\mathsf{NP}\)-completeness of Boolean circuit satisfiability follows from \Cref{thm:satisfiability-import}: formulas are circuits, and a satisfying circuit input is a polynomially checkable witness.
Circuit unsatisfiability is therefore \(\mathsf{coNP}\)-complete.

\begin{theorem}[Circuit-specified targets]
\label{thm:circuit}
Exact synthesis for a Boolean target supplied by a circuit is \(\mathsf{coNP}\)-hard under polynomial-time many-one reductions and belongs to \(\Sigma_2^{\mathsf P}\).
The lower bound holds with \(W=(0)\) and \(n_{\max}=1\).
Both bounds hold for at-most-once use and exactly-once use.
\end{theorem}
\begin{proof}
By the parameter-count argument in \Cref{thm:hardness}, a candidate network has a polynomial-length description.
Its validity and its equality with \(C\) on one Boolean input are polynomial-time checkable.
Let \(z\) range over binary strings of the polynomial length bounding a candidate description, padded as necessary.
The answer is yes precisely when
\[
\exists z\,\forall x\in\{0,1\}^d\,
\left(\text{\(z\) describes a permitted network }N_z
\text{ and }N_z(x)=C(x)\right).
\]
This proves the upper bound.
For a fixed candidate, a disagreeing Boolean input certifies failure in polynomial time, so candidate equivalence belongs to \(\mathsf{coNP}\).

For hardness map a circuit \(C\) to \((C,(0),1)\).
Every permitted network for this inventory has only zero parameters and computes the constant zero function.
Such a network exists: use the single zero as the output bias.
It uses the occurrence exactly once.
Thus synthesis succeeds if and only if \(C\) is unsatisfiable.
This proves both lower bounds.
\end{proof}

Enumerating networks and all Boolean inputs gives a deterministic running time of at most
\[
N_{\mathcal H} 2^d\operatorname{poly}(L),
\]
where \(L\) includes the circuit, inventory, and budget encodings.
Checking a fixed candidate is already \(\mathsf{coNP}\)-hard, by taking that candidate to be the fixed zero network and allowing \(C\) to vary.

Exact reproduction is stronger than agreement with high probability under random inputs.
A Boolean function can differ from zero at one cube vertex, with disagreement probability only \(2^{-d}\) under the uniform distribution on \(\{0,1\}^d\).
The completeness theorem below concerns all vertices, without a promise that a disagreement has substantial probability.

\subsection{Blocking Exact Subset Sums}

The numerical problem used below asks for a small set of occurrences meeting every exact subset sum.
This is a form of subset-sum interdiction, studied by Gr\"une and Wulf \cite{GrueneWulf2026}.
We need additional arithmetic restrictions on the numbers.
The following proof establishes those restrictions directly and treats equal-valued occurrences as distinct choices.

\begin{definition}[Exact-sum blocker]
\label{def:exact-sum-blocker}

For integers \(n\geq1,p\geq0\) and positive integers \(\beta_1,\ldots,\beta_n,T\), an \emph{exact-sum blocker} of size at most \(p\) is a set \(H\subseteq\{1,\ldots,n\}\) such that \(|H|\leq p\) and
\[
\sum_{j=1}^n\beta_jz_j=T
\quad\Rightarrow\quad
\sum_{j\in H}z_j\geq1
\qquad\text{for every }z\in\{0,1\}^n.
\]
The condition is vacuous when no subset sums to \(T\).
An index \(j\) with \(z_j=1\), and the occurrence it labels, is said to be selected by \(z\).
\end{definition}

A prime is an integer \(q\ge2\) whose only positive divisors are one and \(q\).
We will cancel nonzero residues modulo a prime.
A map is injective when equal outputs require equal inputs.
For an integer \(a\) not divisible by \(q\), choose the least positive integer \(g\) of the form \(ua+vq\), with \(u,v\in\mathbb Z\); such positive integers exist because \(q\) is one of them.
Dividing \(a\) and \(q\) by \(g\) with remainders in \(\{0,\ldots,g-1\}\) shows that each remainder is also such an integer combination.
Minimality of \(g\) forces both remainders to be zero, so \(g\) divides both \(a\) and \(q\).
Primality and \(q\nmid a\) then force \(g=1\), giving integers \(u,v\) with \(ua+vq=1\).
If \(q\) divides \(ab\), multiplying this identity by \(b\) shows that \(q\) divides \(b\).
Thus multiplication by a nonzero residue is injective on residues modulo \(q\), and a product of nonzero residues remains nonzero.

Trial division tests whether an integer \(q\ge2\) is prime by checking divisors from two through \(\lfloor\sqrt q\rfloor\).
Indeed, any factorization \(q=c_1c_2\) with \(c_1,c_2\ge2\) has at least one factor at most \(\sqrt q\).
For integers bounded by a polynomial in the input length, this uses polynomially many binary divisions of polynomial bit length.
The numerical calibration uses the following classical prime bound.
\begin{theorem}[Bertrand's bound; Erd\H{o}s {\cite{Erdos1932}}]
\label{thm:bertrand-import}
For every integer \(h\geq2\), there is a prime \(q\) with \(h<q<2h\).
\end{theorem}

The restrictions in the next lemma separate the numerical values that will play different roles in a synthesis instance.
They leave the universal exact-sum condition computationally difficult.

\begin{lemma}[Restricted exact-sum blockers]
\label{lem:restricted-blockers}

Deciding whether an exact-sum blocker exists is \(\Sigma_2^{\mathsf P}\)-complete under deterministic polynomial-time many-one reductions, even under the following restrictions.
An odd prime \(q\geq5\) is supplied, \(p=q-1\), every \(\beta_j\) satisfies \(0<\beta_j<T\) and is not a power of two, and
\[
\beta_j\equiv3\pmod q,
\qquad
T\equiv0\pmod q.
\]
The reduction produces \(p\) bounded by a polynomial in its input length.
Every produced instance has \(n>p\), where \(n\) is the number of its source numbers \(\beta_j\).
\end{lemma}

\begin{proof}

Membership follows by choosing \(H\) existentially and quantifying \(z\) universally in \Cref{def:exact-sum-blocker}.
The size constraint and the implication are checkable by polynomial-time binary arithmetic.
We give the hardness construction in three steps: an encoding of counterexamples, ordinary subset sums with designated occurrences, and protected choices with the required congruences.

\emph{Counterexamples and calibration assignments.}

Let \(\ell,s\geq0\) be integers and let \(\psi\) be a CNF formula on bit vectors \(u\in\{0,1\}^{\ell}\), \(v\in\{0,1\}^{s}\), with clause width at most three.
We start from the quantified formula
\[
\exists u\in\{0,1\}^{\ell}\,\forall v\in\{0,1\}^{s}\,\neg\psi(u,v).
\]
This source is complete for \(\Sigma_2^{\mathsf P}\), by the following encoding of polynomial-time computation as a circuit and then as a formula.
A polynomial-time deterministic computation has polynomially many configuration bits and steps.
The configuration records its state, head positions, and symbols in the polynomially many tape cells it can visit.
One transition is computed by a polynomial-size Boolean circuit: test the head positions, update the state and the scanned cells, and copy every other cell.
Unrolling polynomially many copies of this transition circuit yields a polynomial-size circuit for the predicate, constructible in polynomial time.

To see the required polarity directly, start with any polynomial-time predicate defining a \(\Sigma_2^{\mathsf P}\) problem, and apply the circuit construction just described to its failure predicate, the negation of that predicate.
Introduce one bit for every circuit gate, impose its truth-table constraints as clauses of size at most three, and assert a failure output of one.
These new bits belong to the universally quantified counterexample block.
We call the predicate's own input bits, existential and universal alike, the original input bits, as opposed to the new gate bits.
For fixed original input bits, a satisfying extension of the gate bits exists precisely when the failure circuit outputs one.
Thus forbidding every satisfying counterexample gives the displayed quantified formula.
This is the usual circuit computation encoding underlying satisfiability \cite{Cook1971}.

Choose an odd prime \(q\geq\max\{5,\ell+1\}\) of size polynomial in \(\ell+1\), and put \(p=q-1\).
Such a prime can be found by deterministic search in a polynomial-sized interval: \Cref{thm:bertrand-import} supplies one between an integer \(h\geq2\) and \(2h\), and trial division is polynomial for these polynomially bounded integers.
Add unused existential variables until there are exactly \(p\) of them.
Their unused values do not change the quantified formula.

For each \(i\in\{1,\ldots,p\}\), introduce two marker bits \(m_i^0,m_i^1\).
Construct a Boolean circuit for a relation \(R\) with the following accepting assignments.
A main assignment has exactly one true marker in each pair and satisfies \(\psi(u',v)\), where \(u'_i=m_i^1\).
A calibration assignment for index \(i\) has both \(m_i^0,m_i^1\) true and every other marker false; its \(v\)-bits are unrestricted.
The relation \(R\) is the disjunction of these cases.
It has a polynomial-size circuit.
Encode that circuit by gate constraints and an asserted true output, obtaining a formula \(\Phi\) with clauses of size at most three.
The projections of satisfying assignments of \(\Phi\) onto the markers and \(v\) are exactly the assignments accepted by \(R\).
Gate bits are auxiliary variables, distinct from the markers.

\emph{Subset sums and designated choices.}

Use one base-ten digit for every variable of \(\Phi\) and one for every clause.
The target has digit one in each variable position and digit four in each clause position.
For every variable create a true item and a false item.
Each item has digit one in its own variable position, and digit one in every clause satisfied by that literal.
For each clause add two slack items, with digits one and two in that clause position and zero elsewhere.
All these item values are positive integers.
Only the true items of the \(2p\) marker variables are \emph{designated}: each will receive a single supplied take occurrence below, so that a blocker can attack it.

Remove repeated literals from clauses and discard tautological clauses first.
Every clause then has at most three distinct literal occurrences.
Even an arbitrary subset of all items contributes at most three literal units and three slack units to a clause digit, and at most two units to a variable digit.
Consequently no base-ten carry occurs in any subset sum.
The target variable digits force exactly one of the two items for each variable.
If \(t\in\{0,1,2,3\}\) literals satisfy a clause, its target digit four can be obtained using slack of total at most three if and only if \(t\geq1\).
Thus exact subsets correspond precisely to satisfying assignments of \(\Phi\), with appropriate slack choices.
Write their positive item values as \(a'_1,\ldots,a'_{N_0}\), and their target as \(A'\).

Multiply every item and the target by \(q\), giving \(a_j=qa'_j\) and \(A=qA'\).
Append fewer than \(q\) additional items of value \(q(A'+1)>A\) until the number \(N\) of items is a multiple of \(q\).
The appended items cannot occur in an exact \(A\)-sum.
They are not designated.
There are at least \(2p\) marker variables, so \(N\geq2\).
The scale and padding preserve every original solution and its designated choices.

\emph{Protected classes.}

Choose positive multiples \(b,L\) of \(q-1\) so that
\[
B=2^b>2p+2,
\qquad
K=3\cdot2^L>4\left(A+(p+1)\sum_{j=1}^{N}a_j+1\right).
\]
For every \(j\in\{0,\ldots,N-1\}\), create a take value and a skip value
\[
\tau_j=KB^j+a_{j+1},
\qquad
\sigma_j=KB^j.
\]
If \(a_{j+1}\) is designated, supply one occurrence of \(\tau_j\).
Otherwise supply \(p+1\) occurrences of \(\tau_j\).
Supply \(p+1\) occurrences of every \(\sigma_j\).
The occurrences of \(\tau_j\) and \(\sigma_j\) form \emph{class \(j\)}, and its two \emph{class options} are take, an occurrence of \(\tau_j\), and skip, an occurrence of \(\sigma_j\).
An option supplied with \(p+1\) copies is \emph{protected}, and so are its occurrences; since every skip option is protected, we call these classes \emph{protected classes}.
A \emph{take class} of a subset is a class from which it selects a take occurrence.
These take and skip occurrences are the numbers \(\beta_1,\ldots,\beta_n\) produced by the reduction.
Their target is
\[
T=A+K\sum_{j=0}^{N-1}B^j.
\]

The total possible contribution of all low parts \(a_j\), even when all copies are selected, is at most \((p+1)\sum_j a_j<K/4\).
Also \(A<K/4\).
Reducing an exact \(T\)-sum modulo \(K\) therefore forces its selected low parts to sum exactly to \(A\).
After subtracting those parts and dividing by \(K\), the remaining equality has target \(\sum_j B^j\).
At class \(j\), at most \(2p+2<B\) supplied occurrences can contribute \(B^j\).
There is no carry in base \(B\), so exactly one occurrence is selected from each class.
The selected take classes give an exact \(A\)-sum of original items.
Conversely, every such original sum lifts to a \(T\)-sum by taking its chosen classes and skipping the others.

An occurrence that lies in a candidate blocker \(H\) is said to be attacked by \(H\), and a set of attacked occurrences is an attack.
A class option with \(p+1\) copies cannot be removed completely by attacking at most \(p\) occurrences.
For every fixed original solution, any attack of that size on its protected options can therefore be bypassed by choosing surviving copies independently in each class.
Only designated take occurrences can be removed without such a replacement.

Consider a blocker \(H\) of size at most \(p\).
For each \(i\), a calibration assignment exists whose only designated true items are \(m_i^0,m_i^1\).
If \(H\) contained neither of these two designated take occurrences, lift that calibration solution while avoiding every attacked protected occurrence.
This would produce a \(T\)-sum disjoint from \(H\), contradicting its blocker property.
Hence \(H\) contains at least one designated take occurrence from every marker pair.
There are \(p\) pairs and the budget is \(p\), so it contains exactly one from each pair and no other occurrence.

Let \(\theta_i\in\{0,1\}\) be the label it attacks in pair \(i\).
A main assignment has exactly one true marker in each pair, and it avoids \(H\) exactly when, for every \(i\), that single true marker has label \(1-\theta_i\).
Its encoded existential assignment is therefore \(u'_i=1-\theta_i\).
No main assignment avoids \(H\) if and only if \(\psi(u',v)\) is false for every \(v\).
Conversely, any existential witness \(u'\) to that universal condition gives a blocker by attacking marker \(1-u'_i\) in every pair.
Each calibration is hit, and no main solution avoids these attacks.
This proves both directions of the reduction.

\emph{Arithmetic restrictions and encoding length.}

For a nonzero residue \(\alpha\) modulo the prime \(q\), multiplication by \(\alpha\) permutes the \(q-1\) nonzero residues.
Multiplying those residues and canceling their nonzero product gives \(\alpha^{q-1}\equiv1\pmod q\).
Apply this with \(\alpha=2\).
Since \(b,L\) are multiples of \(q-1\), it gives \(B\equiv1\pmod q\) and \(K\equiv3\pmod q\).
Every low part \(a_j\) and \(A\) is divisible by \(q\), and \(N\equiv0\pmod q\).
Thus every take or skip value is congruent to three modulo \(q\), whereas \(T\equiv0\pmod q\).

Put \(E=L+bj\).
The skip value is \(3\cdot2^E\), strictly between \(2^{E+1}\) and \(2^{E+2}\).
The take value adds \(a_{j+1}<K/4\leq3\cdot2^E/4\), so it remains strictly below \(4\cdot2^E=2^{E+2}\).
Both values lie strictly between consecutive powers of two.
For a take value,
\[
T-\tau_j=A-a_{j+1}+K\sum_{h\ne j}B^h>-K/4+K>0,
\]
since \(N\geq2\).
A skip value is smaller than its corresponding take value.
All output values are consequently positive and strictly below \(T\).
Every class supplies \(p+1\) skip occurrences and at least one take occurrence, and there are \(N\geq2\) classes, so the number of output values is \(n\geq2(p+2)>p\).

Finally, \(q,p,N\) are polynomially bounded in the original formula length.
The exponents \(b,L\) can be chosen polynomially bounded by rounding the relevant binary-length bounds up to multiples of \(q-1\).
Every exponent \(L+bj\) and every output binary length is polynomial.
There are polynomially many supplied copies.
The construction is therefore an explicit deterministic polynomial-time reduction, including when the original existential block is empty.
\end{proof}

\subsection{A Calibrated Threshold Construction}

Fix an instance in the hardness family produced\linebreak by \Cref{lem:restricted-blockers}, with source numbers \(\beta_1,\ldots,\beta_n\), target \(T\), prime \(q\), and budget \(p=q-1\).
In that family the number \(n\) of source numbers satisfies \(n>p\), by \Cref{lem:restricted-blockers}.
Thus \(q=p+1\leq n\) is polynomially bounded in the source encoding.
We construct a Boolean threshold target and a rational inventory.
The construction has two purposes.
A small perturbation of the source coefficients will encode a blocker, while additional input coordinates will constrain every alternative use of the inventory.

Choose an odd prime \(r>q\) large enough that
\[
Q=2^r-1>8\left(T+\sum_{j=1}^{n}\beta_j+2\right)^2.
\]
The multiplicative order of two modulo \(q\) is the least positive integer \(t\) with \(2^t\equiv1\pmod q\).
It exists with \(t\leq q-1\).
Indeed, \(q\nmid2\) because \(q\) is an odd prime, so by the cancellation of nonzero residues, \(q\mid2^k\) with \(k\geq1\) would give \(q\mid2^{k-1}\), and repeating would give \(q\mid1\); thus no power of two is divisible by \(q\).
Hence the \(q\) numbers \(2^0,\ldots,2^{q-1}\) have residues among the \(q-1\) nonzero residues, and two of them, \(2^k\) and \(2^{k'}\) with \(0\leq k<k'\leq q-1\), have the same residue.
Canceling the nonzero residue of \(2^k\) gives \(2^{k'-k}\equiv1\pmod q\) with \(1\leq k'-k\leq q-1\).
If \(q\) divided \(Q\), then \(2^r\equiv1\pmod q\).
Write \(r=ht+k_0\) with integers \(h\geq0\) and \(0\leq k_0<t\).
Then \(1\equiv2^r=(2^t)^h2^{k_0}\equiv2^{k_0}\pmod q\), so the minimality of \(t\) forces \(k_0=0\), and \(t\) divides \(r\).
Since \(r\) is prime, \(t\in\{1,r\}\), and \(t\leq q-1<r\) excludes \(t=r\); hence \(t=1\).
Then \(2\equiv1\pmod q\), that is, \(q\) divides one, which is impossible for a prime, so \(Q\not\equiv0\pmod q\).
Since \(T\equiv0\pmod q\) by \Cref{lem:restricted-blockers}, it follows that \(Q-T\equiv Q\not\equiv0\pmod q\).
These two nonzero residues are used in \Cref{lem:boolean-sole-negative} and \Cref{lem:boolean-finite-biases} of \Cref{app:boolean-placements}, where \(Q-T\) and \(Q\), the multiples of \(F\) in the anchor baseline \(B\) and the control baseline \(C\) defined below, must be nonzero modulo \(q\); up to sign, \(Q\) is also the multiple of \(F\) in each of the two finite negative parameters.

Let
\[
n_{\mathrm a}=p+3,
\qquad K=2n+3p+14,
\qquad J=K+1,
\qquad R=3Jr,
\qquad d=n+n_{\mathrm a}+R+1.
\]
The input coordinates consist of the \(n\) source bits \(z_j\), \(n_{\mathrm a}\) anchor bits \(a_h\), \(J\) query blocks, and one control bit \(c\).
Each query block has three query coordinates, called copies, at every \emph{level} \(i\in\{0,\ldots,r-1\}\).
The level of a query coordinate is a label fixed by the construction, and it determines the baseline that the coordinate receives, defined below.
There are \(R\) query coordinates in total.
Each query coordinate keeps its specified level even if a candidate assigns parameter occurrences of other values to its incident edges.

Choose a power of two \(F\) satisfying
\[
F>\max\{8Q^4,4d+64q^2+8\},
\qquad P=FQ.
\]
Put \(m=K+2R\) and
\[
H_0=16q^4\left(64(m+1)^4P^4+1\right).
\]
Choose a power of two \(M>\max\{2H_0,q(2R+3)\}\), and set
\[
w=M^{-1},
\qquad \Gamma=M^9,
\qquad s_0=\frac{1}{q\Gamma}.
\]
The four baseline families are
\[
F\beta_j,
\qquad B=F(Q-T)-w,
\qquad F2^i+2w,
\qquad C=P-\frac32w.
\]
Assign these to source, anchor, level-\(i\) query, and control coordinates, respectively.
Write \(\lambda_t\) for the baseline of input coordinate \(t\), and define
\[
\Lambda(x)=\sum_{t=1}^{d}\lambda_t x_t,
\qquad f(x)=\begin{cases}1,&\Lambda(x)\geq P-w,\\0,&\Lambda(x)<P-w.\end{cases}
\]
We call \(\Lambda(x)\) the \emph{baseline score} of \(x\).
Every baseline is positive.
The inventory contains two occurrences of every coordinate's baseline, \(p\) occurrences of \(s_0\), two of \(\Gamma\), and the four signed parameters
\[
+M,\qquad -M,\qquad -(P-w),\qquad -P.
\]
Its size is \(m\), because the number of occurrences outside the query-baseline family is
\[
2n+2n_{\mathrm a}+2+p+2+2+2=K.
\]
The computational-node cap is four, including the affine output.
We call \(+M,-M\) the \emph{gain parameters} and the remaining parameters \emph{nongain parameters}.
A \emph{primitive} means a single inventory value, before any path multiplication.
The nongain values of the form \(F\nu+\delta w\), with nonzero integer \(\nu\) and rational \(\delta\), are called \emph{finite primitives}; they are exactly the baseline families and the two negative finite values.
Here \(\delta\) is \(0\) for a source baseline and for \(-P\), \(-1\) for \(B\), \(2\) for a query baseline, \(-3/2\) for \(C\), and \(1\) for \(-(P-w)\).
The other nongain types are \(\Gamma\) and \(s_0\).
We call occurrences of \(s_0\) \emph{selectors}.

All these numbers have polynomial binary length.
The blocker instances produced by \Cref{lem:restricted-blockers} have polynomially bounded \(p\).
\Cref{thm:bertrand-import} permits a polynomially bounded numerical choice of \(r\), larger than \(q\) and a constant times the binary length of \(T+\sum_j\beta_j+2\).
The quantities \(K,J,R,d,m\) are polynomially bounded integers, and the logarithms of the bounds defining \(F,M\) are polynomial.
Clearing the common denominator \(2M\) turns the condition defining \(f\) into an integer threshold comparison.
Binary addition and comparison therefore give a polynomial-size Boolean circuit for \(f\).
The reduction supplies this circuit, rather than a table of its values.

\subsection{Formal Identities and Query Restrictions}

A Laurent polynomial in formal variables \(w,\Gamma\) is a finite sum of terms \(c_{ij}w^i\Gamma^j\), where \(c_{ij}\in\mathbb Q\) and \(i,j\in\mathbb Z\).
A power product \(w^i\Gamma^j\) is a monomial; after collecting equal monomials, \(c_{ij}\) is its coefficient.
Formal equality means equality of all these coefficients, before evaluating the variables at numerical values.
The \emph{formal expansion} of an inventory value is the following Laurent polynomial, in which \(F\), \(P\), \(Q\) and \(q\) are numerical coefficients.
A finite primitive \(F\nu+\delta w\) expands as itself, with coefficients \(F\nu\) and \(\delta\); \(\Gamma\) expands as the monomial \(\Gamma\); \(s_0\) expands as \(q^{-1}\Gamma^{-1}\); and the gains \(\pm M\) expand as \(\pm w^{-1}\).
The specialization \(w=M^{-1}\), \(\Gamma=M^9\) returns each inventory value from its formal expansion.
The choices of \(F\) and \(M\) allow two different separations.
The factor \(F\) distinguishes products from individual finite parameters.
The much larger specialization \(\Gamma=M^9\), \(w=M^{-1}\) makes every exact Boolean output equation a formal polynomial identity.
The argument for the second separation freezes the actual activation pattern at the tested input before treating the expression formally; it does not assert that the same pattern persists under changes to the numerical parameters.

\begin{lemma}[Uniform specialization]
\label{lem:boolean-specialization}

Fix any candidate with at most four computational vertices and any Boolean input.
Replace each hidden ReLU by either its affine preactivation or zero according to its actual activation at that input.
Expand the resulting output as a Laurent polynomial in \(w,\Gamma\), replacing every parameter occurrence by the formal expansion of its value.
If the numerical output at that input is \(b\in\{0,1\}\), the expanded Laurent polynomial is identically \(b\).
This assertion holds uniformly for every legal assignment of inventory occurrences and every Boolean input.
\end{lemma}

\begin{proof}

Under a fixed activation pattern, the output is the sum of the products along the active input-to-output and bias-to-output paths.
A directed path visits distinct computational vertices, and there are at most four of them.
An input-to-output path therefore has at most four edges, and a bias-to-output path consists of one bias and at most three edges, so every path contains at most four parameter occurrences.
The occurrences along a path determine it, since its first occurrence fixes its starting input or bias and each edge fixes its head; hence there are at most \(\sum_{k=1}^4m^k\leq(m+1)^4\) paths.

Each nongain finite parameter \(F\nu+\delta w\) has at most two formal terms; the other parameters are monomials.
Every formal term of a single parameter has the form \(cw^i\Gamma^j\) with \(i,j\in\{-1,0,1\}\): the finite primitives give \(F\nu\) and \(\delta w\), \(\Gamma\) gives \(\Gamma\), \(s_0\) gives \(q^{-1}\Gamma^{-1}\), and the gains give \(\pm w^{-1}\).
Expanding the product along one path, with its input coordinate in \(\{0,1\}\) as a further factor, gives at most \(2^4=16\) terms, and the exponent of either formal variable in each term lies between minus four and four.
The single-parameter coefficients \(F\nu\), \(\delta\), \(1\), \(q^{-1}\) and \(\pm1\) have denominators dividing \(2q\), because \(F\nu\) is an integer and \(\delta\in\{0,-1,2,-3/2,1\}\).
A product of at most four of them therefore has a denominator dividing \((2q)^4=16q^4\), which is a common denominator of all coefficients.
Each single-parameter coefficient has absolute value at most \(P\): every finite primitive has \(|\nu|\leq Q\), so \(|F\nu|\leq FQ=P\), while \(|\delta|\leq2\), \(q^{-1}\) and \(1\) are at most \(P\).
Hence each term coefficient has absolute value at most \(P^4\).

After collecting equal monomials, each coefficient of the output expansion is a sum of at most \(16(m+1)^4\leq64(m+1)^4\) term coefficients, so it has absolute value at most \(64(m+1)^4P^4\).
Subtracting \(b\in\{0,1\}\) changes only the constant coefficient, by at most one.
Multiplying by \(16q^4\) clears all denominators, so the resulting integer coefficients have absolute value at most \(16q^4\left(64(m+1)^4P^4+1\right)=H_0\).

Under the numerical substitution, a monomial \(w^u\Gamma^v\) becomes \(M^{-u+9v}\).
The map \((u,v)\mapsto-u+9v\) is injective on \(\{-4,\ldots,4\}^2\): equality would give \(u-u'=9(v-v')\), whose left side has absolute value at most eight.
The resulting exponents lie between minus forty and forty.
Multiplying by \(M^{40}\) gives an integer polynomial of degree at most eighty.
For any nonzero such polynomial, the highest nonzero term, of degree \(k\), has absolute value at least \(M^k\), while the sum of the absolute values of the lower terms is less than
\[
H_0\frac{M^k}{M-1}<M^k.
\]
Because \(M>2H_0\), the displayed inequality holds and the polynomial cannot vanish.
Consequently the numerical output equation forces every formal coefficient to vanish.
The same calculation applies to a preactivation.
Its actual sign, when nonzero, is the sign of the highest specialized term.
Nothing here asserts that the activation pattern remains fixed when \(w\) and \(\Gamma\) vary independently.
\end{proof}

The next restriction resolves a possible mismatch between input labels and parameter values.
An input called a query coordinate need not receive a query-valued parameter in a candidate.
Replication ensures that, for every candidate, one complete block receives only query-valued parameters on its input edges.

A coordinate face of a Boolean cube is obtained by fixing some coordinates to bits and leaving the others free.
Such a face is also called a subcube.
For a Boolean-valued function, a vertex wins if its value is one and loses if its value is zero.
A coordinate is selected at a vertex of the cube when its value there is one.
A coordinate is essential if changing only that coordinate changes the value at some fixed setting of the other coordinates.
Such a setting is called a background, and the coordinate is pivotal at it.
A background has \emph{(binary) weight} \(W\), and is then a \emph{query background of weight \(W\)}, when the coordinates it sets to one are query coordinates whose values \(2^i\), for their levels \(i\), sum to \(W\), and it sets every other coordinate to zero.
The function is coordinatewise monotone if changing any coordinate from zero to one never decreases its value.
For two free coordinates of a function \(h\), the mixed difference is \(h(1,1)-h(1,0)-h(0,1)+h(0,0)\), with every other coordinate fixed.

\begin{lemma}[A block with only query-valued input edges]
\label{lem:boolean-clean-block}

For every candidate, some query block has no input edge from its coordinates that carries an occurrence outside the query-baseline family.
Fix every coordinate outside that block to zero.
Then each nonzero input edge from a coordinate of the block carries an actual value \(F2^i+2w\) for some level \(i\); this is the level of the occurrence, which need not be the level of the coordinate.
Write \(y_{i,\gamma}\) for the copy-\(\gamma\) coordinate of level \(i\) in the block, where \(i\) is now the level of the coordinate.
The target restriction is
\[
g(y)=1\quad\text{if and only if}\quad
\sum_{i=0}^{r-1}\sum_{\gamma=1}^{3}2^i y_{i,\gamma}\geq Q.
\]
The origin and all unit vectors lose; the all-ones vector and every one- or two-coordinate deletion win.
Every query coordinate is essential.
The function \(g\) cannot be expressed as an affine function plus a scalar multiple of one ReLU of an affine function, and neither can any restriction of \(g\) obtained by fixing up to two coordinates to zero or any one coordinate to one.
\end{lemma}

\begin{proof}

An occurrence outside the query-baseline family can contaminate at most one block when used on an \emph{external-input edge}, an edge leaving an input vertex.
There are \(K\) such occurrences and \(J=K+1\) blocks.
The asserted block follows by counting.
All input edges from the other coordinates contribute zero on its restriction.
The target value there depends on the baseline score, which there equals \(FI+2w\kappa\), where \(I\) is the displayed binary weighted sum and \(\kappa\leq3r\) is the number of selected query coordinates.
Since \(2Rw<1\), inputs with \(I\geq Q\) win and inputs with \(I\leq Q-1\) lose.

A single query weight is at most \((Q+1)/2<Q\).
The all-ones sum is \(3Q\), and deleting two coordinates leaves at least \(3Q-(Q+1)=2Q-1>Q\).
A level-\(i\) coordinate is pivotal against a background of one copy of every other level, whose sum is \(Q-2^i\).
Thus every coordinate is essential.

For two tested level-zero coordinates, backgrounds of weight \(Q-2\) and \(Q-1\) give respectively AND and OR on that same pair.
Their pairwise mixed differences are one and minus one.
The other coordinates are fixed to the stated background in each difference.

Let \(\phi(y)=\ell(y)+a\rho(t+v\cdot y)\), where \(\ell\) is affine.
We show that every mixed difference of \(\phi\) in coordinates \(i\) and \(j\) has the fixed sign of \(av_iv_j\), or is zero.
The affine contribution to this mixed difference is zero.
For an interval length \(u>0\), the ReLU definition gives
\[
\rho(t+u)-\rho(t)=
\begin{cases}
0,&t\le-u,\\
t+u,&-u<t<0,\\
u,&t\ge0.
\end{cases}
\]
This increment is nondecreasing in the starting point \(t\).
Let \(u'>0\) be a second length.
The increment from the starting point \(t+u'\) minus the increment from \(t\) is the mixed difference of \(\rho\) for the lengths \(u\) and \(u'\).
By the monotonicity just noted, this mixed difference is nonnegative.
For arbitrary real lengths write \(\Delta(t;u,u')=\rho(t+u+u')-\rho(t+u)-\rho(t+u')+\rho(t)\) for this mixed difference, which is symmetric in \(u\) and \(u'\).
If \(u<0<u'\), put \(t'=t+u\) and \(\bar u=-u>0\); since \(t'+\bar u=t\), \(t'+u'=t+u+u'\) and \(t'+\bar u+u'=t+u'\), substitution gives
\[
\Delta(t;u,u')=-\Delta(t';\bar u,u')\leq0.
\]
The case \(u>0>u'\) follows by exchanging the two lengths.
If \(u,u'<0\), put \(t''=t+u+u'\); the points \(t'',t''-u,t''-u',t''-u-u'\) are \(t+u+u',t+u',t+u,t\), so \(\Delta(t;u,u')=\Delta(t'';-u,-u')\geq0\).
A zero length gives zero.
At a background \(y\) with \(y_i=y_j=0\), the mixed difference of \(\phi\) in coordinates \(i\) and \(j\) equals \(a\Delta(t+v\cdot y;v_i,v_j)\).
Taking the two lengths to be \(v_i\) and \(v_j\) and multiplying by \(a\) therefore gives the stated sign of \(av_iv_j\), or zero.
Consequently, one ReLU plus an affine function cannot produce the two opposite nonzero differences.

After at most two zero fixings, choose a level \(h<r-1\) with all three copies still free.
Such a level exists because \(r\geq7\), since \(r\) is an odd prime larger than \(q\geq5\).
Reserve two copies for testing.
The binary expansions of \(Q-2^{h+1}\) and \(Q-2^h\) give AND and OR backgrounds, using at most one surviving copy of every other level and a third copy of level \(h\) for the AND background.
Every level still has at least one surviving copy.
If instead a coordinate of level \(k\) is fixed to one, choose \(h<r-1\) with \(k\notin\{h,h+1\}\).
Both backgrounds then include the prescribed fixed coordinate at level \(k\); the two tested coordinates and the third level-\(h\) copy are free.
The same mixed-difference argument applies on each of these faces.
\end{proof}

\Cref{app:boolean-placements} uses the last assertion of \Cref{lem:boolean-clean-block} to exclude placements that leave the restricted output an affine function plus a multiple of one ReLU.

The full target also has the geometric properties needed after the graph has been constrained.
Those properties are independent of the candidate's assignment of parameters.

\begin{lemma}[Full-target tests]
\label{lem:boolean-full-target}

The function \(f\) is monotone, every coordinate is essential, and \(f(0)=f(e_j)=0\) for every coordinate \(j\).
Call a winning vertex nonminimal if deleting one of its selected coordinates gives a winning vertex.
The nonminimal winning vertices affinely span \(\mathbb R^d\).
The function has a pairwise mixed difference of minus one on a Boolean face.
\end{lemma}

\begin{proof}

All baselines are positive, so \(f\) is monotone.
A source coordinate \(z_j\) is pivotal at a query background of weight \(Q-\beta_j\), an anchor at one of weight \(T\), and a level-\(i\) query at one of weight \(Q-2^i\).
These backgrounds lie between zero and \(Q\) and are realized by binary expansions using one available copy per level.
They can avoid the tested query coordinate.
The losing side is separated by at least one integer unit in the \(F\)-scaled sum; its total small correction is less than one.
On the winning side the source and query tests reach at least \(P\), and the anchor test reaches \(P-w\) plus a nonnegative correction.
The control coordinate is pivotal at the background in which one level-zero query coordinate is one and every other coordinate is zero.
Its baseline score without the control is \(F+2w<P-w\), while adding \(C\) gives more than \(P\).
All singleton baseline scores, including \(C=P-3w/2\), are below \(P-w\).

On the slice \(c=1\), every vertex with another selected coordinate wins.
Every vertex of that slice with at least two selected coordinates other than \(c\) is a nonminimal winner.
The all-ones vertex of that slice and its single deletions affinely span the hyperplane \(c=1\).
There is also a nonminimal winner with \(c=0\) and three anchors selected: two anchors already have baseline score greater than \(P\), because \(Q>2T\).
This gives a point outside that hyperplane and proves the full affine span.
The OR background in \Cref{lem:boolean-clean-block} gives the claimed negative mixed difference.
\end{proof}

Write \(D=M=w^{-1}\) for the gain magnitude.
A \emph{positive wire} is a hidden ReLU whose incoming coefficients and present bias are positive; on Boolean inputs it equals its nonnegative preactivation.
The following placement statement is the part of the reduction that addresses arbitrary graph choices.
Its proof, including every internal, output, bias and unused-parameter case, is given in \Cref{app:boolean-placements}.
The argument first restricts the two large gains, then forces the finite negative parameters into hidden biases and identifies the source vertices of the two gain edges.
It next proves that the two \emph{biased hinges}, the hidden ReLUs that receive the finite negative parameters as biases, are not \emph{coupled}: there is no directed path from one to the other, neither a direct edge nor a path through the remaining hidden vertex.
A \emph{coupling edge} is an edge on such a path.
Only after these actual edges are excluded does the full-cube argument identify their common \emph{score}, the preactivation of each biased hinge without its finite negative bias.

\begin{lemma}[Placement for the calibrated construction]
\label{lem:boolean-placement}

Let \(W,f\) be the calibrated inventory and target above, with computational-node cap four and affine output.
If any permitted network represents \(f\) on the Boolean cube, then some permitted network using a submultiset of the same occurrences has the form
\[
D\left(\rho(S(x)-(P-w))-\rho(S(x)-P)\right),
\qquad S(x)=a u(x)+\sum_{j=1}^{d}b_jx_j.
\]
The coefficients \(a,b_1,\ldots,b_d,v_{\mathrm b},v_1,\ldots,v_d\) are nonnegative reals; a zero denotes an absent parameter, and each nonzero coefficient is a single positive actual inventory primitive.
The optional wire is \(u(x)=v_{\mathrm b}+\sum_{j=1}^{d}v_jx_j\), also called the sharing wire.
It enters each hinge through one \emph{sharing edge} of weight \(a\), and each input \(x_j\) enters each hinge through one direct input edge of weight \(b_j\).
Every nonzero \(a\) uses two identical-value occurrences for its two sharing edges, and every nonzero \(b_j\) uses two identical-value occurrences for its two direct input edges.
The two displayed biases and gains use their specified supplied occurrences.
The hinges carry no other bias, so the \emph{score bias}, the constant term of \(S\), is \(a v_{\mathrm b}\).
The output has no \emph{ordinary output contribution}, meaning no output bias, no skip edge and no edge from the wire to the output.
A wire without incoming parameters, or with no sharing edge, is omitted.
\end{lemma}

We call the displayed representation a \emph{common ramp}: its two hinges share one affine score and their thresholds differ by \(w\).

\subsection{Recovering a Blocker from a Common Ramp}

The placement argument ensures a common score without assigning every remaining primitive to its intended numerical role.
We now recover an actual source blocker from any legal common-ramp representation.
A binary threshold identity will first determine every query coefficient.

\begin{lemma}[Three-copy binary reference]
\label{lem:boolean-binary-reference}

Let \(Q=2^r-1\), with \(r\geq2\).
Assign positive integer weights to three copies of every level \(i\in\{0,\ldots,r-1\}\).
If the assigned weights with threshold \(Q\) give exactly the same Boolean function as the weights \(2^i\) with threshold \(Q\), then every assigned weight at level \(i\) equals \(2^i\).
The conclusion also holds with any larger number of copies per level.
\end{lemma}

\begin{proof}

Let \(l_i,u_i\) be the minimum and maximum assigned weights at level \(i\), and put \(L=\sum_i l_i\).
Call a selection of copies an original winner if its sum under the weights \(2^i\) is at least \(Q\), and an original loser otherwise.
Selecting one minimum at every level is an original winner, so \(L\geq Q\).
Selecting one maximum at every positive level and no level-zero coordinate is an original loser, giving \(\sum_{i\geq1}u_i\leq Q-1\).
For \(i\geq1\), start instead with the losing pattern of minima at every positive level.
Remove its level-\(i\) coordinate and add two unused level-\((i-1)\) coordinates.
Its original binary sum remains \(Q-1\), so it is an original loser and its assigned sum is at most \(Q-1\).
Its assigned sum is also at least \(L-l_0-l_i+2l_{i-1}\), because the pattern of minima has assigned sum \(L-l_0\), the removed coordinate has weight \(l_i\), and each of the two added coordinates has weight at least \(l_{i-1}\).
Combining the two bounds with \(L\geq Q\) gives \(L-l_0-l_i+2l_{i-1}\leq Q-1\leq L-1\), that is,
\[
l_i\geq2l_{i-1}+1-l_0.
\]
This uses two additional copies when the lower-level minimum was already selected, which explains the three-copy hypothesis.
Induction on \(i\) gives \(l_i\geq l_0+2^i-1\): equality holds at \(i=0\), and for \(i\geq1\) the inductive hypothesis gives \(l_i\geq2(l_0+2^{i-1}-1)+1-l_0=l_0+2^i-1\).
Summing over the positive levels, using \(u_i\geq l_i\), the maxima bound and \(\sum_{i=1}^{r-1}2^i=2^r-2=Q-1\), gives
\[
Q-1\geq\sum_{i=1}^{r-1}u_i\geq\sum_{i=1}^{r-1}l_i\geq\sum_{i=1}^{r-1}\left(l_0-1+2^i\right)=(r-1)(l_0-1)+Q-1.
\]
The weights are positive integers, so \(l_0\geq1\), and \(r\geq2\); the display therefore forces \(l_0=1\).
Every inequality in the display is then an equality, and since \(u_i\geq l_i\geq2^i\) for each \(i\geq1\), it follows that \(l_i=u_i=2^i\) for all \(i\geq1\).

It remains to identify the weight of any chosen level-zero copy.
If that copy has the minimum weight \(l_0\), its weight is already one.
Otherwise let its weight be \(v\), remove the level-one coordinate from the losing positive-level pattern, and add that copy and a distinct level-zero copy of weight \(l_0\).
The new binary sum is still \(Q-1\), while its assigned sum is \(Q-1-2+1+v\).
Hence \(v\leq1\), and positivity gives \(v=1\).
This proves the three-copy case.
With more copies, restrict to any three per level, including a chosen copy whose weight is to be identified.
\end{proof}

The next statement permits finite sharing values, selector sharing, direct selectors, a positive wire bias, and rearrangements of all baseline occurrences.
Here the sharing value is the weight \(a\) of the sharing edges, and sharing by a value means that \(a\) takes it, so selector sharing, or sharing by \(s_0\), means \(a=s_0\).
Only the common-ramp form and the actual occurrence budget are assumed.

\begin{lemma}[Common-ramp witness extraction]
\label{lem:boolean-extraction}

There exists a legal common-ramp network
\[
N(x)=D\left(\rho(S(x)-(P-w))-\rho(S(x)-P)\right),
\qquad S(x)=a u(x)+\sum_jb_jx_j,
\]
that equals \(f\) on all Boolean inputs if and only if the fixed source instance \(\beta_1,\ldots,\beta_n,T\) has an exact-sum blocker of size at most \(p\).
Here the network follows the positive primitive and paired-occurrence conventions of \Cref{lem:boolean-placement}, \(u\) is the optional sharing wire, and \(a\) and the \(b_j\) are the hinges' sharing and direct input weights, each zero or a single positive inventory primitive.
The statement includes the case in which no source subset sums to \(T\), and allows an omitted sharing wire.
\end{lemma}

\begin{proof}

We first prove extraction, allowing all legal primitive roles.
The ramp is zero exactly for \(S\leq P-w\) and one exactly for \(S\geq P\).
The target is zero at the origin and at every unit vector by \Cref{lem:boolean-full-target}, so at these points the score, and hence each of its nonnegative contributions, is at most \(P-w\).
A direct \(\Gamma\) primitive exceeds this bound at the unit vector of its coordinate.
By the form of \(S\) and of the wire \(u\), the only products in \(S\) are the terms \(av_{\mathrm b}\) and \(av_j\).
A product containing two finite positive primitives is at least \(F^2>P\), because every finite positive primitive is at least \(F\) and \(F>Q\); a finite-\(\Gamma\) product and \(\Gamma^2\) are also too large, being at least \(\Gamma=M^9>P\).
Such products cannot occur at an origin or unit score.
Call a nonfinite term of the score \emph{tiny} when it is at most \(Ps_0\).
The only remaining nonfinite products that are not tiny are \(\Gamma s_0=1/q\).
Every other remaining nonfinite contribution is tiny: this includes direct selectors, finite-selector products and \(s_0^2\).

Consequently the score has the form
\[
S(x)=I_{\mathrm b}/q+\tau_0+
\sum_j\left(\mu_j+I_j/q+\tau_j\right)x_j,
\]
where \(\mu_j\) is an actual finite direct primitive or zero, \(I_{\mathrm b},I_j\in\{0,1\}\), and \(\tau_0,\tau_j\geq0\).
In this proof, an \emph{increment} is a product \(\Gamma s_0\).
The total number of increments satisfies \(I_{\mathrm b}+\sum_jI_j\leq p\).
Recall that the products in \(S\) are the terms \(av_{\mathrm b}\) and \(av_j\).
We therefore consider the sharing values \(a=\Gamma\) and \(a=s_0\) in turn.
If \(a=\Gamma\), the two sharing edges consume both occurrences of \(\Gamma\), and the wire uses at most \(p\) selector inputs or a selector bias.
If \(a=s_0\), the two sharing edges consume two identical occurrences of \(s_0\), by the paired-occurrence convention of \Cref{lem:boolean-placement}.
The wire can then use at most two \(\Gamma\) inputs or biases, and \(2\leq p\).
No other primitive sharing produces an increment \(1/q\).

There are at most \(d\) tiny direct terms, \(d\) tiny shared coordinate terms and one tiny shared bias.
Since
\[
2d+1=2n+2p+2R+9<m=2n+3p+2R+14,
\]
we have
\[
\tau_0+\sum_j\tau_j\leq mPs_0< w/4.
\]
The last inequality is equivalent to \(M^8>4mP/q\), which follows from \(M^8\geq M>2H_0>4mP\geq4mP/q\).
Tiny bias terms remain allowed throughout the argument.
Write \(s_j=\mu_j+I_j/q+\tau_j\) for the effective score coefficient of coordinate \(j\).

\emph{Excluding anchor and control baselines from queries.}

Fix any anchor and any level-\(i\) query coordinate.
Choose a query background of binary weight \(T+1\), avoiding the compared copy; spare copies allow this.
Adding the anchor wins, whereas adding the compared query loses, because \(T+1+2^i<Q\).
In the candidate, the score at the winning point is at least \(P\) and at the losing point at most \(P-w\).
Subtracting their common background and score bias gives
\[
s_{\mathrm{anchor}}-s_{\mathrm{query}}\geq w.
\]
Each nonzero \(\mu_j\) is a finite primitive \(F\nu+\delta w\).
We call \(\mu_j\) the \emph{finite part} of coordinate \(j\), the integer \(\nu\) its \emph{leading integer}, \(F\nu\) its \emph{leading part}, and \(\delta w\) its \emph{finite correction}.
Suppose some query has finite part \(B\) or \(C\).
Every finite family below \(B\), namely the source baselines \(F\beta_j\) and the query baselines \(F2^i+2w\), has leading integer smaller than \(Q-T\), the leading integer of \(B\), by more than one.
Indeed, \(\beta_j\leq T-1\) and \(2^i\leq(Q+1)/2\), while \(Q-T\geq T+2\) and \(Q-T\geq(Q+1)/2+2\) because \(Q>8(T+2)^2\).
The query's effective score coefficient is at least its finite part, hence at least \(B=F(Q-T)-w\), since \(C>B\) and its other contributions are nonnegative.
An anchor whose finite part lies in such a family, or is zero, has effective score coefficient at most
\[
F(Q-T-2)+2w+\frac1q+\frac w4<F(Q-T)-w,
\]
because its finite part is at most \(F(Q-T-2)+2w\), its increment at most \(1/q\), its tiny terms less than \(w/4\), and \(2F>1/q+13w/4\).
Such an anchor would have effective score coefficient below that of the query, contrary to the displayed gap, so every anchor must itself have finite part \(B\) or \(C\).
An anchor with finite part \(B\) but no increment cannot satisfy the displayed gap against a query with finite part at least \(B\): the anchor's effective score coefficient exceeds the query's by at most \(\tau_j\), the sum of its tiny terms, which is smaller than \(w/4\).
Thus at most \(p\) anchors can have finite part \(B\).
At most two anchors can use \(C\), since only two actual occurrences exist; the paired direct convention only strengthens that bound.
This accounts for at most \(p+2<n_{\mathrm a}=p+3\) anchors, a contradiction.
Therefore no query has finite part \(B\) or \(C\).

\emph{Identifying all query primitives.}

All remaining query finite corrections are nonnegative, being zero for source primitives and \(2w\) for query primitives.
On the query-only cube, the \emph{nonleading score} is the score minus the sum of the leading parts of the selected coordinates.
It consists of the finite corrections, the increments and the tiny terms, including those in the score bias, and it lies in
\[
0\leq\text{nonleading score}\leq p/q+2Rw+w/4<1.
\]
The sum of the leading parts is \(F\) times an integer weighted sum.
Thus a leading integer sum at least \(Q\) gives score at least \(P\), while a sum at most \(Q-1\) gives score below \(P-w\).
The assigned integer query coefficients therefore represent precisely the original threshold at quota \(Q\).
None is zero, by query essentiality.
\Cref{lem:boolean-binary-reference}, applied to any three copies per level, identifies each integer coefficient as \(2^i\).
A source primitive cannot have that value by the \emph{nonpower promise} of \Cref{lem:restricted-blockers}, that no \(\beta_j\) is a power of two, so each full finite query part is exactly \(F2^i+2w\).

\emph{Identifying sources, anchors and control.}

For source coordinate \(j\), use query backgrounds of weights \(Q-\beta_j\) and \(Q-\beta_j-1\).
The first wins when that source is selected, and the second loses.
All query-background perturbations, including score bias, are nonnegative and total less than one.
Hence
\[
F\beta_j-1<s_j<F(\beta_j+1).
\]
A smaller finite leading integer cannot bridge the gap of at least \(F\).
A larger source or query leading integer has nonnegative correction and violates the upper bound.
The negative-correction families \(B,C\) have leading integers larger than \(T\), by a gap exceeding one, and cannot lie in this interval either.
Thus \(\mu_j=F\beta_j\) in full; the nonpower promise excludes a query primitive at the same leading value.

For an anchor, backgrounds \(T,T-1\) give the interval
\[
F(Q-T)-1<s_{\mathrm{anchor}}<F(Q-T+1).
\]
Only \(B\) lies in it: every source or query leading value is much smaller, and \(C\)'s is larger by at least \(T\geq3\).
For the control coordinate, its singleton loss and its win with one level-zero query give
\[
F(Q-1)-1<s_c<P.
\]
Only \(C\) lies in this interval.
These comparisons use actual finite baselines and do not assign physical selector roles in advance.

The control singleton now gives
\[
(I_{\mathrm b}+I_c)/q+\tau_0+\tau_c\leq w/2.
\]
Since \(1/q>w/2\), it forces \(I_{\mathrm b}=I_c=0\).
A tiny score bias remains within the already proved bound.

Choose an anchor with \(I_h=0\), possible because there are \(p+3\) anchors and at most \(p\) increments in all.
Let \(H=\{j\leq n:I_j=1\}\) be the incremented source coordinates.
Its size is at most \(p\).
At that anchor together with any source solution \(z\) satisfying \(\sum_j\beta_jz_j=T\), the sum \(\Lambda\) of the selected baselines is \(P-w\), so the target is one.
If that solution avoided \(H\), the complete candidate score would be less than
\[
P-w+w/4<P,
\]
contradicting its required ramp value one.
Thus \(H\) hits every source solution and is an exact-sum blocker.

\emph{The forward network.}

Conversely, fix an exact-sum blocker \(H\) of size at most \(p\).
The output uses the standard difference of two shifted ReLUs \cite{Williams2018}.
For each input coordinate, assign one of its two baseline occurrences to each hinge.
Use the prescribed negative biases and opposite output gains.
If \(H\ne\varnothing\), add
\[
u=\rho\left(s_0\sum_{j\in H}z_j\right)
\]
and use the two \(\Gamma\) sharing occurrences.
On the Boolean cube the common score of this network is \(S=\Lambda+\sum_{j\in H}z_j/q\).
This uses at most four computational vertices and at most \(p\) selectors, with no reused occurrence or parallel edge.
If \(H\) is empty, omit \(u\) and both sharing edges, retaining a legal three-vertex network.

To verify all inputs, we use letters local to this verification, unrelated to the symbols of the extraction above: let \(k\) count selected anchors, \(t\) count selected queries, \(v\) be their binary weighted sum, \(b=\sum_j\beta_jz_j\), and \(h=\sum_{j\in H}z_j\).
Then
\[
\Lambda=FI+w\delta,
\quad I=b+k(Q-T)+v+Qc,
\quad\delta=-k+2t-3c/2,
\quad S=FI+w\delta+h/q.
\]
The perturbations of \(\Lambda\) and \(S\), namely \(w\delta\) and \(w\delta+h/q\), have absolute value less than one, using \(p/q+2Rw<1\) and \(2R>n_{\mathrm a}+3/2\).
If \(I<Q\), both target and ramp are zero; if \(I>Q\), both are one, since the \(F\)-gap exceeds the perturbation.
At \(I=Q\), a selected control forces every other coordinate zero, giving the losing control singleton.
With \(c=0\), two or more anchors are impossible because \(Q>2T\).
With no anchor, some query must be selected since \(\sum_j\beta_j<Q\), and \(\delta=2t\geq2\) gives baseline score above \(P\), hence also \(S>P\).
With one anchor and at least one query, \(\delta=-1+2t\geq1\) again gives baseline score above \(P\), hence also \(S>P\).
With one anchor and no query, \(b=T\) and \(\Lambda=P-w\).
The blocker condition gives \(h\geq1\), so \(S\geq P-w+1/q>P\).
These cases exhaust every Boolean input, including the case in which no source subset sums to \(T\).
\end{proof}

\subsection{Completeness for Boolean Targets}

The calibrated construction separates numerical selection from verification of every Boolean input.
Its graph restrictions are necessary: a network may otherwise assign any occurrence to a bias, an input edge, an internal edge or an output edge, and may leave occurrences unused.
The placement and extraction lemmas address all those choices.

\begin{theorem}[restatement of \Cref{ithm:boolean-complete}]
\label{thm:boolean-complete}

Under either the at-most-once or the exactly-once rule, with an affine output, exact synthesis for a Boolean target supplied by a circuit is \(\Sigma_2^{\mathsf P}\)-complete under deterministic polynomial-time many-one reductions.
Hardness holds with a computational-node budget of four, including the output, even when the circuit computes a monotone threshold with positive rational coefficients.
\end{theorem}

\begin{proof}

Membership is the existential candidate followed by universal Boolean-input verification in \Cref{thm:circuit}.
For hardness, compose the quantified-CNF reduction in \Cref{lem:restricted-blockers} with the calibrated network construction above.
The blocker instances that reduction produces satisfy \(p<n\) by \Cref{lem:restricted-blockers}, as required for the polynomial numerical-size bounds.
Form the calibrated inventory and its polynomial-size threshold circuit.
Every explicit number has polynomial binary length, every coordinate and occurrence is listed in polynomial time, and the budget is the fixed value four.
If the source has a blocker, the forward construction in \Cref{lem:boolean-extraction} gives a legal permitted network for the target.
Conversely, any permitted network for that target has a legal common-ramp representation by \Cref{lem:boolean-placement}.
\Cref{lem:boolean-extraction} then returns an actual source-coordinate blocker of size at most \(p\).
The source and synthesis answers are therefore equivalent under this deterministic polynomial-time transformation.
All target weights are positive, so the target function is monotone.
This proves the restricted hardness and hence the full completeness claim.

For exactly-once use, enlarge a source blocker to cardinality exactly \(p\).
This is possible because \(p<n\), and enlarging a blocker preserves the property of meeting every source solution.
The source construction has \(p\geq2\), so the enlarged blocker is nonempty and the selector neuron keeps an incoming edge, as \Cref{def:network} requires.
Use all \(p\) selector occurrences on the enlarged blocker in the same forward network.
Both baseline copies for every coordinate, both sharing occurrences, the two hinge biases and the two output gains are then used, so every inventory occurrence is assigned exactly once.
The forward proof needs only the blocker property and the bound \(|H|/q\leq p/q\), which remain true after enlargement.
Conversely, every exactly-once witness is an at-most-once witness, so the same placement and extraction lemmas apply.
Membership still follows by guessing a legal occurrence-complete graph and verifying its Boolean outputs universally.
\end{proof}

In the forward network of \Cref{lem:boolean-extraction}, the budget of four counts three hidden ReLUs, namely the two hinges and the sharing wire \(u\), and one affine output.
The theorem does not assume those vertices or any of their parameter roles are prescribed to the synthesis algorithm.
The inventory in \Cref{thm:boolean-complete} contains three negative occurrences, \(-D,-(P-w),-P\).

\section{A ReLU Network as the Target}
\label{sec:relu-target}

Here the target is an explicitly given rational ReLU network \(F:\mathbb R^d\to\mathbb R\), with an affine output.
Its graph and all parameter positions are listed explicitly; its parameters need not belong to \(W\).
The synthesis input is \((F,W,n_{\max})\), and the requirement is
\[
N(x)=F(x)\qquad\text{for every }x\in\mathbb R^d.
\]
This formulation asks whether the supplied computation can be represented using the available inventory and node budget.
Unlike a nonlinear polynomial target, a ReLU target has the same piecewise-affine form as the candidate networks.
This shared form does not ensure that the target is representable with a particular inventory.

\begin{lemma}[A network encoding of unsatisfiability]
\label{lem:cnf-relu}
From a CNF formula \(\Phi\) with \(d\geq1\) variables and at least one clause, every clause nonempty, one can construct in polynomial time a rational ReLU network \(F_\Phi\) with the following two properties.
If \(\Phi\) is unsatisfiable, then \(F_\Phi=0\) on all of \(\mathbb R^d\).
If \(\Phi\) is satisfiable, then some Boolean input \(a\) has \(F_\Phi(a)=1\).
\end{lemma}
\begin{proof}
For a positive literal on variable \(i\), use the affine function \(2x_i-1\); for a negative literal, use \(1-2x_i\).
Denote the function corresponding to literal \(l\) by \(s_l(x)\), and put
\[
F_\Phi(x)=\max\left\{0, \min_{C\text{ a clause of }\Phi} \max_{l\in C}s_l(x)\right\}.
\]
At a satisfying Boolean input every clause has a literal of value one, and every literal value is either one or minus one.
Every clause maximum is therefore one, and the displayed expression is one.

Suppose instead that the expression is positive at some real input \(x\).
Each clause then has a literal whose affine value is positive.
Assign variable \(i\) to true when \(x_i>1/2\), and to false otherwise.
A positive literal of positive affine value has \(x_i>1/2\) and is true.
A negative literal of positive affine value has \(x_i<1/2\) and is true.
Thus this assignment satisfies every clause.
Consequently an unsatisfiable formula gives the zero function everywhere.

It remains to realize the expression by an explicitly listed ReLU network of polynomial size.
Delete repeated literals within each clause and repeated clauses; the displayed expression takes a maximum over the set of literals of a clause and a minimum over the set of clauses, so it does not change.
Write the clauses as \(C_1,\ldots,C_k\), with \(k\geq1\), and the literals of \(C_j\) as \(l_{j,1},\ldots,l_{j,w_j}\), with \(w_j\geq1\).
For all real \(a,b\),
\[
\max\{a,b\}=\rho(a-b)+b,\qquad
\min\{a,b\}=a-\rho(a-b),
\]
because \(\rho(a-b)=a-b\) when \(a\geq b\) and \(\rho(a-b)=0\) otherwise.

For every clause \(C_j\) and every \(t\) with \(2\leq t\leq w_j\), introduce a hidden vertex \(g_{j,t}\).
Put \(G_{j,1}=s_{l_{j,1}}(x)\) and, for \(t\geq2\), \(G_{j,t}=g_{j,t}+s_{l_{j,t}}(x)\), and let the preactivation of \(g_{j,t}\) be \(G_{j,t-1}-s_{l_{j,t}}(x)\).
Then \(G_{j,t}=\rho(G_{j,t-1}-s_{l_{j,t}}(x))+s_{l_{j,t}}(x)=\max\{G_{j,t-1},s_{l_{j,t}}(x)\}\), and induction on \(t\) shows that \(c_j=G_{j,w_j}\) equals \(\max_{l\in C_j}s_l(x)\).
Put \(R_1=c_1\), and for \(2\leq j\leq k\) introduce a hidden vertex \(h_j\) with preactivation \(R_{j-1}-c_j\), and put \(R_j=R_{j-1}-h_j\).
Then \(R_j=R_{j-1}-\rho(R_{j-1}-c_j)=\min\{R_{j-1},c_j\}\), and induction on \(j\) gives \(R_k=\min_jc_j\).
Finally introduce a hidden vertex \(o\) with preactivation \(R_k\), and let the affine output be the value of \(o\), through one edge of weight one and with no bias.
The output equals \(\rho(R_k)=F_\Phi(x)\) at every real input \(x\), since every identity used holds for all real numbers.

Each preactivation above is an affine combination of the input coordinates and of the values of earlier hidden vertices, because \(G_{j,t}\) involves only \(g_{j,t}\) and the input, and \(R_j=c_1-\sum_{i=2}^{j}h_i\).
For each preactivation, collect the coefficient of every input coordinate and of every hidden vertex, and the constant term.
A nonzero collected coefficient becomes the weight of one edge, a nonzero constant term becomes a present bias, and zero coefficients give absent edges and biases, so there is at most one edge per ordered pair.
Every preactivation involves at most two literal functions and hidden coefficients \(\pm1\), so all weights and biases are integers of absolute value at most four.
In the order \(g_{1,2},\ldots,g_{k,w_k},h_2,\ldots,h_k,o\), followed by the output, every edge points forward, so the graph is acyclic.

Every hidden vertex reaches the output: \(g_{j,t}\) with \(t<w_j\) enters \(g_{j,t+1}\) with weight one, \(g_{1,w_1}\) enters \(o\) with weight one, \(g_{j,w_j}\) with \(j\geq2\) enters \(h_j\) with weight minus one, each \(h_j\) enters \(o\) with weight minus one, and \(o\) enters the output.
Each hidden vertex also has an incoming edge.
For \(g_{j,t}\) with \(t\geq3\) it is the edge from \(g_{j,t-1}\), and for \(h_j\) with \(j\geq3\) it is the edge from \(h_{j-1}\).
The preactivation of \(g_{j,2}\) is \(s_{l_{j,1}}-s_{l_{j,2}}\) for two distinct literals, whose input coefficients are \(\pm2\) on two variables, or \(\pm4\) on one variable when the literals are opposite, so it has an input edge.
The preactivation of \(h_2\) is \(c_1-c_2\); it has an edge from \(g_{1,w_1}\) or from \(g_{2,w_2}\) when \(w_1\geq2\) or \(w_2\geq2\), and otherwise it is \(s_l-s_{l'}\) for the single literals \(l\ne l'\) of two distinct clauses, which has an input edge as before.
The preactivation of \(o\) has the edge from \(h_k\) when \(k\geq2\); when \(k=1\) it is \(c_1\), which has the edge from \(g_{1,w_1}\) when \(w_1\geq2\) and is otherwise a literal function with input coefficient \(\pm2\).
The network has \(\sum_j(w_j-1)+(k-1)+1=\sum_jw_j\) hidden vertices and one output, and it is computed in polynomial time.
\end{proof}

This construction reduces unsatisfiability to exact reproduction with the zero network as the only permitted candidate function.
It gives the lower bound in the next theorem.

\begin{theorem}[ReLU-network targets]
\label{thm:relu-target}
Exact synthesis on all real inputs for a target supplied as a rational ReLU network is \(\mathsf{coNP}\)-hard under deterministic polynomial-time many-one reductions and belongs to \(\Sigma_2^{\mathsf P}\).
Both bounds hold for at-most-once and exactly-once use.
The lower bound holds with \(W=(0)\) and \(n_{\max}=1\).
Equality of a supplied candidate and a supplied target belongs to \(\mathsf{coNP}\).
\end{theorem}
\begin{proof}
For candidate verification, guess the activation patterns of both networks and a sign for their nonzero output difference.
Introduce variables for the input and all computational values in both networks.
For an active hidden vertex, set its value equal to its affine preactivation and require that preactivation to be nonnegative.
For an inactive hidden vertex, set its value to zero and require its preactivation to be nonpositive.
The output equations are affine.
These constraints form a rational linear system of polynomial encoding length.
Add the strict inequality \(N(x)-F(x)>0\), or its negative according to the guessed sign.
If the networks disagree, their actual states at a disagreeing point satisfy one of these systems.
Conversely a feasible system gives the actual network values and hence a disagreement.

Such a system has a polynomial-bit rational solution whenever it is feasible.
Indeed, introduce a slack \(s\), constrain \(0\le s\le1\), and replace the strict signed gap by a gap at least \(s\).
The original strict system is feasible exactly when this linear program has a positive maximum of \(s\).

The objective of this linear program is bounded, so a feasible program attains its optimum.
Split unrestricted variables into differences of nonnegative variables.
A slack variable for an inequality is an additional nonnegative variable equal to its right side minus its left side, turning that inequality into an equality.
An inequality is tight at a point when equality holds.
In this standard nonnegative-variable form, a basic feasible solution is a feasible point uniquely determined by a linearly independent selection of tight constraint equations.

To see that an optimal basic feasible solution exists here, write the standard form as \(Az=b,z\geq0\), with objective \(c\cdot z\).
An optimum exists by \Cref{thm:rational-lp-import}.
Choose an optimum \(z\) with the smallest number of positive coordinates, let \(S\) be the set of their indices, and write \(A_S\) for the corresponding column submatrix of \(A\).
If the columns \(A_S\) were dependent, a nonzero vector \(d\) supported on \(S\) would satisfy \(Ad=0\).
Both \(z+td\) and \(z-td\) are feasible for all sufficiently small \(t>0\), so optimality forces \(c\cdot d=0\).
Reverse \(d\) if necessary so it has a positive coordinate, and take \(t=\min_{i:d_i>0}z_i/d_i\).
Then \(z-td\) is another optimum, is nonnegative, and has strictly fewer positive coordinates, a contradiction.
Thus the columns \(A_S\) are independent.

If \(S\) is empty, the optimum is the zero vector.
Otherwise select \(|S|\) independent rows of \(A_S\); the resulting square system, together with the zero coordinates outside \(S\), uniquely determines the optimum.
These are independent tight equations, so the point is a basic feasible solution.

Cramer's rule expresses each nonzero coordinate as a ratio of determinants of this system.
Clear denominators in each row, including its right side.
If its integer entries have absolute value at most \(2^{B_{\mathrm{bit}}}\) and its square size is \(r\), the determinant expansion has \(r!\) terms of absolute value at most \(2^{rB_{\mathrm{bit}}}\).
Every numerator and nonzero denominator therefore has binary length at most \(rB_{\mathrm{bit}}+\lceil\log_2(r!)\rceil+2\), a polynomial bound.

The optimal value of \(s\) is positive when the strict system is feasible, so the optimal point satisfies the strict signed gap.
Discarding \(s\) and the standard-form slack variables, and replacing each split pair of nonnegative variables by its difference, yields a rational point of the original strict system, the required certificate.
By Cramer's rule all nonzero coordinates of the basic feasible solution share the denominator given by the determinant of the square system, so each such difference has that denominator and a numerator of binary length at most one more than the bound above.
The guessed states give actual ReLU values, including either state at a zero preactivation.

Checking the patterns, node values, and signed gap takes polynomial time.
Non-equivalence is therefore in \(\mathsf{NP}\), proving the verification claim.
Guessing a polynomial-size permitted network and applying this \(\mathsf{coNP}\) test gives synthesis in \(\Sigma_2^{\mathsf P}\).

For hardness start with CNF unsatisfiability.
Its \(\mathsf{coNP}\)-completeness follows from circuit satisfiability by the following translation.
Introduce one Boolean variable per gate, exclude every invalid assignment to a gate's input and output variables by a clause, and require the output variable to be true.
Each gate has a constant-size truth table, so the resulting CNF has polynomial size and is satisfiable precisely when the circuit is.
Empty clauses or a formula with no clauses can be recognized directly and replaced by fixed unsatisfiable or satisfiable formulas with nonempty clauses.
Apply \Cref{lem:cnf-relu} and use target \(F_\Phi\), inventory \(W=(0)\), and budget one.
Every permitted network computes zero, and an output with a zero bias is legal and uses the occurrence exactly once.
The lemma shows that synthesis succeeds precisely when \(\Phi\) is unsatisfiable.
This proves the lower bound for both inventory variants.
\end{proof}


\subsection{At Most Two Computational Vertices}

A small computational budget permits reconstruction from finitely many target values even though equality must hold on all real inputs.
The output may have direct input edges, and these edges are included in the reconstruction.

\begin{theorem}[Two-vertex exact synthesis; restatement of \Cref{ithm:fixed-cap}]
\label{thm:two-vertex}
Full-space exact synthesis of a supplied rational ReLU target, with at-most-once inventory, an affine output, and at most two computational vertices, is \(\mathsf{coNP}\)-complete under deterministic polynomial-time many-one reductions.
The input dimension is unrestricted.
\end{theorem}
\begin{proof}
Write \(T\) for the supplied target throughout this proof.
Write \(m=|W|\) for the number of inventory occurrences, and choose an integer \(R\geq2\) bounding the absolute numerators and positive denominators of the reduced inventory values.
The binary length of \(R\) is polynomial in the input length.
Put \(B=4R^2+4\), and let \(v\in\mathbb Q^d\) have coordinates \(v_i=B^i\), for \(1\leq i\leq d\).
These coordinates also have polynomial binary length.
We construct a polynomial list of legal candidates containing an exact witness whenever one exists.

A nonaffine network with at most two computational vertices has exactly one hidden vertex and computes
\[
p\cdot x+q+c\rho(a\cdot x+b),\qquad a\ne0,\quad c\ne0.
\]
Here \(p\) and \(q\) are the direct output coefficients and the output bias, \(a\) and \(b\) are the input weights and bias of the hidden vertex, and \(c\) is its output gain.
Both \(a\ne0\) and \(c\ne0\) hold, because \(a=0\) makes the hidden value the constant \(\rho(b)\) and \(c=0\) removes the hidden term, and either would make the computed function affine.
Every nonzero displayed coefficient is one inventory occurrence; zero positions may be omitted.
If \(k\) is the largest index with \(a_k\ne0\), then \(|a_k|\geq1/R\) and
\[
|a\cdot v|
\geq \frac{B^k}{R}-R\sum_{i<k}B^i
>\frac{3B^k}{4R}>2R.
\]
Indeed the geometric sum is less than \(B^k/(B-1)\), and \(R/(B-1)<1/(4R)\).
Since \(|b|,|a_i|\leq R\), the hidden preactivation has opposite signs at \(v\) and \(-v\), and adding any coordinate unit vector \(e_i\) preserves each sign.
Thus the target values at these points recover both affine pieces of every true nonaffine witness:
\[
p_{s,i}=T(sv+e_i)-T(sv),\qquad
q_s=T(sv)-p_s\cdot(sv),\qquad s\in\{+1,-1\}.
\]
For each choice of the sign \(s\) at which the hidden vertex is inactive and of a nonzero output-gain occurrence \(c\in W\), form
\[
a_i=(p_{-s,i}-p_{s,i})/c,\qquad
b=(q_{-s}-q_s)/c.
\]
Here \(p_s\) is the vector with coordinates \(p_{s,i}\).
The inactive sign is not known in advance, so both values of \(s\) are tried; only at the inactive sign are \(p_s\) and \(q_s\) the direct output coefficients \(p\) and the output bias \(q\).
The candidate graph for the sign \(s\) uses \(p_s\) as its direct output coefficients and \(q_s\) as its output bias.
Omit zero positions, require \(a\ne0\), and retain this graph only if all nonzero positions can be assigned distinct occurrences of their numerical values.
This is a polynomial-time multiset check.
Every retained graph is legal: the hidden vertex has a nonzero input edge and reaches the output through the nonzero gain \(c\).
There are at most \(2m\) such candidates.

It remains to include every possible affine witness.
Evaluate the target at zero and at every coordinate unit vector.
Let \(L_0(x)=\alpha\cdot x+\beta\) be their affine interpolant, where \(\beta=T(0)\) and \(\alpha_i=T(e_i)-T(0)\).
Its realizability with at most two vertices is decidable in polynomial time.
First test a graph with no hidden vertex by matching its nonzero coefficients to the inventory.
If all coefficients vanish, a present zero occurrence is necessary and sufficient, since by \Cref{def:network} the output must have an incoming edge or a present bias to be legal.
Next consider a constant hidden vertex, meaning a hidden vertex whose incoming edges, if any, all carry zero, so that its value is \(\rho(b)\) for its bias \(b\).
When that bias is present, deleting those zero edges changes neither the function nor legality, so we may assume the vertex has no incoming edge.
Enumerate distinct nonzero occurrences \(b,c\) for its bias and output gain, and an optional nonzero output-bias occurrence \(q\), allowing the bias to be absent with value zero.
Test
\[
\beta=q+c\rho(b)
\]
and simultaneously match the nonzero direct slopes \(\alpha_i\), with all numerical multiplicities respected.
These tests cover every witness whose constant hidden vertex has a nonzero bias and a nonzero output gain; the remaining constant cases reduce to the no-hidden test, as shown below.

A hidden vertex with a nonconstant preactivation and a nonzero output gain cannot produce an affine function, even after the affine skip terms are added.
Indeed, write that function as \(p\cdot x+q+c\rho(a\cdot x+b)\), with \(a\ne0,c\ne0\).
On \(a\cdot x+b<0\) its coefficient vector is \(p\), and on \(a\cdot x+b>0\) it is \(p+ca\).
Both open halfspaces are nonempty: the points \(x=(-b-1)a/(a\cdot a)\) and \(x=(-b+1)a/(a\cdot a)\) lie in them, respectively.
Each point admits sufficiently small perturbations in every coordinate while staying in its halfspace, so near each point these perturbations read off that halfspace's coefficient vector.
A globally affine function has a single coefficient vector everywhere, which would then equal both \(p\) and \(p+ca\).
Their difference is \(ca\ne0\), a contradiction.

A zero output gain contributes nothing, so its hidden vertex can be pruned.
If the output is then left with no incoming edge and no present bias, the zero occurrence of that gain can serve as a present zero bias, which makes the output legal.
A hidden vertex whose incoming edge weights and bias are all zero likewise computes zero and can be pruned, leaving a no-hidden witness that may use one of the vertex's zero occurrences as its output bias.
A constant hidden vertex with zero bias likewise computes zero and reduces to the no-hidden test in the same way.
These cases exhaust all affine witnesses.
If any test succeeds, append one legal realization of \(L_0\).

The resulting list has at most \(2m+1\) graphs.
If the supplied cap is one, use only the no-hidden affine feasibility test and discard every reconstructed graph with two vertices.

All target evaluations and candidate coefficients have polynomial bit length.
The bounds in the proof of \Cref{thm:population-risk} are stated for Boolean inputs, so we repeat the argument for the integer probe points used here.
Let \(L\geq2\) be the input length, let \(D\) be the product of the positive denominators of all parameter positions of \(T\), and let \(x\) be one of the probe points \(0\), \(e_i\), \(sv\) or \(sv+e_i\).
Every coordinate of \(x\) is an integer of absolute value at most \(X=B^d+1\).
Freezing the activation pattern of \(T\) at \(x\) expresses \(T(x)\) as a sum of bias-to-output path products and of input-to-output path products, each of the latter multiplied by one coordinate of \(x\).
A path uses each parameter position at most once and the coordinates of \(x\) are integers, so \(D\) times each such term is an integer, and \(DT(x)\) is an integer.

Every parameter of \(T\) has absolute value at most \(2^L\), every preactivation has at most \(L\) terms, \(T\) has at most \(L\) computational vertices, and \(D<2^L\).
Every input value is at most \(X\) in absolute value, and every bias is at most \(2^L\leq2^LX\).
Induction in topological order therefore bounds the absolute value of the \(j\)th computational value by \((L2^L)^jX\), since a ReLU does not increase absolute value.
Hence \(|DT(x)|\leq2^L(L2^L)^LX\), whose binary length is at most \(L+L\log_2L+L^2+\log_2X+1\).
This is polynomial in the input length, because \(\log_2X\leq1+d\log_2B\) and the coordinates of \(v\) have polynomial binary length.
The quantities \(p_{s,i}\), \(q_s\), \(a_i\), \(b\), \(\alpha_i\) and \(\beta\) arise from polynomially many such values by polynomially many subtractions, multiplications by the integers \(\pm B^i\), sums of \(d\) terms and divisions by one inventory value, so they also have polynomial bit length.

If a true witness is nonaffine, its two pieces and gain put it on the list.
If it is affine, it equals \(L_0\), whose feasibility test succeeds.
Conversely each retained graph is a permitted candidate, so synthesis succeeds precisely when one listed graph equals \(T\) globally.
By \Cref{thm:relu-target}, each equality test is in \(\mathsf{coNP}\).
A polynomial disjunction of these tests remains in \(\mathsf{coNP}\): a certificate for failure gives one polynomial-length inequivalence certificate for every listed graph.
An empty list is rejected deterministically.

For hardness, use the \(W=(0)\) construction of \Cref{thm:relu-target} with cap two.
The single occurrence still permits only one computational vertex, so the only candidate function is zero.
The same reduction from CNF unsatisfiability applies.
\end{proof}

The reconstruction relies on having only one hidden vertex.
For example, with cap three the legal network
\[
h_1=\rho(x-\tfrac12),\qquad
h_2=\rho(x-\tfrac14-2h_1),\qquad N=h_2
\]
uses \(W=(1,1,1,-2,-\tfrac12,-\tfrac14)\), with the third occurrence of \(1\) as the unit output edge from \(h_2\), and equals \(\rho(\tfrac14-|x-\tfrac12|)\).
Here \(R=4\) and \(v=68\), and the function vanishes at all six probe points \(0,1,68,69,-68,-67\).
Thus those probes alone do not reconstruct a general cap-three witness.
This example limits the stated reconstruction procedure and does not classify larger budgets.

\subsection{A Fixed Computational-Node Cap}
\label{sec:fixed-cap-synthesis}

A fixed number of computational vertices still permits arbitrarily many input coordinates and inventory occurrences.
Nevertheless, a polynomially constructible set of inputs distinguishes all different functions realizable with that cap.
The set depends on the inventory and the cap, rather than on the target.
Recall that \(\mathsf{DP}\) consists of languages \(A\cap\overline C\) with \(A,C\in\mathsf{NP}\).

\begin{theorem}[Fixed-cap exact synthesis; restatement of \Cref{ithm:fixed-cap}]
\label{thm:fixed-cap-dp}
For each fixed integer \(H\ge1\), full-space exact synthesis of a supplied rational ReLU target with computational-node cap \(H\) belongs to \(\mathsf{DP}\), under either inventory rule.
The dimension, inventory size and target size are unrestricted.
\end{theorem}

We first locate disagreements in low-dimensional coordinate subspaces.
Recall that the support of a vector is the set of coordinates where it is nonzero.
\begin{lemma}[Sparse disagreements]
\label{lem:sparse-disagreement}
If two ReLU networks with affine outputs and respectively \(h_1,h_2\) hidden vertices compute different functions on \(\mathbb R^d\), they disagree at a vector whose support has size at most \(h_1+h_2+1\).
\end{lemma}
\begin{proof}
Choose an input \(x\) where the outputs differ.
For each hidden vertex \(v\) in either network, let \(a_v\in\mathbb R^d\) contain the coefficients of its direct input edges, with zero for an absent edge.
Its preactivation is therefore \(a_v\cdot x+b_v\) plus a linear combination of its hidden predecessors' values.
For network \(i\in\{1,2\}\), let \(p_i\in\mathbb R^d\) be its direct input-to-output coefficient vector.
Form a matrix \(A\) with one row \(a_v\) for every hidden vertex in the two networks and one final row \(p_1-p_2\).
It has \(h_1+h_2+1\) rows.

Choose a set \(J\) of column indices whose columns form a basis of the column space of \(A\), meaning the linear span of its columns, and write \(A_J\) for their submatrix.
Because \(Ax\) is in that column space, there is a vector \(u\) satisfying \(A_Ju=Ax\).
Set \(x'_J=u\) and every coordinate outside \(J\) to zero.
Then \(Ax'=Ax\), and \(x'\) has support of size at most
\[
|J|=\operatorname{rank}A\le h_1+h_2+1.
\]

Every hidden vertex has the same value at \(x'\) and \(x\).
Indeed, process each network in topological order.
At a hidden vertex \(v\), the direct input contribution is unchanged because \(a_v\cdot x'=a_v\cdot x\).
Its bias is unchanged, and the induction hypothesis preserves the values of all hidden predecessors.
Thus its entire preactivation, and hence its ReLU value, is unchanged.
The two output biases and all hidden-to-output contributions are consequently unchanged as well.
Their direct input difference is preserved by the last row of \(A\): \((p_1-p_2)\cdot x'=(p_1-p_2)\cdot x\).
The output difference is therefore the same nonzero number at both inputs.
If \(A\) has rank zero, take \(J=\varnothing\) and \(x'=0\); the same argument applies.
\end{proof}

The sparse disagreement bound permits a finite set of tests for the whole candidate class.
\begin{lemma}[A separating set for a fixed cap]
\label{lem:fixed-cap-probes}
For fixed \(H\), one can construct in deterministic polynomial time a polynomial-size set \(S\subseteq\mathbb Q^d\) of polynomial-bit vectors such that any two legal cap-\(H\) candidates agreeing on \(S\) agree on all of \(\mathbb R^d\).
\end{lemma}
\begin{proof}
Put \(r=2H-1\).
A cap-\(H\) candidate has at most \(H-1\) hidden vertices, since the cap counts the affine output, so the support bound \(h_1+h_2+1\) of \Cref{lem:sparse-disagreement} is at most \(r\) for two candidates.
Enumerate the coordinate subsets \(J\subseteq[d]\) of size at most \(r\), including the empty subset.
There are at most \(\sum_{j=0}^r d^j\) of them.
For each \(J\), enumerate topologically labeled local graphs with input vertices for the coordinates in \(J\) and at most \(H\) computational vertices, the last of which is the affine output.
Assign distinct occurrences of \(W\) to present edges and biases, allowing unused occurrences.
Only for this auxiliary enumeration, allow parameterless vertices and vertices without a path to the output.
A parameterless vertex has value zero.
These relaxed graphs are test functions, not enlarged legal synthesis witnesses.

There are at most \(C=H(r+H+1)\) edge and bias positions in a local graph.
Assigning either absence or one of the \(m\) occurrences to each position, then rejecting repeated occurrence IDs, gives at most a constant multiple of \((m+1)^C\) descriptions.
This is polynomial because \(H\) is fixed.
A legal global candidate restricted to the coordinate subspace supported on \(J\) occurs in this enumeration: delete its edges from coordinates outside \(J\), retaining every other position, vertex and occurrence ID.
On that subspace the deleted edges contribute zero.
A deletion may leave a vertex with no parameters, and such a vertex is one of the relaxed parameterless vertices, whose value is zero, so the restriction is still described locally.

For each pair of local descriptions, enumerate their activation masks and the two signs of an output disagreement.
There are constantly many masks.
Their masked computations and preactivation inequalities are affine; adding a strict signed output gap gives a rational linear feasibility problem.
A weak choice at a zero preactivation gives the actual ReLU value, so these systems cover all degeneracies.
Rational linear programming decides feasibility in polynomial time.
For a strict gap introduce \(0\le\delta\le1\), require the gap to be at least \(\delta\), and maximize \(\delta\).
A strict gap exists exactly when the maximum is positive.
The polynomial-bit witness bound is the same rational determinant argument used in \Cref{thm:relu-target}.
The local affine coefficients have polynomial bit length by path expansion.
Thus a feasible pair supplies a polynomial-bit rational distinguishing input.

Call two local descriptions inequivalent when they compute different functions of the coordinates in \(J\).
For every inequivalent pair, select one distinguishing input from a feasible system, and embed it in \(\mathbb R^d\) by putting zero outside \(J\).
Let \(S\) be the set of these selected distinguishing points.
Its construction and size are polynomial.
If legal global candidates differ, \Cref{lem:sparse-disagreement} gives a coordinate set of size at most \(r\) on which their restrictions differ.
The corresponding pair of local descriptions has a selected distinguishing point in \(S\).
Consequently agreement on \(S\) implies global agreement.
This reasoning also covers an empty separating set.
\end{proof}

The separating set removes ambiguity among consistent candidates, while the supplied target may still require a universal equality test.
\begin{proof}[Proof of \Cref{thm:fixed-cap-dp}]
Construct \(S\) and evaluate the supplied target \(F\) at each of its points.
All these evaluations have polynomial bit complexity by the rational path bound.
Let \(A\) be the language asserting that some legal candidate agrees with \(F\) on \(S\).
Let \(C\) assert that some such consistent legal candidate disagrees with \(F\) at some real input.
Both are in \(\mathsf{NP}\): guess the graph and occurrence assignment, check legality and the sample equalities, and for \(C\) additionally use the polynomial-bit inequivalence certificate of \Cref{thm:relu-target}.
For exactly-once use, the legality check additionally requires every occurrence to be assigned; the same separating set is valid for this subclass.

If an exact witness exists, every consistent candidate computes its function by \Cref{lem:fixed-cap-probes}, so \(A\) holds and \(C\) fails.
Conversely, if \(A\) holds and \(C\) fails, any consistent candidate is an exact witness.
The synthesis language is therefore \(A\cap\overline C\), proving the claimed upper bound.
\end{proof}

\Cref{thm:fixed-cap-dp} is a bound for each fixed cap; it gives no \(\mathsf{DP}\) upper bound for arbitrary supplied caps.
The probe enumeration grows as \((d+m)^{O(H^2)}\).
For cap two, \Cref{thm:two-vertex} gives the stronger at-most-once classification.
The \(\mathsf{coNP}\) lower bound with \(W=(0)\) remains valid for every fixed cap.

\subsection{Three Computational Vertices}
\label{sec:three-vertex-synthesis}

For cap three the preceding \(\mathsf{DP}\) upper bound is attained: \Cref{thm:three-vertex-dp} proves matching \(\mathsf{DP}\)-hardness.
The numerical part uses an explicit affine target; a separate target component enforces unsatisfiability.
We use the distinct-number form of Numerical Matching with Target Sums (NMTS).
Strong \(\mathsf{NP}\)-completeness means that \(\mathsf{NP}\)-completeness persists when the numerical integers are encoded in unary, with encoding length proportional to their values.
\begin{theorem}[Numerical matching {\cite[Appendix, Corollary 8]{HulettWillWoeginger2008}}]
\label{thm:distinct-nmts-import}
Given three lists of positive integers \((a_j)_{j=1}^d,(b_j)_{j=1}^d,(t_i)_{i=1}^d\), deciding whether there are permutations \(\pi,\sigma\) of \(\{1,\ldots,d\}\) such that \(a_{\pi(i)}+b_{\sigma(i)}=t_i\) for every \(i\) is strongly \(\mathsf{NP}\)-complete, even when all \(3d\) integers are pairwise distinct and \(\sum_j a_j+\sum_j b_j=\sum_i t_i\).
\end{theorem}

These promises will ensure distinct inventory values without changing the matching question.
An affine target is homogeneous here when its constant term is zero.
\begin{theorem}[Affine synthesis with cap three]
\label{thm:three-vertex-affine}
Full-space synthesis of an explicitly encoded rational affine target with computational-node cap three is \(\mathsf{NP}\)-complete under either inventory rule.
Hardness holds for homogeneous targets with distinct positive integer slopes and distinct nonzero signed integer inventory values.
\end{theorem}
\begin{proof}
Membership follows by guessing the legal graph and occurrence assignment.
At most two hidden vertices give four activation masks.
For each mask, \Cref{thm:rational-lp-import} tests either strict sign of a disagreement with the affine target by maximizing a slack in \([0,1]\), as in \Cref{thm:relu-target}.
This is polynomial time for the guessed candidate.

For hardness, use Numerical Matching with Target Sums on positive integers \((a_j)_{j=1}^d,(b_j)_{j=1}^d,(t_i)_{i=1}^d\), all \(3d\) numbers distinct.
It asks for permutations \(\pi,\sigma\) with \(a_{\pi(i)}+b_{\sigma(i)}=t_i\).
\Cref{thm:distinct-nmts-import} supplies strong \(\mathsf{NP}\)-hardness with these promises and matching total sums.
We may assume \(d\ge3\): the smaller cases have constantly many matchings to test and can be replaced by fixed distinct-number YES or NO instances.
For example, \(a=(1,2,3)\), \(b=(10,20,30)\) and \(t=(11,22,33)\) form a YES instance, while \(t=(11,21,34)\) gives a NO instance; both have matching total sums.
Put
\[
R=\max\{a_j,b_j,t_i\},\qquad M=10R+10,\qquad B=M+R,
\]
and supply
\[
W=(M+a_j)_j\uplus(M+b_j)_j\uplus(-M-b_j)_j\uplus(1,-1),
\qquad T(x)=\sum_{i=1}^d(2M+t_i)x_i.
\]
The unit occurrences are \(1,-1\); call all other occurrences nonunit.
Here \(m=3d+2\), all values are distinct nonzero integers, every nonunit magnitude lies in \([M+1,B]\), and
\[
2M+t_i>B,\qquad (M+1)^2>3M+2R.
\]
The transformation has polynomial binary length.

A source matching gives
\[
L(x)=\sum_i(M+b_{\sigma(i)})x_i,\quad
h_+=\rho(L),\quad h_-=\rho(-L),\quad
N=\sum_i(M+a_{\pi(i)})x_i+h_+-h_-.
\]
The identity \(\rho(L)-\rho(-L)=L\) proves \(N=T\).
The three input-edge lists and the two unit output gains use every occurrence exactly once.
There are exactly three computational vertices, no biases or parallel edges, and each hidden vertex reaches the output.

Conversely take any at-most-once cap-three witness for \(T\), including possible biases and nested vertices.
Let \(\widetilde N\) be the computation obtained by deleting its biases, before pruning zero vertices.
For all real \(s,t\), \(|\rho(s)-\rho(t)|\le|s-t|\): the same-sign cases follow directly from the definition, and in the opposite-sign case the nonzero output difference is at most the input difference.
Also \(\rho(\lambda t)=\lambda\rho(t)\) for \(\lambda>0\).

For each input vertex \(v\) set \(C_v=0\), and for each computational vertex \(v\), in topological order, put
\[
C_v=|b_v|+\sum_{u:(u,v)\in E}|a_{uv}|C_u.
\]
Write \(z_v\) and \(\widetilde z_v\) for the original and bias-free vertex values.
The preceding ReLU inequality and the triangle inequality prove, by induction in the same topological order,
\[
\left|\frac{z_v(\lambda x)}{\lambda}-\widetilde z_v(x)\right|
\le\frac{C_v}{\lambda}.
\]
At a hidden vertex this uses the already bounded predecessor differences and the divided bias \(b_v/\lambda\); the affine output satisfies the same estimate without applying ReLU.
The target is positively homogeneous, \(T(\lambda x)=\lambda T(x)\), and \(N=T\).
Thus the output estimate gives \(|\widetilde N(x)-T(x)|\le C_{\mathrm{out}}/\lambda\) for every \(\lambda>0\).
A positive difference would violate this bound for sufficiently large \(\lambda\), so \(\widetilde N=T\).

Repeatedly delete newly parameterless hidden zero vertices together with their outgoing edges, until none remains, then prune vertices not reaching the output.
Because \(T\ne0\), the output remains legal.
We obtain a bias-free at-most-once witness with no larger cap and no new occurrence.

Let \(p_i\) be its direct input-output coefficient, or zero if absent, so \(|p_i|\le B\).
The hidden computation must supply the linear residual \(r\cdot x\), where
\[
r_i=2M+t_i-p_i,\qquad 1<M+1-R\le r_i\le3M+2R.
\]
No-hidden synthesis would require the unavailable slope \(2M+t_i\).
With a single hidden vertex, its contribution has a fixed sign on all inputs and cannot equal this nonzero linear residual.
Thus there are two hidden vertices.
Order them and write
\[
h_1=\rho(a\cdot x),\quad h_2=\rho(c\cdot x+u h_1),\quad
N=p\cdot x+v h_1+w h_2.
\]
An absent internal edge is denoted by \(u=0\).
Reachability gives \(w\ne0\), and \(a\ne0\) because otherwise there is only one effective hidden vertex.
Also \(v\ne0\), since \(w h_2\) alone has a fixed sign.

On \(a\cdot x<0\) we have \(r\cdot x=w\rho(c\cdot x)\).
If \(c\) is not proportional to \(a\), set
\[
z=c-\frac{c\cdot a}{a\cdot a}a.
\]
The denominator is positive because \(a\ne0\).
Then \(a\cdot z=0\), while \(c\cdot z=z\cdot z>0\), since nonproportionality makes \(z\ne0\).
The line \(x_0+s z\), with \(a\cdot x_0<0\), stays in that halfspace and makes the right side a nonaffine univariate hinge, contradicting the linear left side.
Therefore \(c=t a\).
For \(t\ge0\), the residual vanishes on this open halfspace, contrary to its nonzero coefficients.
Thus \(t<0\), and equality there gives \(r=w c\) globally.

If \(w\) is nonunit, every nonzero \(c_i\) must be a unit occurrence: otherwise \(|w c_i|\ge(M+1)^2>3M+2R\).
But all \(d\ge3\) coefficients \(c_i\) are nonzero, because \(r=wc\) and every \(r_i\ge M+1-R>1\).
Distinct nonzero \(c_i\) sit on distinct edges and therefore use distinct occurrences, while only two unit occurrences exist.
Consequently \(w=1\) or \(-1\).
Every skip \(p_i\) is now nonzero, since \(p_i=0\) would imply \(|c_i|=2M+t_i>B\).
Moreover every \(a_i\ne0\), since \(c=t a\) and every \(c_i\ne0\).
The witness therefore has \(d\) input edges into each hidden vertex, \(d\) skips, and the two nonzero gains \(v,w\).
These \(3d+2=m\) distinct positions consume the entire inventory.
No occurrence is left for the internal edge, so \(u=0\).

On \(a\cdot x>0\), we now have \(h_2=0\), hence \(r=v a\).
The same magnitude and unit-scarcity argument forces \(v\) to be a unit.
Since \(v=w t\) and \(t<0\), the gains are opposite units and \(t=-1\), giving \(c=-a\).
Both unit occurrences are used as gains.
The two hidden input edges of each coordinate must therefore use a positive and a negative occurrence of equal nonunit magnitude.
The only available pairs are \(M+b_j,-M-b_j\); source distinctness excludes every \(M+a_j\) value from those pairs.
All pairs are used bijectively, leaving exactly the \(M+a_j\) values for skips.
As every residual slope is positive,
\[
2M+t_i=(M+a_{\pi(i)})+(M+b_{\sigma(i)}),
\]
which extracts an NMTS matching.
The converse covered the larger at-most-once class, and the forward witness used every occurrence, so hardness holds under both inventory rules.
\end{proof}

The explicit affine classification combines with a universal target condition.
\begin{theorem}[ReLU-target synthesis with cap three; restatement of \Cref{ithm:fixed-cap}]
\label{thm:three-vertex-dp}
Full-space exact synthesis of a supplied rational ReLU target with computational-node cap three is \(\mathsf{DP}\)-complete under deterministic polynomial-time many-one reductions, under either inventory rule.
\end{theorem}
\begin{proof}
Membership is \Cref{thm:fixed-cap-dp}.
For hardness combine the NMTS instance above with a CNF formula \(\Phi\).
Let \(F_\Phi(z)\) be the nonnegative network of \Cref{lem:cnf-relu}, whose input \(z\) has one coordinate for each variable of \(\Phi\).
With three fresh coordinates \(y\), supply
\[
\widehat T(x,z,y)=T(x)+\min\{F_\Phi(z),1,\rho(y_1),\rho(y_2),\rho(y_3)\},
\]
keeping the inventory and cap three unchanged.
Binary minimum and the signed-pair construction in \Cref{lem:cnf-relu} compile this target in polynomial size.
If \(\Phi\) is unsatisfiable, the added term is zero.
A matching then gives the preceding witness using all occurrences.
Conversely, consider a witness on the inputs \((x,z,y)\) that computes \(T(x)\), which is the target \(\widehat T\) when \(\Phi\) is unsatisfiable.
Restrict it by setting the added inputs \(z,y\) to zero and deleting their edges.
Repeatedly pruning parameterless zero vertices and vertices not reaching the output gives a legal at-most-once witness for the affine target \(T\); the nonzero \(T\) keeps the output legal.
Thus a NO matching instance cannot become YES.
This holds even when the expanded witness uses every occurrence exactly once, because its restriction is still an at-most-once witness, which the converse in the proof of \Cref{thm:three-vertex-affine} excludes.

If \(\Phi\) is satisfiable, fix a Boolean \(z^*\) with \(F_\Phi(z^*)=1\) and fix \(x=0\).
The target slice is
\[
Q(y)=\min\{1,\rho(y_1),\rho(y_2),\rho(y_3)\}.
\]
It has five distinct affine formulas on nonempty open sets: zero, one, and \(y_1,y_2,y_3\).
The zero formula occurs when \(y_1<0\), one when all coordinates exceed one, and \(y_i\) when it is positive, below one, and smaller than both others.
Any cap-three candidate, even after arbitrary inputs are fixed, has at most two hidden vertices and four masks, hence at most four affine output formulas.

On an open set where the target equals an affine form \(L\), some masked formula must equal \(L\) identically.
Otherwise each difference between a masked formula and \(L\) is an affine function that is not identically zero, and the zero sets of these finitely many differences, called proper affine zero sets, would cover that open set.
Each proper affine zero set is a hyperplane or empty.
A hyperplane cannot contain an open ball: moving a point slightly in the direction of its nonzero normal changes the defining affine value.
Inside any open ball one can therefore choose a point off the hyperplane and then a smaller open ball around that point disjoint from it, since its affine value stays nonzero under sufficiently small perturbations.
Repeating this choice for finitely many hyperplanes leaves a nonempty open ball outside their union.
Hence finitely many proper affine zero sets cannot cover the open set in question.
Five distinct formulas therefore cannot be represented by four masks.

The constructed instance is YES precisely when NMTS is YES and \(\Phi\) is unsatisfiable.
For every language \(A\cap\overline C\) with \(A,C\in\mathsf{NP}\), independently reduce \(A\) to NMTS and \(C\) to CNF satisfiability, then apply this construction.
It is a polynomial-time many-one reduction from that language, proving \(\mathsf{DP}\)-hardness directly from the class definition.
Both inventory rules follow from the same forward and converse arguments.
\end{proof}

These classifications distinguish cap two from cap three while leaving unrestricted-budget synthesis open.
They do not automatically prove hardness with any larger fixed cap: extra hidden vertices can invalidate the numerical converse.
\section{Uniform Approximation on a Compact Box}
\label{sec:compact}

By \Cref{thm:fixed-cap-dp}, the full-space exact problem has a \(\mathsf{DP}\) upper bound at every fixed cap.
We now change the requirement to uniform approximation on a supplied compact box.
The target is a supplied rational ReLU network \(F\), and the input also lists rational endpoints \(a_j<b_j\) and a rational tolerance \(\varepsilon\ge0\).
With \(\mathcal B=\prod_{j=1}^d[a_j,b_j]\), the requirement is
\[
|N(x)-F(x)|\le\varepsilon\qquad\text{for every }x\in\mathcal B.
\]
A violation for a fixed candidate is certified by activation masks, the box constraints, and one strict signed error inequality.
The rational linear feasibility argument in \Cref{thm:relu-target} therefore puts candidate verification in \(\mathsf{coNP}\) and synthesis in \(\Sigma_2^{\mathsf P}\).
This \(\Sigma_2^{\mathsf P}\) upper bound for uniform approximation is tight even with no hidden candidate vertex and an inventory of identical positive occurrences, by \Cref{thm:compact-relu-hardness}.

\subsection{A Covering Source}

The target supplies the complexity of the uniform error test.
We first state the succinct covering problem used in the reduction.
For a Boolean circuit \(D\) on \(s\) input bits, its one-set is \(\{y\in\{0,1\}^s:D(y)=1\}\).
A family of sets covers the Boolean cube when every cube point belongs to at least one set.
Such covering formulations are standard sources of second-level hardness \cite{Umans1999}; the circuit version below has a direct proof.

\begin{lemma}[Circuit-specified set cover]
\label{lem:succinct-cover}
Given Boolean circuits \(D_1,\ldots,D_n\) on the same input bits and an integer \(1\leq k<n\), deciding whether some \(k\) or fewer of their one-sets cover the Boolean cube is \(\Sigma_2^{\mathsf P}\)-complete.
Hardness holds when every one-set is nonempty and the full family covers the cube.
\end{lemma}
\begin{proof}
Membership follows by guessing the chosen indices and universally checking the input bits.
For hardness let \(p\geq1,s\geq0\) be integers and let \(C\) be a Boolean circuit on \(p+s\) input bits.
Start with the quantified source sentence
\[
\exists u\in\{0,1\}^p\,\forall y\in\{0,1\}^s\, C(u,y)=1.
\]
An unused existential bit can be added if necessary.
These circuit sentences are complete for the defining existential-universal polynomial-time predicate class.
Indeed, a polynomial-time deterministic computation has polynomially many configuration bits and steps.
One transition is computed by a polynomial-size Boolean circuit, and unrolling the polynomially many transitions yields a polynomial-size circuit for the predicate.
Thus any predicate in the definition of \(\Sigma_2^{\mathsf P}\) has the required circuit representation.

Use universe coordinates \((u',y,w)\), where \(u'\) has \(p\) bits and the \(r=\lceil\log_2(p+2)\rceil\) bits of \(w\) encode an integer.
Each code \(c\) specifies the subcube of universe points with \(w=c\): code zero specifies the main subcube, codes \(1,\ldots,p\) specify pair-control subcubes, and code \(p+1\) specifies one additional control subcube.
Write \(U(w)\) for the predicate \(w>p+1\), covering any unused codes.
For \(1\leq i\leq p\) and \(b\in\{0,1\}\), define
\[
D_{i,b}=(w=0\land u'_i\ne b)\lor(w=i)\lor U(w),
\]
and define
\[
D_*=(w=0\land C(u',y))\lor(w=p+1)\lor U(w).
\]
These are \(2p+1\) circuits of polynomial size; choose budget \(p+1\).
Every set is nonempty because it contains its designated control subcube.
The full family covers every code: at code zero, either \(D_{i,0}\) or \(D_{i,1}\) holds for any fixed \(i\), and the remaining codes have their control or unused-code coverage.

Only \(D_*\) covers code \(p+1\), and code \(i\) is covered only by \(D_{i,0}\) and by \(D_{i,1}\), each of which covers it on its own.
The budget therefore forces \(D_*\) and exactly one choice \(D_{i,b_i}\) for each \(i\).
At code zero, the chosen sets \(D_{1,b_1},\ldots,D_{p,b_p}\) cover every \(u'\ne b\), where \(b=(b_1,\ldots,b_p)\).
They cover no point with \(u'=b\), so coverage of that remaining subcube is exactly \(C(b,y)=1\) for every \(y\).
This proves both directions and polynomial-time many-one hardness.
\end{proof}

A polynomial is \emph{multilinear} if each variable has exponent at most one in every monomial.
Its extrema on a box occur at vertices: with the other coordinates fixed, it is affine in one coordinate, whose maximizing or minimizing endpoint can be chosen in turn.
This fact will control fractional input and gate coordinates without interpreting them as Boolean values.

\begin{theorem}[Compact approximation; restatement of \Cref{ithm:compact-relu}]
\label{thm:compact-relu-hardness}
Uniform-approximation synthesis for a rational ReLU target on a rational box is \(\Sigma_2^{\mathsf P}\)-complete under deterministic polynomial-time many-one reductions.
Hardness holds on the unit box, at tolerance one, with candidate node budget one and an inventory consisting of identical positive rational occurrences.
The same completeness holds for the restriction to explicitly listed multilinear polynomial targets of degree at most five with candidate node budget one.
Both assertions hold for at-most-once and exactly-once use, with affine scalar outputs.
\end{theorem}
\begin{proof}
We first construct a polynomial target with a positive integer tolerance, replace it by a ReLU target preserving the uniform-error condition, and then rescale to tolerance one.
Take circuits \(D_1,\ldots,D_n\) and budget \(1\leq k<n\) from \Cref{lem:succinct-cover}.
Introduce feature coordinates \(x_1,\ldots,x_n\), source input coordinates \(y\), and coordinates \(z\) for all internal gate values of these circuits.
The feature coordinate \(x_i\) is required to equal circuit \(D_i\)'s output in a consistent Boolean assignment.
Two further coordinates \(s,t\) will select calibration conditions.
All coordinates range independently over \([0,1]\).

For every gate, form a multilinear polynomial that is zero at consistent Boolean gate assignments and one at inconsistent ones.
For distinct gate coordinates \(a,b,v\), use
\[
\begin{aligned}
q_{\mathrm{AND}}(a,b,v)&=v+ab-2abv,\\
q_{\mathrm{OR}}(a,b,v)&=(1-v)(a+b-ab)+v(1-a)(1-b).
\end{aligned}
\]
For a NOT gate and an equality constraint, respectively, use
\[
q_{\mathrm{NOT}}(a,v)=1-a-v+2av,
\qquad q_{\mathrm{EQ}}(a,v)=a+v-2av.
\]
A fixed-zero or fixed-one wire uses \(v\) or \(1-v\).
Repeated gate inputs can be simplified or replaced by fresh copy coordinates with equality constraints.
Every local polynomial has degree at most three and, by its Boolean values and multilinearity, lies in \([0,1]\) throughout its local cube.
Let \(Q(x,y,z)\) be their sum, including the output equalities, and let \(g\geq1\) be their number.
Then \(0\leq Q\leq g\) everywhere.
At Boolean vertices, \(Q=0\) means complete gate and output consistency; every other vertex has \(Q\geq1\).
Expanding these constant-size local formulas gives a polynomial-size explicit monomial list.

Put \(S=\sum_i x_i\), choose the integer \(E=n+k+g+1\), and define
\[
P=-E+S+st(2E+1-Q-S).
\]
This polynomial is multilinear of degree at most five and has polynomially many explicitly listed coefficients of polynomial bit length.
Initially use tolerance \(E\), node budget one, and \(k\) occurrences of the value one.
Every candidate is therefore
\[
N=b+\sum_{j\in J}\xi_j,
\]
where \(\xi=(x,y,z,s,t)\), the index set \(J\) lists the distinct coordinates that receive an input edge, and \(b\) is zero for an absent bias or one for a present bias.
At most \(k\) positions are occupied, or exactly \(k\) under the exactly-once rule.
The output must occupy at least one position.

Set \(s=t=0\) and \(x=0\).
The target is \(-E\), independently of \(y,z\), so the error bound forces the nonnegative candidate value to be zero.
This excludes its bias and every edge from \(y,z\): any such coordinate can be set to one at a calibration vertex.
The calibration vertices \((s,t)=(1,0)\) and \((0,1)\), again with \(x=y=z=0\), exclude the two selector edges.
Every successful candidate must consequently be
\[
N=\sum_{i\in I}x_i,\qquad 1\leq|I|\leq k,
\]
with equality in the cardinality bound for exactly-once use.

For any Boolean source input \(y\), choose its unique consistent gate values and feature outputs, and set \(s=t=1\).
Then \(Q=0\) and \(P=E+1\).
The inequality \(P-N\leq E\) requires \(\sum_{i\in I}D_i(y)\geq1\).
Thus a successful candidate extracts a cover.

Conversely, choose a cover \(I\) with \(|I|\leq k\).
For exactly-once use, pad it to \(k\) distinct indices; this is possible because \(k<n\), and adding sets preserves coverage.
Use the one-node candidate \(N=\sum_{i\in I}x_i\).
At a Boolean vertex with \(st=0\), the two error inequalities are
\[
S-2E\leq N\leq S,
\]
which hold because \(0\leq N\leq S\leq n<2E\).
At a vertex with \(s=t=1\), they are
\[
1-Q\leq N\leq2E+1-Q.
\]
For the lower bound, a consistent vertex is covered and an inconsistent vertex has \(Q\geq1\).
For the upper bound, use \(N\leq k\), \(Q\leq g\), and \(2E+1-g\geq k\).
Both signed errors are multilinear, so their vertex bounds extend to the entire unit box.
This proves the reduction for the polynomial target.

To produce a ReLU target, write the multilinear polynomial as \(P(\xi)=\sum_A c_A\prod_{j\in A}\xi_j\), with at most five coordinates in each support \(A\).
Replace it by the standard Lov\'asz extension of its Boolean vertex values \cite[Section 4]{GrabischMarichalRoubens2000}:
\[
F(\xi)=\sum_A c_A\min_{j\in A}\xi_j,
\]
where the minimum for \(A=\varnothing\) is defined to be one.
This is the established min-polynomial representation, not a new extension.
Each nonempty minimum has an explicit constant-size rational ReLU implementation, valid on all of \(\mathbb R^D\) and in particular on the box.
Remove the supports with \(c_A=0\).
For a remaining nonempty support \(A=\{j_1,\ldots,j_h\}\), put \(m_{A,1}=\xi_{j_1}\), and for \(2\leq t\leq h\) introduce a hidden vertex \(v_{A,t}\) with
\[
v_{A,t}=\rho\left(m_{A,t-1}-\xi_{j_t}\right),\qquad m_{A,t}=m_{A,t-1}-v_{A,t}.
\]
The identity \(\min\{a,b\}=a-\rho(a-b)\), valid for all real \(a,b\), and induction on \(t\) give \(m_{A,h}=\min_{j\in A}\xi_j\).
Since \(m_{A,t-1}=\xi_{j_1}-\sum_{t'=2}^{t-1}v_{A,t'}\), the preactivation of \(v_{A,t}\) has an edge of weight one from \(\xi_{j_1}\), an edge of weight minus one from \(\xi_{j_t}\), edges of weight minus one from \(v_{A,2},\ldots,v_{A,t-1}\), and no bias.
The affine output is \(c_\varnothing+\sum_{A\ne\varnothing}c_A(\xi_{j_1}-\sum_{t=2}^{h}v_{A,t})\), with \(j_1\) and \(h\) those of \(A\): its bias is the constant term \(c_\varnothing=-E\ne0\), each \(v_{A,t}\) enters it with weight \(-c_A\ne0\), and each coordinate enters it with its collected coefficient when that is nonzero, so negative coefficients are output weights and there are no parallel edges.
Ordering the vertices support by support and then by \(t\) makes every edge point forward, every hidden vertex has two input edges and an edge to the output, and there are at most four hidden vertices per support.
Thus \(F\) is an explicitly encoded rational ReLU target of polynomial size.

We verify that this replacement preserves every affine candidate's error bound.
Let \(D\) be the total number of coordinates in \(\xi\).
Sort their values as \(r_1\geq\cdots\geq r_D\), and put \(r_0=1,r_{D+1}=0\).
Let \(a^j\) be the Boolean vertex whose first \(j\) coordinates in that order are one, and write \(a^j_i\) for its coordinate \(i\).
The numbers \(\lambda_j=r_j-r_{j+1}\), for \(0\leq j\leq D\), are nonnegative, sum to one, and satisfy \(\xi=\sum_j\lambda_j a^j\).
For every monomial support, including the empty support,
\[
\min_{i\in A}\xi_i
=\sum_{j=0}^D\lambda_j\prod_{i\in A}a_i^j.
\]
Therefore \(F(\xi)=\sum_j\lambda_jP(a^j)\), while an affine candidate satisfies \(N(\xi)=\sum_j\lambda_jN(a^j)\).
The signed difference is a convex combination of its Boolean vertex differences.
Its absolute value is consequently at most \(E\) everywhere if it is at most \(E\) at every vertex.
The reverse implication is immediate because the vertices belong to the box.
This argument includes negative coefficients, constant terms, and ties in the coordinate ordering.

Finally divide the target and every inventory occurrence by \(E\), and set tolerance one.
For a one-node affine candidate this uniformly divides its output by \(E\), so legal descriptions correspond exactly to those above.
The inventory now consists of \(k\) occurrences of \(1/E\), and the cover witness has at least one input edge and no bias.
The reduction is polynomial in the original circuit input length and respects both inventory rules.

For ReLU targets, membership is the activation-pattern argument given at the start of this section.
For the polynomial restriction, the candidate is affine and the explicitly listed target is multilinear, so a strict error violation has a box vertex witness, specified by one endpoint-choice bit per coordinate.
Checking that witness uses polynomial-time rational arithmetic.
Guessing the permitted candidate and universally checking the vertices proves membership in \(\Sigma_2^{\mathsf P}\).
Both completeness assertions follow.\qedhere
\end{proof}

\section{Exact Population Risk on the Boolean Cube}
\label{sec:population-risk}

A target given explicitly also permits an exact population-risk question without a sampling oracle.
A probability distribution specifies the probabilities of possible inputs.
A density is a nonnegative function of integral one; probabilities of regions are its integrals over those regions.
For a random input \(X\) with probability distribution \(\nu\) and a real function \(f\), the expectation \(\mathbb E_{X\sim\nu}[f(X)]\) is the average of \(f\) under that probability distribution.
On a finite input set it is a probability-weighted sum, and for the continuous input distributions used here it is an integral against the specified density.

All continuous integrals below use Lebesgue measure, the usual volume on Euclidean space.
A measurable function is one whose superlevel sets \(\{x:f(x)>t\}\) are measurable for every real \(t\).
For an event \(A\), its indicator \(\mathbf1_A\) is one on \(A\) and zero otherwise, so \(\Pr[A]=\mathbb E[\mathbf1_A]\).
Independent coordinates have the product of their separate probability distributions as their joint probability distribution.
A fair bit takes each value in \(\{0,1\}\) with probability \(1/2\).

A measurable real-valued function \(f\) is integrable under \(\nu\) when \(\mathbb E_\nu[|f|]<\infty\).
A measurable real-valued function \(f\) is square-integrable under a probability distribution \(\nu\) when \(\mathbb E_\nu[f^2]<\infty\); write
\[
\|f\|_{2,\nu}=(\mathbb E_\nu[f^2])^{1/2}.
\]
The pointwise inequality \(2|fg|\leq f^2+g^2\), obtained from \((|f|-|g|)^2\geq0\), shows that \(fg\) is integrable whenever \(f,g\) are square-integrable.

For square-integrable \(f,g\), Cauchy--Schwarz is the inequality
\[
|\mathbb E_\nu[fg]|\leq\|f\|_{2,\nu}\|g\|_{2,\nu}.
\]
To verify it, suppose first that \(\mathbb E_\nu[g^2]=0\).
Then \(g=0\) except on a set of probability zero, and both sides vanish.
Otherwise expand the nonnegative average \(\mathbb E_\nu[(f-tg)^2]\) and take \(t=\mathbb E_\nu[fg]/\mathbb E_\nu[g^2]\).
This gives \((\mathbb E_\nu[fg])^2\leq\mathbb E_\nu[f^2]\mathbb E_\nu[g^2]\).

Expanding the square and applying Cauchy--Schwarz gives
\[
\mathbb E_\nu[(f+g)^2]
\leq\|f\|_{2,\nu}^2+2\|f\|_{2,\nu}\|g\|_{2,\nu}+\|g\|_{2,\nu}^2
=(\|f\|_{2,\nu}+\|g\|_{2,\nu})^2.
\]
Taking nonnegative square roots proves the triangle inequality \(\|f+g\|_{2,\nu}\leq\|f\|_{2,\nu}+\|g\|_{2,\nu}\).

We also use convex averaging, or Jensen's inequality: if \(X\) and \(f(X)\) are integrable and \(f:\mathbb R\to\mathbb R\) is convex, then \(f(\mathbb E[X])\leq\mathbb E[f(X)]\).
For completeness we prove it; in this argument only, \(m=\mathbb E[X]\) denotes the mean, not the inventory size \(|W|\).
Convexity makes every secant slope \((f(m)-f(u))/(m-u)\), with \(u<m\), at most every secant slope \((f(v)-f(m))/(v-m)\), with \(v>m\).
The supremum of the former and infimum of the latter are finite: use \(u=m-1\) and \(v=m+1\) to bound the two collections on the needed sides.
Choose a real \(c\) between them.
Then \(f(t)\geq f(m)+c(t-m)\) for every real \(t\), on either side of \(m\).
Averaging at \(t=X\) proves the inequality.

Jensen's inequality also applies after fixing some independent input coordinates, called observed coordinates below, by using the product probability distribution of the remaining coordinates.
Conditioning on specified observed coordinates means fixing those coordinates and averaging with their conditional probability distribution; the other coordinates keep their original probability distributions when all input coordinates are independent.

Let \(T:\mathbb R^d\to\mathbb R\) be a supplied rational ReLU network with an affine output.
Write \(U_d\) for the uniform distribution on \(\{0,1\}^d\), whose coordinates are independent fair bits.
For a permitted network \(N\), define
\[
R_T(N)=\mathbb E_{X\sim U_d}[(N(X)-T(X))^2]
=2^{-d}\sum_{x\in\{0,1\}^d}(N(x)-T(x))^2.
\]
The decision input consists of \(T,W,n_{\max}\) and a rational threshold \(\tau\).
The question is whether some permitted network satisfies \(R_T(N)\leq\tau\).
Candidate networks have affine outputs.
Unless exactly-once use is specified, occurrences are used at most once.
All input coordinates and the target graph are listed explicitly.

The relevant counting class is \(\mathsf{PP}\).
A language belongs to this class if a polynomial-time machine making independent fair binary choices accepts an input precisely when strictly more than half its computation paths accept.
The class \(\mathsf{NP}^{\mathsf{PP}}\) allows a nondeterministic polynomial-time machine to query a language in \(\mathsf{PP}\).
The input encodings and running times here are ordinary binary encodings and bit running times.

\begin{theorem}[Exact population risk; restatement of \Cref{ithm:population-risk}]
\label{thm:population-risk}
Under either inventory rule, deciding whether \(R_T(N)\leq\tau\) for some permitted network is \(\mathsf{NP}^{\mathsf{PP}}\)-complete under deterministic polynomial-time many-one reductions.
Hardness holds for integer inventory, with computational-node cap two under at-most-once use and cap \(m=|W|\) under exactly-once use.
Both caps include the affine output.
The hard targets take only the values \(-1,+1\) on the Boolean cube.
\end{theorem}

For the lower bound we use an existential choice followed by a strict majority test.
The following is the source problem, including the threshold convention needed below.
\begin{theorem}[Littman, Goldsmith, and Mundhenk {\cite[Theorem 15]{LittmanGoldsmithMundhenk1998}}]
\label{thm:emajsat}
Given a Boolean formula \(\varphi(x,z)\), deciding whether there exists an assignment to \(x\) such that strictly more than half the assignments to \(z\) satisfy \(\varphi(x,z)\) is \(\mathsf{NP}^{\mathsf{PP}}\)-complete.
\end{theorem}

The parity of a list of bits is their sum modulo two.
We use a parity construction from probabilistic explanation complexity.
It prevents a candidate with limited coordinate access from observing the random bits used by the source formula.
\begin{lemma}[Parity replacement; Koriche, Lagniez, and Tran {\cite[Definition 4 and Lemma 2]{KoricheLagniezTran2026}}]
\label{lem:parity-replacement}
Replace each of several independent fair bits by the parity of its own block of \(k+1\) fresh independent fair bits.
Conditioning on the values of at most \(k\) fresh coordinates in total leaves these block parities independent and uniform.
They also remain independent of all other independent coordinates outside the replacement blocks.
\end{lemma}

The cited construction was used for conditional relevance of sparse explanations.
Here the additional ingredient is an integer-output inequality for unconditional squared loss.
The source problem, parity replacement, and counting framework are established ingredients; the occurrence-constrained synthesis reduction is proved below.

\begin{proof}[Proof of \Cref{thm:population-risk}]
\emph{Membership.}
Let \(L\geq2\) be the total binary input length and \(m=|W|\).
A guessed legal network has at most \(m\) computational vertices, because every vertex requires a distinct incoming-edge or bias occurrence.
Its graph and occurrence assignment therefore have polynomial length.
All graph conditions in \Cref{def:network}, including the cap, can be checked in polynomial time.

Let \(D\) be the product of the positive denominators of the occurrences in \(W\) and of all parameter positions in the supplied target.
Repeated target values contribute separate denominator factors for their separate positions.
The binary length of \(D\) is at most \(L\).
Fix a Boolean input \(x\).
Replacing each active hidden ReLU by the identity and each inactive hidden ReLU by zero gives an acyclic affine computation with the same value at \(x\).
At a zero preactivation either choice gives that value.
Its expansion is a sum of input-to-output and bias-to-output path products.
Every path uses a parameter position at most once.
For the candidate, distinct positions also use distinct inventory occurrences.
Every path denominator consequently divides \(D\), so both \(DN(x)\) and \(DT(x)\) are integers.

Exact evaluation has polynomial bit complexity.
For an explicit bound, every parameter has absolute value at most \(2^L\), every affine preactivation has at most \(L\) terms, and each network has at most \(L\) computational vertices.
Induction through its vertices gives absolute value at most \((L2^L)^L\) on Boolean inputs.
A ReLU does not increase this bound.
Together with the common denominator, this bounds every intermediate numerator by a polynomial number of bits.
In particular,
\[
A(x)=(DN(x)-DT(x))^2
\]
is a nonnegative integer with \(A(x)<2^K\), where the uniform choice \(K=8(L+1)^2\) suffices.
Indeed \(\log_2\max\{1,A(x)\}\) is at most \(2L+2+2L\log_2 L+2L^2<K\).

A polynomial-time counting machine guesses \(d\) input bits and \(K\) bits specifying an integer \(t\in\{0,\ldots,2^K-1\}\).
It accepts exactly when \(t<A(x)\).
Its number of accepting paths is
\[
S=\sum_{x\in\{0,1\}^d}A(x),
\]
and its total number of paths is \(Q=2^{d+K}\).

The desired risk condition is
\[
S\leq J,\qquad J=\left\lfloor\tau D^2 2^d\right\rfloor.
\]
Compute \(J\) by exact rational arithmetic.
If \(J<0\), reject; if \(J\geq Q\), accept.
For \(0\leq J<Q\), use a machine with one additional fair selector bit.
On the first selector branch, simulate the counting machine and accept exactly when it rejects, giving \(Q-S\) accepting paths.
On the second, accept exactly when the remaining \(d+K\) bits specify an integer less than \(J+1\), giving \(J+1\) accepting paths.
Of its \(2Q\) paths, exactly \(Q-S+J+1\) accept.
This is strictly greater than \(Q\) precisely when \(S\leq J\).
The fixed candidate can thus be checked with a \(\mathsf{PP}\) query.
Guessing the candidate proves membership in \(\mathsf{NP}^{\mathsf{PP}}\).

\emph{Inventory and candidate witnesses for hardness.}
Take an instance \(\varphi(x,z)\) of \Cref{thm:emajsat}, with \(e\) existential bits and \(n\) random bits.
Adding ignored variables preserves the answer, so assume \(e,n\geq1\).
For \(a\in\{0,1\}^e\), write
\[
p(a)=2^{-n}\#\{z\in\{0,1\}^n:\varphi(a,z)=1\}.
\]
Supply \(e+2\) occurrences of \(+1\), \(e\) occurrences of \(-1\), and one occurrence of each value
\[
1,0,-1,\ldots,1-e.
\]
The last group has \(e+1\) occurrences; thus \(m=3e+3\).
The three groups have separate occurrence identities even when their numerical values agree.
Set the computational cap to two.

In the hardness construction, \(x\in\{0,1\}^e\) denotes the existential coordinates of an input; the share coordinates introduced with the target below make up the rest.
For every \(a\), a permitted hidden neuron can compute
\[
h_a(x)=\rho\left(\sum_{i:a_i=1}x_i-\sum_{i:a_i=0}x_i+1-|a|\right),
\]
where \(|a|\) is the number of its one coordinates.
Use an affine output with the single edge \(+1\) from this neuron.
The positive input edges and output edge consume at most \(e+1\) occurrences from the first group.
The negative input edges consume at most \(e\) from the second.
The bias is assigned its separate occurrence in the last group; a zero bias may instead be omitted.
There are no parallel edges, both computational vertices are legal, and the hidden vertex reaches the output.
On Boolean inputs the preactivation is one minus the number of coordinates where \(x\) differs from \(a\).
Consequently \(h_a(x)=1\) when \(x=a\), and \(h_a(x)=0\) otherwise.

\emph{The target.}
Replace every random bit \(z_j\), and one additional fair bit \(u\), by the parity of a disjoint block of \(m+1\) fresh coordinates, which we call share coordinates.
There are now
\[
d=e+(n+1)(m+1)
\]
Boolean input coordinates: the \(e\) existential coordinates \(x\) and the share coordinates.
In the definition of \(F\) and in the conditioning argument below, \(u\) and every \(z_j\) denote the parities of their blocks.
Define the Boolean function \(F=u\lor\varphi(x,z)\) with these parities, and let the target be \(T=2F-1\) on the cube.
The XOR of two bits is one precisely when the bits differ.
Each parity has a linear-size XOR circuit, and XOR has a constant-size expression using AND and NOT.
The circuit for \(F\) therefore has polynomial size.

This circuit compiles into a polynomial-size rational ReLU target.
For Boolean gate values \(a,b\), NOT is \(\rho(1-a)\) and AND is \(\rho(a+b-1)\).
Replace OR and XOR by their fixed AND/NOT expressions first, and create the corresponding hidden ReLU vertices in circuit order.
If an AND gate has the same predecessor twice, use one edge of weight two instead of parallel edges.
Constants are bias-only vertices when needed, and unused vertices can be removed.
All resulting gate values are Boolean on the cube.
The final affine output \(2F-1\) gives the prescribed signed target.
Target parameters are independently supplied as part of its explicit graph and need not belong to \(W\).
This construction imposes no requirement on target values outside the cube.

For a fixed \(x\), the parities are independent fair bits.
Therefore
\[
\Pr[F=1| x]=\frac{1+p(x)}2,
\qquad
\mathbb E[T| x]=p(x),
\qquad
T^2=1.
\]
The last equality holds at every Boolean input.

\emph{All competing graphs in a no instance.}
Fix any permitted candidate \(N\).
Let \(S_N\) be the set of fresh share coordinates that occur on an input edge of its graph.
Each distinct such coordinate requires a parameter occurrence, so \(|S_N|\leq m\).
Including coordinates on zero-weight edges in \(S_N\) preserves this bound.
Condition on all \(e\) original \(x\)-coordinates and on the share coordinates in \(S_N\).
The candidate value is determined by this conditioning.
Every parity block retains at least one unconditioned coordinate.
By \Cref{lem:parity-replacement} with \(k=m\), the block parities remain independent fair bits, independently of \(x\).
The conditional target mean is consequently \(p(x)\).

Every candidate value is an integer on the cube.
This follows by induction because the supplied parameters and Boolean inputs are integers, integer affine operations preserve integrality, and ReLU preserves integers.
In a no instance, \(0\leq p(x)\leq1/2\) for every \(x\).
For every integer \(b\),
\[
b^2-2bp(x)\geq0.
\]
If \(b\leq0\), both summands \(b^2\) and \(-2bp(x)\) are nonnegative.
If \(b\geq1\), the expression is at least \(b^2-b=b(b-1)\geq0\).
Thus, on every event specified by the conditioning,
\[
\mathbb E[(N-T)^2| x,(X_j)_{j\in S_N}]
=1+N^2-2Np(x)\geq1.
\]
Here \(X_j\) denotes coordinate \(j\) of the random input \(X\sim U_d\).
Averaging shows \(R_T(N)\geq1\) for every permitted graph, including arbitrary sharing, skips, biases, and negative affine outputs.

\emph{A yes instance and the threshold.}
If \(p(a)>1/2\), use the permitted indicator network \(h_a\).
The event \(x=a\) has probability \(2^{-e}\), and \(h_a^2=h_a\) on the cube.
It follows that
\[
R_T(h_a)=1+2^{-e}(1-2p(a)).
\]
Since \(p(a)\) is an integer multiple of \(2^{-n}\), strict majority implies \(p(a)-1/2\geq2^{-n}\).
Hence
\[
R_T(h_a)\leq1-2^{1-e-n}
<1-2^{-e-n}<1.
\]
Choose \(\tau=1-2^{-e-n}\).
This rational has polynomial binary length.
A source yes instance has feasible risk below \(\tau\), while a source no instance has no feasible risk at most \(\tau\).
The dimension, target graph, inventory and threshold are all constructed in polynomial time.
This proves the at-most-once hardness.

\emph{Exactly-once use.}
Membership uses the same graph certificate and counting test, with the additional polynomial-time check that every occurrence is assigned.
For hardness retain the inventory, target, and threshold above, but set the cap to \(m\).
The no-instance argument covered every graph using the inventory and did not require cap two, so it remains valid.

For a yes witness \(a\), omit the indicator neuron's bias when \(|a|=1\).
Let \(\delta\) be one if this bias is nonzero and zero otherwise.
The two-vertex indicator then uses \(e+1+\delta\) occurrences, leaving
\[
\ell=m-e-1-\delta=2e+2-\delta\geq3
\]
unused occurrences.
The sole zero occurrence is unused, because all indicator input and output weights are units and its present bias is nonzero.

List the unused occurrences as \(q_0,q_1,\ldots,q_{\ell-1}\), with \(q_0=0\).
Add hidden vertices
\[
z_0=\rho(q_0),\qquad
z_j=\rho(q_jz_{j-1})\quad(1\leq j\leq \ell-2),
\]
and add the term \(q_{\ell-1}z_{\ell-2}\) to the existing affine output.
The first vertex has a present zero bias and every later vertex has an incoming edge.
Each value is zero, by induction, regardless of the signs of the chain weights.
The output and its risk are unchanged.

The chain consumes all \(\ell\) unused occurrences: one bias, \(\ell-2\) internal edges, and one output edge.
Its \(\ell-1\) new hidden vertices give \(\ell+1=m-e-\delta\leq m\) computational vertices in total.
All new vertices have structural paths to the output.
The added output predecessor differs from the indicator neuron, so no parallel edge is introduced.
This is a legal exactly-once witness attaining the same risk below \(\tau\).
The reduction is polynomial, completing the exactly-once hardness and the theorem.
\end{proof}

The loss threshold in this reduction separates risks by an exponentially small amount.
The theorem is an exact decision classification and gives no constant-gap approximation hardness.
Sampling bounds at a fixed accuracy concern a different requirement.

The signed-label reduction used integer output values.
Allowing a single half occurrence gives a finer output grid and permits ordinary binary labels.
\begin{corollary}[Binary targets]
\label{cor:binary-risk}
Under either inventory rule, population squared-risk synthesis is \(\mathsf{NP}^{\mathsf{PP}}\)-complete even when the target takes values only in \(\{0,1\}\) on the Boolean cube and the inventory consists of integers and exactly one occurrence of \(1/2\).
Hardness holds with cap three under at-most-once use and cap \(m=|W|\) under exactly-once use, counting the affine output.
\end{corollary}
\begin{proof}
The upper bound follows from \Cref{thm:population-risk}.
For hardness take the same existential-majority source \(\varphi(x,z)\), with \(e,n\geq1\), and retain \(p(a)=\Pr_z[\varphi(a,z)=1]\).
Supply \(e+2\) positive units, \(e\) negative units, the separate bias occurrences \(1,0,-1,\ldots,1-e\), and one occurrence of \(1/2\).
There are \(m=3e+4\) occurrences in total.
Replace every random source bit and the additional bit \(u\) by parity blocks of size \(m+1\), as before.
Use the binary target
\[
Y=F=u\lor\varphi(x,z),\qquad
q(x)=\mathbb E[Y| x]=\frac{1+p(x)}2.
\]
The preceding Boolean-gate compilation supplies this target as a polynomial-size rational ReLU network with an affine output, without the final signed transformation.

Every candidate value on the cube belongs to \(\tfrac12\mathbb Z\), the set of integer multiples of one half, whose elements we call half-integers.
Indeed, freeze its actual activation pattern and use the path expansion from the preceding proof.
Every path product contains only integers and at most one factor \(1/2\), since there is just one such occurrence and a path cannot reuse it.
Every product is a half-integer, and so is their sum.
This argument permits sharing, reconverging paths, and any placement or nonuse of the half occurrence.

Fix any candidate and condition on all existential coordinates and its observed fresh coordinates.
At most \(m\) fresh coordinates are observed, so \Cref{lem:parity-replacement} leaves the conditional target mean equal to \(q(x)\).
The candidate value is fixed by the conditioning and is a half-integer.
In a source no instance, \(1/2\leq q(x)\leq3/4\).
For every half-integer \(b\), the conditional squared loss minus \(1/4\) is
\[
b^2+(1-2b)q(x)-\frac14.
\]
There is no half-integer strictly between \(1/2\) and one.
If \(b\leq1/2\), the coefficient \(1-2b\) is nonnegative, and the expression is at least
\[
b^2+\frac{1-2b}{2}-\frac14=(b-1/2)^2\geq0.
\]
If \(b\geq1\), that coefficient is nonpositive, and the expression is at least
\[
b^2+\frac34(1-2b)-\frac14=(b-1)(b-1/2)\geq0.
\]
Averaging proves risk at least \(1/4\) for every legal candidate, independently of its cap.

For a source yes witness \(a\), use the earlier indicator \(g=h_a\), then a second hidden vertex \(h=\rho(1+g)\), and affine output \(N=h/2\).
This has three computational vertices and equals \((1+h_a)/2\), since \(h_a\geq0\).
The positive input edges, the edge \(g\to h\), and the bias of \(h\) use \(|a|+2\leq e+2\) positive unit occurrences.
The negative input edges use at most \(e\) negative units; the bias of \(g\) uses its separate \(1-|a|\) occurrence and is omitted if zero.
The half occurrence is the output edge.
Every vertex is legal and reaches the output, and there are no parallel edges.

Outside \(x=a\), prediction \(1/2\) has squared error exactly \(1/4\) for either binary label.
At \(x=a\), prediction one has conditional mean squared error \(1-q(a)\).
Changing the prediction there therefore changes its conditional risk by
\[
1-q(a)-\frac14=\frac34-q(a)=\frac{1-2p(a)}4.
\]
Consequently
\[
R_Y(N)=\frac14+2^{-e}\frac{1-2p(a)}4
\leq\frac14-2^{-e-n-1}.
\]
The inequality uses \(p(a)-1/2\geq2^{-n}\), exactly as before.
Choose the polynomial-length rational threshold \(\tau=1/4-2^{-e-n-2}\).
The yes risk is strictly below it, and every no risk is at least \(1/4\).
This proves at-most-once hardness at cap three.

For exactly-once use, set the cap to \(m\).
Let \(\delta\) indicate whether the nonzero bias of \(g\) is present.
The displayed witness consumes \(e+3+\delta\) occurrences, leaving
\[
\ell=m-e-3-\delta=2e+1-\delta\geq2
\]
unused, including the sole zero.
The half occurrence is already assigned to the output edge.
Use the zero-chain construction in the proof of \Cref{thm:population-risk} on these \(\ell\) leftovers.
It consumes them all and adds \(\ell-1\) hidden vertices without changing the output.
When \(\ell=2\), it consists of a zero-bias hidden vertex and its edge to the output; no internal chain edge is needed.
The resulting computational count is \(3+\ell-1=\ell+2=m-e-1-\delta\leq m\).
The chain is legal and has a distinct final output predecessor, as in that proof.
The cap-independent no-instance inequality still excludes every exactly-once candidate.
This proves the second hardness assertion.
\end{proof}

The corollary does not assert binary-target hardness for an entirely integer inventory, or cap three under exactly-once use.
It retains an exponentially small decision gap and does not establish fixed-gap approximation hardness.

\section{Exact Population Risk under Continuous Uniform Inputs}
\label{sec:continuous-risk}

The distribution changes the arithmetic of exact risk.
Let \(T,N:\mathbb R^d\to\mathbb R\) be rational ReLU networks with affine outputs, and define
\[
R_T^{\mathrm{cont}}(N)=\int_{[0,1]^d}(N(x)-T(x))^2\,dx.
\]
The cube has volume one, so this is squared population risk for continuous uniform inputs.
For analysis, the same notation applies to measurable real-valued functions whose squared difference has finite integral.
In the synthesis problems, the supplied target and candidate descriptions remain networks.
In fixed-network comparison both networks are supplied.
In synthesis only \(T\) is supplied, and \(N\) must obey the inventory and the computational cap.
The threshold \(\tau\) is a binary rational in both problems.
Fixed-network comparison asks whether the risk of the supplied \(N\) is at most \(\tau\), and synthesis asks whether some legal candidate has risk at most \(\tau\).

The appropriate comparison class permits different denominators on different counting paths.
Following de Campos, Stamoulis, and Weyland \cite[Definition 9]{DeCamposStamoulisWeyland2019}, write \(\mathsf{PP}_{\mathbb Q}^{(pt)}\) for the languages expressible as
\[
F(a)=\sum_{w\in\{0,1\}^{p(|a|)}}\frac{u(a,w)}{v(a,w)}\geq t(a).
\]
Here \(a\) is the ordinary binary input, \(p\) is a polynomial, \(u\) is a signed integer, \(v\) is a positive integer, and the weights and rational threshold \(t\) are computed exactly in polynomial time.
Their individual binary lengths are polynomially bounded.
The definition requires the weights to be computed exactly, not merely approximated.
In particular, it does not include the undecidable comparisons possible in the broader approximation-defined model of that source.
Taking integer counting weights shows that the class contains \(\mathsf{PP}\).

We use the following recent arithmetic result with its version stated explicitly.
The notation \(\mathsf{BPP}\) means polynomial-time computation with independent fair random bits and error at most \(1/3\) on every input.
Oracle superscripts have the same meaning as for \(\mathsf{NP}^{\mathsf{PP}}\).
A dyadic rational is an integer divided by a power of two.

\begin{theorem}[Meiburg {\cite[Theorem 12]{Meiburg2026}}]
\label{thm:rational-weight-import}
\leavevmode\par\nopagebreak
For exactly poly\-nomial-time computable rational weights,
\[
\mathsf{PP}_{\mathbb Q}^{(pt)}
\subseteq \mathsf{BPP}^{\mathsf{PP}}
\cap\mathsf P^{\mathsf{NP}^{\mathsf{PP}}}.
\]
\end{theorem}

This is an import from the cited preprint, rather than a new general arithmetic bound.
Our contribution here is the exact ReLU integration characterization, including a one-dimensional realization of arbitrary rational weighted sums.
Previous work already studies ReLU integration hardness, including fixed dimension \cite[Proposition 2.7 and Theorem 3.2]{FeischlGrassleZehetgruber2026}, and probabilistic neural verification through weighted model integration \cite{MorettinPasseriniSebastiani2024}.

\begin{theorem}[Continuous exact risk; restatement of \Cref{ithm:population-risk}]
\label{thm:continuous-risk}
Comparing the continuous squared risk of two supplied networks with a rational threshold is \(\mathsf{PP}_{\mathbb Q}^{(pt)}\)-complete under deterministic poly\-nomial-time many-one reductions.
Hardness holds with \(d=1\), \(N=0\), dyadic target parameters, and \(0\leq T\leq1\).
Under either inventory rule, continuous-risk synthesis is \(\mathsf{PP}_{\mathbb Q}^{(pt)}\)-hard even with \(W=(0)\) and cap one, and belongs to
\[
\mathsf{NP}^{\mathsf{PP}_{\mathbb Q}^{(pt)}}
\cap\mathsf{BPP}^{\mathsf{NP}^{\mathsf{PP}}}
\subseteq\mathsf{PSPACE}.
\]
The randomized synthesis upper bound uses \Cref{thm:rational-weight-import}.
\end{theorem}

We first establish the geometric encoding needed for membership.
Recall the affine hyperplanes, halfspaces and open balls defined in \Cref{sec:model}.
A set is closed when the points outside it form an open set.
A set is convex when it contains every segment between two of its points: \(\lambda x+(1-\lambda)y\) belongs to the set whenever \(x,y\) do and \(0\le\lambda\le1\).
Affine halfspaces are convex by linearity, and intersections preserve this property.
A polytope is a bounded intersection of finitely many closed affine halfspaces.

A hyperplane supports a polytope when it meets the polytope and the polytope lies wholly in one of its two closed halfspaces.
A face is the intersection of the polytope with a supporting hyperplane, and the whole polytope is also a face.
The polytope and its faces are therefore convex.

An inequality is tight at a point when equality holds.
The affine hull of a nonempty set consists of all finite affine combinations of its points.
The dimension of the affine hull is the dimension of the linear span of the differences of the set's points.
A set's boundary consists of the points whose every open ball contains both a point in the set and a point outside it.
The relative boundary is defined in the same way, with balls taken within the set's affine hull and outside points taken from that hull.
The relative interior consists of the points of the set that have a ball within the affine hull contained in the set.
A polytope is full-dimensional in \(\mathbb R^d\) when it contains an open ball in \(\mathbb R^d\).
A vertex is a zero-dimensional face, identified with its unique point.
A facet of a face is a subface of dimension one smaller than that face.
A face flag is a strictly nested sequence of faces, complete when it contains one face of every dimension.

Points \(v_0,\ldots,v_k\) are affinely independent when \(v_1-v_0,\ldots,v_k-v_0\) are linearly independent.
Their simplex is the set of convex combinations \(\sum_{i=0}^k\lambda_i v_i\), where \(\lambda_i\ge0\) and \(\sum_i\lambda_i=1\); these weights are its barycentric coordinates.
A triangulation decomposes a polytope into simplices such that any two distinct simplices intersect, if at all, in a common face.
Lexicographic tie breaking compares the first coordinate at which two vectors differ.

A bounded set has measure zero, or volume zero, if for every \(\epsilon>0\) it can be covered by boxes whose total volume is less than \(\epsilon\).
A finite union of such sets has the same property, by dividing \(\epsilon\) among their covers.

Each bounded part of a proper affine hyperplane has volume zero.
To see this in a bounded cube, choose one nonzero normal coefficient and solve for that coordinate as an affine function of the others.
Partition the other coordinates into boxes of side \(1/n\).
There are \(O(n^{d-1})\) such boxes, and the affine graph above each has height variation at most \(C/n\), where \(C\) is the sum of the absolute remaining coefficient ratios.
Cover that piece of the graph by a box with that base and height at most \((C+2)/n\).
The sum of these box volumes is \(O(1/n)\), and hence can be made smaller than any positive \(\epsilon\).
This also covers \(d=1\), with an empty base and a shrinking interval.

A bounded lower-dimensional polytope lies in a proper affine hyperplane and consequently has volume zero.
For a full-dimensional polytope, a boundary point must make some nonconstant defining inequality tight, since finitely many strictly satisfied inequalities admit a small ball.
Its boundary is therefore contained in finitely many proper hyperplanes and has volume zero.
Changing a bounded integrand on these sets does not affect its integral: the absolute change is at most its bound times the volume of a covering set.
For polynomials restricted to bounded polytopes, integration by successive one-dimensional slices agrees with volume integration.
Their boundaries have volume zero, and polynomial oscillations on fine boxes tend uniformly to zero, so upper and lower box sums give the same limit.
These are the integration conventions used below.

\begin{lemma}[Polynomial-length rational integration leaves]
\label{lem:risk-leaves}
From two supplied rational ReLU networks one can construct a polynomial-time algorithm whose input is a binary word of one fixed length, polynomial in the size of the two networks, and whose output is a nonnegative rational of polynomial binary length, such that the sum of these outputs over all words of that length equals the continuous squared risk of the two networks.
\end{lemma}
\begin{proof}
Let \(L\geq2\) bound the full input encoding length, and let \(p\) be the total number of hidden vertices of both networks.
Freeze one of the \(2^p\) activation patterns: replace each active ReLU by the identity and each inactive ReLU by zero.
Every computed quantity becomes a rational affine form.
Recursively substituting these affine computations expresses each coefficient as a sum of input-to-vertex or bias-to-vertex path products, as in the earlier exact-evaluation argument of \Cref{thm:population-risk}.

Let \(D\) be the product of all positive position denominators in both supplied networks.
Each directed path uses a position at most once, so multiplication by \(D\) turns its term into an integer product of used numerators and unused denominators.
The full encoding length \(L\) bounds the number of parameter positions and the sum of their numerical bit lengths; in particular each such scaled product has absolute value at most \(2^{2L}\).

An acyclic path is determined by its set of positions, so the two networks together supply at most \(2^{L+1}\) terms for any coefficient of their difference.
Thus each scaled numerator has at most \(3L+3\) bits, including its sign, and the common positive denominator has at most \(L+1\) bits.
These bounds apply to frozen preactivations as well as the output difference, and are smaller than \(b_0=4(L+1)^3\) for \(L\ge2\).

Intersect the frozen preactivation sign inequalities with the cube.
Write the resulting polytope as \(P=\{x:a_i\cdot x\leq b_i,\quad 1\leq i\leq M\}\), with \(a_i\in\mathbb Q^d\), \(b_i\in\mathbb Q\), and \(M\) polynomial.
Discard patterns for which \(P\) is not full-dimensional.
This test is exact rational linear programming: reject any false constant row \(0\leq b_i\) with \(b_i<0\), ignore true constant rows in this test, and maximize a common slack \(\delta\in[0,1]\) in all remaining inequalities.
The maximum is positive precisely when an open ball lies in \(P\).
All coefficients have polynomial binary length and \(\delta\leq1\) bounds the objective.
Apply \Cref{thm:rational-lp-import} first to reject infeasibility and otherwise to compute the maximum exactly, including when the feasible region is lower-dimensional.
Also discard any pattern declaring active a gate whose frozen preactivation is the zero affine form.

The retained polytopes cover the cube up to a set of measure zero and have disjoint interiors.
To see coverage, exclude the cube boundary and all zero hyperplanes of nonconstant affine forms arising in any frozen pattern.
There are finitely many such hyperplanes.
At any remaining point, choose the active state for each gate whose actual preactivation is positive and the inactive state otherwise.
Nonzero frozen preactivations are then strictly signed in a neighborhood, while identically zero ones are inactive, so the resulting polytope is full-dimensional and retained.
For disjointness, suppose two retained patterns overlap on an open set.
At their first differing gate the earlier computations agree, so its preactivation is zero on that open set.
The frozen affine form is therefore identically zero, contradicting retention of the pattern that declares it active.

On every activation polytope, including those discarded as lower-dimensional, the frozen computation gives the actual network values in topological order.
Indeed, each labeled ReLU substitution agrees with \(\max\{0,t\}\) under its sign constraint, and either substitution gives zero when \(t=0\).
Each frozen affine output difference has \(d+1\) coefficients of absolute value at most \(2^{b_0}\).
Thus \(|N(x)-T(x)|\le(d+1)2^{b_0}\) throughout the cube, and its square is bounded.

This squared difference is also measurable.
For \(t\ge0\), its superlevel set is a finite union of activation regions intersected with either \(a(x)>\sqrt t\) or \(a(x)<-\sqrt t\), where \(a\) is the corresponding frozen affine output difference.
For \(t<0\), the superlevel set is the whole cube.
These sets have boundaries in finitely many proper affine hyperplanes, so the box covers above make the upper and lower sums of their indicators agree; in particular, they are measurable for volume.
The boundedness and measurability just established show that lower-dimensional pieces and boundaries do not affect integration.

On a retained \(P\), let \(g\) be the frozen affine output difference.
We enumerate a rational triangulation without storing it.
Retain every original inequality row, including constant rows, for this enumeration.
A nonempty face is uniquely described by its full tight set \(J\): the rows that are equalities throughout the face.
We justify this description and the facet coverage used below.

Every nonempty convex subset of \(\mathbb R^d\) has a relative-interior point.
Choose one point \(v_0\) in the set, then successively add a point outside the current affine span while possible.
Each addition increases the rank of the differences from \(v_0\), so at most \(d\) points are added.
For the resulting family \(v_0,\ldots,v_k\), these difference columns are independent and the affine span is the set's affine hull.
The simplex of these points lies in the set by convexity.

The equal-weight mean \((v_0+\cdots+v_k)/(k+1)\) has all barycentric coordinates equal to \(1/(k+1)\).
For \(k\ge1\), select \(k\) coordinate rows making the \(k\) difference columns a nonsingular square matrix, as proved in \Cref{sec:model}.
Solving this system expresses \(\lambda_1,\ldots,\lambda_k\) as affine functions of the point's coordinates within this affine hull; \(\lambda_0=1-\sum_{i=1}^k\lambda_i\) is affine too.
Each coordinate changes by at most a fixed sum of absolute coefficients times the maximum coordinate change of the point.
Thus all these finitely many positive coordinates remain positive on a sufficiently small relative ball about the mean, placing that ball in the simplex and in the set.
For \(k=0\), the set is the single point \(v_0\), which is its own relative interior.

Choose such a point \(z\) for a nonempty face \(F\), and let \(I\) be the original rows tight at \(z\).
If a row tight at \(z\) were slack at some \(y\in F\), the point \(z+\varepsilon(z-y)\), for sufficiently small \(\varepsilon>0\), would still lie in \(F\), because \(z\) is a relative-interior point, but would violate that row.
Hence every row in \(I\) is tight throughout \(F\).

Conversely, let \(x\in P\) satisfy equality in all rows of \(I\).
For sufficiently small \(\varepsilon>0\), the point \(y=z+\varepsilon(z-x)\) also lies in \(P\): rows in \(I\) stay equalities, and the other finitely many rows have positive slack at \(z\).
Write \(F=\{u\in P:c\cdot u=\alpha\}\), with \(c\cdot u\le\alpha\) on \(P\); for \(F=P\), take \(c=0\), \(\alpha=0\).
Since \(z\) is a convex combination of \(x\) and \(y\) with both coefficients positive, equality \(c\cdot z=\alpha\) forces \(c\cdot x=c\cdot y=\alpha\).
Thus \(x\in F\), proving that the rows in \(I\) describe exactly \(F\) and are its full tight set.
In the other direction, any nonempty intersection obtained by making selected original rows tight is a face: equality in the sum of those valid inequalities forces equality in each selected row.
The empty selection gives \(P\).

We also need that each relative boundary point of a positive-dimensional face lies in a facet.
Work within that face's affine hull, and remove redundant inequalities, meaning rows whose removal leaves the set unchanged.
Every remaining row is nonconstant in this affine hull.
For one such row, nonredundancy supplies a point satisfying all the other rows but violating that row.
Join it to a relative-interior point of the face, where all remaining rows are strict.
Their segment meets the chosen row's equality hyperplane before its violating endpoint, with every other row still strict.
A small neighborhood of this meeting point within that hyperplane still satisfies the other rows.
The row therefore defines a face of dimension one smaller, a facet.

Any point at which every remaining row is strict has a relative open ball inside the face.
Thus every relative boundary point is tight in a remaining row and belongs to its facet.
These facet cuts are faces of \(P\) by the preceding tight-row argument, so they occur among the enumerated faces.
This geometric justification does not change the enumeration's original row set.

For a proposed subset \(J\subseteq\{1,\ldots,M\}\), solve the following linear program in the variables \((x,\delta)\), which we call the center program of \(J\):
\[
\begin{split}
\text{maximize }&\delta,\qquad 0\leq\delta\leq1,\quad x\in P,\\
a_j\cdot x&=b_j\quad(j\in J),\\
a_i\cdot x&\leq b_i-\delta\quad(i\notin J).
\end{split}
\]
A positive optimum occurs exactly when \(J\) is the full tight set of a nonempty face.
Indeed, such a point is in that face and strictly satisfies all other rows; conversely a relative-interior point has a positive minimum of those finitely many slacks.
If there are no other rows, the independent bound \(\delta\leq1\) gives the same test.

For each \(J\) with positive optimum, choose one optimal vertex of this bounded program, with lexicographic tie breaking, and denote its \(x\)-coordinate by \(z_J\).
This choice can be made with polynomially many exact rational linear programs.
First fix \(\delta\) to its computed maximum, then minimize \(x_1\), fix that value, and continue through \(x_d\).
All these objectives are bounded because \(x\in[0,1]^d\) and \(0\le\delta\le1\).

The final point is unique.
If it lay strictly between two different feasible points of the original program, optimality would force both endpoints to have its \(\delta\)-value, then its \(x_1\)-value, and successively every coordinate, a contradiction.
Consider the constraints of the center program of \(J\), without the coordinate-fixing constraints, that are tight at this point.
They contain a full independent square subsystem in the variables \((x,\delta)\).
Otherwise the coefficient vectors of these tight constraints span a proper subspace of \(\mathbb R^{d+1}\), so some nonzero direction \(y\in\mathbb R^{d+1}\) is orthogonal to all of them.
Moving from the final point by \(\varepsilon y\) or \(-\varepsilon y\) keeps every tight constraint tight, and for sufficiently small \(\varepsilon>0\) it keeps every other constraint of the center program satisfied, since those finitely many constraints have positive slack at the final point.
The final point would then be the midpoint of two different feasible points of the center program, which the preceding paragraph excludes.
Those independent tight equations expose a single point, so the final point is a vertex, in the sense defined earlier in this section, of the feasible set of the center program in the variables \((x,\delta)\).

Cramer's rule using these rows of the center program gives a polynomial bit bound for every coordinate.
All values fixed in the successive programs are coordinates of this final point, so their encodings have the same uniform polynomial bound.
There are only \(d+1\) optimization calls.
The point \(z_J\) is a relative-interior point of the face and has polynomial-bit coordinates by Cramer's rule.
The face dimension is \(d-\operatorname{rank}\{a_j:j\in J\}\), since all other inequalities are strict there.
Thus all tests and this choice of point are polynomial-time exact computations.

For a nonempty face \(F\), let \(J(F)\) be its full tight set of original inequality rows and write \(z_F=z_{J(F)}\).
Enumerate complete face flags \(F_0\subsetneq\cdots\subsetneq F_d=P\), where \(\dim F_i=i\), by their full tight sets.
The full tight-row description gives \(F_i\subseteq F_{i+1}\) exactly when \(J(F_{i+1})\subseteq J(F_i)\): inclusion of faces forces every row tight on the larger face to be tight on the smaller, and the reverse row inclusion forces every point of the smaller face into the larger.
Test these row-set inclusions and the prescribed dimensions in polynomial time; the distinct dimensions ensure strict nesting.
A face of dimension \(d\) is \(P\), because every proper face lies in a proper supporting hyperplane.

For each flag form the simplex with vertices \(z_{F_0},\ldots,z_{F_d}\).
These points are affinely independent.
For each \(i\ge1\), the faces \(F_{i-1}\subsetneq F_i\) are distinct, so their full tight sets are distinct, and the row-set inclusion criterion above gives \(J(F_i)\subsetneq J(F_{i-1})\).
Choose a row \(j\in J(F_{i-1})\setminus J(F_i)\).
That row is an equality throughout \(F_{i-1}\), hence at every affine combination of its points, so the affine hull of \(F_{i-1}\) lies in \(\{x:a_j\cdot x=b_j\}\).
The point \(z_{F_i}\) is feasible for the center program of \(J(F_i)\) with a positive value of \(\delta\), and \(j\notin J(F_i)\), so \(a_j\cdot z_{F_i}\le b_j-\delta<b_j\).
Thus \(z_{F_i}\) lies outside the affine hull of \(F_{i-1}\).
That hull contains the preceding points \(z_{F_0},\ldots,z_{F_{i-1}}\), because each \(z_{F_l}\) with \(l<i\) lies in \(F_l\subseteq F_{i-1}\).
Suppose \(z_{F_0},\ldots,z_{F_{i-1}}\) are affinely independent, and take a vanishing linear combination of the differences \(z_{F_l}-z_{F_0}\), \(1\le l\le i\).
A nonzero coefficient on \(z_{F_i}-z_{F_0}\) would place \(z_{F_i}\) in the affine hull of the preceding points, so that coefficient is zero, and the remaining coefficients vanish by the assumed independence.
Induction on \(i\), starting from the single point \(z_{F_0}\), proves the claim.

These simplices triangulate \(P\) up to their boundaries: they cover \(P\) and have pairwise disjoint interiors.

We prove this by induction on the dimension \(k\) of a nonempty face \(F\) of \(P\), with the following assertion.
The simplices with vertices \(z_{F_0},\ldots,z_{F_k}\), over all complete face flags \(F_0\subsetneq\cdots\subsetneq F_k=F\) with \(\dim F_i=i\), cover \(F\) and have pairwise disjoint relative interiors in the affine hull of \(F\).
The case \(F=P\) is the claim.
For the induction step, \(k\ge1\), the induction hypothesis triangulates each facet of \(F\) in this way, using the same selected points on shared subfaces.
Let \(c=z_F\) be the selected relative-interior point of \(F\).
For a facet simplex \(S\), its convex hull with \(c\) is \(\{(1-t)y+tc:y\in S,0\leq t\leq1\}\).
Coning \(S\) to \(c\) means taking this set.
For a positive-dimensional face, a ray from \(c\) in its affine span is a set \(\{c+tv:t\ge0\}\), where \(v\ne0\) is a direction parallel to that span.

The face is closed, bounded and convex, and \(c\) is in its relative interior, so its intersection with this ray is a closed segment from \(c\) to a unique boundary endpoint.
The segment has positive finite length, and an interior endpoint would permit extension.
Every point strictly between \(c\) and the endpoint is in the relative interior: a small ball about \(c\) inside the face scales to a small ball about that point under convex combination with the endpoint.
Thus only the endpoint is on the boundary within the face's affine span.

Each point of the face other than \(c\) lies on one such segment.
By the facet coverage just proved, the facet simplices cover these boundary endpoints and have disjoint relative interiors, consistently on shared subfaces.
Their convex hulls with \(c\) therefore cover the face and have disjoint relative interiors in the face's affine span; two simplex relative interiors could overlap only if the corresponding ray endpoint lay in both facet-simplex relative interiors.
This proves the assertion in every dimension, starting with a single point.

We use affine changes of variables only on bounded polytopes with polynomial integrands.
A coordinate swap preserves the integral after permuting the variables.
Adding a multiple of one coordinate to another translates the corresponding one-dimensional slices and also preserves it.
Scaling one coordinate by a nonzero scalar multiplies slice integrals by the absolute value of that scalar.
Translation has factor one.
Every invertible linear matrix reduces to the identity by these elementary row operations.
The determinant row laws proved in \Cref{sec:model} show that their accumulated volume factor is the absolute determinant of that matrix.

For an invertible \(d\times d\) matrix \(A\), a vector \(b\), a bounded polytope \(Q\) and a polynomial \(h\), define the affine image \(AQ+b=\{Ay+b:y\in Q\}\).
The preceding coordinate substitutions give
\[
\int_{AQ+b}h(x)\,dx
=|\det A|\int_Q h(Ay+b)\,dy.
\]
The standard simplex is \(\Delta_d=\{(\lambda_1,\ldots,\lambda_d):\lambda_i\geq0,\sum_{i=1}^d\lambda_i\leq1\}\), with \(\lambda_0=1-\sum_{i=1}^d\lambda_i\).
The slice and box-sum conventions above justify composing these substitutions.
For a simplex, use \(A=(v_1-v_0,\ldots,v_d-v_0)\), \(b=v_0\), and the standard simplex as \(Q\).
Its columns are independent by the simplex hypothesis.
For \(d=0\), the empty determinant and the point-volume convention are both one.

For simplex vertices \(v_0,\ldots,v_d\), put \(g_i=g(v_i)\), where \(g\) is the frozen affine output difference on the retained polytope.
The integral of \(g^2\) over the simplex is
\[
\frac{|\det(v_1-v_0,\ldots,v_d-v_0)|}{d!}
\frac{(\sum_{i=0}^d g_i)^2+\sum_{i=0}^d g_i^2}{(d+1)(d+2)}.
\]
This formula follows from the moments of the barycentric coordinates under the uniform distribution on the simplex, which we prove below: \(\mathbb E[\lambda_i^2]=2/((d+1)(d+2))\) and \(\mathbb E[\lambda_i\lambda_j]=1/((d+1)(d+2))\) for \(i\ne j\).

For integers \(a,b\geq0\), integration by parts gives
\[
\int_0^1 t^a(1-t)^b\,dt=\frac{a!b!}{(a+b+1)!}.
\]
Indeed, for \(a=0\) the integral is \(1/(b+1)\); for \(a\geq1\), the boundary term vanishes and integration by parts multiplies the integral with exponents \((a-1,b+1)\) by \(a/(b+1)\).
Repeating this \(a\) times gives the displayed factorial ratio.

For nonnegative integers \(k_0,\ldots,k_d\), we now prove
\[
\int_{\Delta_d}\prod_{i=0}^d\lambda_i^{k_i}\,d\lambda
=\frac{\prod_{i=0}^d k_i!}{(d+\sum_{i=0}^d k_i)!}.
\]
For \(d=0\), the single point has zero-dimensional volume one and both sides are one.
For \(d\geq1\), put \(\lambda_d=t\) and \(\lambda_i=(1-t)\mu_i\) for \(0\le i<d\), where \((\mu_1,\ldots,\mu_{d-1})\in\Delta_{d-1}\) and \(\mu_0=1-\sum_{i=1}^{d-1}\mu_i\), so that \(\sum_{i=0}^{d-1}\mu_i=1\).
The substitution of the \(d-1\) free coordinates \(\mu_1,\ldots,\mu_{d-1}\) scales their volume by \((1-t)^{d-1}\).
The induction hypothesis gives the factor \(\prod_{i<d}k_i!/(d-1+\sum_{i<d}k_i)!\).
The remaining integral is
\[
\int_0^1 t^{k_d}(1-t)^{d-1+\sum_{i<d}k_i}\,dt
=\frac{k_d!(d-1+\sum_{i<d}k_i)!}{(d+\sum_{i=0}^d k_i)!},
\]
proving the formula.

At zero exponents the volume is \(1/d!\), so multiplying by \(d!\) converts these integrals to uniform-simplex expectations and gives the two stated moments.
Expanding \((\sum_i\lambda_i g_i)^2\) gives the squared-affine integral formula above.
Determinants, factorials, and affine evaluations here use polynomially many polynomial-bit rationals, so each contribution has polynomial bit length.

The following chain of explicit bounds makes this precise: it ends with one polynomial that bounds the numerator and denominator lengths of every contribution, so each leaf output is a rational of polynomial bit length.
Let \(b_0=4(L+1)^3\) bound frozen numerator and denominator lengths and set
\[
\begin{split}
b_1&=\lceil\log_2((d+1)!)\rceil+(d+1)(d+4)b_0+2,\\
b_2&=4(d+2)(b_0+b_1+1),\\
b_3&=10(d+2)^2(b_2+1),\qquad
b_{\mathrm{vol}}=10(d+2)^3(b_1+1).
\end{split}
\]
Write \(n=d+1\).
A row of the center program of a subset \(J\), displayed earlier in this proof, has \(d+2\) rational entries, including its right-hand side.
Clearing their denominators bounds each integer entry in absolute value by \(2^{(d+2)b_0}\).
The determinant expansion therefore bounds each Cramer numerator and denominator by \(n!2^{n(d+2)b_0}\), whose binary length is smaller than \(b_1\).

An affine evaluation \(g(z_J)\) is a sum of \(n\) terms: the constant coefficient of \(g\), and each other coefficient of \(g\) times the corresponding coordinate of \(z_J\).
Each term has numerator and positive denominator length at most \(b_0+b_1\).
Putting the \(n\) terms of an affine evaluation over the product denominator gives numerator length at most
\[
n(b_0+b_1)+\lceil\log_2 n\rceil+2
\]
and denominator length at most \(n(b_0+b_1)+1\).
Both bounds are smaller than \(b_2=4(n+1)(b_0+b_1+1)\).

Next put the \(n\) values \(g_i\) over their product denominator.
The squared-value bracket \((\sum_i g_i)^2+\sum_i g_i^2\) then has numerator length at most \(2nb_2+\lceil\log_2(n^2+n)\rceil+2\) and denominator length at most \(2nb_2+1\).
Both bounds are smaller than \(b_3=10(n+1)^2(b_2+1)\).

For \(d\ge1\), each coordinate difference has numerator magnitude below \(2^{2b_1+1}\) and positive denominator below \(2^{2b_1}\).
Clear all \(d^2\) such denominators in the volume determinant and divide by \(d!\).
The determinant expansion gives respective numerator and denominator length bounds
\[
2d^3b_1+d+\lceil\log_2(d!)\rceil+2,
\qquad
2d^3b_1+\lceil\log_2(d!)\rceil+1.
\]
They are smaller than \(b_{\mathrm{vol}}\).
For \(d=0\), the volume is one and obeys the same bound.

Multiplying the volume by the squared-value bracket and dividing by \(n(n+1)\) therefore shows that
\[
b_{\mathrm{vol}}+b_3+\lceil\log_2((d+1)(d+2))\rceil+4
\]
bounds each contribution's numerator and positive denominator length.

An activation pattern together with the \(d+1\) tight sets of a face flag is a binary word of length \(p+(d+1)M\).
Use a fixed polynomial bound on this length and accept only zero padding after these fields, so each flag is counted once.
Every invalid word contributes \(0/1\).
The valid contributions sum to the risk by the preceding partition and triangulation arguments.
\end{proof}

The hardness direction uses narrow triangular bumps whose widths encode reciprocals.
The folding itself is classical \cite[Lemmas 3.10--3.11]{Telgarsky2016}; we give the exact weighted construction and its integral.

\begin{proof}[Proof of \Cref{thm:continuous-risk}]
Membership of fixed comparison follows from \Cref{lem:risk-leaves}: negate each risk leaf and compare their sum with \(-\tau\) in the definition of \(\mathsf{PP}_{\mathbb Q}^{(pt)}\).
A guessed legal synthesis graph has polynomial description length, so one such oracle query gives the synthesis upper bound \(\mathsf{NP}^{\mathsf{PP}_{\mathbb Q}^{(pt)}}\).

For hardness, fix a language represented by \(F(a)\geq t(a)\) as above and one input \(a\).
Choose a polynomial integer \(q\geq1\) such that \(|u(a,w)|\leq2^q\) and \(1\leq v(a,w)\leq2^q\).
Put \(K=2^q\), \(b=2q+2\), and
\[
U(w)=Kv(a,w)-u(a,w),\qquad 0\leq U(w)\leq2^{2q+1}<2^b.
\]
With \(a\) fixed, write \(v(w)=v(a,w)\).
Boolean circuits of polynomial size compute the digits of \(v(w)-1\) and the predicate \(j<U(w)\), where \(j\) is a \(b\)-bit unsigned integer.
Such circuits are obtained by unrolling the fixed polynomial-time weight evaluator and polynomial-bit integer arithmetic.
Let \(n=p(|a|)+b\geq1\), so that an \(n\)-bit word encodes the pair \((w,j)\).
No padding of the source counting word is needed.

Write
\[
\theta(t)=\rho(2t)-2\rho(2t-1)+\rho(2t-2).
\]
The power \(\theta^i\) denotes iteration: \(\theta^0(t)=t\) and \(\theta^{i+1}=\theta\circ\theta^i\).
Let \(z_i=\theta^i(x)\), put \(\eta=2^{-n}\), and define \(h=\eta\theta^{n+1}(x)\).
For \(0\leq i<n\), set \(s_i=2z_i-1\).
The \(i\)th itinerary bit is \(\mathbf1_{\{s_i>0\}}\), so a zero value of \(s_i\) gives bit zero.

For a real \(x\) and a nonempty set \(A\subseteq\mathbb R\), write \(\operatorname{dist}(x,A)=\inf_{a\in A}|x-a|\).
The map \(\theta\) is zero outside \([0,1]\) and has the two branches \(2t,2-2t\) inside.
The itinerary bit is one exactly when \(z_i>1/2\), so for \(z_i\in[0,1]\) it records that the next iterate \(z_{i+1}=\theta(z_i)\) is computed on the branch \(2-2t\).
Composing its inverse branches \(t/2,1-t/2\) shows that the \(2^n\) open dyadic cells of width \(\eta\) realize all strict sign words of \((s_0,\ldots,s_{n-1})\) exactly once.
On each cell \(h\) has slopes \(2,-2\) and is twice the distance to its endpoints.
The zero set of \(s_i\) consists of the odd multiples of \(2^{-(i+1)}\), all of which are fine-cell endpoints.
On the unit interval,
\[
|s_i(x)|=2^{i+1}\operatorname{dist}(x,\{s_i=0\})\geq h(x).
\]

Set \(r=1-\theta^{n+1}(x)\), and let \(\epsilon=\eta/4\).
For each itinerary coordinate use the clipped ramp
\[
\beta_i(x)=\rho((s_i+\epsilon)/(2\epsilon))-
\rho((s_i-\epsilon)/(2\epsilon)).
\]
It lies in \([0,1]\) everywhere.
Where \(r\leq1/2\), which is the central half of each fine cell, we have \(h\geq\eta/2=2\epsilon\), so each \(\beta_i\) is exactly its Boolean itinerary bit.

Compile the source circuit with \(\operatorname{AND}(u,v)=\rho(u+v-1)\) and \(\operatorname{NOT}(u)=\rho(1-u)\).
Every gate preserves \([0,1]\) on fractional inputs, while on the central half of a fine cell, where \(r\le1/2\), all gates compute the correct Boolean values.
For \(0\leq k<q\), let \(c_k\in[0,1]\) be the compiled circuit's outputs for the binary digits of \(v(w)-1\), and let \(A\in[0,1]\) output the predicate \(j<U(w)\).
On Boolean circuit inputs the convention is \(v(w)-1=\sum_{k=0}^{q-1}2^kc_k\), with \(c_k\in\{0,1\}\).

Define
\[
P=r+\sum_k2^k\min\{r,c_k\},\qquad
B=\rho(1-2P),\qquad T=\min\{B,A\}.
\]

Since \(P\geq r\), the function \(B\) is zero wherever \(r\geq1/2\).
Within the central half, \(c_k\) is Boolean, so \(\min\{r,c_k\}=r c_k\) and \(P=v(w)r\).
There \(A\) is either zero or one.
It follows on the entire cell, including its boundary, that
\[
T(x)=
\begin{cases}
\rho(1-2v(w)r(x)),&j<U(w),\\
0,&j\geq U(w).
\end{cases}
\]
Both sides vanish outside the central half.
Outside \([0,1]\), \(r=1\) and the same vanishing holds.
Thus no fractional decoding region contributes to the integral.

A selected cell has a triangular bump of height one and slope magnitude \(4v(w)/\eta\) on each side.
Each side's squared integral is its height cubed divided by three times that magnitude.
The two sides therefore contribute \(\eta/(6v(w))\).
There are exactly \(U(w)\) selected integers \(j\) for each \(w\), giving
\[
\int_0^1 T(x)^2\,dx
=\frac{2^{-n}}6\sum_w\frac{U(w)}{v(w)}
=\frac{2^{-n}}6\left(K2^{p(|a|)}-F(a)\right).
\]
Comparison with \(\tau=2^{-n}(K2^{p(|a|)}-t(a))/6\) is consequently equivalent to \(F(a)\geq t(a)\), including equality.
All threshold data have polynomial bit length.
If only nonnegative risk thresholds are allowed and this threshold is negative, \(F\leq K2^p<t\), so send the instance to the fixed NO instance \(T=1,\tau=0\).

All displayed functions have polynomial-size ordinary ReLU realizations with dyadic parameters.
For each fold, create its three hinge gates and a ReLU of their nonnegative affine sum.
The nonnegative quantity \(r\) can similarly be a ReLU of \(1-z_{n+1}\).
Use \(\min(a,b)=a-\rho(a-b)\), storing positive and negative parts of signed intermediate quantities when needed.
All circuit outputs and final quantities declared nonnegative may be passed through a ReLU unchanged.

Powers of two used in the slopes and clipped ramps have polynomial-bit exponents.
Repeated predecessors in an affine expression have their coefficients combined, so no parallel edges are required.
Remove unused target vertices; a constant zero vertex, if needed, has a present zero bias.

The resulting target satisfies the conditions of \Cref{def:network} other than the computational-vertex bound and the inventory, lies in \([0,1]\), and is supplied independently of the candidate inventory.
Taking \(W=(0)\) and cap one forces the candidate to be zero under either inventory rule, and a zero output bias supplies a legal witness.
This proves both comparison completeness and synthesis hardness.

Finally, apply \Cref{thm:rational-weight-import} to a fixed-candidate comparison.
Let \(s\) bound a padded candidate certificate's length.
Amplify its \(\mathsf{BPP}^{\mathsf{PP}}\) algorithm to error at most \(2^{-s-3}\) using polynomially many independent repetitions.
A union bound over all \(2^s\) certificates makes the same random tape correct for every legal candidate with probability at least \(7/8\).
For a fixed tape, asking whether some legal certificate accepts is an \(\mathsf{NP}^{\mathsf{PP}}\) query.
This gives the stated randomized synthesis upper bound.
The containment in \(\mathsf{PSPACE}\) also follows directly.
Polynomial-time nondeterministic or randomized paths can be enumerated with polynomial-size counters, and a \(\mathsf{PP}\) query can be evaluated by enumerating its polynomial-length random paths and counting acceptances in polynomial space.
Reusing this space through the nested oracle simulations preserves a polynomial space bound.
\end{proof}

This result distinguishes fixed-network comparison from existential synthesis.
It does not show that continuous-risk synthesis is complete for its upper-bound class.
It also does not reduce exact comparison to an additive approximation with only polynomially many precision bits.

\subsection{The Length of the Exact Rational Value}

A threshold comparison need not print the value it compares.
Lawrence \cite{Lawrence1989} already exhibited rational polytopes with compact descriptions and volumes whose reduced denominators have no polynomial bit bound.
The following construction gives an explicit one-dimensional squared-risk counterpart with dyadic network parameters.

\begin{theorem}[Long risk denominators]
\label{thm:risk-denominator}
For every sufficiently large integer \(n\), an explicit dyadic ReLU network \(T_n:\mathbb R\to[0,1]\) has \(O(n\log n)\) binary encoding length, while the reduced denominator of\linebreak \(\int_0^1 T_n(x)^2\,dx\) has \(\Omega(2^n)\) bits.
Thus no polynomial in network encoding length bounds all such exact risk denominators, even with \(W=(0)\) and cap one.
\end{theorem}

The number-theoretic input is the following established asymptotic consequence of explicit prime estimates.
Write \(\pi(t)\) for the number of primes at most \(t\), and \(\log\) for the natural logarithm in this estimate.

\begin{theorem}[Dusart {\cite[Theorem 6.9, inequality (6.7)]{Dusart2010}}]
\label{thm:prime-estimate}
As \(t\to\infty\),
\[
\pi(t)=\frac{t}{\log t}+\frac{t}{\log^2t}
+O\left(\frac{t}{\log^3t}\right).
\]
\end{theorem}

The cited bounds are unconditional.
We use only their asymptotic form and require no explicit finite starting value of \(n\).

\begin{proof}[Proof of \Cref{thm:risk-denominator}]
Use the tent map \(\theta\) from the proof of \Cref{thm:continuous-risk}.
Put
\[
\eta=2^{-n},\quad B_n(x)=\sum_{i=1}^n\theta^i(x),\quad
a=4\cdot2^{\lceil\log_2(n+1)\rceil},\quad c=n+1,\quad C=2a,
\]
and define
\[
T_n(x)=C^{-1}\rho\left(a\theta^{n+1}(x)+B_n(x)-c\right).
\]
Both \(a,C\) are powers of two, with \(4(n+1)\leq a<8(n+1)\).
On the unit interval, \(0\leq B_n\leq n\) and \(0\leq\theta^{n+1}\leq1\), so \(0\leq T_n\leq(a-1)/C<1\).
Outside the interval all the positive iterates vanish, hence \(T_n=0\).

Each fold uses four ReLUs as in the preceding proof.
Its output is shared by the next fold and the final sum, with all parameter positions explicitly listed.
The final hidden preactivation uses the \(n\) coefficients one, the coefficient \(a\), and bias \(-c\); its affine output uses \(C^{-1}\).
There are \(4(n+1)+2\) computational vertices and \(O(n)\) positions, no parallel edges, and every vertex reaches the output.
The parameter lengths and vertex-index lengths are \(O(\log n)\), giving the asserted encoding bound.

For each integer \(0\leq j<2^n\), let \(I_j=(j\eta,(j+1)\eta)\).
On this open cell the function \(B_n\) is affine with slope
\[
s_j=\sum_{i=1}^n\varepsilon_{i,j}2^i,\qquad \varepsilon_{i,j}\in\{-1,1\}.
\]
Every sign vector occurs once.
Indeed if the tent branch signs are \(\beta_1,\ldots,\beta_n\), then \(\varepsilon_i=\beta_1\cdots\beta_i\); conversely \(\beta_1=\varepsilon_1\) and \(\beta_i=\varepsilon_i\varepsilon_{i-1}\).
The inverse-branch cell correspondence is bijective, so the numbers \(s_j/2\) are precisely all odd integers from \(-(2^n-1)\) to \(2^n-1\).
The identity \(\theta^i(1-x)=\theta^i(x)\) for \(i\geq1\) also shows that reflected cells have opposite slopes and equal midpoint values.

At a fine-cell endpoint the last preactivation is \(B_n-c\leq-1\).
At its midpoint it is at least \(a-c>0\).
Write
\[
u_j=c-B_n(j\eta),\qquad D=a2^{n+1},\qquad
H_j=a-u_j+s_j\eta/2,\qquad p_j=(D+s_j)/2.
\]
Here \(1\leq u_j\leq n+1\), and \(H_j\) is the positive midpoint height before multiplication by \(C^{-1}\).
The two half-cell slopes are \(D+s_j\) and \(-(D-s_j)\), both with positive magnitudes.
There is therefore exactly one zero in each half, and the positive portions are two linear triangles of height \(H_j\).
Their squared integrals give
\[
I_n=\int_0^1T_n^2
=\frac1{3C^2}\sum_j H_j^3
\left(\frac1{D+s_j}+\frac1{D-s_j}\right)
=\frac1{3C^2}\sum_j\frac{H_j^3}{p_j}.
\]
For the last equality, reindex the terms \(H_j^3/(D-s_j)\) by the reflection \(x\mapsto1-x\), which permutes the cells.
Reflected cells have opposite slopes and equal midpoint values, hence equal heights \(H_j\), so these terms sum to \(\sum_jH_j^3/(D+s_j)\).
The sum over \(j\) is therefore \(2\sum_jH_j^3/(D+s_j)\), which equals \(\sum_jH_j^3/p_j\) because \(D+s_j=2p_j\).

Set \(U_j=2^{n+1}u_j\) and \(E_j=2^{n+1}H_j\).
Iterate values at dyadic endpoints have denominators dividing \(2^n\), so these are integers, and
\[
E_j=2p_j-U_j,\qquad
I_n=\frac1{3C^2\cdot2^{3(n+1)}}\sum_j\frac{E_j^3}{p_j}.
\]
Moreover,
\[
0<U_j\leq2^{n+1}(n+1)<(a-1)2^n<p_j.
\]
The \(p_j\) run once through all odd integers in \(((a-1)2^n,(a+1)2^n)\).

Let \(p>3\) be a prime in that interval.
Exactly one denominator \(p_j\) equals \(p\).
Every other denominator is less than \(2p\) and differs from \(p\), because the ratio of the interval endpoints is \((a+1)/(a-1)<2\).
None is divisible by \(p\).
The corresponding numerator obeys \(E_j\equiv-U_j\not\equiv0\pmod p\), and since \(p\) is prime, \(p\) does not divide \(E_j^3\).
Combining the other terms over a denominator coprime to \(p\) therefore cannot cancel this term's factor \(p\) in the denominator.
The common prefactor has denominator divisible only by two and three, so it cannot cancel it either.
Thus every prime in the interval divides the reduced denominator of \(I_n\).

To count them, put \(X=a2^n\), \(Y=2^n\), and \(F(t)=t/\log t+t/\log^2t\).
In this proof, \(f=\Theta(g)\) means both \(f=O(g)\) and \(f=\Omega(g)\), and \(o(1)\) denotes a quantity tending to zero as \(n\to\infty\).
Then \(\log X=\Theta(n)\), \(a=\Theta(n)\), and
\[
F'(t)=\frac1{\log t}-\frac2{\log^3t}.
\]
Uniformly for \(X-Y\leq t\leq X+Y\), the derivative is \((1+o(1))/\log X\), since \(Y/X=1/a\to0\).
By \Cref{thm:prime-estimate},
\[
\pi(X+Y)-\pi(X-Y)
=(1+o(1))\frac{2Y}{\log X}
+O\left(\frac{X}{\log^3X}\right)
=\Omega(2^n/n).
\]
The error relative to \(2Y/\log X\) is \(O(a/\log^2X)=O(1/n)\), so it does not remove the positive main term.

Each counted prime has \(\Theta(n)\) binary logarithm.
Their product, which divides the reduced denominator, consequently has \(\Omega(2^n)\) bits.
The interval endpoints are even, so endpoint conventions do not change the count for large \(n\).
This proves the claimed output-size bound.
\end{proof}

The exponential bound is in the construction parameter \(n\); the ordinary indexed encoding has length \(O(n\log n)\).
The conclusion is a failure of every polynomial output-length bound, not a proof that threshold comparison lies outside \(\mathsf{PP}\) or another decision class.

\subsection{Synthesis with a Binary Accuracy Gap}
\label{sec:continuous-gap-synthesis}

By \Cref{prop:binary-gap-risk}, synthesis with a binary accuracy gap has a separator in \(\mathsf{NP}^{\mathsf{PP}}\), and the matching hardness holds already for continuous uniform inputs on the cube.
The reduction below supplies an integer inventory for which every legal candidate under cap two can be analyzed, whatever role each occurrence plays in it.
Balanced corners, with equally many zeros and ones, give all intended corner functions the same bias, so the inventory needs only one bias occurrence.
This leaves no second bias occurrence for a candidate to reassign to a steeper input edge.

For a prescribed input distribution, a binary accuracy gap supplies a rational threshold \(\tau\) and a positive rational gap \(\delta\), both in binary.
The synthesis promise is YES if some legal candidate has squared risk at most \(\tau\), and NO if every legal candidate has risk at least \(\tau+\delta\).
These are disjoint promised input sets; inputs in neither set have no required answer.
No unary encoding of \(1/\delta\) and no bounded-loss assumption are imposed.
A separator is an ordinary language containing every promised YES instance and no promised NO instance.
A polynomial-time many-one promise reduction maps promised YES inputs to YES inputs and promised NO inputs to NO inputs.

\begin{theorem}[Continuous uniform synthesis with a binary gap; restatement of \Cref{ithm:continuous-gap-complete}]
\label{thm:continuous-gap-complete}
For continuous uniform inputs on \([0,1]^d\), the squared-risk synthesis promise problem of \Cref{prop:binary-gap-risk} is \(\mathsf{NP}^{\mathsf{PP}}\)-complete under deterministic polynomial-time many-one promise reductions, under either inventory rule.
Hardness holds with integer inventory and computational-node cap two, including the affine output.
The target is an independently supplied rational ReLU network and the positive gap is encoded in ordinary binary.
\end{theorem}

Membership follows from \Cref{prop:binary-gap-risk}; we prove hardness.
The dimension in the reduction is explicit, and may include auxiliary coordinates that every competing graph is permitted to use.
We write \(\|f\|_2^2=\int f^2\) for uniform-cube integrals in the dimensions in question.

\begin{lemma}[Balanced corner inventory]
\label{lem:balanced-corner-inventory}
Let \(e\ge2\) be an integer, and set \(D=2e\), \(K=D+2\), \(M=2^{10K^2}\), and \(b=1-eM\).
Supply \(e\) occurrences of \(M\), \(e\) occurrences of \(-M\), one occurrence of \(b\), and one occurrence of \(1\), with cap two.
On \([0,1]^D\) define
\[
U(x)=\rho\left(1-M\sum_{i=1}^D\min\{x_i,1-x_i\}\right).
\]
For every corner \(a\in\{0,1\}^D\), define
\[
h_a(x)=\rho\left(1-M\sum_{i=1}^D|x_i-a_i|\right),
\]
and write \(|a|=\sum_{i=1}^D a_i\) for its Hamming weight.
Then the following two statements hold.
First, for any integer \(d_{\mathrm{aux}}\ge0\), additional independent uniform coordinates \(z\in[0,1]^{d_{\mathrm{aux}}}\), legal at-most-once candidate \(N(x,z)\), and measurable \(\mu(x)\) with \(0\le\mu\le U/2\),
\[
\int_{[0,1]^{D+d_{\mathrm{aux}}}}\left(N^2-2N\mu\right)\,dx\,dz\ge0.
\]
Second, if \(|a|=e\), then \(h_a\) agrees on \([0,1]^D\) with a legal candidate consuming every supplied occurrence.
\end{lemma}
\begin{proof}
Here the positive support of a nonnegative function \(h\) means the set \(\{x:h(x)>0\}\).
For \(M\ge4\), the positive supports of the \(h_a\), over all corners, are disjoint and \(U=\sum_a h_a\), with \(0\le U\le1\).
The elementary simplex integral is
\[
\int_{\substack{y_i\ge0\\\sum_{i=1}^j y_i\le1}}
(1-\textstyle\sum_i y_i)^q\,dy=\frac{q!}{(j+q)!}
\qquad(j,q\in\mathbb Z_{\ge0}).
\]
For \(j=0\) both sides equal one.
For larger \(j\), integrating one coordinate and rescaling the remaining simplex proves the formula inductively, using \(\int_0^1(1-t)^{j+q-1}\,dt=1/(j+q)\).
Consequently, with
\[
I_j=\frac{2}{(j+2)!M^j},\qquad E_0=2^D I_D,
\]
we have \(\|h_a\|_2^2=I_D\) and \(\|U\|_2^2=E_0<1/256\).

On the cube, \(|x_i-a_i|\) equals \(1-x_i\) if \(a_i=1\) and \(x_i\) if \(a_i=0\).
Hence, if \(|a|=e\), then \(1-M|a|=1-eM=b\) and
\[
h_a(x)=\rho\left(b+M\sum_{i:a_i=1}x_i-M\sum_{i:a_i=0}x_i\right).
\]
This preactivation has \(e\) positive \(M\)-edges, \(e\) negative \(M\)-edges and bias \(b\); the output edge uses the unit occurrence.
It has two legal computational vertices and uses all \(m=D+2\) occurrences.

We now examine every competing graph.
A candidate with no hidden vertex is integer-affine.
Otherwise it has exactly one hidden vertex and can be written
\[
N=A+g\rho(\ell),
\]
where \(A\) is the direct affine output part, \(\ell\) is the hidden integer-affine preactivation, and \(g\) is its nonzero output-edge occurrence.
Both formal affine branches \(A\) and \(A+g\ell\) have integer coefficients.

Suppose neither branch is identically zero.
The candidate equals \(A\) on the halfspace \(\{\ell\le0\}\) and \(A+g\ell\) on \(\{\ell>0\}\), and one of these two halfspaces has cube measure at least \(1/2\).
On it, \(N\) equals a nonzero integer-affine form \(Q=c+\sum_{i=1}^{D+d_{\mathrm{aux}}}\alpha_i\xi_i\), where \(\xi=(x,z)\) and \(c,\alpha_i\) are integers.
If \(Q\) is constant, its magnitude is at least one.
Otherwise choose \(j\) with \(\alpha_j\ne0\); then \(|\alpha_j|\ge1\), and conditioning on the other coordinates gives
\[
\Pr[|Q|<1/8]\le1/4.
\]
Indeed the admissible interval in that coordinate has length at most \(1/(4|\alpha_j|)\).
Thus the integral of \(Q^2\) on any set of measure at least \(1/2\) is at least \((1/2-1/4)(1/8)^2=1/256\).
This proves \(\|N\|_2\ge1/16\), also for any nonzero integer-affine candidate with no hidden vertex.

Cauchy--Schwarz then gives
\[
\int(N^2-2N\mu)\ge\|N\|_2^2-\int|N|U
\ge\|N\|_2(\|N\|_2-\|U\|_2)\ge0.
\]
A zero candidate is immediate.
This argument includes halfspaces of measure zero or one.

If a formal branch is identically zero, then \(N=g\rho(q)\), where \(q=\ell\) if \(A=0\), and \(q=-\ell\) if \(A+g\ell=0\).
The latter identity follows from \(\rho(\ell)-\ell=\rho(-\ell)\); the sign of \(g\) is unchanged.
For \(g<0\), the candidate is nonpositive and soundness holds pointwise.
The only positive occurrences in the inventory are \(M\) and \(1\), since \(-M\) and \(b=1-eM\) are negative, so only \(g=1\) or \(g=M\) remains.
Consider first the case \(q=-\ell\), in which the active branch \(A+g\ell\), the formula for \(N\) where \(\ell>0\), is identically zero.
Then \(A=-g\ell\), so each nonzero hidden coefficient \(a\), including a bias, requires the output coefficient \(-ga\) on the same input or as the output bias.
That output coefficient is a single inventory occurrence, because \Cref{def:network} allows at most one edge for each ordered pair of vertices and each vertex has at most one bias.
For \(g=M\), a hidden coefficient of magnitude at least \(M\) would require an output coefficient of magnitude at least \(M^2>eM\), which is unavailable in the inventory.
Only the sole unit occurrence remains; then \(\ell\) is either an input coordinate or the constant one and \(\rho(-\ell)=0\) on the cube.
For \(g=1\), the hidden value \(b\) would require the output coefficient \(-b=eM-1\), which is absent from the inventory.
All hidden coefficients are therefore multiples of \(M\), and a positive vertex maximum of \(q=-\ell\) is at least \(M\).

Consider next \(A=0\), and write \(B=\max\ell\), attained at a cube vertex.
For \(g=1\), the unit occurrence is already the gain.
Without bias \(b\), the \(b\)-occurrence is either unused or a negative input coefficient, contributing zero to the vertex maximum; all other contributions to the maximum are multiples of \(M\).
Thus \(B>0\) implies \(B\ge M\).
With bias \(b\), let \(p\le e\) be the number of positive input edges.
Then \(B=1-M(e-p)\), positive only when \(p=e\).
In that case \(B=1\).
Write \(J\) for the set of coordinates selected by the hidden vertex and \(k=|J|\), so \(e\le k\le D\).
For \(j\in J\), the distance \(\operatorname{dist}_j\) is the selected coordinate or one minus that coordinate, and exactly \(e\) distances are of the latter type.
The candidate is then
\[
N=\rho\left(1-M\sum_{j\in J}\operatorname{dist}_j\right).
\]
Selected coordinates may be original or auxiliary.

For \(g=M\), one positive \(M\) occurrence is consumed.
With bias \(b\), the vertex maximum \(B\) is \(b\) plus the sum of the positive input coefficients.
Those are at most the \(e-1\) remaining occurrences of \(M\) and possibly the unit, so
\[
B\le(1-eM)+(e-1)M=1-M,
\]
or \(B\le2-M<0\) if the unit is also an input coefficient.
Hence a nonzero candidate has no such bias.
Its maximum is a multiple of \(M\) or a multiple plus one.
If \(0<B<M\), then \(B=1\).
The unit is either the bias, with no positive \(M\)-input, or an input coefficient.
In the latter case there are no positive \(M\)-inputs if the bias is absent, or exactly one if the bias is \(-M\).
Counting the remaining negative \(M\) occurrences and the optional negative \(b\)-input gives at most \(e+2\le D\) nonzero input coordinates in every such case.
In the bias-\(-M\) case one negative occurrence has already been consumed by that bias.

For the following norm estimates, write \(B=\max q\), with the maximum taken over the full input cube.
We first show that every case with \(B\ge M\) has a norm large enough for the Cauchy--Schwarz estimate above; the cases with \(B=1\) follow.
Every hidden input coefficient has magnitude at most \(eM\), with at most \(K\) nonzero input coordinates.
If \(B\ge M\), a one-sided cube of side \(r=1/(2K(e+1))\) at a maximizing vertex has \(q\ge M/2\).
It follows that
\[
\|g\rho(q)\|_2^2\ge M^2r^K/4\ge1/256.
\]
The final inequality follows from
\(\log_2((2K(e+1))^K)\le3K^2\) and \(M^2=2^{20K^2}\).

For \(g=M,B=1\), there are at most \(D\) input coordinates.
The same argument with side \(1/(2D(e+1)M)\) gives
\[
\|N\|_2^2\ge
\frac{M^{2-D}}{4(2D(e+1))^D}\ge E_0.
\]
The last inequality follows from
\(M^2\ge2^{D+3}(2D(e+1))^D\), whose right-hand side has binary logarithm at most \(4K^2\).
Fewer or no input coordinates only improve these weaker bounds.
Cauchy--Schwarz proves soundness in both cases.

Only the case \(g=1\), \(B=1\) remains, in which the candidate is \(N=\rho(1-M\sum_{j\in J}\operatorname{dist}_j)\), as displayed above.
Let \(s\) selected coordinates be original and \(t\) be auxiliary, so \(s+t=k\le D\), and put \(l=D-s\).
The squared norm of \(N\) is \(I_k\).
For each selected original coordinate \(i\), the distance \(\operatorname{dist}_i\) equals \(|x_i-a_i|\) for one value \(a_i\in\{0,1\}\), its corner sign.
Since \(U=\sum_ah_a\), the product \(NU\) can be nonzero only on the positive support of a summand \(h_a\) whose corner \(a\) has these corner signs on the selected original coordinates.
Writing \(V=\rho(1-M\sum_{i=1}^s\operatorname{dist}_i)\), integration of \(N\) over the selected auxiliary coordinates gives \(V^{t+1}/((t+1)!M^t)\).
Call the \(l\) original coordinates that \(N\) does not select missing.
Integration of \(U\) over the missing coordinates gives \(2^lV^{l+1}/((l+1)!M^l)\), one contribution for each choice of the missing corner signs.
Since \(N\) does not depend on the missing coordinates and \(U\) does not depend on the auxiliary ones, \(\int NU\) is the integral, over the \(s\) selected original coordinates, of the product of these two partial integrals.
Applying the simplex integral to this product, \(2^lV^{t+l+2}/((t+1)!(l+1)!M^{t+l})\), gives
\[
C:=\int NU=
\frac{2^{D-s}(t+D-s+2)!}
{(t+1)!(D-s+1)!(t+D+2)!M^{t+D}}.
\]

If \(s=D,t=0\), then \(C=I_D=\|N\|_2^2\).
Otherwise \(l\ge1\), and
\[
\frac C{I_k}=
\frac{2^{l-1}(t+l+2)!(s+t+2)!}
{(t+1)!(l+1)!(D+t+2)!}M^{-l}
\le\frac{2^{3D+1}}M\le\frac12.
\]
Here \((s+t+2)!\le(D+t+2)!\).
The remaining factorial ratio is the binomial coefficient \(\binom{t+l+2}{t+1}=(t+l+2)!/((t+1)!(l+1)!)\), which counts subsets of size \(t+1\) in a set of size \(t+l+2\).
It is at most \(2^{t+l+2}\), the number of all subsets of that set.
Since \(N\ge0\) and \(2\mu\le U\), we have \(2N\mu\le NU\) pointwise, so
\[
\int(N^2-2N\mu)\ge\|N\|_2^2-\int NU=I_k-C.
\]
In the case \(s=D\), \(t=0\), the right side is exactly zero, which the lemma permits, since it asserts only a nonnegative integral.
Otherwise \(C\le I_k/2\), and the right side is at least \(I_k/2>0\).
This proves the required nonnegative integral.
All possible affine branches, gains and occurrence roles have been covered.
\end{proof}

We next compile the probabilistic source into an actual continuous target.
The discontinuous rounded variables used in the analysis are not supplied as network operations.
\begin{proof}[Proof of \Cref{thm:continuous-gap-complete}]
Take the strict-majority source of \Cref{thm:emajsat}, with existential bits \(w\in\{0,1\}^e\) and random bits \(z\in\{0,1\}^n\).
Adding ignored variables, we may assume \(e\ge2\) and \(n\ge1\).
Replace its existential encoding by \(a=(w,1-w)\in\{0,1\}^{D}\), where \(D=2e\).
Let \(\psi(a,z)\) assert the original formula on the first \(e\) bits and all complementary-pair tests.
Put \(p(a)=\Pr_z[\psi(a,z)=1]\).
Every assignment with positive acceptance probability is balanced.
A YES source has some \(p(a)>1/2\), whereas a NO source has \(0\le p(a)\le1/2\) for every \(a\).
Use precisely the inventory and cap of \Cref{lem:balanced-corner-inventory}.

Represent each source random bit and one further bit \(u\) by the parity of its own block of \(m+1\) fresh uniform continuous coordinates rounded at \(1/2\), where \(m=D+2\) is the number of inventory occurrences.
There are \(S=(n+1)(m+1)\) share coordinates and total dimension \(d=D+S\).
For analysis let \(a(x)\) round the original coordinates, let
\(F^*=u\lor\psi(a(x),z)\), and set \(T^*=U(x)(2F^*-1)\).
A candidate reads at most \(m\) distinct share coordinates because each input edge consumes an occurrence.
Condition on the original \(x\) and on those read continuous values.
Each parity block retains an independent unobserved uniform coordinate, whose rounded bit is fair.
Its parity is therefore fair independently of all other blocks and of the conditioning.
This is the independent-block argument of \Cref{lem:parity-replacement}, applied directly to the rounded unseen coordinates.

Consequently
\[
\mathbb E[T^*| x,\text{read share values}]=U(x)p(a(x)),
\qquad (T^*)^2=U(x)^2.
\]
In a NO instance, \Cref{lem:balanced-corner-inventory} gives \(R_{T^*}^{\mathrm{cont}}(N)\ge E_0\) for every legal candidate.
In a YES instance use the balanced witness \(h_a\).
Its support lies at the selected corner and has squared norm \(I_D\), so
\[
R_{T^*}^{\mathrm{cont}}(h_a)=E_0+(1-2p(a))I_D\le E_0-2\eta,
\qquad \eta=I_D2^{-n},
\]
since a strict majority exceeds one half by at least \(2^{-n}\).

Let \(0<\epsilon\le1\) be a rational transition width, to be chosen below.
To obtain the actual target, decode each original coordinate by
\(\operatorname{clip}(2x_i-1/2)\), where \(\operatorname{clip}(t)=\min\{1,\max\{0,t\}\}\).
This equals the corner bit wherever \(U>0\), since those coordinates are within \(1/M\le1/4\) of a cube corner.
Decode a share coordinate \(y\) by
\[
\operatorname{clip}\left((y-1/2)/\epsilon+1/2\right).
\]
It equals the rounded bit outside a transition interval of length \(\epsilon\).
Compile the parities and \(F=u\lor\psi\) by the AND/NOT ReLU identities in \Cref{thm:population-risk}, decomposing OR and XOR first.
Every intermediate value remains in \([0,1]\), and agrees with its Boolean value outside the share transitions wherever \(U>0\).

Supply the target
\[
T=2\rho(U+F-1)-U.
\]
This has a polynomial-size rational ReLU graph.
Indeed \(\min\{x_i,1-x_i\}=x_i-\rho(2x_i-1)\), so \(U\) itself has such a graph, and clipping and the final displayed gate use affine operations and ReLUs.
Signed intermediate values can be passed by the signed-pair construction of \Cref{lem:cnf-relu}.
No inventory constraint is imposed on target parameters.

We have \(|T|\le U\), and \(T=T^*\) except on the union of the share transitions, whose probability is at most \(S\epsilon\).
Where \(U=0\), both targets are zero regardless of the original decoders.

A uniform bound for all candidates on the cube is
\[
|N|\le H:=\left((m+1)(eM+1)\right)^{m+1}.
\]
Topological induction gives this bound, since each affine preactivation has at most \(m\) terms, each parameter has magnitude at most \(eM\), and there are at most \(m\) computational vertices.
Thus this weaker bound suffices and has polynomial binary length.
Since \(|T|,|T^*|\le1\), their squared losses differ by at most \(4(H+1)\) on any input.

Choose
\[
\epsilon=\frac{\eta}{16S(H+1)}.
\]
The two risks then differ by at most \(\eta/4\), simultaneously for every candidate.
Hence actual YES risk is at most \(E_0-7\eta/4\), whereas every actual NO risk is at least \(E_0-\eta/4\).

Set
\[
\tau=E_0-3\eta/2,\qquad \delta=\eta.
\]
These are positive rational inputs with polynomial binary encodings; factorials and \(M^D\) have polynomial bit length.
The YES risk is below \(\tau\), and every NO risk is above \(\tau+\delta=E_0-\eta/2\).
All graphs, dimensions, occurrences and decoder coefficients are constructed in polynomial time.
The YES witness consumes every occurrence and the NO argument covered at-most-once use, proving hardness for both inventory rules with cap two.
Combining this reduction with \Cref{prop:binary-gap-risk} proves completeness.
\end{proof}

Discarding the gap field also gives \(\mathsf{NP}^{\mathsf{PP}}\)-hardness of exact continuous-risk synthesis.
The binary gap above may be exponentially small.
The result does not classify the inverse-gap clipped-loss problem, and it does not identify the exact-threshold upper bound of \Cref{thm:continuous-risk} with \(\mathsf{NP}^{\mathsf{PP}}\).

\subsection{Exact Continuous-Risk Synthesis}
\label{sec:continuous-exact-synthesis}

Exact synthesis under continuous uniform inputs is classified by the exact comparison class of \Cref{sec:continuous-risk} preceded by one discrete existential choice.
For a class \(\mathcal C\), write \(\exists\!\cdot\mathcal C\) for the languages \(L\) for which there are a polynomial \(s\) and a language \(B\in\mathcal C\) such that
\[
a\in L\quad\leftrightarrow\quad
\exists y\in\{0,1\}^{s(|a|)}\,(a,y)\in B.
\]
This is one existential string followed by one test in \(\mathcal C\).
The definition does not identify this class with an adaptive oracle class.
We use \(\mathcal C=\mathsf{PP}_{\mathbb Q}^{(pt)}\), with the exact rational-weight convention specified in \Cref{sec:continuous-risk}.

\begin{theorem}[Exact continuous-risk synthesis; restatement of \Cref{ithm:population-risk}]
\label{thm:continuous-exact-synthesis}
Under continuous uniform inputs on \([0,1]^d\), squared-risk ReLU-network synthesis with a rational threshold is complete for
\[
\exists\!\cdot\mathsf{PP}_{\mathbb Q}^{(pt)}
\]
under deterministic polynomial-time many-one reductions, under either inventory rule.
The target is an independently supplied rational ReLU network and the dimension is listed explicitly.
Hardness holds with integer inventory and computational-node cap two, including the affine output.
The hard targets have dyadic parameters and satisfy \(0\le T\le1\) on the integration cube.
\end{theorem}

The proof needs a stronger version of the separation estimate in \Cref{lem:balanced-corner-inventory}.
It allows a target mean \(\mu\) up to \(U\) instead of \(U/2\), and its only exceptions are the balanced corner functions.

\begin{lemma}[Separation from the balanced corner functions]
\label{lem:balanced-corner-exceptions}
Use the inventory, \(e,D,K,M,b,U\), and corner functions of \Cref{lem:balanced-corner-inventory}.
For any number of additional independent uniform coordinates \(z\), any legal at-most-once candidate \(N(x,z)\), and any measurable \(0\le\mu(x)\le U(x)\),
\[
\int(N^2-2N\mu)\,dx\,dz\ge0,
\]
unless the function computed by \(N\) is \(h_a(x)\) for a corner \(a\in\{0,1\}^D\) of Hamming weight \(e\).
\end{lemma}
\begin{proof}
We retain the complete graph and occurrence-role classification in the proof of \Cref{lem:balanced-corner-inventory}.
Write \(I_j=2/((j+2)!M^j)\) and \(E_0=\|U\|_2^2=2^DI_D\).
For the larger allowed mean, the Cauchy--Schwarz estimate of that proof becomes
\[
\int(N^2-2N\mu)
\ge\|N\|_2(\|N\|_2-2\|U\|_2).
\]
A nonzero integer-affine candidate, a candidate with neither formal affine branch zero, and each of the classified large-maximum cases have \(\|N\|_2\ge1/16\).
Here \(E_0\le2^{D+1}/M^D<1/1024\), so that estimate remains nonnegative.
Zero functions and negative-gain functions satisfy the assertion pointwise.
The active-branch-zero cases either vanish or have a maximum at least \(M\), as established there.

For the remaining gain-\(M\), maximum-one case, that proof gives
\[
\|N\|_2^2\ge
\frac{M^{2-D}}{4(2D(e+1))^D}.
\]
This bound is at least \(4E_0\) whenever
\[
M^2\ge2^{D+5}(2D(e+1))^D.
\]
The binary logarithm of the right side is at most \(4K^2\), using \(D\ge4\), \(K=D+2\), and \(2D(e+1)=D(D+2)\le K^2\); the left side has logarithm \(20K^2\).
Thus Cauchy--Schwarz applies in this case as well.

Only the unit-gain candidates with maximum one remain; we call them \emph{templates}.
In that proof each template has the form \(\rho(1-M\sum_{j\in J}\operatorname{dist}_j)\) for its set \(J\) of selected coordinates.
If a template selects \(s\) original and \(t\) auxiliary coordinates, put \(k=s+t\le D\) and \(l=D-s\).
Its squared norm is \(I_k\).
The exact mixed-coordinate calculation in that proof gives, unless \(s=D\) and \(t=0\),
\[
\frac{\int NU}{I_k}
=\frac{2^{l-1}(t+l+2)!(s+t+2)!}
{(t+1)!(l+1)!(D+t+2)!}M^{-l}
\le\frac{2^{3D+1}}M\le\frac12.
\]
Since these templates are nonnegative and \(\mu\le U\), their integral \(\int(N^2-2N\mu)\) is at least \(I_k-2\int NU\ge0\).
The exceptional templates use all \(D\) original coordinates and no auxiliary coordinate.
In the occurrence classification of that proof, such a template is \(\rho(1-M\sum_{i=1}^D\operatorname{dist}_i)\), where exactly \(e\) of the distances have the form \(1-x_i\) and the others have the form \(x_i\).
Setting \(a_i=1\) for the former coordinates and \(a_i=0\) for the others gives \(\operatorname{dist}_i=|x_i-a_i|\) on the cube, so the template computes \(h_a\) with \(|a|=e\).
This exhausts every case of that classification, including direct skips, unused occurrences, biases, and arbitrary auxiliary-coordinate access.
\end{proof}

We next implement a uniform variable that stays independent of the raw coordinates a candidate can read.
The underlying modular-sharing mechanism is classical \cite[Section 4]{ChorKushilevitz1993}; its independent-share formulation is stated explicitly in \cite[Scheme 3.1]{DibertCsirmaz2013}.
We give a continuous triangular realization and prove the conditional identity needed here.

\begin{lemma}[A continuous uniform variable hidden from a bounded number of coordinates]
\label{lem:continuous-uniform-sharing}
For an integer \(r\ge1\), define
\[
\Theta_r(t)=2\rho(t)+4\sum_{j=1}^{2r-1}(-1)^j\rho(t-j/2)+2\rho(t-r).
\]
If \(Y_1,\ldots,Y_r\) are independent uniform variables on \([0,1]\), then
\(\Theta_r(\sum_jY_j)\) is uniform on \([0,1]\) and independent of any selected collection of at most \(r-1\) of them.
The function \(\Theta_r\) has a rational ReLU realization of size \(O(r)\).
\end{lemma}
\begin{proof}
For each integer \(k\in\{0,\ldots,r-1\}\), the displayed expression on \([k,k+1]\) is the triangle with values zero, one, and zero at \(k,k+1/2,k+1\).
It vanishes outside \([0,r]\).
The slopes follow by adding the changes at successive half-integers; their magnitudes inside are two.
Condition on every variable except one unobserved \(Y_j\).
The sum of the conditioned variables is a fixed \(c\in[0,r-1]\).
Because \(\Theta_r\) has period one on \([0,r]\) and each triangle consists of two branches, every bounded measurable \(f\) satisfies
\[
\int_0^1f(\Theta_r(c+t))\,dt
=\int_0^1f(\Theta_1(t))\,dt
=\int_0^1f(u)\,du.
\]
Averaging over the other unobserved variables shows that, given the observed variables, \(\Theta_r(\sum_jY_j)\) is conditionally uniform on \([0,1]\), as asserted, and independent of them.
The defining sum of \(2r+1\) ReLU terms gives the stated realization.
\end{proof}

\begin{proof}[Proof of \Cref{thm:continuous-exact-synthesis}]
For membership, existentially specify a padded candidate graph and its occurrence assignment.
Its length is polynomial, and legality is polynomial-time decidable under either rule.
For a legal graph, \Cref{lem:risk-leaves} provides short rational leaves \(r_w\) summing to its risk.
In a bitstring equality, \(0\) denotes the all-zero word of the required length, including the empty word when that length is zero.
The single weighted comparison
\[
\sum_w\left(-r_w+\tau\mathbf1_{\{w=0\}}\right)\ge0
\]
holds exactly when the candidate's risk is at most the threshold \(\tau\).
For an illegal certificate use constant negative weights.
This proves membership directly in the stated existential closure.

For hardness, fix an input to any language in that closure.
Its condition is an existential string \(y\) followed by a comparison of an exactly polynomial-time computable rational weighted sum with a rational threshold \(t=t_1/t_2\), where \(t_2>0\).
If the counting word has length \(p\), replace each weight \(u/v\) by
\[
\frac{2^pt_2u-t_1v}{2^pt_2v}.
\]
Summing the new weights over all \(2^p\) counting words subtracts \(t\) from the sum, and every new numerator and denominator still has polynomial binary length.
Pad ignored existential coordinates so their number \(e\) is at least two.
An ignored counting bit, with each weight divided by two, ensures \(p\ge1\) without changing the sum.
Choose a polynomially bounded integer \(q\ge1\) such that the normalized evaluators satisfy
\[
|u(y,w)|\le Q:=2^q,\qquad 1\le v(y,w)\le Q.
\]
We must therefore encode
\[
\exists y\in\{0,1\}^e\,F_y\ge0,
\qquad F_y=\sum_{w\in\{0,1\}^p}\frac{u(y,w)}{v(y,w)}.
\]
No positive lower bound on a nonzero \(|F_y|\) is assumed.

Use the balanced inventory of \Cref{lem:balanced-corner-inventory}, with \(D=2e\) and \(m=D+2\) occurrences.
For \(a\in\{0,1\}^D\), use the source evaluators if \(a=(y,1-y)\); otherwise set \(u=-1,v=1\).
Write the resulting sum as \(F_a\), and put \(A=Q2^p\).
Every invalid code has \(F_a=-2^p<0\), and \(|F_a|\le A\) for every code.
The source is YES precisely when some balanced code has \(F_a\ge0\).

Write \(\operatorname{clip}(t)=\rho(t)-\rho(t-1)\).
On the positive support of \(U\), the functions
\[
d_i(x)=\operatorname{clip}(2x_i-1/2)
\]
are exactly the bits of the unique corner supporting \(U\).
Indeed every coordinate is within \(1/M\le1/4\) of that corner.
In this proof, \(I_2\) and \(I_3\) denote the second and third moments of a corner function \(h_a\), displayed next; they are not the quantities \(I_j=2/((j+2)!M^j)\) of \Cref{lem:balanced-corner-exceptions} at \(j=2,3\), whose subscript counts selected coordinates.
The simplex formula used in \Cref{lem:balanced-corner-inventory} gives
\[
I_2=\int h_a^2=\frac2{(D+2)!M^D},\qquad
I_3=\int h_a^3=\frac6{(D+3)!M^D},\qquad
\frac{I_3}{I_2}=\frac3{D+3}\le\frac37.
\]
For \(j=2,3\), disjoint corner supports give \(\int U^j=2^DI_j\).

We first construct a payload using a local uniform coordinate \(t\in[0,1]\).
Set \(b_1=2q+2\), \(n=p+b_1\), and \(\eta=2^{-n}\).
Use the tent decoder in the proof of \Cref{thm:continuous-risk}: the \(n\) itinerary bits, recalled below, encode \((w,j)\), where \(j\) is a \(b_1\)-bit unsigned integer.
For completeness, recall the tent map \(\theta(t)=\rho(2t)-2\rho(2t-1)+\rho(2t-2)\) of that proof, whose power \(\theta^i\) denotes iteration.
Let \(z_0=t\), \(z_i=\theta^i(t)\), and \(s_i=2z_i-1\) for \(0\le i<n\), and put \(r_t=1-\theta^{n+1}(t)\).
As in that proof, the \(i\)th itinerary bit is \(\mathbf1_{\{s_i>0\}}\).
With \(\epsilon=\eta/4\), the ramps
\[
\beta_i=\rho((s_i+\epsilon)/(2\epsilon))-
\rho((s_i-\epsilon)/(2\epsilon))
\]
lie in \([0,1]\) and equal the itinerary bits wherever \(r_t\le1/2\).
The fine cells are the \(2^n\) open dyadic cells of width \(\eta\) from that proof, which realize all strict sign words of \((s_0,\ldots,s_{n-1})\) exactly once.
Each fine cell thus realizes one word and has two affine branches on which \(r_t\) traverses \([0,1]\).
Thus integration on that cell is \(\eta\) times integration against a uniform \(r_t\).

Compile the source evaluator, its validity test, the digits \(c_k\) of \(v-1\) for \(0\le k<q\), and the selection predicate
\[
J:\quad j<Qv+u
\]
by Boolean AND and NOT gates, realized as in \Cref{thm:continuous-risk}.
All gate values stay in \([0,1]\), even outside exact decoding regions.
Since \(0\le Qv+u\le2Q^2<2^{b_1}\), this predicate selects exactly \(Qv+u\) integers for each \(w\).
Define
\[
P=r_t+\sum_{k=0}^{q-1}2^k\min\{r_t,c_k\},\qquad
B_U(x,t)=\min\{\rho(U(x)-4P),J\}.
\]
No product of two variable signals occurs in these formulas.
If \(\rho(U(x)-4P)>0\), then \(r_t\le P<U/4\le1/4\).
Both the itinerary and original-coordinate decoders are then Boolean, so \(P=vr_t\) and \(J\) is the exact selection bit.
Consequently the payload on a fine cell is \(\rho(U-4vr_t)\) if selected, and zero otherwise.
Where \(\rho(U(x)-4P)=0\) the payload vanishes; if \(U=0\) it vanishes independently of every decoded value.
It is continuous and zero at \(t=0,1\).

A selected cell contributes
\[
\int_{\mathrm{cell}}B_U\,dt=\frac{\eta U^2}{8v},\qquad
\int_{\mathrm{cell}}B_U^2\,dt=\frac{\eta U^3}{12v}.
\]
These follow by integrating \(U-4vr\) and its square over \(0\le r\le U/(4v)\).
On the support of the corner \(a\), summing selected words therefore gives
\[
\int_0^1B_U\,dt=\frac{\eta U^2}{8}(A+F_a),\qquad
\int_0^1B_U^2\,dt=\frac{\eta U^3}{12}(A+F_a).
\]
In particular \(3\int B_U^2=2U\int B_U\), exactly.

For \(x\in[0,1]^D\) and \(s\in[0,1]\), fix \(\lambda=1/8\) and introduce the background
\[
\mathcal E_U(x,s)=\rho((s-2\lambda)/\lambda)
-\rho((s-2\lambda)/\lambda-U(x)).
\]
As a function of \(s\), it is zero on \([0,2\lambda]\), rises linearly to \(U\), and equals \(U\) on \([3\lambda,1]\).
Place three copies of the payload at
\[
\begin{split}
B_1(x,s)&=B_U(x,\operatorname{clip}(s/\lambda)),\\
B_2(x,s)&=B_U(x,\operatorname{clip}((s-\lambda)/\lambda)),\\
B_3(x,s)&=B_U(x,\operatorname{clip}((s-3\lambda)/\lambda)),
\end{split}
\]
and put \(T_0=\mathcal E_U+B_1+B_2-B_3\).
Since the payload vanishes at \(t=0\) and \(t=1\), these copies are supported in \([0,\lambda]\), \([\lambda,2\lambda]\), and \([3\lambda,4\lambda]\), respectively.
The background is zero on the first two intervals and equals \(U\) on the third.
Thus \(0\le T_0\le U\).
Relative to the background alone, the copies change the energy \(\int_0^1T_0^2\,ds\) by
\[
\lambda\left(3\int B_U^2-2U\int B_U\right)=0.
\]
Direct integration of the capped ramp gives
\[
\begin{split}
\int_0^1T_0^2\,ds&=(1-2\lambda)U^2-\frac{2\lambda}{3}U^3,\\
\mu(x):=\int_0^1T_0\,ds
&=(1-2\lambda)U+cU^2+\frac{\lambda\eta}{8}F_aU^2,\\
c&=-\frac\lambda2+\frac{\lambda\eta A}{8}.
\end{split}
\]
The signed multiplicity of the payload copies in the first moment is \(2-1=1\).
The formula for \(\mu\) applies on each corner support; it is zero elsewhere, and \(0\le\mu\le U\).

Supply \(m+1\) fresh raw uniform coordinates \(Y\), and define the actual target
\[
T(x,Y)=T_0\left(x,\Theta_{m+1}\left(\sum_{j=1}^{m+1}Y_j\right)\right).
\]
Every candidate reads at most \(m\) distinct raw share coordinates, since each read requires an input-edge occurrence.
By \Cref{lem:continuous-uniform-sharing}, conditioning on all original coordinates and on every raw share value read by a candidate leaves the virtual coordinate uniform.
Both displayed moments of \(T_0\) therefore remain exact for \(T\) under this conditioning.
The target's energy is the known rational
\[
\mathcal R_0=\int T^2=2^D\left(\frac34I_2-\frac1{12}I_3\right),
\]
independent of all source sums \(F_a\).

Choose the rational threshold
\[
\begin{split}
\tau&=\mathcal R_0+I_2-2((1-2\lambda)I_2+cI_3)\\
&=\mathcal R_0-\frac12I_2+\lambda(1-\eta A/4)I_3.
\end{split}
\]
Here \(\eta A=1/(4Q)\le1/8\), so the last correction is positive.
Moreover,
\[
\mathcal R_0-\tau\ge\frac{25}{56}I_2>0,\qquad
\mathcal R_0\ge2^D\frac57I_2\ge\frac{80}{7}I_2.
\]
In particular \(0<\tau<\mathcal R_0\).

For every candidate, the exact conditional first moment gives
\[
R_T^{\mathrm{cont}}(N)=\mathcal R_0+\int(N^2-2N\mu).
\]
Unless its function is a balanced original-coordinate \(h_a\), \Cref{lem:balanced-corner-exceptions} makes this risk at least \(\mathcal R_0>\tau\).
For such a corner, \(U=h_a\) on its support, and the first-moment formula instead gives
\[
R_T^{\mathrm{cont}}(h_a)=\tau-\frac{\lambda\eta}{4}I_3F_a.
\]
A YES source therefore has a corner witness of risk at most \(\tau\); if its \(F_a=0\), the corner risk is exactly \(\tau\), preserving the non-strict convention without a margin assumption.
For a NO source, every valid and invalid code has \(F_a<0\), so each corner has risk strictly above \(\tau\); every other candidate was already excluded.
The YES witness consumes every occurrence, whereas the NO argument covered the at-most-once superset.
This proves hardness under both rules.

Finally, the source evaluators unroll into polynomial-size Boolean circuits, and their ReLU compilation uses only polynomially many positions.
The fold depth \(n\), decoder slopes, \(Q,M\), factorials in the moments, and the threshold all have polynomial binary encoding length.
The dimension is \(D+m+1\), listed explicitly.
The three payload copies and continuous sharing use polynomially many rational target parameters.
Minimum, clipping, and signed intermediate expressions have the ordinary ReLU implementations described in \Cref{thm:continuous-risk}; repeated predecessors are combined, so no parallel edge is needed.
The target inventory is independent of the supplied candidate inventory.
Thus this is a polynomial-time many-one reduction to the exact synthesis problem.
\end{proof}

\section{Synthesis under Gaussian Inputs}
\label{sec:gaussian-risk}

Changing the input probability distribution preserves the occurrence constraint but changes the arithmetic of exact loss.
Let \(\gamma_d\) be the standard Gaussian distribution on \(\mathbb R^d\), with density
\[
\psi_d(x)=(2\pi)^{-d/2}\exp(-\|x\|_2^2/2).
\]
We use the vector norms defined in \Cref{sec:model}, so \(\|x\|_2^2=\sum_i x_i^2\).
The one-dimensional density and its upper tail are
\[
\phi(t)=\psi_1(t)=(2\pi)^{-1/2}e^{-t^2/2},
\qquad
\overline\Phi(t)=\int_t^\infty\phi(u)\,du.
\]
For a real random variable \(V\) with \(\mathbb E[V^2]<\infty\), its mean is \(\mathbb E[V]\) and its variance is \(\mathbb E[(V-\mathbb E[V])^2]\).
It is centered when its mean is zero.
For \(v>0\), a centered normal of variance \(v\) has density
\[
f_v(t)=(2\pi v)^{-1/2}\exp(-t^2/(2v)).
\]
This is the density of \(\sqrt v Z\) for a standard normal \(Z\); the moment calculation below gives its stated mean and variance.

Independent coordinates have product density.
If independent centered normals have positive variances \(s,t\), the density of their sum at \(r\) is \(\int_{\mathbb R}f_s(x)f_t(r-x)\,dx\), by substituting \(r=x+y\) in that product density.
The identity
\[
\frac{x^2}{s}+\frac{(r-x)^2}{t}
=\frac{s+t}{st}\left(x-\frac{sr}{s+t}\right)^2+\frac{r^2}{s+t}
\]
shows that this integral is \(f_{s+t}(r)\): the remaining Gaussian integral is \(\sqrt{2\pi st/(s+t)}\).
Scaling a standard normal by a nonzero real \(a\) gives density \(f_{a^2}\), also for negative \(a\) because \(\phi\) is even.
For independent standard normal \(X_i\), repeatedly adding these independent summands proves that \(\sum_i a_iX_i\) is a centered normal of variance \(\sum_i a_i^2\) whenever the coefficient vector is nonzero.
If every coefficient is zero, the sum is identically zero.
Functions of disjoint sets of independent coordinates remain independent, because their joint probabilities factor under the product probability distribution.

Evenness of \(\phi\) makes every integrable odd moment of \(Z\) zero.
Since \(\phi'(z)=-z\phi(z)\), integration by parts gives
\[
\int_{\mathbb R}z^m\phi(z)\,dz
=(m-1)\int_{\mathbb R}z^{m-2}\phi(z)\,dz
\qquad(m=2,4).
\]
The boundary term \(z^{m-1}\phi(z)\) vanishes at both infinities, since Gaussian exponential decay dominates each fixed power.
Starting from \(\int\phi=1\), this yields \(\mathbb E[Z^2]=1\) and \(\mathbb E[Z^4]=3\), and hence the claimed variances after scaling.

For square-integrable functions, meaning \(\mathbb E_{X\sim\gamma_d}[f(X)^2]<\infty\), write
\[
\|f\|_{2,\gamma_d}^2=\mathbb E_{X\sim\gamma_d}[f(X)^2],
\qquad
\langle f,g\rangle_{\gamma_d}=\mathbb E_{X\sim\gamma_d}[f(X)g(X)].
\]
We abbreviate the function norm to \(\|f\|_2\) within the Gaussian arguments; vector norms retain their coordinate definitions above.
Functions are orthogonal when this inner product is zero.

For a supplied rational ReLU target \(T\) and a permitted candidate \(N\), define
\[
R_T^{\mathrm G}(N)=\mathbb E_{X\sim\gamma_d}[(N(X)-T(X))^2].
\]
Gaussian-risk synthesis asks whether some legal occurrence assignment has risk at most a supplied rational threshold.
All statements below concern this synthesis question; calculations for a fixed candidate are supporting algorithms.
Gaussian agnostic learning of a freely chosen ReLU has different input and output requirements \cite{GoelKarmalkarKlivans2019}.
Weighted model integration also distinguishes its exact integration procedures from Gaussian Monte Carlo approximation \cite[Section IV-B]{MorettinPasseriniSebastiani2024}.

\subsection{Exact Comparison and Effective Approximation}

Finite candidate enumeration does not by itself decide a non-strict inequality between computable real risks.
A real number is computable here if an algorithm produces a rational interval of arbitrarily prescribed width containing it.
A language is recursively enumerable if an algorithm halts and accepts exactly its members, with no termination requirement on other inputs.
A language whose complement is recursively enumerable is called co-recursively enumerable.
We use these notions only for the following exact synthesis bound, without claiming a running-time bound.

A function \(f\) is \(K\)-Lipschitz for a vector norm \(\|\cdot\|\) if \(|f(x)-f(y)|\leq K\|x-y\|\) for every pair of inputs.
The following growth estimate is used both for the semidecision bounds below and for the accuracy-gap result of \Cref{prop:binary-gap-risk}.
\begin{lemma}[Uniform growth and Lipschitz bounds]
\label{lem:risk-global-growth}
From a synthesis input and its rational ReLU target, one can compute a positive integer \(H\) of polynomial binary length such that every legal candidate and the target satisfy
\[
|N(x)|,|T(x)|\le H(1+\|x\|_\infty),
\]
and both functions have Lipschitz constant at most \(H\) for the infinity norm.
Thus \(F=N-T\) has Lipschitz constant at most \(2H\) and \(|F(x)|\le2H(1+\|x\|_\infty)\).
\end{lemma}
\begin{proof}
Let \(L\ge2\) bound the complete input length, all vertex and parameter-position counts, and the explicit dimension.
Candidate counts are bounded by the inventory length, even if the supplied cap is larger.
Every target and inventory parameter has absolute value at most \(2^L\).
Put \(A=1+(L+1)2^L\) and \(H=A^{L+1}\).
In topological order, if previous growth and Lipschitz bounds are at most \(A^j\), each next affine preactivation has at most \(L\) incoming terms and one bias, and has bounds at most \(A^{j+1}\).
The ReLU is one-Lipschitz and maps zero to zero, so it preserves these bounds.
The affine output obeys the same estimate.
There are at most \(L\) computational vertices in each graph; inputs start with bound one.
The chosen \(H\) therefore suffices uniformly, and \(\log_2 H=O(L^2+L\log(L+1))\).
The difference bounds follow by the triangle inequality.
\end{proof}

The growth bound makes Gaussian tail error effectively controllable.
This yields terminating certificates for strict inequalities on the appropriate side.
\begin{proposition}[Semidecision bounds for Gaussian synthesis]
\label{prop:gaussian-semidecision}
Under either inventory rule, Gaussian-risk synthesis with a non-strict rational threshold is co-recursively enumerable; the version with a strict threshold is recursively enumerable.
The zero-threshold non-strict problem is exactly full-space exact synthesis.
\end{proposition}
\begin{proof}
There are effectively finitely many legal graphs and occurrence assignments: each computational vertex consumes a distinct occurrence through an incoming position or bias, so a legal graph has at most \(\min(m,n_{\max})\) computational vertices, where \(n_{\max}\) is the cap.
They can all be enumerated and their legality tested.
We show that each risk is uniformly computable.

Put \(A_0=2H\) from \Cref{lem:risk-global-growth}.
The squared difference is bounded by \(2A_0^2(1+\|x\|_2^2)\).
For an integer \(B\ge1\), integration by parts gives
\[
\Pr[|Z|>B]\le\frac{2\phi(B)}B,\qquad
\mathbb E[Z^2\mathbf1_{\{|Z|>B\}}]\le2\phi(B)(B+B^{-1})
\]
for a standard normal \(Z\).
A union bound over a coordinate with \(|x_i|>B\), using independence for the squares of the other coordinates, bounds the contribution to the risk from outside \([-B,B]^d\) by
\[
4dA_0^2\phi(B)\left(B+\frac{d+1}B\right).
\]
This bound tends to zero as \(B\) grows, at a computable rate.
On the compact box, the squared difference and Gaussian density have computable Lipschitz constants, so rational grid sums with certified density approximations provide arbitrarily small error.
The explicit implementation in \Cref{prop:binary-gap-risk} below supplies such approximations without any exact transcendental comparison.
Together these give rational upper and lower bounds of arbitrary precision for every risk.

For non-strict synthesis, run these approximations on the finite list until every lower bound exceeds the threshold.
This accepts exactly the complement; an empty candidate list is accepted immediately.
For strict synthesis, accept when some upper bound falls below the threshold.
These procedures terminate on precisely the claimed sides, including negative thresholds.
At threshold zero, continuity and the strictly positive Gaussian density imply that zero risk is equivalent to equality everywhere: a nonzero difference at one point has nonzero magnitude on a neighborhood of positive Gaussian measure.
\end{proof}

\subsection{A Binary-Encoded Accuracy Gap}

We use the binary-gap synthesis promise defined in \Cref{sec:continuous-gap-synthesis}, for either the continuous uniform probability distribution or the standard Gaussian probability distribution.
Let \(\tau\) and \(\delta>0\) be its binary rational threshold and gap.
Exact threshold equality is avoided by the separation between its promised YES and NO risks.

\begin{proposition}[Synthesis with a binary gap]
\label{prop:binary-gap-risk}
Under either inventory rule, the stated risk promise problem has a separator in \(\mathsf{NP}^{\mathsf{PP}}\), for continuous uniform inputs on \([0,1]^d\) and for standard Gaussian inputs on \(\mathbb R^d\).
\end{proposition}
\begin{proof}
Use the uniform bound \(H\) of \Cref{lem:risk-global-growth}, which is computable before a candidate is guessed.
In this proof, \(R(N)\) denotes \(R_T^{\mathrm{cont}}(N)\) or \(R_T^{\mathrm G}(N)\), according to the specified input probability distribution.
For a fixed legal candidate \(N\), set \(F=N-T\) and \(f=F^2\).
We construct a polynomial-time fair-bit experiment with an acceptance probability \(p_N\) and a known positive rational multiplier \(D\) such that
\[
|Dp_N-R(N)|\le\delta/4.
\]
Both \(D\) and its inverse have polynomial binary length.
The test \(p_N\le(\tau+\delta/2)/D\) is a total \(\mathsf{PP}\) predicate by the exact counting-threshold transformation in \Cref{thm:population-risk}; thresholds outside \([0,1]\) are decided directly.
A YES witness makes this predicate accept, and every candidate on a NO instance makes it reject.
Guessing the legal graph and using that total predicate proves the separator claim, without making a promise about individual oracle queries.

For uniform inputs, \(f\le M=16H^2\) on the cube and its infinity-norm Lipschitz constant is at most \(16H^2\).
Couple a uniform point \(X\) to the grid point with coordinates \(Y_i=2^{-b}\lfloor2^bX_i\rfloor\).
Then \(|\mathbb E[f(X)]-\mathbb E[f(Y)]|\le16H^2 2^{-b}\).
Choose \(b\) to make this at most \(\delta/8\).
The grid point is sampled with \(bd\) fair bits, and \(f(Y)\) is exactly computable in polynomial time.
With an independent uniform \(s\)-bit integer \(V\), accept when
\[
V<\left\lfloor\frac{2^s f(Y)}M\right\rfloor.
\]
Taking \(M2^{-s}\le\delta/8\) gives the required approximation with \(D=M\).
The chosen \(b,s\) are polynomial in the input length because \(H\) has polynomial binary length and \(\log_2(1/\delta)\) is at most polynomial in the binary length of the positive rational \(\delta\).

For Gaussian inputs put \(A_0=2H\), choose \(k\) with \(A_0\le2^k\), and let \(b_\delta\ge1\) be a bit-length bound giving \(\delta\ge2^{-b_\delta}\).
Choose
\[
B=8(k+b_\delta+d+2).
\]
The tail estimate in \Cref{prop:gaussian-semidecision} is at most \(\delta/8\).
Indeed \(\phi(B)\le2^{-B^2/2}\), and the prefactor \(4dA_0^2(B+(d+1)/B)\) has base-two logarithm at most \(2k+2d+B+3\).
The chosen \(B\) makes this logarithm minus \(B^2/2\) at most \(-b_\delta-3\).

On \([-B,B]^d\), put \(U=2H(B+1)\) and \(V_B=(2B)^d\).
The function \(g=f\psi_d\) is bounded by \(U^2\) and has infinity-norm Lipschitz constant at most
\[
K_G=4HU+U^2dB.
\]
Here \(f\) has Lipschitz constant at most \(4HU\) and \(\psi_d\le1\).
The gradient \(\nabla\psi_d=(\partial\psi_d/\partial x_i)_{i=1}^d\) of the density satisfies \(\|\nabla\psi_d\|_1\le dB\) on the box, where \(\|v\|_1=\sum_i|v_i|\) for a vector \(v\).
A uniform dyadic grid has mesh \(2B2^{-b}\); choose \(b\) so that \(V_BK_G2B2^{-b}\le\delta/32\).
At every sampled point, approximate \(g\) by a rational \(\widetilde g\in[0,U^2]\) to error at most \(\delta/(32V_B)\).
It suffices to approximate the density to that error divided by \(U^2\), then multiply by the exactly computed \(f\); clipping to the stated interval does not increase the error.
Finally, as in the uniform case, draw the grid point \(Y\) with fair bits and an independent uniform \(s\)-bit integer \(V\), and accept when
\[
V<\left\lfloor\frac{2^s\widetilde g(Y)}M\right\rfloor,\qquad M=U^2+1,
\]
with \(V_BM2^{-s}\le\delta/32\).
Since \(0\le\widetilde g<M\), the acceptance probability \(p_N\) differs from \(\mathbb E[\widetilde g(Y)]/M\) by at most \(2^{-s}\).
The multiplier is now \(D=V_BM\).
The tail, grid, density, and vertical errors total at most \(\delta/8+3\delta/32=7\delta/32<\delta/4\).
All bounds and precision choices have polynomial binary length.

For clarity, the density approximations themselves have polynomial bit running time.
At a grid point the rational \(q=\|Y\|_2^2/2\) has polynomial numerical magnitude, at most \(dB^2/2\).
For \(x\ge0\) and an integer \(n\ge0\), let \(S_n(x)=\sum_{i=0}^n(-x)^i/i!\).
Repeated integration by parts gives the exact remainder
\[
\exp(-x)-S_n(x)
=\frac{(-1)^{n+1}}{n!}\int_0^x\exp(-t)(x-t)^n\,dt.
\]
Its absolute value is at most \(x^{n+1}/(n+1)!\).
The sign of the remainder shows that \(S_{2k+1}(x)\le\exp(-x)\le S_{2k}(x)\) for integers \(k\ge0\), and the two bounds differ by \(x^{2k+1}/(2k+1)!\).
When \(0\le x\le1/2\), this difference is at most \(2^{-2k-1}\).
Choose the least integer \(j\ge0\) with \(q/2^j\le1/2\), apply these bounds there, and clip their endpoints to \([0,1]\).
Squaring the endpoints \(j\) times encloses \(\exp(-q)\).
Each squaring increases their distance by at most a factor of two, so the total amplification is at most \(2^j\le\max\{1,4q\}\).
Polynomially many terms and polynomial extra precision therefore suffice; exact rational numerator and denominator lengths stay polynomial, including after the \(j\) squarings.

For completeness, \(n!\ge(n/3)^n\) for integers \(n\ge1\).
The binomial expansion gives
\[
(1+1/n)^n
=\sum_{i=0}^n\frac{n(n-1)\cdots(n-i+1)}{i!n^i}
\le\sum_{i=0}^n\frac1{i!}<3,
\]
where \(i!\ge2^{i-1}\) for \(i\ge2\) bounds the tail by a geometric sum.
The factorial estimate now follows by induction from \(1!\ge1/3\): multiplication by \(n+1\) preserves it because
\((n+1)(n/3)^n\ge((n+1)/3)^{n+1}\).
This supplies the factorial estimate used in the later Gaussian calibration.

The density constant \((2\pi)^{-d/2}\) also requires rational enclosures of \(\pi\).
Use the integral definition
\[
\arctan x=\int_0^x(1+t^2)^{-1}\,dt
\]
and the usual analytic normalization \(\pi=4\arctan1\).
For \(0\le x<1\), integrating the finite geometric identity gives, for integers \(m\ge1\),
\[
\arctan x
=\sum_{i=0}^{m-1}\frac{(-1)^ix^{2i+1}}{2i+1}
+(-1)^m\int_0^x\frac{t^{2m}}{1+t^2}\,dt.
\]
The last term has absolute value at most \(x^{2m+1}/(2m+1)\).
Machin's identity states that
\[\pi=16\arctan(1/5)-4\arctan(1/239).\]
To verify it, write \(A(x)=\arctan x\).
For nonnegative \(x,y\) with \(xy<1\),
\(A(x)+A(y)=A((x+y)/(1-xy))\): differentiating both sides in \(y\) gives \(1/(1+y^2)\), and the sides agree at \(y=0\).
Consequently,
\[
2A(1/5)=A(5/12),\quad
4A(1/5)=A(120/119),\quad
A(1)+A(1/239)=A(120/119).
\]
Thus \(\pi=16A(1/5)-4A(1/239)\), and the two series tails decrease geometrically.
They give rational enclosures of \(\pi\) to any prescribed binary precision in polynomial time.

Since \(1/2\le(1+t^2)^{-1}\le1\) on \([0,1]\), the normalization gives \(2\le\pi\le4\).
Rational bisection encloses the square root of a positive rational argument by comparing squared rational midpoints with that argument.
Apply this operation to both rational endpoints of each enclosing interval.
Starting with a rational interval containing the root, each comparison halves its width, so polynomially many steps give the required binary precision.
Integer powering and reciprocal operations then give \((2\pi)^{-d/2}\) with polynomially many extra precision bits: on the interval \(2\le t\le4\), the absolute value of the derivative of \(t\mapsto(2t)^{-d/2}\) is at most \(d\).
All interval endpoints have polynomial bit length, and exact rational arithmetic preserves that bound.
Thus the whole acceptance predicate is deterministic polynomial time in its fair bits and the guessed graph.
No exact Gaussian sampling or transcendental equality test is used.
\end{proof}

The proposition does not classify exact synthesis at an arbitrary threshold.
It also differs from the model of \Cref{op:gap-risk} in the Introduction, which uses uniform Boolean inputs, clips the squared loss at one, and supplies \(1/\delta\) in unary, so that its gap is at least inverse-polynomial.
In the proposition, the loss may be unbounded and the binary gap may be exponentially small.


\subsection{Counting Hardness with a Forced Zero Candidate}

Standard Gaussian inputs also support a counting lower bound.
The following restriction separates the difficulty of risk comparison from the choice of an occurrence assignment: its prescribed multiset forces every candidate to compute zero.
Boolean-to-Gaussian lifting is an established technique in learning theory \cite{DanielyVardi2021,ChenGollakotaKlivansMeka2022}.
Here an independent Gaussian coordinate supplies a homogeneous amplitude with second moment one; this permits an explicit rational threshold for a majority-counting reduction.

\begin{theorem}[Forced-zero Gaussian synthesis]
\label{thm:gaussian-counting}
Under either inventory rule, standard Gaussian\linebreak squared-risk synthesis with positive binary rational threshold \(\tau\) and positive binary rational gap \(\delta\) is \(\mathsf{PP}\)-complete in the separator sense when restricted to \(W=(0)\) and cap one.
Hardness holds for bias-free rational ReLU targets, and also when candidate graphs are required to be bias-free.
For the same restrictions, the exact non-strict positive-threshold problem is \(\mathsf{PP}\)-hard.
These hardness statements use deterministic polynomial-time many-one reductions, with promise reductions for the gap problem.
\end{theorem}
\begin{proof}
We reduce from strict-majority circuit acceptance.
Given a Boolean circuit \(C\) on \(n\ge1\) inputs, the source question is whether
\(\Pr[C(B)=1]>1/2\) for a uniform Boolean vector \(B\).
This problem is \(\mathsf{PP}\)-complete: the acceptance predicate of a polynomial-time machine using polynomially many fair bits has a polynomial-size Boolean circuit, and unused bits pad its random tape to a fixed length.
An unused bit ensures \(n\ge1\).
Set
\[
\epsilon=\frac{2^{-n-5}}n,\qquad
\tau=\frac12-2^{-n-1},\qquad
\delta=2^{-n-2}.
\]
These are positive rationals of polynomial binary length.

Use \(n+1\) independent standard normal coordinates \((Z,X_1,\ldots,X_n)\).
The target first computes
\[
r=|Z|=\rho(Z)+\rho(-Z),\qquad
a_i=\min\{r,\epsilon^{-1}\rho(X_i)\}.
\]
All these quantities have bias-free ReLU implementations.
For nonnegative stored signals \(u,v\), the formulas
\[
\min(u,v)=\rho(u-\rho(u-v)),\qquad
\max(u,v)=\rho(v+\rho(u-v))
\]
use two ReLU vertices each.
A nonnegative sum, including \(r\), can be stored using one additional ReLU.

Simulate \(C\) on the signals \(a_i\) with AND implemented by minimum, OR by maximum, and NOT by \(u\mapsto\rho(r-u)\).
The constants FALSE and TRUE are represented by zero and \(r\).
Let \(v_C\) be the final signal and give the target the affine output
\[
T=r-v_C.
\]

Induction over the circuit shows that every simulated circuit signal and \(T\) lies in \([0,r]\).
Each gate needs only constantly many vertices and edges.
The nonunit scaling is \(\epsilon^{-1}=n2^{n+5}\), of \(O(n+\log n)\) bits, so the target has polynomial binary encoding.
Repeated gate arguments may be simplified, or coincident affine terms combined, so no parallel edges are necessary.
All target biases are absent.
Prune unused computations; a zero target can instead be represented by one zero-weight input-output edge.
Thus the target satisfies the graph conventions and is positively homogeneous, meaning \(T(tx)=tT(x)\) for every \(t\ge0\).

For analysis only, put
\[
B_i=\mathbf1_{\{X_i>0\}},\qquad
J=r(1-C(B)),\qquad
E=\bigcup_{i=1}^n\{|X_i|<\epsilon r\}.
\]
Outside \(E\), \(a_i=rB_i\), and gate induction gives \(T=J\).
Everywhere \(T,J\in[0,r]\), whence
\[
|T^2-J^2|\le r^2\mathbf1_E.
\]
Conditionally on \(r\), the bound \(1/\sqrt{2\pi}\) on the standard Gaussian density gives
\[
\Pr[|X_i|<\epsilon r| r]\le\frac{2\epsilon r}{\sqrt{2\pi}}.
\]
The union bound therefore yields
\[
\left|\mathbb E[T^2]-\mathbb E[J^2]\right|
\le\frac{2n\epsilon}{\sqrt{2\pi}}\mathbb E[|Z|^3]
=\frac{4n\epsilon}{\pi}\le4n\epsilon=2^{-n-3}.
\]
Indeed, substituting \(u=z^2/2\) in the Gaussian integral gives
\(\mathbb E[|Z|^3]=2\sqrt{2/\pi}\).
The signs \(B_i\) are independent fair bits and are independent of \(r\), while \(\mathbb E[r^2]=1\).
Writing \(q=\Pr[C(B)=0]\), we obtain
\[
|\mathbb E[T^2]-q|\le e,\qquad e=2^{-n-3}.
\]
This bound includes the unbounded Gaussian amplitude; no separate truncation error is omitted.

Supply \(W=(0)\) and cap one, including the output.
Every legal candidate computes zero, and a single zero-weight input-output edge is a legal candidate under either inventory rule, also with biases prohibited.
Thus every legal candidate has risk \(R=\mathbb E[T^2]\).
On a source YES instance, \(q\le1/2-2^{-n}\), so
\[
R\le\tfrac12-2^{-n}+e<\tau.
\]
For a source NO instance, \(q\ge1/2\), so
\[
R\ge\tfrac12-e>\tfrac12-2^{-n-2}=\tau+\delta.
\]
Every source circuit maps into the promise, proving \(\mathsf{PP}\)-hardness for both rules.
Dropping \(\delta\) from the output of this same reduction proves exact positive-threshold \(\mathsf{PP}\)-hardness.

For gap membership, use the fixed-candidate construction in \Cref{prop:binary-gap-risk} with the legal zero network.
It produces a polynomial-time fair-bit experiment and a known positive polynomial-bit rational \(D\).
Here \(b\) denotes the total number of fair bits of that experiment, its grid and vertical bits, not the grid precision \(b\) of \Cref{prop:binary-gap-risk}.
The experiment accepts on \(A\) of the \(2^b\) outcomes, so its acceptance probability is \(p=A/2^b\), satisfying
\( |Dp-R|\le\delta/4\).
There is no graph guess because every candidate has the same function.
Accept when \(p\le(\tau+\delta/2)/D\).
The approximation margins give a separator for every promised instance.
To see directly that this is a total \(\mathsf{PP}\) predicate, let
\(K=\lfloor2^b(\tau+\delta/2)/D\rfloor\).
If \(K<0\) reject, and if \(K\ge2^b\) accept.
Otherwise a machine with \(b+1\) fair bits has \(2^b-A\) accepting paths in one half, by complementing the original experiment, and \(K+1\) accepting paths in the other half.
Its acceptance count exceeds \(2^b\) exactly when \(A\le K\).
This proves the claimed \(\mathsf{PP}\) separator and completes the classification.
\end{proof}

The theorem gives no exact \(\mathsf{PP}\) upper bound or exact equality procedure.
Its \(\mathsf{PP}\) lower bound also applies to the unrestricted synthesis problem, and \Cref{thm:gaussian-choice} attains the \(\mathsf{NP}^{\mathsf{PP}}\) upper bound under either rule when the candidate function can vary.
The construction uses an exponentially small binary gap; it does not settle the clipped inverse-polynomial-gap question in \Cref{op:gap-risk}.


\subsection{A Matching Gaussian Classification with Two Vertices}

Allowing a nonzero candidate adds an existential choice to Gaussian counting.
We obtain a matching classification by representing each witness bit with opposite unit weights on a pair of Gaussian coordinates.
Each witness then defines a rare hinge: the ReLU of the resulting signed sum of coordinates minus a large bias, which is positive only far in the Gaussian tail.
Every such intended hinge uses the complete supplied multiset.
Different signed coordinate choices have much smaller cross moments than their individual squared norms.
A second Gaussian moment calculation excludes every competing cap-two graph outside this family.

\begin{theorem}[Gaussian binary-gap synthesis; restatement of \Cref{ithm:continuous-gap-complete}]
\label{thm:gaussian-choice}
Under either inventory rule, standard-Gaussian squared-risk ReLU synthesis with a positive binary rational threshold and a positive binary rational gap is \(\mathsf{NP}^{\mathsf{PP}}\)-complete in the separator sense, under deterministic polynomial-time many-one promise reductions.
Hardness holds with integer inventory and cap two, including the affine output.
With the same restrictions, the exact non-strict positive-threshold problem is \(\mathsf{NP}^{\mathsf{PP}}\)-hard under deterministic polynomial-time many-one reductions.
\end{theorem}
\begin{proof}
Membership for the gap problem is \Cref{prop:binary-gap-risk}.
For hardness use E-MAJSAT: given a Boolean circuit \(\varphi\) with \(e\) existential inputs and \(n\) random inputs, put \(p(a)=\Pr_{z\in\{0,1\}^n}[\varphi(a,z)=1]\) for each \(a\in\{0,1\}^e\), and ask whether
\[
\exists a\in\{0,1\}^e\,p(a)>\tfrac12.
\]
This is \(\mathsf{NP}^{\mathsf{PP}}\)-complete with exactly this fixed-witness, strict-majority convention \cite[Appendix A, Theorem 15]{LittmanGoldsmithMundhenk1998}.
Ignored variables ensure \(e,n\ge1\).
Put \(q=2^{-n}\).
A YES instance has some \(p(a)\ge1/2+q\); a NO instance has every \(p(a)\le1/2\).

Set
\[
k=2e,\qquad E=k+n+10,\qquad B=16E^2,\qquad d=k+n+1,\qquad \eta=2^{-10E}.
\]
The independent standard normal coordinates are \((Y_1,Y'_1,\ldots,Y_e,Y'_e,R_1,\ldots,R_n,U)\).
Recall from \Cref{sec:gaussian-risk} that a centered random variable has mean zero.
Write \(X\) for this raw input vector and put \(D_i=Y_i-Y'_i\); these differences are independent centered normals of variance two, independent of the verifier coordinates \(R_1,\ldots,R_n,U\).
Supply cap two and the multiset
\[
W=(+1)^{e+1}\uplus(-1)^e\uplus\{-B\}.
\]

For a signed witness \(a\in\{-1,+1\}^e\), write \(\bar a=(a+\mathbf1)/2\in\{0,1\}^e\) for its Boolean encoding and define
\[
h_a=\rho\left(\sum_{i=1}^e a_iD_i-B\right),\qquad
H=\rho\left(\sum_{i=1}^e|D_i|-B\right)=\max_a h_a.
\]
The equality \(H=\max_ah_a\) holds because \(\sum_ia_iD_i\le\sum_i|D_i|\), with equality when each \(a_i\) is a sign of \(D_i\), and \(\rho\) is nondecreasing.
The envelope \(H\) has an integer-weight ReLU implementation with \(2e+1\) hidden vertices.
For each \(i\), two hidden vertices compute \(\rho(Y_i-Y'_i)=\rho(D_i)\) and \(\rho(Y'_i-Y_i)=\rho(-D_i)\), whose sum is \(|D_i|\).
One further hidden vertex receives these \(2e\) values with weight one and has bias \(-B\), so it computes \(\rho(\sum_i(\rho(D_i)+\rho(-D_i))-B)=H\).

Every intended \(h_a\) uses all occurrences exactly once: the two input weights \((a_i,-a_i)\) consume one positive and one negative unit per pair, the hidden bias consumes \(-B\), and the output edge consumes the remaining positive unit.
Thus it is a legal cap-two candidate under either rule.
Its raw coefficient vector has \(k=2e\) nonzero unit entries, so its hidden preactivation has variance \(k\).

We first provide quantitative bounds for these functions.
For integers \(1\le j\le2k\), let
\[
s_j=\mathbb E[\rho(\sqrt{j}Z-B)^2],\qquad s=s_k,
\]
where \(Z\) is standard normal.
Using the Gaussian density \(\phi\) and its upper tail \(\overline\Phi\), we show
\begin{equation}
\label{eq:gaussian-tail-moment}
\frac{\exp(-B^2/(2j))}{200B^3}\le s_j\le2j(B+1)\exp(-B^2/(2j)).
\end{equation}

For the lower bound put \(t=B/\sqrt{j}\ge1\) and integrate over \([t+t^{-1},t+2t^{-1}]\).
The squared excess is at least \(j^2/B^2\), the interval has length \(\sqrt{j}/B\), and the density of \(Z\) on it is at least
\[\exp(-4)\exp(-t^2/2)/\sqrt{2\pi}>\exp(-t^2/2)/200.\]
Their product is at least the stated lower bound because \(j\ge1\).

For the upper bound, integration by parts and \(\overline\Phi(t)\le\phi(t)/t\) yield
\[
s_j\le j\mathbb E[Z^2\mathbf1_{\{Z>t\}}]
=j(t\phi(t)+\overline\Phi(t))\le2j(B+1)\exp(-t^2/2).
\]

For completeness, the same upper bound holds for any positive real variance \(v\) with \(t=B/\sqrt v\ge1\).
Shifting the integral gives
\[
\mathbb E[\rho(\sqrt v Z-B)^2]
=v\phi(t)\int_0^\infty u^2\exp(-tu-u^2/2)\,du
\le\tfrac v2\exp(-t^2/2),
\]
since \(\int_0^\infty u^2\exp(-u^2/2)\,du=\sqrt{\pi/2}\).
This is smaller than the stated upper estimate.

The quantities \(s_j\) increase with \(j\): add an independent centered Gaussian increment and apply Jensen's inequality to the convex function \(x\mapsto\rho(x-B)^2\).

Identify an intended witness with its raw coefficient vector \(\widetilde a=(a_1,-a_1,\ldots,a_e,-a_e,0,\ldots,0)\in\mathbb R^d\).
For any signed coordinate vector \(w\in\{-1,0,1\}^d\) with exactly \(k\) nonzero entries, distinct from an intended vector \(\widetilde a\), we have
\(\|w+\widetilde a\|_2^2\le4k-2\).
Writing \(h_w=\rho(w\cdot X-B)\), the arithmetic-geometric mean inequality gives pointwise
\[
h_wh_a\le\tfrac14\rho((w+\widetilde a)\cdot X-2B)^2.
\]
If one factor on the left is zero this is immediate; otherwise both preactivations are positive and their sum is the argument on the right.
Applying the bound above for a general variance \(v\), or using zero if \(w+\widetilde a=0\), shows
\[
\mathbb E[h_wh_a]\le c:=12kB\exp(-B^2/(2k-1)).
\]
This also covers vectors \(w\) with nonzero entries on the verifier coordinates.

Put
\[
\kappa=2400kB^4\exp(-B^2/(4k^2)).
\]
Taking ratios in \Cref{eq:gaussian-tail-moment} gives
\[
s_{k-1}/s\le\kappa\quad(k\ge2),\qquad c/s\le\kappa,\qquad s/s_{k+1}\le\kappa.
\]
Each ratio divides the upper bound in \Cref{eq:gaussian-tail-moment}, or the definition of \(c\), by the lower bound there.
For the first ratio,
\[
\frac{s_{k-1}}s
\le400(k-1)(B+1)B^3\exp\left(-\frac{B^2}{2(k-1)}+\frac{B^2}{2k}\right)
=400(k-1)(B+1)B^3\exp\left(-\frac{B^2}{2k(k-1)}\right).
\]
For the second,
\[
\frac cs
\le2400kB^4\exp\left(-\frac{B^2}{2k-1}+\frac{B^2}{2k}\right)
=2400kB^4\exp\left(-\frac{B^2}{2k(2k-1)}\right).
\]
For the third, where \(k+1\le2k\) keeps \(s_{k+1}\) within the range of \Cref{eq:gaussian-tail-moment},
\[
\frac s{s_{k+1}}
\le400k(B+1)B^3\exp\left(-\frac{B^2}{2k}+\frac{B^2}{2(k+1)}\right)
=400k(B+1)B^3\exp\left(-\frac{B^2}{2k(k+1)}\right).
\]
The denominators \(2k(k-1)\), \(2k(2k-1)\) and \(2k(k+1)\) are at most \(4k^2\) because \(k\ge1\), so each exponential is at most \(\exp(-B^2/(4k^2))\).
Since \(B+1\le2B\), each prefactor is at most \(2400kB^4\).

Our deliberately large choice of \(B\) ensures
\begin{equation}
\label{eq:gaussian-tail-scales}
\kappa\le\eta,\qquad s_{2k}\le2^{-20E},\qquad 2^ks\le\eta.
\end{equation}
Here \(\log_2(2400kB^4)\le12E\) for \(E\ge12\), whereas \(B^2/(4k^2)\ge64E^2\) and \(\exp(-x)\le2^{-x}\).
For the other two bounds, \(B+1\le2^E\), \(4k\le2^E\), and the exponents \(B^2/(4k)\ge64E^3\), \(B^2/(2k)\ge128E^3\) dominate the respective prefactors.

A convenient explicit lower bound from \Cref{eq:gaussian-tail-moment} is
\[
0<L:=2^{-B^2-4B-10}\le s.
\]
Indeed \(\exp(-B^2/(2k))\ge2^{-B^2}\), using \(\exp(1)<4\), and \(200B^3\le2^{4B+10}\).

We next construct the independently supplied target.
Set
\[
\epsilon=\frac{\eta L}{256de^3}.
\]
Here \(e\) is the number of existential inputs of \(\varphi\), not Euler's number.
The target reads a bit from each coordinate \(t\) among the differences \(D_i\) and the verifier coordinates \(R_j,U\), using the signal
\[
\min\{H,\epsilon^{-1}\rho(t)\}.
\]
Simulate the Boolean circuit \(F(a,z,u)=u\lor\varphi(a,z)\), whose arguments \(a\), \(z\) and \(u\) are the bits read from the \(D_i\), the \(R_j\) and \(U\), respectively, with amplitude \(H\): AND and OR are minimum and maximum, NOT is \(v\mapsto\rho(H-v)\), and constants are zero and \(H\).
Use the two-ReLU minimum and maximum implementations from the proof of \Cref{thm:gaussian-counting}.
If \(v_F\) is the final signal, take the affine target output \(T=2v_F-H\).
Every simulated circuit signal is in \([0,H]\), and \(|T|\le H\).

Absolute values of the rational differences, the sum defining \(H\), and the gate simulation all have polynomial size.
The integers \(B\) and \(\epsilon^{-1}\) have polynomial binary length, as do all target coefficients.
Coincident terms may be combined to avoid parallel edges, and unused vertices may be pruned.
The target has biases; no homogeneity restriction is claimed in this theorem.

For analysis put \(\operatorname{bit}(t)=\mathbf1_{\{t>0\}}\), applied coordinatewise to a vector, and define
\[
J=H\left(2F(\operatorname{bit}D,\operatorname{bit}R,\operatorname{bit}U)-1\right).
\]
In particular, \(\operatorname{bit}(0)=0\).
For fixed witness signs \(a\), independence and the extra fair bit \(u\) give
\[
\mathbb E_{R,U}[2F(\bar a,\operatorname{bit}R,\operatorname{bit}U)-1]=p(\bar a).
\]
Let \(\mathcal E\) be the union of the events \(|D_i|<\epsilon H\), \(|R_j|<\epsilon H\), and \(|U|<\epsilon H\) over the decoded coordinates.
Outside this event the simulation equals \(J\), and everywhere \(|T-J|\le2H\mathbf1_{\mathcal E}\).
Put \(S=\mathbb E[H^2\mathbf1_{\mathcal E}]\).
We claim
\[
S\le64d\epsilon e^3\le\eta s/4.
\]

Indeed \(H\le Q=\sum_i|D_i|\), and convexity gives
\(\mathbb E[Q^3]\le e^2\sum_i\mathbb E[|D_i|^3]\le6e^3\), since \(\mathbb E[|D_i|^3]=8/\sqrt\pi<6\).
Every decoded coordinate has density at most \(1/2\).
For an independent verifier coordinate, conditioning on \(H\) bounds its contribution by \(\epsilon\mathbb E[H^3]\le6\epsilon e^3\).

For \(D_i\), put \(Q_{-i}=\sum_{j\ne i}|D_j|\).
On \(|D_i|<\epsilon H\), since \(\epsilon\le1/2\), we have \(|D_i|<2\epsilon Q_{-i}\) and \(H^2\le4Q_{-i}^2\).
Conditioning on the other differences bounds the contribution by
\(8\epsilon\mathbb E[Q_{-i}^3]\le48\epsilon e^3\).
This includes \(e=1\), when the dependent event has zero contribution.
There are \(e+n+1\le d\) events; the union bound and the chosen \(\epsilon\) prove the claim.

Put \(C=\mathbb E[T^2]\) and \(C_0=2^es\).
Since \(H=\max_a h_a\), choosing a maximizing hinge at each point bounds the sum of the remaining squared hinges by their products with that maximizer.
Summing the cross-moment bounds over ordered pairs gives
\[
0\le C_0-\mathbb E[H^2]\le2^{2e}c.
\]
Next, \(|T|\le H\) everywhere, and outside \(\mathcal E\) we have \(T=J\), so \(T^2=J^2=H^2\) because \(2F-1\in\{-1,1\}\).
Hence
\[
0\le\mathbb E[H^2]-C\le S.
\]
Adding these two displays gives
\[
0\le C_0-C\le \Delta:=2^{2e}c+S.
\]

For a signed witness \(a\), its difference-sign orthant is \(\{X:a_iD_i\ge0\text{ for every }i\}\).
On this orthant the intended hinge satisfies \(H=h_a\).
Define the random signed label \(\ell_a=2F(\bar a,\operatorname{bit}R,\operatorname{bit}U)-1\), whose mean over \(R,U\) is \(p(\bar a)\).
Each \(D_i\) has a continuous normal density, so the union of the events \(D_i=0\) has probability zero and may be omitted in these expectations.
Off these events, \(\operatorname{bit}D=\bar a\) on the orthant, so there \(J=h_a\ell_a\) and \(h_aJ=h_a^2\ell_a\).
Outside the difference-sign orthant, the pointwise difference between \(h_aJ\) and \(h_a^2\ell_a\) is at most \(2h_aH\).
There \(H=h_{a'}\), where the signed witness \(a'\ne a\) records the signs of the \(D_i\), so the difference is at most \(2h_ah_{a'}\).
Since \(h_a\) depends only on the differences and \(\ell_a\) only on \(R,U\), averaging the labels gives \(\mathbb E[h_a^2\ell_a]=p(\bar a)\mathbb E[h_a^2]=p(\bar a)s\).
Summing the bounds \(\mathbb E[h_ah_{a'}]\le c\) over the \(2^e-1\) witnesses \(a'\ne a\) therefore yields
\[
|\mathbb E[h_aJ]-p(\bar a)s|\le2^{e+1}c.
\]

Moreover \(h_a\le H\), so \(|\mathbb E[h_a(T-J)]|\le2S\).
Thus
\[
|\mathbb E[h_aT]-p(\bar a)s|\le A:=2^{e+1}c+2S.
\]

The preceding bounds imply
\begin{equation}
\label{eq:gaussian-tail-errors}
\Delta\le qs/128,\qquad A\le qs/128,\qquad
2\sqrt{2^e\kappa}s\le qs/128.
\end{equation}
For explicit comparison, \(\Delta/s\le(2^{2e}+1/4)2^{-10E}\), \(A/s\le(2^{e+1}+1/2)2^{-10E}\), and the last left side divided by \(s\) is at most \(2^{1+e/2-5E}\).
Each is at most \(2^{-n-7}\) because \(E=k+n+10\) and \(e\le k\).

It remains to exclude every other legal graph.
With cap two a candidate is affine, or has one hidden vertex and the form
\[
N(X)=g\rho(w\cdot X+b)+A_0(X),
\]
where \(A_0\) is affine and the displayed present coefficients use distinct occurrences of \(W\).
The hidden-to-output edge is compulsory by reachability, and all coefficients are integers.
If the hidden input vector is zero, or if there is no hidden vertex, the output is an integer affine function.
It is either zero or has squared Gaussian norm at least one.
In the latter case the triangle inequality and \(C_0\le\eta\) make its risk greater than \(C_0\).

Suppose \(w\ne0\), and \(-B\) is not the hidden bias.
Then \(b\in\{-1,0,1\}\), \(\sigma=\|w\|_2\ge1\), and \(|g|\ge1\).
The linear-combination rule established at the start of \Cref{sec:gaussian-risk} makes \(Z=w\cdot X/\sigma\) standard normal.
Integration by parts twice on \((-b/\sigma,\infty)\) gives
\[
\mathbb E[\rho(\sigma Z+b)(Z^2-1)]=\sigma\phi(b/\sigma).
\]
Indeed \(\phi''=(z^2-1)\phi\); the first boundary term vanishes because the hinge is zero at its breakpoint, and the remaining integral is \(-\sigma\int_{-b/\sigma}^\infty\phi'(z)\,dz\).

The polynomial \(Z^2-1\) is the second Hermite polynomial in the normalization used here.
The preceding moment calculation gives \(\mathbb E[Z^2]=1\) and \(\mathbb E[Z^4]=3\).
Thus constants are orthogonal to \(Z^2-1\).
The reflection \(X\mapsto-X\) preserves the standard Gaussian density and \(Z^2-1\), while changing the sign of every linear function of \(X\); the inner product of every linear function of \(X\) with \(Z^2-1\) is therefore zero.
Every affine function of \(X\) is consequently orthogonal to \(Z^2-1\), whose squared norm is \(3-2+1=2\).
Cauchy--Schwarz therefore gives
\[
\|N\|_2^2\ge\frac{g^2\sigma^2\phi(b/\sigma)^2}{2}
\ge\frac{\phi(1)^2}{2}>\frac1{100}.
\]

Since \(\|T\|_2^2\le C_0\le\eta\), every such candidate has risk greater than \(C_0\), by the triangle inequality.
This covers \(-B\) as an input weight, output gain, output bias or unused occurrence, and permits arbitrary affine skips and cancellations.

If \(-B\) is the hidden bias, all other coefficients are units and \(g=\pm1\).
After the output edge is assigned, at most \(2e=k\) units remain for hidden input edges.
The hidden term is then \(\pm h_w\), and its squared norm \(\mathbb E[h_w^2]\) is at most \(s_k\le s_{2k}\).
If \(A_0\ne0\), it has norm at least one, so \(\|N\|_2\ge1-\sqrt{s_{2k}}\); its risk again exceeds \(C_0\).
Otherwise \(A_0=0\) and \(N=\pm h_w\), with \(w\) having \(j\le k\) nonzero unit entries.
Under exactly-once use \(j=k\) is forced: every remaining unit must occupy a distinct hidden input edge.
When \(A_0=0\), no direct input-to-output edge and no output bias can be present.
Such an edge or bias would contribute a nonzero coefficient to \(A_0\), and cancelling it would require a parallel edge or a second output bias.
This accounts for every exactly-once occurrence without padding or increasing the cap.

Under at-most-once use, for \(1\le j<k\), monotonicity and Cauchy--Schwarz give
\[
R_T^{\mathrm G}(N)\ge C-2\sqrt{Cs_j}\ge C-2\sqrt{2^e\kappa}s\ge C-qs/128.
\]
For \(j=0\) the candidate is the zero function.
For \(j=k\), if \(w\) is outside the intended paired family, the cross-moment bound applies to every intended \(a\), so
\[
|\mathbb E[h_wT]|\le\mathbb E[h_wH]\le\sum_a\mathbb E[h_wh_a]\le2^ec.
\]
The risk is then at least \(C-2^{e+1}c\ge C-2A\), for either output sign.
For an intended vector, on a source NO instance \(p(\bar a)\le1/2\) gives \(R_T^{\mathrm G}(h_a)\ge C-2A\).
Also \(p(\bar a)\ge0\), so \(R_T^{\mathrm G}(-h_a)\ge C+s-2A\ge C-2A\), covering even any sign placements excluded by inventory balance.
Every zero function has risk \(C\).
Combining all cases proves, under either inventory rule,
\[
R_T^{\mathrm G}(N)\ge C_0-3qs/128
\quad\hbox{for every legal candidate on a source NO instance}.
\]
On a source YES instance choose \(p(\bar a)\ge1/2+q\).
Its exactly-once legal hinge satisfies
\[
R_T^{\mathrm G}(h_a)=C+s-2\mathbb E[h_aT]
\le C_0-2qs+2A\le C_0-(2-1/64)qs.
\]

Finally we supply rational thresholds, including a finite-bit method of obtaining their calibration.
One can compute a positive dyadic \(\widehat s\) with
\[
|\widehat s-s|\le\theta:=2^{-k-n-12}L
\]
in polynomial time.
Here is an explicit implementation, so no exact Gaussian comparison is assumed.
Write
\[
s=(2\pi k)^{-1/2}\int_B^\infty(y-B)^2\exp(-y^2/(2k))\,dy.
\]
Choose an integer \(b\ge1\) with \(2^{-b}\le\theta\), and put \(M=8(k+b+B+10)\).
In this calibration, \(b\) is a precision, \(M\) a truncation point and \(N\) below a polynomial degree; they are unrelated to the hidden bias \(b\) and the candidate \(N\) above.
The omitted tail is at most
\(2k(M+1)\exp(-M^2/(2k))<2^{-b-4}\): use the Gaussian second-moment tail, \(M\ge8k\), and \(\log_2(2k(M+1))\le k+M+2\).
On \([B,M]\), approximate the exponential uniformly to
\(\xi=2^{-b-6}/M^3\) by its Taylor polynomial in \(y^2/(2k)\).
A degree \(N\ge8M^2+4(b+\lceil\log_2(M^3)\rceil+10)\) suffices: the integral remainder and factorial estimate in \Cref{prop:binary-gap-risk} give an error at most \((3M^2/(2k(N+1)))^{N+1}\le(3/16)^{N+1}\le\xi\).
All degrees and numerical bounds are polynomial in \(k,B,b\).

Approximate \((2\pi k)^{-1/2}\) to \(\xi\), using the \(\pi\) series and square-root bisection already described in \Cref{prop:binary-gap-risk}.
Exact integration of the rational polynomial times \((y-B)^2\), followed by multiplication by that approximation, has error at most \(3\xi M^3\).
Rounding the result dyadically to error at most \(2^{-b-4}\), and adding the tail bound, gives total error below \(2^{-b}\le\theta\).
Polynomially many rational coefficients have polynomial bit length, their common denominator has polynomial bit length, and exact integration and polynomial multiplication preserve polynomial bit complexity.
Since \(\theta<s/2\), the resulting \(\widehat s\) is positive.
The requested \(b\) is polynomial in \(e+n\), by the explicit formula for \(L\).

Take
\[
\tau=(2^e-q)\widehat s,\qquad \delta=(q/4)\widehat s.
\]
Both rationals are positive and have polynomial binary encoding.
The calibration errors for \(\tau\) and \(\tau+\delta\), compared respectively with \(C_0-qs\) and \(C_0-3qs/4\), are at most \(2^e\theta\le qs/4096\).
The displayed YES and NO risk bounds therefore imply respectively \(R_T^{\mathrm G}(h_a)<\tau\) and \(R_T^{\mathrm G}(N)>\tau+\delta\) for every legal NO candidate.
This is the required promise reduction.
Every zero candidate has risk \(C\ge C_0-qs/128>\tau\).
Thus the reduction requires a nonzero occurrence choice.
Dropping \(\delta\) proves the exact positive-threshold hardness claim.
\end{proof}

The exponentially small binary gap and unbounded squared loss also leave \Cref{op:gap-risk} open.
For the exact Gaussian problem, \Cref{thm:gaussian-choice} strengthens the lower bound but supplies neither an equality algorithm nor an exact \(\mathsf{NP}^{\mathsf{PP}}\) upper bound.


\subsection{A Fixed Absolute Gap with Unbounded Loss}

The exponentially small gap in the preceding reduction can be enlarged while retaining the node cap and both inventory rules.
This uses large parameter values and a large threshold; the loss remains unbounded.
The distinction matters for comparison with the clipped-loss model of \Cref{op:gap-risk}.

\begin{corollary}[Fixed absolute Gaussian gap]
\label{cor:gaussian-fixed-gap}
With the absolute gap fixed to \(1/4\), standard-Gaussian unbounded squared-risk synthesis is \(\mathsf{NP}^{\mathsf{PP}}\)-complete in the separator sense under either inventory rule.
Hardness holds with integer inventory and cap two.
The forced-zero restriction \(W=(0)\), cap one, is \(\mathsf{PP}\)-complete for this fixed gap under either rule, even for bias-free rational targets.
The positive rational threshold is part of the input and is not bounded by a numerical constant.
Both hardness statements use deterministic polynomial-time many-one promise reductions.
\end{corollary}
\begin{proof}
We first strengthen \Cref{thm:gaussian-choice} by rescaling the reduction in its proof, which we call the base reduction.
Use its parameters \(e,n,k,E,B,q,s,\widehat s,\eta,L,\tau,\delta\), target \(T\), and reference energy \(C_0=2^es\).
In particular, \(\eta=2^{-10E}\) is the error scale of the base reduction and \(L=2^{-B^2-4B-10}\) is its positive lower bound on \(s\).
Choose \(K\) as the least nonnegative integer power of two satisfying
\[
K^4q\widehat s\ge1.
\]
The base estimates ensure \(q\widehat s<1\), so minimality gives
\[
1\le K^4q\widehat s<16.
\]
Supply target \(T_*=K^2T\), cap two, inventory
\[
W_*=(+K)^{e+1}\uplus(-K)^e\uplus\{-KB\},
\]
and threshold \(\tau_*=K^4(2^e-q)\widehat s\).
Each intended paired hinge becomes \(K^2h_a\): scaling its hidden parameters multiplies its hidden value by \(K\), and scaling its output gain contributes another \(K\).
It uses every occurrence exactly once.

We must control all other graphs because output skips and biases use a single scaled occurrence, so their coefficients lie in \(K\mathbb Z\) rather than \(K^2\mathbb Z\).
The base precision gives \(s/2\le\widehat s\le3s/2\), and hence
\[
K^4<\frac{32}{qs},\qquad
K^2C_0\le6\cdot2^e\sqrt{s/q}
\le6\cdot2^{-5E+n/2}<\frac1{100}.
\]
Here \(k=2e\), \(2^ks\le2^{-10E}\), and \(E=k+n+10\).
Thus \(K\sqrt{C_0}<1/10\).
Every candidate with no input-dependent hidden vertex is affine with coefficients in \(K\mathbb Z\).
Such a candidate is zero or has Gaussian norm at least \(K\).
In the latter case its distance from \(T_*\) is at least
\(K-K^2\sqrt{C_0}>9K/10\), so its risk exceeds \(K^4C_0\).
This class includes a candidate with a constant hidden gate, since the contribution of such a gate has coefficients in \(K^2\mathbb Z\subseteq K\mathbb Z\).

For an input-dependent hidden vertex, if \(-KB\) is not its bias, the hidden bias is absent or \(\pm K\), its input-vector norm is at least \(K\), and its output gain has magnitude at least \(K\).
The calculation using the second Hermite polynomial \(Z^2-1\) in the proof of \Cref{thm:gaussian-choice} therefore gives
\[
\|N\|_2^2\ge K^4\phi(1)^2/2>K^4/100.
\]
This bound remains valid with every affine skip and bias, including all other placements of \(-KB\).
Since \(\|T_*\|_2^2\le K^4C_0\) and \(C_0\le\eta\), its risk is greater than \(K^4C_0\).

If \(-KB\) is the hidden bias, write
\[
N=\pm K^2h_w+KA,
\]
where \(A\) is integer affine and \(w\) has \(j\le k\) nonzero unit entries.
If \(A\ne0\), its norm is at least one; monotonicity gives \(\|h_w\|_2\le\sqrt s\) (or zero for \(j=0\)).
Consequently
\[
\|N-T_*\|_2
\ge K\left(1-K(\sqrt s+\sqrt{C_0})\right)>4K/5,
\]
using \(s\le C_0\).
Its risk again exceeds \(K^4C_0\).
With \(A=0\), the candidate \(\pm K^2h_w\) is \(K^2\) times the base candidate \(\pm h_w=\pm\rho(w\cdot X-B)\), whose hidden bias is \(-B\), or zero when \(j=0\).
Under exactly-once use, the same occurrence count forces \(j=k\); under at-most-once use, the proof of \Cref{thm:gaussian-choice} already handles every \(j<k\).
All corresponding risk bounds therefore multiply exactly by \(K^4\).
The base YES witness has risk below \(\tau_*\), and every base NO candidate has risk above
\[
\tau_*+K^4(q/4)\widehat s\ge\tau_*+1/4.
\]
The discarded large-risk cases also exceed this threshold, because \(C_0>\tau+\delta\) for the unscaled base reduction.
Indeed \(\tau+\delta=(2^e-3q/4)\widehat s\), and the base calibration \(|\widehat s-s|\le\theta\), with \(2^e\theta\le qs/4096\), gives
\[
\tau+\delta\le\left(2^e-\frac{3q}4\right)s+2^e\theta
\le C_0-\left(\frac34-\frac1{4096}\right)qs<C_0,
\]
since \(q,s>0\).
Each discarded risk therefore exceeds \(K^4C_0>K^4(\tau+\delta)=\tau_*+K^4(q/4)\widehat s\).
Thus the reduction has the required fixed gap under both rules.

All parameters have polynomial binary length.
Indeed \(\widehat s\ge L/2=2^{-B^2-4B-11}\), so
\[
\log_2K\le\left\lceil\frac{B^2+4B+n+11}{4}\right\rceil.
\]
Rational comparisons find \(K\) in polynomial time; \(KB,K^2\), the scaled target, and \(\tau_*\) then have polynomial encodings.
The separator in \Cref{prop:binary-gap-risk} applies with the fixed gap \(1/4\), completing this classification.

For the forced-zero restriction, start with the \(\mathsf{PP}\) reduction in \Cref{thm:gaussian-counting}, whose gap is \(\delta=2^{-n-2}\).
Multiply its target by \(\lambda=2^{\lceil(n+2)/2\rceil}\) and its positive threshold by \(\lambda^2\).
The sole candidate remains zero, all risks multiply by \(\lambda^2\), and the resulting NO margin is at least \(\lambda^2\delta\ge1>1/4\).
Target size and coefficient bit lengths remain polynomial, and bias-freeness is preserved.
The fixed-candidate \(\mathsf{PP}\) separator still applies, proving the second classification.
\end{proof}

This is an absolute-gap result with no bound on the numerical threshold.
In the cap-two reduction, writing \(t=K^4q\widehat s\in[1,16)\) gives
\[
\tau_*=t(2^{e+n}-1).
\]
Thus the threshold can be exponentially large even though its encoding is polynomial.
Clipping the loss destroys the scaling identity, so the corollary does not resolve \Cref{op:gap-risk} or give a bounded-loss constant-gap classification.

\phantomsection
\addcontentsline{toc}{section}{References}
\begingroup
\makeatletter
\let\addcontentsline\@gobblethree
\makeatother
\newcommand{\etalchar}[1]{$^{#1}$}

\endgroup

\appendix

\section{Parameter Placements in the Boolean Construction}
\label{app:boolean-placements}

This appendix proves the placement statement used in the Boolean hardness reduction.
The case analysis has four structural stages.
First, coefficient and mixed-difference obstructions restrict gain positions; \Cref{lem:boolean-input-gains,lem:boolean-internal-edge-gains,lem:boolean-internal-bias-gains,lem:boolean-positive-mixed,lem:boolean-negative-mixed} exclude the unintended external and internal cases.
Second, \Cref{lem:boolean-finite-biases,lem:boolean-gain-sources} place the two finite negative parameters at hidden biases and the gains at those biased vertices' output edges.
Third, \Cref{lem:boolean-parallelization} excludes actual coupling edges, using coefficient identities on the clean query cube.
Finally, \Cref{lem:boolean-local-sharing,lem:boolean-global-ramp} use local coefficient comparisons and the full target to identify one shared affine score.
The final subsection assembles these statements into \Cref{lem:boolean-placement}; no later global identification is used to justify an earlier edge exclusion.

All parameters and inputs are those of the calibrated construction.
For the candidate at hand, the clean block is a query block supplied by \Cref{lem:boolean-clean-block}, with every coordinate outside it fixed to zero.
Its restricted cube is the clean query cube, also called the clean cube or the query cube, and the restriction of \(f\) to it is the clean query target, or the query target.
A clean query input is a query coordinate of the clean block.
Unless a face or the full target is named, a winner, clean winner or clean query winner is a winning vertex of the clean query cube; in \Cref{lem:boolean-global-ramp}, which uses the full target, a winner is a winning vertex of \(f\).
At a tested input, a coordinate is selected when its value there is one, and an input edge, direct query term or input primitive is selected when its coordinate is selected.
The selected query sum, or selected power-of-two sum, of a hidden vertex is the sum of the integers \(2^i\) over its selected direct query inputs, where \(F2^i+2w\) is the actual value on that input edge.
A coordinate is a supported coordinate of a hidden vertex, and of its direct query sum or direct affine form, when its direct input edge into that vertex is present, that is, when it lies in the support of that vertex's direct input coefficient vector.

We retain the actual activation patterns and the formal identities justified by \Cref{lem:boolean-specialization}.
A missing edge or bias has coefficient zero and consumes no occurrence.
Hidden vertices are numbered in topological order, with at most three of them.
For the fixed tested input, \(\eta_v\in\{0,1\}\) is the actual activation indicator of hidden vertex \(v\), as defined in \Cref{sec:model}.
The same convention gives \(\eta_h,\eta_g\) when the vertices are named \(h,g\); the input is suppressed from the notation.
The tuple of these indicators is the actual activation pattern at the input, also called its activation mask, or simply its mask; the mask of a single hidden vertex is its indicator.
Freezing the pattern replaces each hidden ReLU by its affine preactivation or by zero according to its indicator, as in \Cref{lem:boolean-specialization}.

A frozen transfer from a vertex \(u\) to a later vertex \(v\) is the sum over directed paths from \(u\) to \(v\) of the edge-weight product times the activation indicators of the hidden vertices after \(u\) on that path.
It is zero if there is no contributing path.
This is the coefficient transmitting the value at \(u\) through the frozen computation, by expansion in topological order.
The transfer is short when each of its paths passes through at most one hidden vertex strictly between \(u\) and \(v\).
A frozen expression is a vertex value, preactivation, transfer or output computed with the pattern frozen in this way and treated as a Laurent polynomial in the formal variables \(w,\Gamma\) recalled below.
The frozen pattern, or frozen mask, is the actual pattern used for it.

The source vertex of a gain edge is its gain source, and the gain is said to originate at that vertex.
The prefix ending at a hidden vertex consists of that vertex and the earlier hidden vertices from which it can be reached, with the edges and biases entering them.
Its value is the value of that vertex, and we also speak of the prefix ending at \(h\) when that value is named \(h\).
The direct affine form of a hidden vertex is the sum of its direct query terms and its bias.
A first affine query prefix is a prefix ending at a hidden vertex that, under the frozen pattern, receives no contribution from any earlier hidden vertex, for instance because every hidden predecessor is inactive.
Its preactivation is then its direct affine form, an affine function of the clean query inputs called its first affine form; its value is a first affine value, and its final vertex is a first affine source when a gain originates there.

A gain relay is a hidden-to-hidden gain edge; it carries the value of its source vertex, multiplied by the gain, into its target vertex.
The ordinary output coefficient of a hidden vertex is the coefficient of its hidden-to-output edge when that edge does not carry a gain, and zero when that edge is absent.
A hinge, also called a gate, is a hidden ReLU vertex; the first, middle and last gates are the hidden vertices in topological order.
On a set of inputs, a hinge is effective, or genuine, unless it equals its preactivation throughout that set or vanishes throughout it.

Recall that \(w=M^{-1}\) and \(\Gamma=M^9\) are the two formal variables of the construction.
Put \(D=w^{-1}\), so that the two gain occurrences \(+M,-M\) are \(+D,-D\).
When \(+D\) is a hidden-to-output edge, its gain source is the \emph{positive-output source}; when \(-D\) is one, its gain source is the \emph{negative-output source}.
A path, transfer or frozen expression is gain-free when no gain occurrence enters it, that is, when no edge or bias contributing to it carries \(+D\) or \(-D\); a vertex is gain-free when its value is.
When both gains are hidden-to-output edges, the signed gain contribution at an input is \(D\) times the value of the source of \(+D\) minus \(D\) times the value of the source of \(-D\), and its value there is the signed gain value.

By \Cref{lem:boolean-specialization}, the frozen output at a tested input is identically its value zero or one, so its coefficient of every nonzero power of \(w\) vanishes.
The coefficient of \(w^{-1}\) in the frozen output, or in a term, path or transfer contributing to it, is called its inverse coefficient.
The inverse equation, or inverse-\(w\) equation, is the requirement that the inverse coefficient of the frozen output vanish; inverse-square terms and equations refer in the same way to \(w^{-2}\).
For a gain-free expression \(h\) under a frozen pattern, write \(h^0\) for its coefficient of \(w^0\), still a Laurent polynomial in \(\Gamma\).
This notation does not assert that \(h^0\) numerically dominates every positive power of \(w\), since \(\Gamma=M^9\).
Signs are determined after the prescribed specialization.
For a gain-free frozen expression \(h\), write \(h^{(1)}\) for its \emph{derivative}: its formal derivative at \(w=0\), holding \(\Gamma\) fixed, or equivalently its coefficient of \(w^1\).
This does not differentiate the activation pattern or the numerical specialization.
The leading value of a gain-free expression \(h\) is \(h^0\).

For a nongain primitive \(c\), its leading value \(c^0\) is also called a nongain leading primitive, and \(c-c^0\) is its correction.
More generally, the correction of a gain-free expression \(h\) is \(h-h^0\); when it is a scalar multiple of \(w\), that scalar is also called its correction.
Every nongain leading primitive is a signed finite value \(Fn\), a value \(\Gamma\), or a value \(s_0=1/(q\Gamma)\).
For a finite primitive, the integer \(n\) satisfies \(1\leq|n|\leq Q\); this generic \(n\) is not the number of source numbers.
The values \(\Gamma\) and \(s_0\) are the nonfinite primitives.
When an ordinary scalar coefficient \(c\) is finite or absent, write \(c^0=Fn_c\), where \(n_c\in\mathbb Z\) and \(n_c=0\) for an absent coefficient.
The same indexed convention applies to other scalar coefficient names, such as \(n_t,n_a,n_b\).
Dividing such a finite leading value by \(F\) normalizes it, so the normalized value of \(Fn_c\) is \(n_c\).
For a direct affine form \(U\), write \(L_U\) for the normalized finite part of \(U^0\), so that \(U^0\) is \(FL_U\) plus its \(\Gamma\) and \(s_0\) terms; for a first affine form it is the first affine leading sum.
The finite negative primitives both have leading value \(-P\).
A positive finite primitive has correction zero, \(2w\), \(-w\), or \(-3w/2\); a finite negative has correction zero or \(w\).

The \(\Gamma\)-degree and \(w\)-degree of a monomial \(c\,\Gamma^jw^k\), with \(c\) a nonzero scalar, are \(j\) and \(k\), so a factor \(s_0=1/(q\Gamma)\) has \(\Gamma\)-degree \(-1\).
An expression has \(\Gamma\)-degree \(j\), or \(w\)-degree \(k\), when all its monomials do, and it has nonnegative, nonzero, positive or negative such degree when all its monomials do; its constant \(\Gamma\)-degree part is the sum of its monomials of \(\Gamma\)-degree zero.
Since \(\Gamma s_0=1/q\), one \(\Gamma\) factor and one \(s_0\) factor in a product multiply to the constant \(1/q\), and such a pair of factors is called a balanced pair, or a balanced \(\Gamma,s_0\) pair.
A product or path contribution is balanced when it contains at least one \(\Gamma\) or \(s_0\) factor and its \(\Gamma\) and \(s_0\) factors are equal in number, so that they form balanced pairs and its \(\Gamma\)-degree is zero.
It is pure balanced when it has no other factor, so that its value is \(1/q\) or \(1/q^2\).

The \emph{winning stencil} of the clean query cube, also called its ordinary winning stencil, consists of its all-ones vertex and its single-coordinate deletions.
These vertices win by \Cref{lem:boolean-clean-block} and affinely span the query cube.
We use the corresponding stencil, the winning stencil of the face or face stencil, on faces obtained by fixing at most two coordinates to zero or one coordinate to one.
Those vertices also win: even deleting three coordinates leaves a weighted sum at least \(3Q-3(Q+1)/2>Q\).
Winning span, or winning affine span, is the consequence of this spanning used throughout: an affine function of the free query coordinates, whose coefficients may be Laurent polynomials in \(\Gamma\), that takes one value at every vertex of the winning stencil of the cube or face in question equals that constant, so its query coefficients vanish; in particular, one that vanishes at every winner vanishes identically.
When two-coordinate deletions are used in addition, they are specified explicitly.

The robust face statement of \Cref{lem:boolean-clean-block} is its last assertion, which covers exactly these faces.
The robust variation argument is the paragraph of the proof of \Cref{lem:boolean-negative-mixed} headed ``A first finite-negative-biased hinge has a robust leading variation''; it shows that a first hidden vertex with a finite negative bias takes three distinct leading values at winners of the clean cube.
For a hinge with a finite negative bias and direct query sum \(U\), a leading-zero boundary is a tested input at which the normalized leading value of \(U\) equals \(Q\), so that the leading preactivation vanishes and the hinge value is its correction.
The score of a hinge is its preactivation without its negative bias.
When the score is affine, the effective coefficient of a coordinate \(x_j\) is its total coefficient there: the direct input coefficient plus the sharing weight times the wire's coefficient of \(x_j\), with the sharing weight and wire as in \Cref{lem:boolean-placement}.

\subsection{Coefficient and Transfer Comparisons}

Several comparisons recur in the placement cases.
We prove them once with the full numerical ranges, rather than infer distinctness solely from genericity.
In particular, equality of leading finite values does not identify the two negative primitives.

\begin{lemma}[Positive primitives and short products]
\label{lem:boolean-short-products}

Two positive nongain primitives with the same zeroth \(w\)-coefficient have the same full value.
A nonzero product of two nongain leading primitives cannot equal a sum or difference of at most two nongain leading primitives.
The latter assertion permits signed finite factors and signed finite summands.
\end{lemma}

\begin{proof}

Among positive finite leading integers, the source values \(\beta_j\) are below \(T\), the query values are powers of two at most \((Q+1)/2\), and the anchor value \(Q-T\) is larger than both ranges.
The control value is \(Q\).
The source values are not powers of two.
Thus equality can occur only within a single numerical family, whose full correction is fixed.
Equal \(\Gamma\) or \(s_0\) values are already identical.

For products, the six unordered type pairs have leading forms
\[
F^2n_1n_2,\quad Fn\Gamma,\quad Fns_0,\quad
\Gamma^2,\quad 1/q,\quad s_0^2.
\]
A sum of two finite primitives has absolute value at most \(2P<F^2\), so it cannot equal the first form.
At exponent \(\Gamma\), or at exponent \(\Gamma^{-1}\) after scaling by \(q\), a sum of two primitives has coefficient of absolute value at most two.
The corresponding finite-times-nonfinite product has coefficient of absolute value at least \(F>2\).
A square \(\Gamma^2\) or \(s_0^2\) has an exponent unavailable to a primitive sum.
Finally, a sum with no nonzero \(\Gamma\)-exponent has coefficient an integer multiple of \(F\), and hence cannot equal \(1/q\).
These cases also cover absent summands and every permitted sign.
\end{proof}

The next obstruction applies at one winning point, before any global graph normalization.
It also applies after gain terms cancel exactly in a frozen expression.
It does not require deleting their edges from the network.

\begin{lemma}[The gain-free coefficient obstruction]
\label{lem:boolean-coefficient-obstruction}

In an actual frozen output expression, the paths avoiding both gain occurrences have \(\Gamma^0w^0\)-coefficient
\[
\frac{FZ}{q^2}+\frac{k_1}{q}+\frac{k_2}{q^2},
\qquad Z\in\mathbb Z,\quad  0\leq k_1,k_2\leq p.
\]
This coefficient cannot equal one.
Consequently, at an output equal to one, the sum of all gain-containing path contributions cannot be identically zero.
If all gain-containing contributions vanish or cancel exactly, output one is impossible.
\end{lemma}

\begin{proof}

A gain-free path has nonnegative \(w\)-degree.
For its zeroth coefficient, every finite factor therefore contributes its \(Fn\) term, giving a factor \(F\).
A balanced product contains at most two \(\Gamma,s_0\) pairs, so its denominator divides \(q^2\).
This accounts for the term \(FZ/q^2\), with arbitrary signs coming only from finite negative factors.

The length of a path is its number of parameter factors, counting the input edge or bias at which it starts.
That input edge or bias is the origin of the path, and origin counts count such starting parameters.
Elsewhere, as in ``at the origin'' or ``the origin test'', the origin is the all-zero input.
The \emph{origin test} is the requirement that the candidate output the target value zero at the origin, and a \emph{unit test} is the same requirement at a unit vector, where the target is also zero by \Cref{lem:boolean-full-target}.
Without a finite factor, a constant \(\Gamma\)-degree requires either one balanced pair on a length-two path or two pairs on a length-four path.
Their values are \(1/q\) and \(1/q^2\), respectively, and their signs are positive.
For a length-two path, a selector used as an input edge or bias enters a hidden vertex, and the \(\Gamma\) factor must then be that vertex's output edge.
Such a selector therefore has only one possible final \(\Gamma\) edge.
A selector used as the output edge can instead meet two \(\Gamma\) origins, but this consumes both available \(\Gamma\) occurrences and supplies at most two paths in total.
Otherwise, when no selector output edge meets two \(\Gamma\) origins, each pair can be charged to a distinct selector occurrence.
Thus the number \(k_1\) of such paths is at most \(p\), using \(p\geq4\).

A length-four path must use all three hidden vertices in their topological order.
Its last three edges are then fixed.
If the initial occurrence is a selector, at most \(p\) such initial occurrences are available.
If it is a \(\Gamma\) occurrence, the last three edges contain one \(\Gamma\) and two selectors, leaving at most one possible initial \(\Gamma\) occurrence.
These alternatives cannot occur together for the same fixed last three edges.
This proves \(k_2\leq p\).

Equality to one would give
\[
FZ+qk_1+k_2=q^2.
\]
However, \(0\leq qk_1+k_2\leq q^2-1<F\) and \(q^2<F\), which contradicts this equation modulo \(F\).
The cancellations in the statement concern full formal contributions of gain-containing paths, so they leave the above calculation of the gain-free \(\Gamma^0w^0\)-coefficient valid under the original pattern.
\end{proof}

The query target also requires negative parameters at two distinct hidden preactivations.
This elementary allocation statement restricts signed output terms in the subsequent cases.

\begin{lemma}[Negative allocation]
\label{lem:boolean-negative-allocation}

Let a permitted candidate represent the clean query target\linebreak of \Cref{lem:boolean-clean-block} on its restricted cube.
Then at least two distinct hidden vertices must receive negative preactivation contributions.
At most one of the three supplied negative occurrences can therefore be an output parameter.
If all output parameters are positive, the output bias is absent, every direct input-to-output term vanishes on the restriction, and every hidden value with a direct edge to the output is zero at the origin and at the unit vectors of the restricted cube.
If the candidate represents the full target \(f\) and all its present output parameters are positive, all direct input-to-output edges are absent globally.
\end{lemma}

\begin{proof}

A hidden vertex with only nonnegative incoming coefficients, a nonnegative bias and nonnegative predecessor values applies \(\rho\) to a nonnegative expression.
Its ReLU is the identity on the restricted cube.
If negative contributions enter at most one hidden vertex, substitute these positive identity gates in topological order.
The result is an affine function plus a scalar multiple of the sole remaining ReLU of an affine function.
This contradicts \Cref{lem:boolean-clean-block}.
A negative input edge fixed to zero contributes nothing and cannot create an additional effective nonlinearity.
Only three negative occurrences are supplied, so at most one can be assigned to the output.

With positive output parameters every output summand is nonnegative.
A positive output bias would contradict the target value zero at the origin.
A present positive edge from a query input directly to the output would contradict its value zero at the corresponding restricted unit vector.
Other external inputs are fixed to zero and their direct output terms vanish on the restriction.
For the full target, every unit vector also loses.
For a source coordinate, \(\beta_j\leq Q-1\) gives \(F\beta_j\leq P-F<P-w\).
An anchor baseline is \(F(Q-T)-w<P-w\), since \(T>0\).
A query baseline satisfies \((P-w)-(F2^i+2w)=F(Q-2^i)-3w>0\), since \(Q-2^i\geq1\) and \(F>3w\).
The control baseline is \(P-3w/2<P-w\).
Thus \(f(e_t)=0\) for every coordinate \(t\), and the same positive-summand argument excludes every direct input-to-output edge globally.
The remaining nonnegative hidden summands must individually vanish at all those tests.
\end{proof}

Helmbold and Long proved a sign restriction for networks with nonnegative edge weights \cite[Lemma 12]{HelmboldLong2018}.
For a Boolean function, call it supermodular when every pairwise mixed difference, with all other coordinates fixed, is nonnegative.
We record the Boolean-cube form of their restriction and its consequence for the query target.

\begin{lemma}[Nonnegative-edge restriction]
\label{lem:boolean-supermodular}

A ReLU network with nonnegative edge weights, arbitrary hidden biases and an affine output with nonnegative edge weights is monotone and supermodular on the Boolean cube.
An arbitrary output bias does not affect the supermodular conclusion.
Such a network cannot represent the query target.
\end{lemma}

\begin{proof}

Monotonicity follows by induction through the nonnegative edges.
Positive sums preserve supermodularity, and affine input terms have zero mixed difference.
It remains to check that taking positive parts preserves the mixed-difference inequality.
On one two-coordinate face, let \(A\) and \(E\) be the preactivation values at the corners where both free coordinates are zero and where both are one, respectively.
Let \(B\) and \(C\) be the values at the two mixed corners, with the first free coordinate equal to one at \(B\).
Monotonicity gives \(A\leq B,C\leq E\), and supermodularity gives \(A+E\geq B+C\).
If \(A\geq0\), taking positive parts changes no term.
If \(E\leq0\), all four positive parts are zero.
Otherwise \(A<0<E\).
When both \(B,C\leq0\), the required inequality is \(E\geq0\).
When exactly one is positive, it is bounded above by \(E\).
When both are positive, \(E\geq B+C-A\geq B+C\).
Thus positive part preserves the mixed-difference inequality in every case.
Induction proves the assertion for hidden nodes and their nonnegative output sum.
The query target has an \emph{OR face}, a two-coordinate face on which it is the OR of the free coordinates, given by the weight-\((Q-1)\) background in the proof of \Cref{lem:boolean-clean-block}; its mixed difference is negative, so the target cannot belong to this class.
\end{proof}

The remaining short-transfer equalities require slightly more than a primitive-versus-product comparison.
They occur when two gain paths could meet at a later vertex.
All missing coefficients below are interpreted as zero.

\begin{lemma}[Two short-transfer equations]
\label{lem:boolean-short-transfers}

Let \(b\) be a positive nongain leading primitive and \(c_3\) a nonzero nongain leading primitive.
Let \(c_2,e\) be nongain leading primitives or zero, with signed finite values permitted.
Then
\[
c_2=c_3(b-e)
\]
forces \(c_2=0\) and \(b=e\).
Also
\[
b(c_2+c_3e)=c_3
\]
is impossible.
In the first conclusion, \(b=e\) identifies their full positive primitive values.
\end{lemma}

\begin{proof}

Suppose first that \(c_2\ne0\), and divide the first equality by \(c_3\).
If both \(c_2,c_3\) are finite, their ratio is a nonzero rational of absolute value at most \(Q\).
A nonzero finite difference \(b-e\) is an integer multiple of \(F>Q\); a difference involving nonfinite primitives has a nonzero \(\Gamma\)-exponent unless the identical terms cancel to zero.
Neither possibility gives that ratio.

If exactly one of \(c_2,c_3\) is finite, the ratio is a single \(\Gamma\) or \(s_0\) monomial whose coefficient, relative to that primitive, is either \(qFn\) or \(1/(Fn)\).
It cannot match the unit coefficient of one nonfinite primitive in \(b-e\); an additional finite term or a second exponent cannot help.
If both are nonfinite, their ratio is one or has exponent two or minus two in \(\Gamma\).
A difference of primitives cannot give these values: its exponents are only minus one, zero and one, and its nonzero constant coefficient is divisible by \(F\).

Thus \(c_2\ne0\) is impossible.
With \(c_2=0\), the equation and \(c_3\ne0\) give \(b=e\).
The equal value is positive, so \Cref{lem:boolean-short-products} identifies the full primitives.

For the second equation, first take \(c_3\) nonfinite and compare the coefficient at its own \(\Gamma\)-exponent, dividing by its coefficient.
The term \(bc_2\) can contribute only an integer multiple of \(F\) at this exponent.
The term \(bc_3e\) contributes either an integer multiple of \(F\), or the single positive balanced value \(1/q\), or zero.
The resulting equation is \(1=FZ+I/q\) with integer \(Z\) and \(I\in\{0,1\}\).
Clearing \(q\) contradicts reduction modulo \(F\).

It remains that \(c_3=Fn_3\) is finite, with \(n_3\ne0\).
If \(b\) is finite, every constant-degree contribution to \(bc_2+bc_3e\) is divisible by \(F^2\).
Equality to \(Fn_3\) would make \(n_3\) divisible by \(F\), although \(0<|n_3|\leq Q<F\).
If \(b\) is nonfinite, the only constant-degree contributions to \(bc_2\) and \(be\) are balanced products.
Write their indicators as \(I_1,I_2\in\{0,1\}\).
The constant-degree equation becomes
\[
I_1+Fn_3I_2=qFn_3.
\]
Its right side after rearrangement has absolute value \(|Fn_3(q-I_2)|\geq F(q-1)>1\), whereas \(I_1\leq1\).
This contradiction exhausts the types and signs.
\end{proof}

Neither gain may be placed at the affine output bias.
Unlike the later graph cases, its proof uses only the origin.

\begin{lemma}[Gain output biases]
\label{lem:boolean-gain-bias}

A candidate for \(f\) cannot assign either \(+D\) or \(-D\) to its output bias.
\end{lemma}

\begin{proof}

Suppose first that the output bias is \(-D\).
Since the target is zero at the origin, the other output paths must contribute coefficient one at \(w^{-1}\Gamma^0\) there.
Only paths containing \(+D\) can supply this coefficient.
Finite factors contribute an integer multiple of \(F/q\).
A path without a finite factor must contain one balanced \(\Gamma,s_0\) pair in addition to \(+D\).

At the all-zero input there are at most two such paths in a graph with at most three hidden vertices.
Indeed, suppose first that \(+D\) is either the bias of the first hidden vertex or an output edge, in the latter case the final edge of its path.
Then the other two factors must form the balanced pair, and the path passes through two hidden vertices, one of which carries \(+D\) as its bias or output edge.
The other hidden vertex has at most two choices, so there are at most two such paths.
If \(+D\) is an internal edge, the only possible origin of the path is the bias of the source vertex of \(+D\).
The path then ends at the output edge of the target vertex of \(+D\), so there is at most one such path.
If \(+D\) is the bias of the middle hidden vertex, at most one path has its other two factors forming a balanced pair.
If \(+D\) is the bias of the last hidden vertex, there is no such path.
An input gain contributes zero at the origin.
Two balanced pairs and a gain need five factors and do not occur.

Thus the coefficient equation has the form
\[
FZ+k=q,\qquad Z\in\mathbb Z,\quad 0\leq k\leq2,
\]
after clearing \(q\).
It is impossible modulo \(F\), since \(2<q<F\).

For output bias \(+D\), the negative-gain paths must instead supply coefficient one.
Finite-factor paths give an \(F/q^2\) multiple.
Every gain path with no finite factor again has only one balanced pair, with nonnegative count at most \(p\).
The coefficient equation after clearing \(q^2\) is
\[
FZ+qk=q^2,\qquad 0\leq k\leq p.
\]
Since \(qk\leq q(q-1)<q^2<F\), this too is impossible modulo \(F\).
If the other gain is unused, there is no inverse coefficient to cancel the bias at all.
\end{proof}

\subsection{A Single Effective Gain}

We first exclude candidates in which one gain is unavailable on the query restriction.
The two signs require different statements.
The negative-gain statement is pointwise; the positive internal-gain statement uses equality on the whole restricted cube.
This distinction is needed when a gain is assigned to an input coordinate fixed to zero.

\begin{lemma}[The pointwise negative-gain obstruction]
\label{lem:boolean-sole-negative}

At a clean query winner, output one is impossible if every individual active path contribution containing \(+D\) is zero.
The occurrence \(-D\) may occupy any permitted position.
In particular this applies when \(+D\) is unused, lies on a fixed-zero input edge, or is an output edge whose source is inactive.
\end{lemma}

\begin{proof}

Fix the actual pattern at the tested point.
The \(+D\) contributions vanish by hypothesis.
The paths avoiding both gains have constant coefficient \(FZ/q^2+r_0\), where
\[
r_0=\frac{k_1}{q}+\frac{k_2}{q^2},\qquad 0\leq r_0<1,
\]
by \Cref{lem:boolean-coefficient-obstruction}.
There must therefore be a nonzero \(-D\) contribution.
All selected external-input primitives are actual query values, even though their levels need not match their coordinate labels.
Every nonzero path under discussion also avoids \(+D\), since individual path contributions containing that occurrence vanish by hypothesis.
Terms with that occurrence are discarded only as zero contributions in the actual pattern.

\emph{An output edge.}\par\nopagebreak

Here \(-D\) is the hidden-to-output edge from a hidden vertex \(h\), and \(h\) also denotes that vertex's frozen value.
Write the frozen output as \(A-Dh\), where \(A\) contains all other output terms, including any signed finite ones.
The inverse-\(w\) equation forces \(h^0=0\) as a Laurent polynomial in \(\Gamma\).
The next display groups the terms of a constant coefficient by their number of finite factors, and its symbols are explained after it in the order of its terms.
For a prefix ending at \(h\) with at most three hidden vertices, the constant \(\Gamma\)-coefficient of \(h^0\) has the form
\[
F^3A_3+F^2A_2+FL+\frac{FN_0}{q}+\frac{k}{q}=0,
\qquad A_2,A_3,L,N_0\in\mathbb Z,\quad 0\leq k\leq p.
\]
A term with three finite factors belongs to \(F^3A_3\), and one with two finite factors belongs to \(F^2A_2\).
The integer \(L\) is the normalized sum of direct query leading values at the final prefix vertex and its finite bias.

Only one three-factor path shape passes through all three hidden vertices: an origin at the first hidden vertex, followed by the edge from the first to the second hidden vertex and the edge from the second to the third.
We call the paths of this shape the three-factor chain.
If a path of that chain has a balanced \(\Gamma,s_0\) pair and one finite factor, the contribution of such paths to the display is \(FN_0/q\), which defines the integer \(N_0\); otherwise \(N_0=0\).
The finite factor of such a path lies in one of two places.
Either it is the origin, and these origins together give the finite-first sum, the finite direct query inputs and bias of the first hidden vertex, times the two nonfinite edges.
Or it is one of the two internal edges, and that finite edge is multiplied by a nonnegative origin count: the number of origins of the chain that carry the \(\Gamma\) or \(s_0\) value completing the balanced pair.
Input-origin nonfinite terms are absent on the clean block: every input edge from a coordinate of the block carries a query value, and every other input is fixed to zero.
The only possible nonfinite origin of the chain is therefore the bias of the first hidden vertex, so this count is zero or one, and in particular at most \(p\).
All remaining balanced two-factor paths give \(k/q\).

The direct and finite-first sums have absolute value at most \((3r+1)Q\), and an internal normalized product has absolute value at most \(pQ\).
Since \(q<r<Q\) and the calibrated construction takes \(F>8Q^4\), it follows that \(|qL+N_0|<F\).
Clearing \(q\) and reducing modulo \(F\) first gives \(k=0\).
Dividing by \(F\) and reducing modulo \(F\) then gives \(qL+N_0=0\); the higher powers do not affect this reduction.
Thus \(N_0\equiv0\pmod q\).

The \(\Gamma^0w^1\)-coefficient of \(h\) has the following form for an integer \(Z_1\):
\[
\frac{FZ_1}{2q}+\delta+\frac E q.
\]
The term \(\delta=2a+\delta_{\mathrm{bias}}\) is the correction of its direct selected-query sum and finite bias; here \(a\in\{0,\ldots,3r\}\) counts the direct query inputs equal to one at the tested point.
The half-integer \(E\) is \(q\) times the correction of the same possible balanced three-factor chain.
Every other differentiated term retains a finite factor and belongs to \(FZ_1/(2q)\).
If \(E<0\), that chain's finite factor must be the anchor baseline \(B=F(Q-T)-w\) or the control baseline \(C=P-\frac32w\) of the calibrated construction, the only primitives with a negative correction.
For a finite-first sum, one selected query already makes its correction positive, contrary to \(E<0\).
With none selected, its normalized leading value is \(Q-T\) or \(Q\), which is nonzero modulo \(q\) and contradicts \(N_0\equiv0\pmod q\).
For a finite internal factor, its leading value is nonzero modulo \(q\) and is multiplied by a positive count less than \(q\), which again contradicts \(N_0\equiv0\pmod q\).
Hence \(E\geq0\).

The output-one equation consequently gives, for an integer \(Z_2\),
\[
\frac{FZ_2}{2q^2}+r_0-\delta-\frac E q=1.
\]
After clearing \(2q^2\), the part not visibly divisible by \(F\) is
\[
2qk_1+2k_2-2q^2\delta-2qE-2q^2,
\]
an integer, since \(\delta\) and \(E\) are half-integers.
Here \(|\delta|\leq6r+2\), because \(a\leq3r\) and a finite bias has correction of absolute value at most two.
Also \(|E|\leq6r+2\).
When the finite factor of the chain is its origin, \(E\) is the correction of the finite-first sum, which is twice its number of selected query inputs, at most \(3r\), plus the correction of the first finite bias, of absolute value at most two.
When the finite factor is an internal edge, \(E\) is the correction of that edge, of absolute value at most two, times the origin count, which is at most one.
These two shapes cannot occur together, since in the first both internal edges of the chain are nonfinite and in the second exactly one is.
With \(0\leq k_1,k_2\leq p=q-1\), so that \(2qk_1+2k_2\leq2q^2-2\), and with \(q,r<Q\), the displayed integer has absolute value at most
\[
2q^2(6r+4)+2q(6r+2)<12Q^3+20Q^2+4Q<50Q^3<8Q^4<F,
\]
where \(50Q^3<8Q^4\) holds because \(Q=2^r-1\geq127\).
Reduction modulo \(F\) therefore gives literal equality of this small part.

We name these quantities for later use.
The period is \(F/(2q^2)\): before \(2q^2\) is cleared, every term containing a factor \(F\) is an integer multiple of it.
The small residue of a constant coefficient is the sum of its terms without a factor \(F\).
Here the small residue is \(r_0-\delta-E/q\), and it is smaller in absolute value than the period.
The gain residue and the gain-free residue are the parts of the small residue contributed by gain paths and by gain-free paths, respectively.
Here they are \(-\delta-E/q\) and \(r_0\).
Since \(r_0<1\) and \(E\geq0\), the literal equality \(r_0-\delta-E/q=1\) requires \(\delta<0\).
Only a positive \(B\) or \(C\) final bias has a negative correction, so \(L\geq Q-T\).
Also \(N_0\geq-pQ\), including every signed internal finite choice.
It follows that
\[
0=qL+N_0\geq q(Q-T)-pQ=Q-qT>0.
\]
The final strict inequality follows from \(Q>q^2\) and \(Q>8T^2\).
This is a contradiction.

\emph{An edge entering the last hidden vertex.}\par\nopagebreak

Let \(z\) be that last vertex's nonzero output coefficient and \(u\) the source of the gain edge.
The inverse coefficient forces \(u^0=0\).
If \(z\) is finite, the constant gain coefficient is an \(F/(2q)\) multiple, which changes only the divisible part of the preceding residue calculation.
Indeed a two-hidden nongain prefix has half-integer \(\Gamma^0w^1\)-coefficient; a balanced pair without a finite factor has no \(w\)-correction.
The correction of \(z\) multiplies \(u^0=0\).
Thus the small residue remains \(r_0<1\), contradicting output one.

For a nonfinite \(z\), write the two-hidden prefix as \(u=\rho(V+t h_1)\), allowing a missing first vertex or edge.
Here \(h_1\) is the value of the first hidden vertex, \(t\) is the coefficient of its edge into \(u\), and \(V\) collects the direct inputs and bias of \(u\).
The constant gain coefficient then needs the \(w\)-coefficient of \(u\) at the \(\Gamma\)-degree opposite to that of \(z\), and this coefficient can arise in only two ways: from the correction of a finite \(t\), or from the derivative of \(h_1\) carried by a matching nonfinite \(t\).
If \(t\) is finite, its correction multiplies the count of opposite-degree origins entering \(h_1\).
The corresponding coefficient of \(u^0\) has equation \(Fn_t k+k'=0\), where \(k\) and \(k'\) are the numbers of opposite-degree origins entering \(h_1\) and \(V\), respectively, \(k'\leq p\), and \(n_t\ne0\).
As \(F>p\), this forces both counts to be zero, and the derivative vanishes.

In the second way, \(t\) is the matching \(\Gamma\) or \(s_0\) primitive, and the coefficient of \(t\) in \(u^0\) satisfies \(FL_1+k'=0\), where \(FL_1\) is the \(\Gamma^0\)-coefficient of \(h_1^0\).
Again \(L_1=k'=0\).

In this second way, the affine preactivation of \(h_1\) is a sum of positive query inputs and a bias with zero finite leading value.
Its finite bias is therefore absent or negative, since a positive finite bias could not cancel those inputs.
Its derivative \(d=2a_1+\delta_{\mathrm{bias}}\) is nonnegative, where \(a_1\) counts its query inputs equal to one and \(\delta_{\mathrm{bias}}\) is the correction of its finite bias.
The gain then contributes a small constant residue \(-d/q\), or zero if the first vertex is inactive.
These are all possibilities, because a two-factor path has no other opposite-degree source.
The full small residue is therefore at most \(r_0<1\), which again cannot be one after clearing \(2q^2\) and reducing modulo \(F\).

\emph{The first-to-middle edge.}\par\nopagebreak

All gain paths have combined form \(-Dh_1\Theta\), with \(\Theta=b+\eta cs\), where \(\eta\in\{0,1\}\) is the actual last-vertex indicator.
Here \(b\) and \(c\) are the output coefficients of the middle and last hidden vertices, \(s\) is the coefficient of the edge from the middle to the last hidden vertex, and absent edges are zero.
If a gain path is present, \(\Theta^0\ne0\) by \Cref{lem:boolean-short-products}.
The inverse coefficient thus forces \(h_1^0=0\).
The derivative \(d\) of its affine preactivation is nonnegative by the preceding clean-input argument.
The constant \(\Gamma\)-degree of \(\Theta^0\) is an \(F\)-multiple plus either zero or \(1/q\); only a positive balanced pair supplies the latter.
Hence the small gain residue is zero or \(-d/q\).
It cannot supplement \(r_0<1\) to give one.

\emph{Biases and external-input edges.}\par\nopagebreak

A negative gain on an external-input edge is fixed to zero on the clean block and leaves no gain contribution.
A middle hidden bias has inverse coefficient \(-\Theta^0\), with nonzero short transfer whenever a gain path is present.
For a bias at the last hidden vertex, the transfer from that vertex to the output is its nonzero output-edge coefficient.
A first hidden bias \(-D\) makes that gate inactive throughout the clean cube: its finite query inputs sum to at most \(3rP+6rw<M\), and no nonfinite input edge is active there.
An output bias \(-D\) has inverse coefficient minus one with no remaining gain path to cancel it.
These cases exhaust every possible position.
\end{proof}

The next obstruction has the opposite sign but applies to the whole query function.
Positive output coefficients permit the origin and unit tests that its proof needs.

\begin{lemma}[A sole positive internal gain]
\label{lem:boolean-sole-positive-internal}

Suppose \(-D\) is unused in a legal candidate representing the entire clean query target.
Then \(+D\) cannot be a hidden-to-hidden edge or a hidden bias.
\end{lemma}

\begin{proof}

By \Cref{lem:boolean-negative-allocation}, both finite negatives enter distinct hidden preactivations.
Thus all output parameters are positive.
The output bias and skips are absent, and directly output-connected hidden values vanish at the origin and unit vectors.
At every winner the positive gain must contribute to the constant coefficient, since otherwise \Cref{lem:boolean-coefficient-obstruction} applies.

\emph{An edge into the last hidden vertex.}\par\nopagebreak

Let \(z>0\) be the output coefficient of the last hidden vertex, and \(u=\rho(V+t\rho(U))\) the at-most-two-hidden prefix, with \(u\) the source of the gain edge and \(V\), \(t\) as in the case ``An edge entering the last hidden vertex'' of the proof of \Cref{lem:boolean-sole-negative}, and \(U\) the affine preactivation of the first hidden vertex, so that \(\rho(U)\) is the value \(h_1\) there.
At every winner this last vertex is active and \(u^0=0\).
If \(z\) is finite, the \(\Gamma^0w^1\)-coefficient of \(u\) is a half-integer, so the gain contribution is an \(F/2\) multiple.
It cannot change the small residue in \Cref{lem:boolean-coefficient-obstruction} to one.

Let \(z=\Gamma\).
A nonzero constant gain coefficient requires a \(\Gamma^{-1}w^1\) term in \(u\), so the first vertex is active at every winner; otherwise the remaining direct affine terms supply no such derivative.
For finite \(t\), the \(\Gamma^{-1}w^0\) equation has the form \(Fn_t k+k'=0\), with origin counts bounded by \(p<F\).
Both counts vanish and the required derivative is zero.

For \(t=s_0\), the equation is \(FL_U+k'=0\), where \(L_U\) is the finite first affine leading sum.
It forces \(L_U=0\) at every winner and \(k'=0\).
The all-ones query vertex and its single deletions win and affinely span the query cube.
Since the first vertex is active there, its affine form \(L_U\) vanishes identically.
All its finite input coefficients and finite bias are absent, and so is their \(w\)-correction.

For \(t=\Gamma\) or a missing \(t\), no opposite-degree derivative exists.
Thus no type gives the required constant coefficient.
For \(z=s_0\), interchange \(\Gamma\) and \(s_0\) in this argument; the origin count is then at most two, still below \(F\).

\emph{The first-to-middle edge and later biases.}\par\nopagebreak

For the first-to-middle edge, the gain contribution is \(D\rho(U)(b+\eta cs)\), with positive output coefficients \(b,c\) when present and an arbitrary nongain internal \(s\).
The transfer's zeroth coefficient is nonzero whenever any gain path is present, by \Cref{lem:boolean-short-products}.
The first source is active at every winner and its affine \(U^0\) vanishes there by the inverse equation.
Winning affine span makes \(U^0\) identically zero, removing every query input and its bias, hence the entire source.
This contradicts its required activity.
A middle gain bias has that same nonzero short transfer and an uncancelled inverse coefficient.
A last gain bias has its nonzero output coefficient and is excluded in the same way.

\emph{A first hidden bias.}\par\nopagebreak

The first gate is \(h_1=D+I_1\), with \(I_1\) a positive query sum.
It is active throughout the clean cube and has no direct output edge, because \(h_1(0)>0\).
The two remaining vertices have the form
\[
h_2=\rho(V+t h_1),\qquad h_3=\rho(W+u h_1+s h_2),\qquad N=b h_2+c h_3.
\]
All output coefficients are positive when present.
Its gain transfer under a pattern is
\[
Z=\eta_2bt+\eta_3cu+\eta_2\eta_3cst.
\]
Since \(+D\) enters only as the bias of \(h_1\) and \(-D\) is unused, the inverse coefficient is \(Z^0\).
Suppose that only one output branch is present, or that only one of \(h_2,h_3\) is active.
Then \(Z^0\) is one of \(b^0t^0\), \(c^0u^0\) and \(c^0s^0t^0\), except when only the \(c\) branch is present and both \(h_2,h_3\) are active, where \(Z^0=c^0(u^0+s^0t^0)\).
A single present term is a product of nonzero leading primitives, each a nonzero scalar times a power of \(\Gamma\), and is therefore nonzero.
In the exceptional case, with \(u\) and \(st\) both present, \(u^0+s^0t^0=0\) would make the nonzero product \(s^0t^0\) equal to the negative of one primitive, contrary to \Cref{lem:boolean-short-products}; with only one of them present, a single nonzero term remains.
If all gain paths vanish, the coefficient obstruction applies.
Thus \(b,c\) are present and both \(h_2,h_3\) are active at every winner, with
\[
b^0t^0+c^0u^0+c^0s^0t^0=0.
\]

A negative finite \(t\) makes \(h_2\) inactive unless \(V\) has a \(\Gamma\) bias; that bias instead makes it positive at the origin, contradicting its positive output edge.
There are no active \(\Gamma\) input edges on the clean block.
A positive finite or \(\Gamma\) value of \(t\) also makes \(h_2(0)>0\), even against a finite negative bias.
A missing \(t\) leaves a single nonzero inverse transfer, or no gain path.
Therefore \(t=s_0\).
Since \(s_0D>0\), the origin test forces one finite negative to be the bias of \(h_2\).
The other must be an edge \(u\) or \(s\); otherwise every term of the displayed transfer has positive sign.

If \(s=-P_a\), where \(P_a\in\{P-w,P\}\), then \(u\) and the bias of \(W\) are nonnegative.
The origin test forces both absent.
The transfer would require \(b^0=Pc^0\), contrary to \Cref{lem:boolean-short-products}.

If \(u=-P_a\), divide the leading transfer by \(c^0s_0\) to get
\[
\frac{b^0}{c^0}+s^0=Pq\Gamma.
\]
Every term on the left is positive.
The ratio can have the required \(\Gamma\)-exponent only as \(qFn\Gamma\), from finite \(b\) and selector \(c\), or \(\Gamma/(Fn)\), from \(b=\Gamma\) and finite \(c\).
The optional \(s\) must be absent or \(\Gamma\).
In the first ratio case, a present \(s\) would give \(qFn+1=Pq\), impossible modulo \(q\).
With \(s\) absent, \(b\) must be the control primitive \(C\) and \(c=s_0\).
The full gain transfer is then \(s_0(C-P_a)\), whose contribution after multiplication by \(D\) has no constant \(\Gamma\)-degree.
The gain-free coefficient obstruction still excludes output one.
In the second ratio case, either \(1/(Fn)\) or \(1+1/(Fn)\) would have to equal \(Pq\), which is impossible in magnitude.
This exhausts the first-bias case and all internal gain positions.
\end{proof}

The internal-gain analysis leaves the case of a positive gain on an output edge.
Its exclusion will force the negative gain to contribute at every clean winner.

\begin{lemma}[A positive output gain without a negative-gain path]
\label{lem:boolean-sole-positive-output}

At a clean query winner, a candidate cannot have \(+D\) on a hidden-to-output edge if every individual active path contribution containing \(-D\) is zero.
Other output coefficients may be signed finite primitives.
\end{lemma}

\begin{proof}

Let \(h\) be the source of the positive output gain.
Its inverse-\(w\) coefficient forces \(h^0=0\) as an entire Laurent polynomial in \(\Gamma\).
The other nonzero paths avoid both gains, so their constant coefficient has the form in \Cref{lem:boolean-coefficient-obstruction}.
For a prefix containing three hidden vertices, expand the constant-degree prefix equation as
\[
F^3A_3+F^2A_2+FL+\frac{FN_0}{q}+\frac{k}{q}=0.
\]
The definitions and bounds of \(L,N_0,k\) are those in the output-edge case of \Cref{lem:boolean-sole-negative}.
They follow from the unique possible three-factor chain and the direct input/bias terms, irrespective of the signs of the other output coefficients.
The same successive reductions give \(k=0\), \(qL+N_0=0\), and \(N_0\equiv0\pmod q\).
The \(\Gamma^0w^1\)-coefficient has form \(FZ/(2q)+\delta+E/q\).
After combining it with all other output paths, clearing \(2q^2\) and reducing modulo \(F\), the small-residue equation is
\[
\delta+\frac E q+r_0=1,
\qquad 0\leq r_0<1.
\]
Thus \(0<\delta+E/q\leq1\).
This argument uses no positivity assumption on the other output terms.

If the balanced three-factor chain has an internal finite factor, \(N_0\) is a nonzero-modulo-\(q\) normalized finite value times a positive origin count less than \(q\).
It cannot be zero modulo \(q\), so such a chain is absent.
The remaining possible chain has its finite factor first.
Its first vertex has only query inputs and one bias.
When the chain is active, the normalized finite leading sum of its first vertex is the integer \(N_0\) of the display; otherwise \(N_0=0\).
If that bias is finite and the chain is active, this sum is nonnegative, since an active vertex has nonnegative leading preactivation.
If its bias is nonfinite, that finite sum consists only of positive query terms and is also nonnegative.
As in \Cref{lem:boolean-sole-negative}, a negative derivative \(E\) would require a \(B\) or \(C\) bias with no selected queries, contradicting \(N_0\equiv0\pmod q\).
Therefore \(N_0\geq0\), \(E\geq0\), and \(L=-N_0/q\leq0\).

A positive finite final bias would make \(L>0\), impossible.
With a negative finite final bias and any selected direct query, \(\delta\geq2\), contrary to the small-residue interval.
With that bias and no direct query, \(L=-Q\), hence \(N_0=qQ\).
The normalized finite first bias is at most \(Q\), including zero when absent or nonfinite.
Its selected query sum is therefore \(qQ\) minus that bias, and is at least \((q-1)Q\).
Every query weight is at most \((Q+1)/2\), requiring at least \(2q-2\) selected inputs.
Its derivative is at least \(2(2q-2)-3/2>q\), which again makes \(\delta+E/q>1\).

If the final bias is absent or nonfinite, \(L\geq0\); thus \(L=N_0=0\) and \(\delta=0\).
A nonzero finite-first derivative then requires a negative finite first bias and a selected power-of-two sum equal to \(Q\).
The integer \(Q=2^r-1\) has \(r\) binary ones, so such a sum uses at least \(r\) terms.
Its derivative satisfies \(E\geq2r>q\).
Otherwise \(E=0\).
Both possibilities contradict \(0<\delta+E/q\leq1\).

If the prefix ending at \(h\) contains at most two hidden vertices, there is no finite-balanced three-factor chain, so \(N_0=E=0\).
The same reduction gives \(L=0\) and \(0<\delta\leq1\).
An absent or nonfinite final bias gives \(\delta=0\), and a positive finite one gives \(L>0\).
A negative finite bias requires selected query sum \(Q\), hence \(\delta\geq2r\).
This excludes the shorter prefixes too.
\end{proof}

The pointwise lemmas also show why input-gain positions can be eliminated without assuming that deleting a path preserves a ReLU's activation pattern.
Only one case needs an actual graph pruning operation.

\begin{lemma}[Input gains and legal pruning]
\label{lem:boolean-input-gains}

A candidate representing \(f\) cannot put either gain on an external-input edge, and cannot leave one gain unused.
\end{lemma}

\begin{proof}

Choose the block in \Cref{lem:boolean-clean-block}.
Every gain input edge is fixed to zero there.
If \(+D\) is unused or an input edge, \Cref{lem:boolean-sole-negative} excludes a winning output; when both gains are absent the coefficient obstruction also applies directly.
If \(-D\) is unused or an input edge and \(+D\) is an output edge, \Cref{lem:boolean-sole-positive-output} applies.
Output gain biases are excluded by \Cref{lem:boolean-gain-bias}.

It remains that \(+D\) is internal and \(-D\) is unavailable on the restriction.
Delete all external-input edges fixed to zero.
Repeatedly remove a hidden vertex with no incoming edge or present bias, together with its outgoing edges: it is identically zero, so each such deletion preserves the restricted function.
Remove vertices without any path to the output.
All these operations discard occurrences, create no parallel edges and increase no node count.
The final output cannot be parameterless, because the query target is not identically zero.
The remaining network is legal, represents the whole query target and has \(-D\) unused.
If \(+D\) remains, its internal role is unchanged and \Cref{lem:boolean-sole-positive-internal} applies.
If it has been discarded too, \Cref{lem:boolean-coefficient-obstruction} applies.
This pruning is justified by identically zero contributions, rather than by preserving an activation pattern after an arbitrary deletion.
\end{proof}

\subsection{Both Gains on Internal Edges}

After the preceding exclusions, every used gain has a hidden source or is a hidden bias.
For two hidden-to-hidden gain edges, the three topological pairs below exhaust the possible shapes.
Ordinary output coefficients may still be signed finite values.

\begin{lemma}[Two internal gain edges]
\label{lem:boolean-internal-edge-gains}

No candidate for the clean query target places both gains on hidden-to-hidden edges.
\end{lemma}

\begin{proof}

Number the hidden vertices \(1,2,3\), allowing missing vertices and edges.
Here \(ij\) denotes the edge from hidden vertex \(i\) to hidden vertex \(j\).
The gain-edge pairs are serial \((12,23)\), fanout \((12,13)\), or fanin \((13,23)\).
In the case names below, in headings and in the text, bias one, two or three (or the last bias) is a gain bias at that hidden vertex, output one, two or three is the hidden-to-output edge from that vertex carrying a gain, a sign word such as positive gives that gain's sign, and a slash, a comma or the word and joins the positions of the two gains.
For instance, edge \(12\)/bias one places one gain on the edge from vertex one to vertex two and the other at the bias of vertex one.
Either sign order is permitted.
We apply the coefficient obstruction only to the actual frozen expression, or after full gain cancellation.

\emph{Serial edges.}\par\nopagebreak

If vertex one is inactive at a winner, only the last gain relay remains, with a first affine query source.
Its inverse equation forces that source to be \(wK\) for an integer \(K\).
The constant gain contribution is an \(F/2\) multiple for a finite output coefficient, or has nonzero \(\Gamma\)-degree for a nonfinite one.
The gain-free small residue cannot then become one.
The same argument applies if vertex three is inactive and only the first gain relay remains.
If vertex two is inactive, both gain paths disappear.
Thus all three are active at every winner.

The unique inverse-square path forces the first affine source's zeroth coefficient to vanish.
Its active value is therefore \(wK\), where \(K\) is a positive integer at most \(6r+1<F\).
A nonfinite bias could not occur in a zero zeroth coefficient, and a positive finite bias could not cancel the selected query inputs.
The derivative consequently comes only from query corrections and a negative finite bias, giving the stated integer bound.
At vertex two the gain contributes \(+K\) or \(-K\) to its zeroth coefficient.
All its other scalar terms are multiples of \(F\), and any nonfinite bias has a distinct \(\Gamma\)-degree.
Its zeroth coefficient is therefore nonzero.
The last gain edge creates an unmatched inverse-\(w\) term at vertex three and its nonzero output edge.
No earlier output can cancel it, since the first gain has already been multiplied by \(wK\).
This excludes both serial sign orders.

\emph{Fanout edges.}\par\nopagebreak

Write \(c_2,c_3\) for the output coefficients and \(e\) for the possible edge \(2\to3\).
Up to overall sign, the gain contribution is
\[
Dh_1\left(\eta_2c_2-\eta_3c_3+\eta_2\eta_3c_3e\right).
\]
Each winner requires \(h_1>0\) and a nonzero gain contribution.
If \(e\) is present and both branches active, the leading bracket cannot vanish by \Cref{lem:boolean-short-products}.
If a branch is missing or inactive, the same comparison excludes cancellation between its one primitive and one product; a single surviving term is nonzero.
With \(e\) absent, nontrivial leading cancellation requires both branches active and \(c_2^0=c_3^0\).
Two negative output occurrences are forbidden by \Cref{lem:boolean-negative-allocation}.
The equal output primitives are therefore positive and identical in full.
The entire gain contribution then cancels, and \Cref{lem:boolean-coefficient-obstruction} excludes one.

In every remaining winning pattern, the inverse equation forces \(h_1^0=0\).
The first vertex is active at all those winners, so this is an affine identity on the all-ones vertex and its single deletions.
They affinely span the query cube.
Every query coefficient and its bias leading term must vanish, removing all corresponding full primitives and their corrections.
The source is identically zero, contrary to its required activity.

\emph{Fanin edges.}\par\nopagebreak

The last vertex must be active at every winner.
A finite last output coefficient again makes the constant gain contribution an \(F/(2q)\) multiple, and the small gain-free residue cannot give one.
Let its output coefficient be \(\Gamma\) or \(s_0\).
If one gain prefix is inactive, the other is a first affine query prefix with zero zeroth coefficient.
Its \(w\)-derivative has no opposite \(\Gamma\)-degree, so the constant gain coefficient is zero.
The coefficient obstruction excludes that pattern.
Both prefixes are therefore active at every winner.

Write their values as \(h_1=\rho(U)\), \(h_2=\rho(V+t h_1)\).
The inverse equation gives
\[
V^0=(1-t^0)U^0
\]
on the winning stencil.
Since the points of the winning stencil affinely span the query cube, this is an affine identity on the entire query cube.
If \(U\) has no query coefficient, this identity removes every query coefficient of \(V\) too.
Both prefix vertices are then constant functions, leaving only the last effective hinge, which contradicts \Cref{lem:boolean-clean-block}.
Otherwise compare a positive query coefficient of \(U^0\) with the corresponding nonnegative coefficient of \(V^0\).
A positive finite \(t^0>1\) gives a negative right side.
A \(\Gamma\) or \(s_0\) value gives an unmatched negative nonzero \(\Gamma\)-degree.
A negative finite value \(-P\) gives a multiplier \(1+P\), larger than every available ratio of direct query coefficients.
Thus \(t\) is absent.

The direct leading query coefficients then agree in full.
The prefix biases also have equal leading values; their only possible unequal full values are the two finite negatives, differing by \(w\).
Consequently \(h_1-h_2\) on the winning patterns is zero or \(\pm w\), with no \(\Gamma\)-dependence.
Multiplying by the nonfinite output coefficient gives no constant \(\Gamma\)-degree gain contribution.
The gain-free coefficient obstruction supplies the final contradiction.
\end{proof}

\subsection{Internal Gain Biases}

The remaining all-internal placements contain at least one gain bias.
Two elementary consequences of the preceding coefficient comparisons will be used in their case analysis.
A first affine query prefix cannot have zeroth coefficient one.
An at-most-two-hidden nongain prefix cannot have it either: its constant-degree coefficient is \(FZ+I/q\), with \(I\in\{0,1\}\), because only its single internal edge can complete a balanced nonfinite pair from a bias origin.
The equation \(FZ+I/q=1\) is impossible modulo \(F\) after clearing \(q\).

We will also use the following single-relay consequence at a winner.
If the only surviving gain is an edge from a first affine query prefix to a final active hidden vertex, its inverse equation forces that source to be \(wK\), with integer \(K\).
A finite output coefficient makes its constant contribution an \(F/2\) multiple; a nonfinite output coefficient gives nonzero \(\Gamma\)-degree.
Other paths have the residue of \Cref{lem:boolean-coefficient-obstruction}, so they cannot complete the output to one.
A relay whose source or target vertex is inactive has no gain contribution and is excluded directly.
These observations concern actual patterns, not a replacement by a numerically similar graph.

\begin{lemma}[Gain biases and mixed internal placements]
\label{lem:boolean-internal-bias-gains}

No candidate for the clean query target places both gains internally with at least one as a hidden bias.
\end{lemma}

\begin{proof}

Let \(c_i\) be the ordinary output coefficient from hidden vertex \(i\), and \(e_{ij}\) an ordinary internal edge when it is not itself a gain.
Absent coefficients are zero.
The finite signs of these coefficients remain unrestricted except that at most one output occurrence is negative, by \Cref{lem:boolean-negative-allocation}.
The short-product and short-transfer lemmas apply with these signs.

\emph{Biases at vertices one and two.}\par\nopagebreak

If the first bias is \(-D\), its gate is identically zero on the clean cube.
The second gate with bias \(+D\) is then affine and active throughout that cube, leaving only the third effective hinge.
If the first bias is \(+D\), write \(h_1=D+I_1\).
The second preactivation has the form
\[
(e_{12}-1)D+e_{12}I_1+I_2.
\]
For a positive finite or \(\Gamma\) value of \(e_{12}\), it is positive throughout the cube.
For a negative finite value, a selector, or a missing edge, it is negative throughout the cube.
Here finite query sums are bounded by \(3rP+6rw\), \(\Gamma D\) is larger than that scale, and \(s_0<1\).
Thus the first two vertices are affine or zero, and only the third can remain nonlinear.
Both sign orders contradict \Cref{lem:boolean-clean-block}.

\emph{Biases at vertices two and three.}\par\nopagebreak

Up to sign, the inverse gain transfer is
\[
\eta_2c_2+\eta_3c_3(\eta_2e_{23}-1).
\]
A single active term cannot cancel.
When both are active and \(e_{23}\) is present, cancellation would make one product equal a sum or difference of two primitives, contrary to \Cref{lem:boolean-short-products}.
When \(e_{23}\) is absent, it requires \(c_2^0=c_3^0\).
They cannot both be negative output occurrences, and positive equality identifies their full values.
The gain-bias terms then cancel exactly.
The coefficient obstruction excludes output one in this last possibility as well.

\emph{Biases at vertices one and three.}\par\nopagebreak

A first bias \(-D\) is inactive, so the only possible last gain bias has an uncancelled output transfer, or no gain path survives.
Take a first bias \(+D\), giving \(h_1=D+I_1\).
The middle gate is affine or zero except when its incoming edge from \(h_1\) is absent or a selector and its own bias is finite negative.
We verify this classification.
A positive finite or \(\Gamma\) incoming edge makes the middle gate active everywhere, even against a finite negative bias.
A negative finite incoming edge makes it inactive everywhere unless its bias is \(\Gamma\); that bias makes it active everywhere.
An absent or selector edge leaves only a finite negative bias capable of producing a genuine hinge; every other bias gives an everywhere nonnegative preactivation or a constant zero.
If the middle gate is affine or zero, only the last gate remains nonlinear, contradicting the query target.

In the exceptional case the middle gate is zero at the origin.
If \(e_{13}\) is positive finite or \(\Gamma\), the last gate, whose bias is \(-D\), is active at the origin.
The inverse equation is
\[
c_1^0+c_3^0(e_{13}^0-1)=0,
\]
impossible by \Cref{lem:boolean-short-products}.

For negative finite, selector or absent \(e_{13}\), the last gate is zero at the origin, forcing \(c_1\) absent.
At each winner both middle and last gates must be active; otherwise there is one nonzero transfer, or no gain contribution.
A missing \(e_{12}\) cannot cancel the last bias transfer, so \(e_{12}=s_0\).
The inverse equation becomes
\[
s_0c_2^0+c_3^0(e_{13}^0-1+s_0e_{23}^0)=0.
\]

Put \(H=1+P\) when \(e_{13}\) is finite negative, and \(H=1\) otherwise.
Any selector \(e_{13}\) has different degree from the coefficient compared below.
If \(c_3^0=Fn\) is finite, its constant-degree equation is
\[
Fn(H-I/q)=J/q,\qquad I,J\in\{0,1\}.
\]
The absolute value of the left side is at least \(F(1-1/q)>1\), while the right side is at most \(1/q\).
If \(c_3=\Gamma\), comparison at degree one gives \(H=I/q\), again impossible.
If \(c_3=s_0\), comparison at degree minus one gives \(H=FZ+I/q\).
Clearing \(q\) and reducing modulo \(F\) gives \(q=I\), since \(P\) is divisible by \(F\).
This too is impossible.
These comparisons exhaust the output types, including signed finite output coefficients.

\emph{A gain edge whose source vertex has the other gain as its bias.}\par\nopagebreak

Consider edge \(12\) and bias one.
A negative first gain bias is inactive and removes every gain path.
A positive first bias is active; an active gain-edge path then has a nonzero inverse-square transfer proportional to \(c_2+\eta_3c_3e_{23}\).
This transfer cannot vanish by \Cref{lem:boolean-short-products}.
If those paths are absent, the remaining first-bias transfer is \(c_1+\eta_3c_3e_{13}\), likewise nonzero unless all gain paths disappear.
Thus edge \(12\)/bias one is excluded.

For edge \(13\)/bias one, a positive first bias and active last vertex give a unique nonzero inverse-square path; with the last vertex inactive, the remaining transfer is \(c_1+\eta_2c_2e_{12}\).
A negative first bias is again inactive.

For edge \(23\)/bias two, an inactive middle vertex removes both gains, while an active middle and last vertex give a unique inverse-square bias-to-edge path.
With only the middle active, its gain-bias output coefficient is uncancelled or every gain path disappears.
This covers the three source-bias placements, with either gain sign on the edge.

\emph{A gain edge whose target vertex has the other gain as its bias.}\par\nopagebreak

These are edge \(12\)/bias two, edge \(13\)/bias three, and edge \(23\)/bias three.
The gain-dependent part of the target preactivation is \(\sigma D(h-1)\), with \(\sigma\in\{-1,1\}\) and a nongain source prefix \(h\).
An active gain contribution has a nonzero short transfer to the output.
The inverse coefficient therefore forces \(h^0=1\).
The first- and two-hidden-prefix obstruction proved before the lemma excludes this.
Inactive contributions leave the gain-free coefficient obstruction.

\emph{Edge \(12\) and bias three.}\par\nopagebreak

If the last vertex is inactive, the single-relay argument applies.
If the middle is inactive and the last active, its gain-bias output transfer is uncancelled.
Thus both are active at every winner and
\[
h_1^0(c_2^0+c_3^0e_{23}^0)=c_3^0.
\]
The transfer is a nonzero constant Laurent polynomial.
Winning affine span removes every query coefficient of the first source.
It is a positive primitive bias \(b\), and the equation becomes
\[
b^0(c_2^0+c_3^0e_{23}^0)=c_3^0,
\]
contrary to \Cref{lem:boolean-short-transfers}.

\emph{Edge \(13\) and bias two.}\par\nopagebreak

An inactive middle vertex gives the single relay; an inactive last vertex leaves an uncancelled bias transfer or no gain contribution.
Both are active at every winner, with
\[
c_3^0h_1^0=c_2^0+c_3^0e_{23}^0.
\]
The right side is nonzero: otherwise \Cref{lem:boolean-short-products} forces \(c_2=e_{23}=0\), leaving the middle vertex without a path to output.
The first source is active, and winning affine span makes it a positive primitive bias \(b\).
The first short-transfer equation forces \(c_2=0\) and \(b=e_{23}\) in full.
The two gain paths then cancel exactly in the winning pattern, so the coefficient obstruction applies.

\emph{Edge \(23\) and bias one.}\par\nopagebreak

A negative first bias leaves the single relay.
A positive first bias is active throughout the cube.
If the last vertex is inactive, its remaining transfer is \(c_1+\eta_2c_2e_{12}\), which cannot cancel unless all paths disappear.
If the middle is inactive, the transfer \(c_1+c_3e_{13}\) has the same property.
Both middle and last are therefore active at every winner.
Their unique inverse-square path forces \(e_{12}\) absent.
The middle source is now query-affine and satisfies
\[
c_3^0h_2^0=c_1^0+c_3^0e_{13}^0.
\]
If the right side were zero, its two terms would both be absent, leaving the first vertex with no output path after \(e_{12}=0\).
Thus winning affine span makes \(h_2\) a positive primitive bias \(b\).
The first short-transfer equation gives \(c_1=0\) and \(b=e_{13}\) in full.
Again the gain terms cancel exactly, contradicting the coefficient obstruction.

The three bias pairs, the three source-bias placements, the three target-bias placements and the three remaining mixed placements exhaust all cases.
\end{proof}

\subsection{A Positive Output Gain and an Internal Negative Gain}

We next place \(+D\) on an output edge and \(-D\) internally.
At every clean winner the positive-output source is active by \Cref{lem:boolean-sole-negative}.
A negative-gain path is also active there by \Cref{lem:boolean-sole-positive-output}.
These are pointwise conclusions and allow signed finite coefficients on other output edges.
Write \(z_i\) for the ordinary output coefficient of hidden vertex \(i\), that is, the coefficient of its hidden-to-output edge when that edge does not carry a gain.
Likewise, \(e_{ij}\) denotes the edge from hidden vertex \(i\) to hidden vertex \(j\) when it does not carry a gain; absent coefficients are zero.

\begin{lemma}[The positive-output mixed orientation]
\label{lem:boolean-positive-mixed}

No candidate for the clean query target places \(+D\) on a hidden-to-output edge and \(-D\) on a hidden edge or hidden bias.
\end{lemma}

\begin{proof}

Case names follow the key in the proof of \Cref{lem:boolean-internal-edge-gains}.
All coefficient equations below use winning patterns.
Whenever a first affine source is active at all winners and its zeroth coefficient vanishes there, winning affine span makes that entire affine expression zero.
Its query coefficients and bias are then absent in full, so the source cannot be active.
We use this implication repeatedly.

We also record the two-hidden-prefix comparison needed in several cases.
For \(h_1=\rho(U)\), \(h_2=\rho(V+t h_1)\), suppose both are active and \(V^0+t^0U^0=0\).
Write
\[
U^0=FL_U+g_U\Gamma+s_U s_0,
\qquad V^0=FL_V+g_V\Gamma+s_V s_0.
\]
Here \(L_U\) and \(L_V\) are integers, the normalized finite parts.
The nonfinite counts, \(g_U,g_V\) of \(\Gamma\) and \(s_U,s_V\) of \(s_0\), are zero or one, since the active input edges are finite queries and each direct affine form has one bias.
For finite \(t^0=Fn_t\), its nonfinite equations are \(Fn_tg_U+g_V=0\) and \(Fn_ts_U+s_V=0\), forcing all counts zero because \(F>p\).
The scalar equation then forces \(L_U=L_V=0\), since the direct normalized sums are smaller than \(F\).
For \(t=\Gamma\), the degree-two equation gives \(g_U=0\), the degree-minus-one equation gives \(s_V=0\), the degree-one equation \(FL_U+g_V=0\) gives both terms zero, and the scalar equation \(FL_V+s_U/q=0\) gives both zero.
For \(t=s_0\), interchange the two nonfinite types and obtain the same conclusion.
With \(t\) absent, only \(V^0=0\) is required.

\emph{A \(-D\) edge on a path to the positive-output source.}\par\nopagebreak

For edge \(12\)/output two or edge \(13\)/output three, the source first gate and the positive-output gate are active at every winner.
The unique inverse-square coefficient forces the first affine source's zeroth coefficient to vanish.
Winning span gives the contradiction described above.

For edge \(23\)/output three, write \(h_2=\rho(V+th_1)\), and let \(X\) be the direct affine form of hidden vertex three, so that its preactivation is \(X+e_{13}h_1-Dh_2\).
At every winner vertex three is active by \Cref{lem:boolean-sole-negative}, and \(h_2\) is active by \Cref{lem:boolean-sole-positive-output}, since every path containing \(-D\) passes through \(h_2\).
Every path containing a gain ends with the output edge \(+D\) of vertex three, so the gain contribution is
\[
D\left(X+e_{13}h_1\right)-D^2h_2,
\]
with \(h_1,h_2\) gain-free.
Its \(w^{-2}\) coefficient is \(-h_2^0\), so the inverse-square equation gives \(h_2^0=0\).
Its \(w^{-1}\) coefficient is \(X^0+e_{13}^0h_1^0-h_2^{(1)}\), which the inverse-\(w\) equation requires to vanish.

If \(h_1\) is inactive at a winner, then \(h_2=V\) with \(V^0=0\), so \(h_2=w\delta\), where \(\delta=V^{(1)}\) denotes the \(w\)-coefficient of \(V\) at that winner.
A nonfinite bias would leave a term of nonzero \(\Gamma\)-degree in \(V^0\), so the bias of \(V\) is finite or absent.
If no query input of \(V\) were selected, \(V\) would be its bias alone, whose leading value is nonzero when present, and \(V^0=0\) would give \(V=0\), contrary to the activity of \(h_2\).
The selected query inputs have positive leading values, so the bias of \(V\) is a finite negative, of normalized leading value \(-Q\), and the selected query sum is \(Q\).
A sum of \(k\) powers of two has at most \(k\) binary ones, and \(Q=2^r-1\) has \(r\) of them, so at least \(r\) and at most \(3r\) query inputs are selected.
The correction of a finite negative bias is zero or one, so \(2r\leq\delta\leq6r+1<F\).
Since \(h_1=0\) and \(h_2^{(1)}=\delta\), the inverse-\(w\) equation now reads \(X^0=\delta\).
The \(\Gamma^0\)-coefficient of \(X^0\) is \(F\) times the normalized finite leading sum of \(X\), because a nonfinite bias contributes only at \(\Gamma\)-degree one or minus one.
It cannot equal the integer \(\delta\), which lies strictly between zero and \(F\).
Thus \(h_1\) is active at every winner.

The two-hidden-prefix comparison removes both leading affine forms when \(t\) is present, or \(V^0\) alone when it is absent.
Winning span makes the requisite active source identically zero.

For edge \(12\)/output three with edge \(23\) present, the same unique inverse-square path forces the first source zero.
If edge \(23\) is absent, vertices two and three can be interchanged; this is the edge \(13\)/output-two case below.
These cases cover every placement of \(-D\) on an edge that lies on a path to the vertex whose output edge carries \(+D\).

\emph{Edge \(13\), positive output one.}\par\nopagebreak

The inverse equation is
\[
U^0(1-z_3^0)=0.
\]
No primitive has leading value one, so \(U^0\) vanishes at every winner and winning span excludes the active first source.

\emph{Edge \(12\), positive output one.}\par\nopagebreak

The inverse equation has factor
\[
U^0(1-z_2^0-\eta_3z_3^0e_{23}^0)=0.
\]
The second factor cannot be zero.
Its constant \(\Gamma\)-coefficient is \(1-FZ-I/q\), with \(Z\in\mathbb Z\) and \(I\in\{0,1\}\), because the only product without a finite factor is a positive balanced pair.
Clearing \(q\) and reducing modulo \(F\) excludes equality to zero.
Again the first source vanishes on the winning span, a contradiction.

\emph{Edge \(23\), positive output one.}\par\nopagebreak

All three gates are active at every winner, and the inverse identity is
\[
U^0=z_3^0(V^0+t^0U^0).
\]
It is affine on the winning stencil.

If a query coefficient \(U_j=Fu_j>0\) is present and \(z_3\) is finite, a finite or absent \(t\) makes the right coefficient divisible by \(F^2\).
After division by \(F\), it would make \(u_j\) a multiple of \(F\), although \(u_j\leq Q<F\).
A nonfinite \(t\) instead introduces an unmatched \(\Gamma\)-degree.
For nonfinite \(z_3\), a finite \(t\) leaves the right side at the degree of \(z_3\), so it cannot equal the nonzero scalar left side.
Equal nonfinite types give a degree-two term.
Opposite types give the balanced product \(1/q\), but then the direct \(V_j\) term must be zero and the remaining equality is \(U_j=U_j/q\), impossible.
For nonfinite \(z_3\) and missing \(t\), the right side \(z_3^0V_j\) is zero or has the nonzero \(\Gamma\)-degree of \(z_3\).
Thus \(U\) has no query coefficient.
The affine identity then removes every query coefficient of \(V\) too, leaving only the last effective hinge and contradicting the query target.

\emph{Edge \(13\), positive output two.}\par\nopagebreak

All three gates are active at every winner and
\[
V^0=(z_3^0-t^0)U^0.
\]
If \(U\) has no query coefficient, neither has \(V\), and again only one effective hinge remains.
Otherwise compare a positive direct query coefficient of \(U\).
A nonzero finite difference \(z_3^0-t^0\) has an extra factor \(F\), larger than any available normalized direct coefficient.
Different nonfinite types or one finite and one nonfinite type leave unmatched degrees.
Thus \(z_3^0=t^0\) and every query coefficient of \(V\) is zero; the full affine identity also removes its bias.
If \(t\) is negative, \(h_2=\rho(t h_1)=0\), contrary to positive-output activity.
If \(t\) is positive, equality of positive leading primitives is full equality \(z_3=t\).
The gain contribution then cancels exactly in the winning pattern, which the coefficient obstruction excludes.

\emph{Edge \(23\), positive output two.}\par\nopagebreak

The inverse equation \((1-z_3^0)h_2^0=0\) forces \(h_2^0=0\).
If \(h_1\) is inactive at a winner, then \(h_2=w\delta\), where, as in the edge \(23\)/output-three case, \(\delta\geq2r\) is the \(w\)-coefficient of \(h_2\).
The gain contribution has small constant residue \(\delta\): the \(z_3\) term is an \(F\)-multiple when finite and has nonzero \(\Gamma\)-degree otherwise.
The equation \(\delta+r_0=1\), where \(r_0\) is the gain-free residue, with \(0\leq r_0<1\) as in the proof of \Cref{lem:boolean-sole-negative}, is impossible.
Therefore \(h_1\) is active at every winner.
When \(t\) is present, the two-hidden-prefix comparison forces both leading affine forms zero, and winning span excludes the active source.
When \(t=0\), only \(V^0=0\) is needed; winning span makes \(V\) identically zero, contradicting the required activity of \(h_2=V\).
This finishes all nine negative-edge/positive-output placements.

\emph{A negative bias at the first vertex or at the output source.}\par\nopagebreak

A first hidden bias \(-D\) is inactive on the clean cube, contrary to the necessary negative-gain path.
If the negative bias is at the positive-output source itself, its direct inverse-square coefficient is nonzero.
If a middle negative bias feeds a last positive-output source through a present edge, the unique inverse-square transfer is that nonzero edge coefficient.
If the edge is absent, interchange the two incomparable vertices and use the last-bias/output-two case below.

\emph{Bias three, positive output one.}\par\nopagebreak

The inverse equation gives \(U^0=z_3^0\) on every winner.
Winning span removes all query coefficients of the first source and matches its bias leading value with \(z_3^0\).
A negative finite match would make that first gate inactive.
A positive match is full equality by \Cref{lem:boolean-short-products}, including the nonfinite types.
Thus the two gain terms cancel exactly, contradicting the coefficient obstruction.

\emph{Bias three, positive output two.}\par\nopagebreak

At every winner \(h_2^0=z_3^0\), with \(h_2=\rho(V+t\rho(U))\).
If \(t,z_3\) are finite, with \(z_3^0=Fn_z\) and \(L_U,L_V\) as in the two-hidden-prefix comparison, the constant-degree equation is
\[
FL_V+\eta_1F^2n_tL_U=Fn_z.
\]
The same equation has right side zero for nonfinite \(z_3\).
For nonfinite \(t\), the only additional constant-degree term is an optional positive balanced bias pair \(1/q\).
Reduction modulo \(F\) after clearing \(q\) eliminates that pair.
In every case \(L_V=n_z\), or zero for nonfinite \(z_3\), at all winners.
Winning span removes every direct query input of \(V\) and identifies its finite bias leading value when present.

If \(z_3\) is finite negative, \(V\) uses the other finite negative bias.
Both finite negatives are then consumed, and the first gate is positive-only.
If \(t\) is absent, \(h_2=V\) is a negative constant and is inactive.
If \(t\) is present, the identity \(h_2^0=z_3^0\), compared in every \(\Gamma\)-degree and not only in the constant degree, forces any active first gate's leading value to be zero, whether \(t\) is finite or nonfinite.
A nonzero positive-only first affine gate cannot have that property.
An inactive first gate leaves negative \(V\) and makes \(h_2\) inactive too.
Hence a finite \(z_3\) must be positive, and its matching finite bias is equal in full.
Their \(w\)-corrections agree.
If \(z_3\) is nonfinite, both finite corrections are zero.

For finite \(t\), the remaining \(\Gamma^0w^1\) contribution to \(h_2\) through the first gate is an \(F/2\) multiple.
For nonfinite \(t\), it has nonzero \(\Gamma\)-degree, since the derivative \(U^{(1)}\) of the first gate's affine form \(U\) is a scalar.
Thus the constant gain coefficient is the difference of the matching final corrections, namely zero, plus an \(F/2\) multiple, so its gain residue is zero.
It cannot complete the gain-free residue below one to the required output.

\emph{Bias two, positive output one.}\par\nopagebreak

Here the first gate is hidden vertex one, which carries the positive output gain, and \(U\) is its affine preactivation.
The inverse equation makes \(U^0\) equal to one of at most two fixed Laurent values,
\[
U^0=z_2^0+\eta_3z_3^0e_{23}^0.
\]
If it had two positive query coefficients, the all-ones vertex, one deletion and two deletions would give three distinct values, all at winners.
Thus it has at most one clean query input.
Fix that input to zero if it exists, so that the first gate is constant on the resulting face.
If that constant is not positive, the first gate is inactive, so the source of \(+D\) is inactive and \Cref{lem:boolean-sole-negative} excludes the winners of this face.
Otherwise let that constant be the positive primitive \(b\).
The middle preactivation is
\[
-D+t b+I_2,
\]
where \(I_2\) is a direct query sum.

Under \(\Gamma=M^9\), the leading term of a nongain primitive is a multiple of \(M^{-9}\), \(M^0\) or \(M^9\), according as it is a selector, a finite primitive or \(\Gamma\).
A product of two nongain primitives therefore has leading specialized \(M\)-exponent in \(\{-18,-9,0,9,18\}\); in particular this exponent is never one, the \(M\)-exponent of \(D=M\).
At exponent zero its coefficient is bounded in terms of \(P^2\); every query sum is at most \(3rP+6rw\).
The chosen \(M\) therefore makes the middle gate's sign constant on the entire face: a positive exponent-nine or exponent-eighteen product dominates, a negative such product keeps it off, and every smaller product leaves \(-D\) dominant.
The first gate is constant and the middle gate affine or zero, leaving only the third effective hinge.
This contradicts the robust face statement of \Cref{lem:boolean-clean-block}.
The stated bias cases exhaust all positions and complete the proof.
\end{proof}

\subsection{A Negative Output Gain and an Internal Positive Gain}

The opposite mixed orientation has positive ordinary output branches, but those branches need not vanish at the origin and the unit vectors of the restricted cube before this is proved.
We use the pointwise negative-gain obstruction to ensure that the positive gain contributes on every query winner.

\begin{lemma}[The negative-output mixed orientation]
\label{lem:boolean-negative-mixed}

No candidate for the clean query target places \(-D\) on a hidden-to-output edge and \(+D\) on a hidden edge or hidden bias.
\end{lemma}

\begin{proof}

Let \(P_*\) denote either positive magnitude \(P-w\) or \(P\) of a finite negative occurrence.
Both have \(w^0\) value \(P=FQ\).
As at the start of this appendix, \(h^0\) denotes the \(w^0\)-coefficient of a gain-free frozen expression \(h\), retaining its Laurent dependence on \(\Gamma\).
Every positive finite primitive has \(w^0=Fk\), with integer \(1\le k\le Q\).
A direct query coefficient is at most \(FQ\).
The total leading query sum at a vertex is at most \(3rFQ<F^2\).

We will also need bounds on actual values, after numerical specialization.
Every actual query primitive \(F2^i+2w\) is at most \(P=FQ\), since \(Q-2^i\geq1\) and \(F>2w\).
Thus a direct query sum is at most \(A=3rP\).
Put \(U_0=5rP\) and \(U_1=8rP^2\).
A first gain-free vertex without a \(\Gamma\) bias has value at most \(A+P\leq U_0\): all finite positive primitives are at most \(P\), a selector is below one, and a negative bias only lowers its ReLU value.
Here \(P\geq6\), \(r<Q<P\), and \(F>8Q^4>10r\); the choice of \(M\) gives \(D=M>2H_0\), and the definition of \(H_0\) gives \(2H_0>2\cdot16\cdot64P^4=2048P^4\).
Consequently
\[
PU_1+U_0=8rP^3+5rP<13P^4<D/64.
\]
Also \(P<D\) and \(D\geq2\), so
\[
2PD+D+PU_1+U_0<3D^2<D^9=\Gamma,
\qquad P-w>P/2.
\]
These explicit inequalities will justify each comparison with the positive-gain bias.

Both finite negatives enter distinct hidden vertices, by \Cref{lem:boolean-negative-allocation}: at least two distinct hidden vertices must receive negative contributions on the clean cube, \(-D\) is an output edge, and the two finite negatives are the only other negative occurrences.
Thus precisely two hidden vertices can supply effective ReLU nonlinearities; the remaining hidden vertex is a positive wire.
Call a hidden vertex negative-receiving when a finite negative occurrence enters it.
Making either negative-receiving vertex globally linear or inactive leaves at most one effective hinge, contradicting the target.
The same argument applies after fixing at most two query coordinates to zero.

\emph{Elementary facts used.}\par\nopagebreak

\emph{Short internal transfers do not vanish at \(w^0\).}\par\nopagebreak

We call the comparisons of this paragraph the short-transfer comparison.
Between two distinct hidden vertices there are at most a direct path and a two-edge path.
If either exists in the frozen mask, their combined gain-free transfer has nonzero \(w^0\).

Cancellation would make the \(w^0\) value of the direct edge the negative of that of the two-edge product.
At equal \(\Gamma\) exponents, with signs absorbed into the integers, it would require one of the following, which compare a finite direct edge with two finite edges, a \(\Gamma\) or \(s_0\) direct edge with the same primitive times a finite edge, and a finite direct edge with a balanced pair, respectively:

\begin{itemize}
\item \(Fk=F^2m\), with \(0<|k|\le Q<F\);
\item \(1=Fm\);
\item \(Fk=1/q\).

\end{itemize}
None is possible.
The same reasoning applies to a transfer from one hidden vertex to positive output edges through at most one other hidden vertex.

Also, a transfer from a hidden vertex to positive outputs through at most one other vertex cannot have \(w^0\) identically \(1\).
Its \(\Gamma^0\) coefficient is
\[
FZ+\frac n q,\qquad Z\in\mathbb Z,\quad n\in\{0,1\}.
\]
The term \(n/q\) can arise only from \(\Gamma s_0=1/q\), the only two-edge \(\Gamma^0\) path without a factor \(F\).
Equality with one would give \(qFZ=q-n\), a multiple of \(F\) lying strictly between zero and \(F\), which is impossible.

\emph{A first finite-negative-biased hinge has a robust leading variation.}\par\nopagebreak

We call the argument of this paragraph the robust variation argument.
Such a first hidden vertex is
\[
H(x)=\rho\left(F(L(x)-Q)+w(2k(x)+\delta_b)\right),
\qquad \delta_b\in\{0,1\},
\]
where \(L(x)\) is the sum of the supported direct query powers of two times their input bits, and \(k(x)\) counts those inputs equal to one.
The bias is \(-P+w\delta_b\), with \(\delta_b=0\) or \(1\) according to which finite negative occurrence is used.

If no query coordinate is supported, this hinge is identically zero and leaves at most one effective nonlinearity.
Thus at least one is supported.
Put \(E_0=L(\mathbf{1})-Q\).
Then
\[
E_0>\max\{\text{nonzero coefficients of }L\}.
\]
Otherwise, fixing a coordinate with the largest coefficient in \(L\) and one further supported coordinate to zero makes this hinge identically inactive.
If it has only one supported coordinate, fixing that coordinate suffices.
Either outcome leaves at most one effective hinge on a subcube whose target still requires two by \Cref{lem:boolean-clean-block}.

There are at least two supported coordinates, since one alone cannot have excess larger than its own weight.
Let \(i\) and \(j\) be two distinct supported coordinates of \(H\), and let \(2^{i_0},2^{j_0}\) be their coefficients in \(L\), so that \(i_0,j_0\) are the levels of the occurrences on their input edges into \(H\).
Consequently \(H^0\) takes three distinct values at suitable points
\[
\mathbf{1},\qquad \mathbf{1}-e_i,\qquad \mathbf{1}-e_i-e_j,
\]
all of which win.
The first two leading values are \(FE_0\) and \(F(E_0-2^{i_0})>0\).
The third is \(\max\{0,F(E_0-2^{i_0}-2^{j_0})\}\), strictly smaller than the second, whether it is zero or positive.
Moreover, \(H^0>0\) at \(\mathbf{1}\) and at every one-coordinate deletion.

We will also use the resulting strict variation at a later active gate \(G\) whose only possible hidden predecessor is such a first hinge \(H\); we call the following calculation the strict-variation calculation.
We call such a hinge independent, since it has no hidden predecessor of its own.
Suppose \(G\) has a finite negative bias and is active on the winning stencil.
There its leading value has the form
\[
G^0(x)=FZ(x)-P+c^0H^0(x),
\]
where the edge \(c\) from \(H\) to \(G\) is a positive ordinary edge or zero, and \(Z\) is the direct query sum of this gate normalized by \(F\).

If \(c\ne0\), choose a supported coordinate \(j\) of \(H\), of occurrence level \(i_j\) on its input edge into \(H\).
Both stencil values of \(H\) are active, so their leading difference is \(F2^{i_j}>0\).
Let \(d_j\geq0\) be the direct normalized query coefficient of \(G\).
Deleting that coordinate changes \(G^0\) by
\[
Fd_j+c^0F2^{i_j}\ne0.
\]
Both terms have nonnegative Laurent coefficients and the second is nonzero.
If \(c=0\), any present direct query instead gives a strictly positive difference by itself.
With neither a predecessor edge nor a direct query, the negative bias makes \(G\) inactive, contradicting the hypothesis.
Thus such a \(G\) cannot have constant leading value on the winning stencil.
The same difference calculation, without the term \(-P\), shows that a positive wire constant on this stencil has no direct query and no edge from a varying \(H\).
It is then its actual constant bias.

\emph{Positive primitives with the same \(w^0\) value are identical.}\par\nopagebreak

The only possible concern is among finite primitives.
The construction's size separation makes \(Q-T\) larger than \(2^{r-1}\) and every \(\beta_j\), while \(Q\) is distinct from all these values.
The \(\beta_j\) are not powers of two.
Thus equal leading values occur only between identical numerical primitive types, including duplicate occurrences.

The following constant-only exclusions follow directly by comparing \(\Gamma\) exponents and finite magnitudes.
For positive primitive leading values \(b,t\), a positive primitive or absent \(c\), and a positive primitive \(b'\), which occurs only in the last two equations, none of
\begin{equation}
\begin{gathered}
(c-t)b=P,\qquad
(c-1/t)b=P,\\
b'=(P+t)b,\qquad
b'=(P+1/t)b
\label{eq:boolean-reverse-constant-types}
\end{gathered}
\end{equation}
holds.

For the first equality, finite \(b=F\ell\) would require \(c-t=Q/\ell\), a positive scalar smaller than \(F\).
A nonzero scalar difference of primitives is an integer multiple of \(F\), so this is impossible.
For \(b=\Gamma\), the required coefficient of \(\Gamma^{-1}\) is \(P\), while a difference supplies only zero or \(\pm1/q\).
For \(b=s_0\), the required coefficient of \(\Gamma\) is \(Pq\), while a difference supplies only zero or \(\pm1\).

For the reciprocal equality and finite \(b\), a finite \(t=Fn\) makes \(c-1/t\) either negative, when \(c=0\), or at least \(F-1/F>Q\), when \(c\) is finite positive.
A nonfinite \(c\) leaves a nonzero degree.
A nonfinite \(t\) cannot give a scalar: its reciprocal can only meet the opposite primitive with coefficient \(1/q-1\) or \(1-q\), neither zero.
For nonfinite \(b\), a finite \(t\) leaves a nonzero scalar that no finite \(c\) can cancel, again by \(F>1\).
If \(t\) has the same nonfinite type as \(b\), its reciprocal has the degree needed on the right, but its coefficient is negative and adding the only available opposite primitive leaves it negative.
If \(t\) has the opposite type, cancellation of its unwanted reciprocal degree would require one of the same impossible equalities \(1/q=1\) or \(1=q\).
These cases exhaust the second equality.

For each of the last two equalities, all contributions on the right are positive.
A finite \(b\) makes its \(Pb\) contribution exceed the largest finite primitive \(P\), and a nonfinite candidate on the left cannot contain that positive scalar term.
A nonfinite \(b\) gives a coefficient of at least \(P>1\) at its own degree, whereas the corresponding available primitive has coefficient one relative to that type.
Additional positive terms cannot cancel it or reduce that coefficient.

\emph{The positive gain \(+D\) on an internal edge.}\par\nopagebreak

Let that edge be \(u\to v\).

First, there is a winner with \(h_u^0\ne0\).
This is immediate if \(u\) is a first positive wire: since an active \(+D\) path is necessary, \(u\) is active at every winner, and an active first positive wire has a nonzero \(w^0\) coefficient, because its value is a sum of positive primitives, whose positive leading values cannot cancel.
If \(u\) is a first negative-biased hinge, it follows from the robust variation argument.

If \(u\) is second, let \(g\) be its sole possible hidden predecessor.
Suppose \(h_u^0=0\) at every winner.
Since an active \(+D\) path is necessary, \(u\) is active at every winner.

\begin{itemize}
\item If \(g\) is a first negative-biased hinge, its leading value changes by a positive multiple of \(F\) between \(\mathbf{1}\) and a suitable one-coordinate deletion.
A positive edge from \(g\) gives a nonzero change in \(h_u^0\).
A negative finite edge gives a change of magnitude at least \(PF\), which cannot be canceled by the direct query change, at most \(FQ\).
A \(\Gamma\) or \(s_0\) edge gives a nonzero change in its separate \(\Gamma\) exponent.
If the edge is absent, vanishing on the whole stencil forces all direct queries and the bias of \(u\) absent, making \(u=0\).
\item If \(g\) is a positive wire, \(u\) must receive a finite negative.
With a negative bias, all query coefficients are positive.
Vanishing on the stencil removes every query, leaving an impossible primitive-product equality with \(P\).
With a negative edge from \(g\), a coefficient involving a query of \(g\) has magnitude at least \(PF\) and cannot cancel a direct query coefficient.
Again all queries disappear.
Let \(B_{\mathrm g},B_{\mathrm u}\) be the actual optional positive bias values at \(g,u\), with zero for absence.
The remaining equation is \(B_{\mathrm u}^0=P B_{\mathrm g}^0\).
If \(B_{\mathrm g}\ne0\), \Cref{lem:boolean-short-products} excludes this equality, including an absent \(B_{\mathrm u}\).
If \(B_{\mathrm g}=0\), then \(B_{\mathrm u}^0=0\), so \(B_{\mathrm u}=0\), since every present positive primitive has a nonzero leading value.
Both vertices are then exactly zero throughout the clean cube, contradicting activity of \(u\) at every winner.

\end{itemize}
This proves that the claimed winner exists.

Fix such a winner.
Every path containing \(+D\) passes through the edge \(u\to v\) and then through \(v\), so \(v\) is active there, since an active \(+D\) path is necessary; and \(u\) is active, since \(h_u^0\ne0\).
The value \(h_u\) is gain-free, because \(+D\) leaves \(u\) and \(-D\) is an output edge.
If the \(-D\) source equals \(v\), every path containing both gains consists of a path into \(u\), the edge \(u\to v\) and the output edge \(-D\), so the \(w^{-2}\) coefficient of the output is \(-h_u^0\ne0\), which the output identity excludes.
If the source equals \(u\), let \(T_v\) be the frozen transfer from \(v\) to the output; it is gain-free, since no path from \(v\) uses the output edge of \(u\).
No path contains both gains, and the gain contribution is \(Dh_u(T_v-1)\), whose inverse coefficient is \(h_u^0(T_v^0-1)\).
Since \(h_u^0\ne0\) and Laurent polynomials in \(\Gamma\) have no zero divisors, the \(w^{-1}\) equation forces \(T_v^0=1\).
The transfer \(T_v\) passes through at most the remaining hidden vertex and ends at positive output edges, so this contradicts the short-transfer comparison.

The negative-output source is therefore the third vertex \(a\), distinct from \(u,v\).
At such a winner it is active.
Otherwise every \(-D\) contribution vanishes, and the inverse coefficient is \(h_u^0T_v^0\), where \(T_v\) is now the frozen transfer from \(v\) to the positive output edges.
This transfer contains the active \(+D\) path, so \(T_v^0\ne0\) by the short-transfer comparison, and the \(w^{-1}\) term is nonzero.
Nor can an active path run from \(v\) to \(a\).
Since \(u\) precedes \(v\), such a path is the edge \(v\to a\), and with \(v\) and \(a\) active it gives the \(w^{-2}\) coefficient \(-h_u^0\) times the nonzero leading value of that edge, which is nonzero.
Hence, if \(a\) is after \(v\), the edge \(v\to a\) must be absent.

Let \(t>0\) be the output edge from \(v\).
It must exist.
On every winning mask the \(w^{-1}\) equation is
\begin{equation}
\label{eq:boolean-reverse-transfer}
h_a^0=t^0h_u^0.
\end{equation}

In the following case equations, \(FZ\) denotes the corresponding direct query sum, \(b\) an optional positive primitive bias, and \(c\) a positive ordinary edge or zero.
Their values may differ between cases.
There are two structural possibilities.

\emph{First possibility: \(u\) is first and \(a\) is later.}\par\nopagebreak

First exclude a positive wire \(u\) with no bias and exactly one clean query input \(x_j\).
Fix \(x_j=1\).
On this face \(u\) is the positive constant equal to its actual input primitive.
At all-ones and every single deletion of a free coordinate, the face target wins by \Cref{lem:boolean-clean-block}.
\Cref{eq:boolean-reverse-transfer} gives the same nonzero leading value \(t^0u^0\) for \(a\) at all those points, so \(a\) is active there.
The edge \(v\to a\) has already been excluded, and the only other possible hidden predecessor is the now constant \(u\).
Thus the leading value of the preactivation of \(a\) is affine in the free query coordinates.
The face stencil affinely spans their coordinate space, so every free query coefficient in that preactivation is zero.
Every present such direct edge has a positive leading query primitive \(F2^i\); hence all those edges are absent.
Consequently \(a\), like \(u\), is an actual constant throughout the face.

Only \(v\) remains an effective ReLU, with affine preactivation, and the output is affine plus a scalar multiple of that one ReLU.
This contradicts the assertion of \Cref{lem:boolean-clean-block} for a face with one coordinate fixed to one.

A positive wire with neither a query nor a bias is identically zero on the clean cube, contradicting the previously established winner with \(h_u^0\ne0\).
Every remaining first positive wire has either a positive bias or at least two query inputs, so \(u^0>0\) at all-ones and every one-coordinate deletion.
The same positivity for a first negatively biased hinge was proved above.
\Cref{eq:boolean-reverse-transfer} now makes \(a\) active on this entire stencil.
Its frozen leading affine formula can therefore be compared coefficient by coefficient there.
In this case \(c\) is the edge from \(u\) to \(a\).

If \(u\) is a negative-biased hinge, write
\[
u^0=F(L-Q).
\]

\begin{itemize}
\item If \(a\) has a negative bias, then
\[
FZ-P=(t^0-c^0)F(L-Q).
\]
Comparing constants gives \(t^0-c^0=1\), impossible for positive primitives or an absent \(c\).
\item If \(a\) has a negative edge from \(u\), then
\[
FZ+b^0=(t^0+P)F(L-Q).
\]
Its constant equation equates a nonnegative bias with \(-(t^0+P)P<0\).
\item If \(a\) is the positive wire, then
\[
FZ+b^0=(t^0-c^0)F(L-Q).
\]
Query coefficients and the constant force \(t^0=c^0\), with no direct queries or bias at \(a\).
\Cref{lem:boolean-short-products} gives \(t=c\) exactly and hence \(h_a=c h_u\) exactly.
At a winner, \(v\) is active, so the two gain-containing contributions cancel \emph{exactly}, not merely at leading order.
\Cref{lem:boolean-coefficient-obstruction} rules out output \(1\).

\end{itemize}
If \(u\) is a positive wire, \(a\) must receive a finite negative.

\begin{itemize}
\item A negative bias gives
\[
FZ-P=(t^0-c^0)u^0.
\]
If \(u\) has any query, the query equations make the proportionality factor nonnegative, contradicting the negative constant.
If \(u\) is constant, this is \((c^0-t^0)b=P\), excluded by \Cref{eq:boolean-reverse-constant-types}.
\item A negative edge gives
\[
FZ+b^0=(t^0+P)u^0.
\]
Any query of \(u\) would produce a coefficient at least \(PF>FQ\).
With no query, the constant equation is excluded by \Cref{eq:boolean-reverse-constant-types}.

\end{itemize}
This covers both orders \(u,a,v\) and \(u,v,a\).

\emph{Second possibility: \(a\) is first, followed by \(u,v\).}\par\nopagebreak

If \(a\) is a first negative-biased hinge, write \(a^0=Fn\), where \(n\) is an integer and at all-ones \(0<n<F\).
In this subcase allow \(b\) to be an optional positive primitive bias or the remaining finite negative bias, and \(c\) to be a positive ordinary edge, zero, or the remaining finite negative edge from \(a\) to \(u\).
At most one of \(b,c\) can be negative, since the other finite negative has already been used at \(a\).
These are the exceptions to the default positive \(b,c\) convention above.

Since \(u\) is active,
\[
Fn=t^0(FZ+b^0+c^0Fn).
\]
If \(t^0=Fn_t\) is finite, only finite or absent \(b^0=Fn_b,c^0=Fn_c\), with signed integers \(n_b,n_c\), contribute to the \(\Gamma^0\) coefficient of the right side.
That contribution is \(F^2n_t(Z+n_b+Fn_c n)\), divisible by \(F^2\), whereas \(Fn\) is not because \(0<n<F\).
If \(t^0=\Gamma\) or \(s_0\), only an opposite positive nonfinite primitive in \(b\) or \(c\) contributes to that scalar coefficient.
It is therefore
\[
(\eta+\theta Fn)/q,\qquad \eta,\theta\in\{0,1\}.
\]
Equality with \(Fn\) would require \((q-\theta)Fn=\eta\), impossible since \(q\geq5\), \(F(q-1)>1\), and \(n\geq1\).

If \(a\) is a positive wire, \(u\) receives a finite negative.
Comparing affine coefficients in \Cref{eq:boolean-reverse-transfer}:

\begin{itemize}
\item A negative bias gives
\[
FZ-P=(1/t^0-c^0)a^0.
\]
Queries give the same sign contradiction as above; the constant-only case is \((c^0-1/t^0)b=P\), excluded by \Cref{eq:boolean-reverse-constant-types}.
\item A negative edge gives
\[
FZ+b^0=(P+1/t^0)a^0.
\]
Queries violate \(PF>FQ\); the constant-only case is excluded by \Cref{eq:boolean-reverse-constant-types}.

\end{itemize}
Thus every internal-edge placement is excluded.

\emph{The positive gain \(+D\) as a hidden bias.}\par\nopagebreak

Let \(v\) be its vertex and \(a\) the negative-output source.
If \(v=a\), its bias and output edge give a unique nonzero inverse-square term.
With at most two hidden vertices, a first positive-gain bias leaves only one negative-receiving vertex, contrary to \Cref{lem:boolean-negative-allocation}.
Indeed, the preactivation of that first vertex \(v\) is its bias \(+D\) plus input terms, and on the clean cube every input edge either carries a positive query value or is fixed to zero, so \(v\) receives no negative contribution there; the only other hidden vertex is \(a\).
If the positive-gain bias is second, negative allocation requires a finite-negative-biased first hinge and a negative finite edge into the second vertex.
That first hinge is bounded by \(U_0\), so the second preactivation is at least \(D-PU_0>0\) throughout the cube.
Only one effective nonlinearity remains, which is impossible.
The bias cases below therefore have three hidden vertices.

At every winner, \(v\) has an active path to output.
If \(a\) were reachable from \(v\) in that mask, the nonzero short transfer of the short-transfer comparison would give a nonzero \(w^{-2}\) coefficient.
Thus it is not reachable.

If \(a\) were inactive, the remaining \(+D\)-to-positive-output transfer would have nonzero \(w^0\), giving a nonzero \(w^{-1}\) term.
Therefore the source \(a\) is active at every winner, and no active path goes from \(v\) to \(a\).

The \(w^{-1}\) equation gives
\[
h_a^0=T_v^0,
\]
where \(T_v\) is the gain-free transfer from the bias injection to positive output edges.
Excluding \(a\), there is at most one intervening hidden vertex.
Consequently \(h_a^0\) assumes at most two fixed Laurent-polynomial values over all winners.
If \(v\) is last, or the other vertex cannot lie on that transfer, it assumes one fixed value.

Consider the position of \(v\).

\emph{The vertex \(v\) is first.}
It is a positive wire.
The second vertex must be a genuine negative-receiving hinge.
Here \(D\leq v\leq D+A<2D\).
With a negative edge from \(v\) and no \(\Gamma\) bias, the second preactivation is at most \(U_0-(P/2)D<0\).
With a \(\Gamma\) bias it is at least \(\Gamma-P(D+A)>0\).
With a negative bias, a positive finite edge from \(v\) gives preactivation at least \(FD-P>0\), and a \(\Gamma\) edge gives at least \(\Gamma D-P>0\).
Only an \(s_0\) edge can retain a genuine hinge; its contribution is at most \(2D/(q\Gamma)<1\).

If \(a\) is this second hinge, the absence of an active path from \(v\) to \(a\) removes the possible \(s_0\) edge.
It is then a first finite-negative-biased hinge, whose three distinct leading values from the robust variation argument contradict the at-most-two-values conclusion.

Otherwise \(a\) is last.
The direct \(v\to a\) edge is absent by the active-path restriction.
If the second hinge feeds \(a\), it is active at all-ones.
Indeed, its finite negative bias is its only negative parameter on the clean cube and \(v\) is a positive wire, so its preactivation is nondecreasing in every clean query coordinate; inactivity at all-ones would make it inactive throughout the clean cube, and it would not be a genuine hinge.
Since \(v\geq D>0\) everywhere and \(a\) is active at the winner all-ones, the absence of an active path from \(v\) to \(a\) removes the possible incoming \(s_0\) edge of the second hinge from \(v\).
Thus both hinges are independent of \(v\), and \(h_a^0\) must be constant over winners.
A negative-biased \(a\) cannot be constant on the winning stencil, by the strict-variation calculation for a later active gate \(G\) in the elementary facts above.
A negative edge from the finite first hinge makes \(a\) inactive at all-ones unless its bias is \(\Gamma\): the hinge has value at least \(F\) there, giving preactivation at most \(U_0-(P/2)F<0\).
A \(\Gamma\) bias instead gives preactivation at least \(\Gamma-PU_0>0\) throughout the cube.
If the second hinge does not feed \(a\), \(a\) itself is an independent first finite-negative-biased hinge, again contradicting the robust variation argument.

\emph{The vertex \(v\) is second.}
It cannot receive a finite negative.
Its only possible negative is an edge from the first vertex.
Without a \(\Gamma\) bias that first vertex is at most \(U_0\), so \(v\)'s preactivation is at least \(D-PU_0>0\).
With a \(\Gamma\) bias the first vertex is at least \(\Gamma\), and the preactivation of \(v\) is at most \(D+A-(P/2)\Gamma<0\).
Either way \(v\) cannot supply the required second nonlinearity.

Hence \(v\) is the positive wire; the first and last vertices receive the finite negatives.
If \(a\) is first, the robust variation argument contradicts the at-most-two-values conclusion.
If \(a\) is last, the absence of an active path from \(v\) to \(a\) removes \(v\to a\), and \(h_a^0\) equals the fixed output primitive of \(v\).
A negative-biased \(a\) cannot have this constant value on the stencil, by the same strict-variation calculation.
A negative edge from the first finite hinge gives preactivation at most \(U_0-(P/2)F<0\) at all-ones unless the bias is \(\Gamma\); that bias gives the global lower bound \(\Gamma-PU_0>0\).

\emph{The vertex \(v\) is last.}
Here \(h_a^0\) equals its fixed positive output primitive \(t^0\).

If both earlier vertices receive finite negatives, the first is a finite-negative-biased hinge.
If \(a\) is first, the robust variation argument gives a contradiction.
If \(a\) is second, the strict-variation calculation excludes a negative bias, and the preceding negative-edge argument excludes that alternative.

Otherwise one earlier vertex is the positive wire \(W\), the other the sole earlier negative hinge \(H\).

\begin{itemize}
\item If \(H\) is first and \(a=H\), the robust variation argument excludes it.
\item If \(H\) is first and \(a=W\), the positive-wire version of the strict-variation calculation forces it to have no queries and no edge from \(H\).
It is a constant primitive bias, with \(W^0=t^0\).
If the negative edge entering \(v\) comes from the first finite hinge \(H\), its preactivation is at least \(D-PU_0>0\).
If it instead comes from the constant \(W\), a finite or \(s_0\) bias there has value at most \(P\), giving the lower bound \(D-P^2>0\).
The remaining case is \(W=\Gamma\).
Its bias and the output primitive \(t=\Gamma\) consume both \(\Gamma\) occurrences, so no positive \(\Gamma\) edge from \(H\) remains to overcome \(-P_*\Gamma\).
The preactivation of \(v\) is then at most \(D+A+PU_0-(P-w)\Gamma<0\).
\item If \(W\) is first and \(a=H\), the active hinge \(H\) has constant leading value on the whole winning stencil.
With a negative bias, its nonnegative query coefficients must all vanish.
With a negative edge from \(W\), cancellation of a query coefficient would require a direct coefficient at least \(PF>FQ\).
Thus again every query vanishes.
In either case \(H\) is constant on the entire cube and cannot supply a nonlinearity.
\item If \(W\) is first and \(a=W\), constancy forces \(W\) to be a constant primitive bias equal to \(t\).
\end{itemize}
In the case where \(W\) is first and \(a=W\), suppose first that the constant \(W\) is finite or \(s_0\), so \(0<W\leq P\).
A \(\Gamma\) bias at \(H\) would give preactivation at least \(\Gamma-P^2>0\).
A positive \(\Gamma\) edge from a constant finite \(W\geq1\) also gives positive preactivation, at least \(\Gamma-P\); from \(W=s_0\) it contributes only \(1/q\).
In every remaining genuine-hinge case the direct query sum is at most \(A\), an ordinary finite or selector bias is at most \(P\), and a positive finite or selector product with \(W\) is at most \(P^2\).
Including the balanced contribution gives
\[
0\leq H\leq A+P+P^2+1\leq U_1.
\]
A negative edge to \(v\) from either \(W\) or \(H\) consequently leaves preactivation at least \(D-PU_1>0\).

Still in the case where \(W\) is first and \(a=W\), suppose \(W=\Gamma\).
A finite or \(\Gamma\) positive edge to \(H\) gives preactivation at least \(\Gamma-P>0\).
A negative edge gives preactivation at most \(A+\Gamma-(P-w)\Gamma<0\), even allowing a \(\Gamma\) bias.
Thus a genuine nonlinear \(H\) can retain only an \(s_0\) edge or no edge from \(W\), giving \(H\leq A+1\leq U_1\).
If the negative edge entering \(v\) comes from \(H\), the lower bound is again \(D-PU_1>0\).
If it comes from \(W=\Gamma\), the bias at \(W\) and the output primitive \(t=\Gamma\) consume both \(\Gamma\) occurrences.
The positive edge from \(H\), if present, is therefore at most \(P\), and \(v\)'s preactivation is at most
\[
D+A+PU_1-(P-w)\Gamma<0.
\]

Every case contradicts either the formal output identity or the need for two effective nonlinearities.

Absent edges throughout have coefficient zero.
The coefficient obstruction in the cancellation case uses equality of the full gain contributions, not merely their leading terms.\qedhere
\end{proof}

\subsection{Finite Negative Edges with Both Gains at Output}

The previous lemmas leave both gains on distinct hidden-to-output edges.
Negative allocation puts the two finite negatives into distinct hidden preactivations.
The optional third hidden vertex is a positive wire.
We first prove that neither finite negative can be an internal edge.
The argument allows either gain to originate at the positive wire, because the negatively biased vertices have not yet been identified as the gain sources.

For this analysis put \(P_a=P-\delta_a w\), \(P_b=P-\delta_b w\), with \(\{\delta_a,\delta_b\}=\{0,1\}\).
The notation \(U,V,W\) denotes direct query sums, with a positive optional bias only when explicitly allowed.
Every ordinary internal or output coefficient is positive or absent, because all negative occurrences have been assigned.
The inverse output equation equates the zeroth \(w\)-coefficients of the two gain sources at every Boolean point, irrespective of its target label.
At every clean winner both gain sources are active: an inactive positive source contradicts \Cref{lem:boolean-sole-negative}, and an inactive negative source contradicts \Cref{lem:boolean-sole-positive-output}.

We will use a minimal leading-positive query coalition.
For \(h=\rho(U-P_a)\), a \emph{leading-positive query coalition}, or simply a leading-positive coalition, is a selected set whose normalized query sum exceeds \(Q\).
Choose one with no proper subset still having this property.
Such a set is called inclusion-minimal, and exists whenever any selected set exceeds \(Q\), by repeatedly deleting coordinates while the property persists.
Let \(2^{\ell}\) be its smallest selected power of two and \(e\) its excess above \(Q\).
Deleting the coordinate of weight \(2^{\ell}\) leaves sum at most \(Q\), so \(e\leq2^{\ell}\); positive integer excess gives \(1\leq e\).
The selected sum is divisible by \(2^{\ell}\), while \(Q=2^r-1\equiv-1\pmod{2^{\ell}}\), so \(e\equiv1\pmod{2^{\ell}}\).
Therefore \(e=1\), and
\[
h=F+wK_1,\qquad K_1=2k+\delta_a\geq4,
\]
where \(k\) is the number of selected query inputs and at least two are selected.

If no leading-positive coalition exists, \(h\) is zero or \(wK_1\), with every positive \(K_1\) between two and \(6r+1\).
A direct query sum equal to \(Q\) uses at least \(r\) terms, so at that boundary the stronger lower bound \(K_1\geq2r\) holds.
The gate \(h\) is exactly zero at the origin and unit vectors.

All residue comparisons below are literal after clearing \(2q^2\) and reducing modulo \(F\).
Every undisplayed constant-degree term contains a factor \(F\), and the displayed residues and the remainder formed by the pure balanced contributions are smaller than the resulting period \(F/(2q^2)\).
This follows from \(F>8Q^4\), \(q,r<Q\), and the bound \(2k+\delta\leq6r+1\) for a direct correction with \(0\leq k\leq3r\) selected query inputs and finite-bias correction \(\delta\in\{0,1\}\).
In particular a residue of magnitude at least two, or zero, cannot be turned into output one by a gain-free remainder in \([0,1)\).
We continue to retain every actual activation pattern.

\begin{lemma}[The finite negative parameters are biases]
\label{lem:boolean-finite-biases}

If both gains are hidden-to-output edges in a candidate for the clean query target, both finite negative occurrences are hidden biases at distinct vertices.
\end{lemma}

\begin{proof}
At every clean query winner both gain sources are active, by \Cref{lem:boolean-sole-negative,lem:boolean-sole-positive-output}.
Negative allocation puts the two finite negatives at distinct hidden receivers.
If a receiver vanished identically on a face obtained by fixing at most two query coordinates to zero, all other positive gates could be substituted by their affine preactivations, leaving at most one effective hinge on that face.
\Cref{lem:boolean-clean-block} excludes this possibility.
We assume for a contradiction that at least one finite negative is an internal edge.
The remaining hidden vertex, when present, is a positive wire.
We first find one frozen expression valid on an affinely spanning winning stencil, then compare its affine constant term.

\emph{Gains at the first two vertices.}
Suppose the gains originate at the first two hidden vertices.
At every winner their active values have the form
\[
h=U+a,\qquad g=V+b+th,
\]
where \(U,V\) are direct query sums, \(a,b\) are biases or zero, and \(t\) is their internal edge or zero.
Write \(u_j,v_j\) for the respective direct query coefficients of \(U,V\).
The inverse-\(w\) equation gives \(h^0=g^0\) on the winning stencil.
Affine spanning therefore gives, for every query coordinate \(j\),
\[
v_j^0=(1-t^0)u_j^0.
\]
If \(t\ne0\) and \(u_j\ne0\), this equates the product \(t^0u_j^0\) to the difference \(u_j^0-v_j^0\), contrary to \Cref{lem:boolean-short-products}.
Thus a present \(t\) forces all coefficients of \(U,V\) to vanish.
Both gain sources are then constant, and the possible third vertex supplies only one effective hinge, again impossible.

Consequently \(t=0\), and the affine identities give \(U^0=V^0\) and \(a^0=b^0\).
Equal positive leading primitives have equal full values, and absent or nonfinite equal biases also agree in full.
The only unequal biases with equal leading value are the two finite negatives.
If they occur at \(h,g\), the lemma already holds.
Otherwise \(h=g\) on the whole query cube, so the gains cancel exactly; \Cref{lem:boolean-coefficient-obstruction} excludes a winner.
This treats fewer than three hidden vertices as well.

\emph{One mask on a winning stencil.}
We may now assume that there are three hidden vertices, one gain originates at the last, and the other originates earlier.
Only the remaining hidden vertex, which is not a gain source, can have a varying mask at winners.
A positive wire can be replaced by its affine preactivation on the whole query cube.
At a point where its value is zero, its nonnegative summands are zero; their frozen expressions are zero there too by the specialization argument in \Cref{lem:boolean-specialization}.
Thus this replacement preserves the formal output identities even at such a point.
It suffices to make the negative-receiving vertices' masks constant on a winning stencil, possibly after fixing at most two coordinates to zero.

We use the following consequences of the chosen scales:
\[
F>4(d+2),\qquad P-w>P/2,\qquad
\Gamma>2(d+2)P^2,
\qquad (P-w)F/2>(d+2)P.
\]
The first two follow from the displayed bounds defining \(F,M\).
For the third, \(d<m\), \(P\ge6\), and \(M>2H_0>2048(m+1)^4P^4\), whereas \(\Gamma=M^9>M\).
The fourth follows from the first two by multiplying \((P-w)>P/2\) by \(F/2\).
Every direct query sum is at most \(dP\), and an additional finite positive bias is at most \(P\).

If the first vertex receives a finite negative, it is
\[
h=\rho(U-P_a)
\]
with \(U\) a direct query sum.
It is active at the all-ones vertex and at every vertex obtained from it by deleting up to two coordinates.
For a gain source this follows from winner activity.
Otherwise inactivity there, together with monotonicity, would make it vanish on the entire corresponding zero-fixed face, which was excluded above.
Its leading preactivation is nonnegative at all those points: a negative integer multiple of \(F\) cannot be overcome by the correction, bounded by \((6r+1)w<1\).
The direct query coefficient vector of \(U\) has support on at least two coordinates, since deleting a smaller support would leave the negative bias alone.
At a single deletion choose a different coordinate in that support and delete it too.
Restoring that coordinate adds at least \(F\) to the leading preactivation.
Thus \(h^0\ge F\) on the ordinary winning stencil, and \(h\ge F\) there because its correction is nonnegative.

If the second vertex is positive, its ReLU is already redundant.
Otherwise the internal-negative assumption makes it \(\rho(V+b-P_bh)\), with \(b\) positive or absent.
For \(b\ne\Gamma\), its preactivation is at most \((d+2)P-(P-w)F<0\).
For \(b=\Gamma\), it is at least \(\Gamma-(d+2)P^2>0\), since \(h\le(d+2)P\).
Its mask is therefore constant on the stencil in either case.

Suppose instead that the first vertex is the positive wire \(A=U+a\).
If the second vertex has a negative bias, its form is
\[
\rho(V-P_a+tA),\qquad t\ge0.
\]
It is monotone.
If it is a gain source, it is active at all winners.
If it is not, inactivity at a winning deletion would make it vanish on that zero-fixed face, leaving only one effective negative receiver.
Thus it is uniformly active on the winning stencil.

The remaining second-vertex form is
\[
\rho(V+b-P_aA),\qquad A=U+a.
\]
If this vertex is a gain source, its mask is already active on the stencil.
Otherwise the wire itself is a gain source and is active at all winners.
For \(a=\Gamma\), the second preactivation has the following upper bound on the whole cube:
\[
\Gamma+(d+2)P-(P-w)\Gamma<0.
\]
For a positive finite \(a\), or when \(U\) has at least two supported coordinates, we have \(F/2\le A\le(d+2)P\) on the ordinary winning stencil.
The preceding scale inequalities then make the second gate active when \(b=\Gamma\), and inactive otherwise.
If \(a=0\) and \(U\) has at most one supported coordinate, a winning deletion makes the gain-bearing wire zero, impossible.

The only remaining case is \(a=s_0\) with at most one supported query coordinate in \(U\).
Fix that coordinate to zero if it exists; then \(A=s_0\).
For a positive finite \(b\) or \(b=\Gamma\), the second gate is active everywhere on this face.
For \(b=0\) or \(b=s_0\), it is active precisely when \(V>0\), because
\[
0<P_as_0<1<F.
\]
If \(V\) has at least two free supported coordinates, it is active on the face's winning stencil.
If it has one, fix that coordinate to zero too; if it has none, no further fixing is needed.
It is then uniformly inactive.
At most two zero fixings have been used.

These cases exhaust the orderings: the first vertex is either a negative-biased hinge or the positive wire; a negative edge into the second can only come from the first; and every last gain source is already active at winners.

We have obtained one frozen affine output expression valid at all points of a winning stencil on such a face.
Those points affinely span its free query coordinates.
Specialization makes the expression formally one at each stencil point, so each Laurent coefficient is the corresponding constant affine function on the whole face.
In particular, the affine constant term is identically one.
Zero fixings introduce no new constant terms.

\emph{The inverse coefficient's affine constant term.}
Every finite leading primitive is \(Fn\) with
\[
1\le |n|\le Q,\qquad q\nmid n.
\]
For sources this uses \(\beta_j\equiv3\pmod q\); for queries it uses that \(q\) is odd; for anchors, controls and the two negatives it uses \(q\nmid(Q-T)\) and \(q\nmid Q\).
The two nonfinite leading types are \(\Gamma,s_0\), with \(\Gamma s_0=1/q\).
Let \(H_+,H_-\) be the frozen values of the positive- and negative-gain sources.
The inverse coefficient says \(H_+^0-H_-^0=0\).
Take its affine constant term and then its \(\Gamma^0\) coefficient.
A constant path starts with a bias and uses at most two internal edges.
In the difference of the two gain sources there are at most two direct-bias terms, at most three two-factor terms, and just one possible three-factor term, the chain through all three hidden vertices.
Indeed, number the hidden vertices \(1,2,3\) in topological order, so that the last gain source is vertex three and the other is vertex one or two.
The bias-origin paths to vertex three are its own bias, the bias of vertex one times edge \(13\), the bias of vertex two times edge \(23\), and the bias of vertex one times edges \(12\) and \(23\).
Those to vertex two are its own bias and the bias of vertex one times edge \(12\), and the only one to vertex one is its own bias.
The earlier source therefore adds one direct-bias term and at most one two-factor term, and inactive vertices only remove terms.

In the difference \(H_+^0-H_-^0\), write \(FN_1\) for the signed difference of the direct finite-bias terms.
The quantities \(F^2N_2\) and \(F^3N_3\) collect the signed all-finite path contributions with two and three factors, respectively.
Let \(k\) be the signed count of balanced two-factor paths, each contributing \(1/q\).
Let \(N_q\) be zero if the unique three-factor term with one finite factor and one of each nonfinite type is absent.
Otherwise let \(N_q\) be the signed integer \(n\) in its finite leading factor \(Fn\), so that this term contributes \(FN_q/q\).
Hence
\[
0=FN_1+F^2N_2+F^3N_3+\frac{FN_q}{q}+\frac{k}{q},
\]
and these integers satisfy
\[
|N_1|\le2Q,\quad |N_2|\le3Q^2,\quad |N_3|\le Q^3,\quad |k|\le3.
\]
If the mixed three-factor term is present, \(N_q\) is nonzero modulo \(q\).

Multiply by \(q\) and reduce modulo \(F\).
Since \(|k|<F\), this gives \(k=0\).
After division by \(F\), reduction modulo \(q\) gives \(N_q\equiv0\pmod q\), so that three-factor term is absent and \(N_q=0\).
The remaining equation is \(N_1+FN_2+F^2N_3=0\).
Because \(|N_1|<F\) and \(|N_2|<F\), successive reduction and division by \(F\) give \(N_1=N_2=N_3=0\).
In particular the direct finite biases of the two gain sources have equal leading value.
Under the internal-negative assumption they cannot be the two finite negatives.
They are therefore both absent or nonfinite, or are equal positive finite primitives.
Their full \(w\)-derivatives agree in each case.

\emph{The constant output coefficient.}
The constant \(w^0\Gamma^0\) contribution of the gains is the difference of the \(w\)-derivatives of these two source constant expressions.
The direct-bias derivatives cancel.
A product with at least two finite factors has derivative divisible by \(F/2\), since every finite correction is a half-integer multiple of \(w\) and one finite factor remains after differentiation.
A three-factor product with one finite factor and the balanced nonfinite pair was just excluded.
All-nonfinite products have derivative zero.
Thus the gain-derived constant contribution belongs to \((F/2)\mathbb Z\).

For gain-free constant paths, every finite factor contributes \(F\) at degree \(w^0\), and any denominator divides \(q^2\).
The pure balanced paths contribute \(1/q\) or \(1/q^2\), positively.
The path counts in \Cref{lem:boolean-coefficient-obstruction} also bound this bias-origin subset: taking the affine constant term only discards input-origin paths, and every pure balanced contribution is positive.
The constant output identity consequently has form
\[
1=\frac{FZ}{2q^2}+\frac{k_1}{q}+\frac{k_2}{q^2},
\qquad Z\in\mathbb Z,\quad 0\le k_1,k_2\le p=q-1.
\]
It implies
\[
FZ=2(q^2-qk_1-k_2),\qquad
2\le2(q^2-qk_1-k_2)\le2q^2<F,
\]
which is impossible for an integer multiple of \(F\).
No internal finite-negative edge is therefore possible.
Both finite negatives are biases at their distinct hidden receivers, as claimed.
\end{proof}

\subsection{Output Gains Cannot Originate at a Positive Wire}

By \Cref{lem:boolean-finite-biases}, the finite negatives are now biases at two distinct hidden vertices.
The third vertex, if present, is positive and hence a wire.
We still allow one gain to originate at that wire.
In a topological order of the hidden vertices, the wire is \emph{upstream} when it precedes both biased vertices, \emph{downstream} when it follows both, and \emph{between} them, a between wire, when it follows one biased vertex and precedes the other.
The following proof excludes all three positions of the wire.

\begin{lemma}[The gain sources are the biased vertices]
\label{lem:boolean-gain-sources}

Let a permitted candidate represent the clean query target of \Cref{lem:boolean-clean-block} on its restricted cube.
If both finite negatives are hidden biases and both gains are output edges, neither gain can originate at the remaining positive wire.
\end{lemma}

\begin{proof}

Let \(P_1,P_2\) be the positive magnitudes of the first and second negatively biased vertices in topological order, so \(\{P_1,P_2\}=\{P-w,P\}\).
For \(h=\rho(U-P_1)\), with direct query sum \(U\), set \(K_1=0\) at a tested input where \(h\) is inactive.
When it is active, write \(h=F\kappa+wK_1\), where \(\kappa\in\mathbb Z_{\ge0}\) and \(0<K_1\leq6r+1\).
Thus \(K_1\) is the frozen derivative contribution of the first hinge, including its activation indicator.
At a leading-zero boundary the normalized sum of \(U\) is \(Q\), so \(K_1\geq2r\).
A minimal leading-positive coalition gives \(\kappa=1\), as shown before \Cref{lem:boolean-finite-biases}.
For the second hinge, \(K_2\) denotes only the correction of its direct query sum and finite bias, excluding coupling terms.
Its full derivative may be \(K_2+t^0K_1\) or the coupled expression displayed below; only the direct quantity is bounded by \(6r+1\).
All output-gain sources are active at clean winners by \Cref{lem:boolean-sole-negative,lem:boolean-sole-positive-output}.
The gain-free residues and the period comparison are those stated before \Cref{lem:boolean-finite-biases}.

\emph{An upstream wire.}\par\nopagebreak

A positive bias in the wire makes its zeroth coefficient nonzero everywhere.
A positive query input does so on that input's fixed-one face.
Inverse equality activates the other gain source on that whole cube or face.
Both gain sources are then active, hence equal to their affine preactivations, throughout it, so substitution leaves at most one effective hinge, contradicting the robust face statement of \Cref{lem:boolean-clean-block}.
A wire with neither bias nor clean query input is zero on the restriction and cannot carry a gain at a clean winner.
Thus an upstream gain-bearing wire is excluded.

\emph{A downstream wire.}\par\nopagebreak

Write
\[
h=\rho(U-P_1),\quad g=\rho(V+th-P_2),\quad u=W+ah+bg,
\]
with direct query sums \(U,V\) and positive affine \(W\).
Both hinges are exactly zero at origin and units: there \(U\) and \(V\) are zero or a single query value, at most \(F(Q+1)/2+2w<P-w\), and \(h=0\) removes the term \(th\).
Inverse equality for either pair involving \(u\) forces \(W^0=0\) at those tests, hence \(W=0\) identically, since every present positive primitive has a nonzero leading value.
All three hidden values are therefore zero at origin and units, and the output there is its bias plus at most one skip term.
These ordinary output parameters are positive or absent and the target is zero at these tests, so the output bias and every skip edge from a clean query coordinate are absent; the other skip edges are fixed to zero on the restriction.
Thus the ordinary output contribution, here the output bias and the skip edges, vanishes on the clean restriction.

First put the gains on \(h,u\).
Their inverse equation is
\[
(1-a^0)h^0=b^0g^0.
\]
Suppose \(h^0=F\kappa>0\) at a point.
An active \(g\) then has \(g^0\geq0\) at the prescribed specialization \(\Gamma=M^9\).
A finite \(t\) makes its positive product dominate the finite negative bias.
A \(\Gamma\) edge gives a positive highest degree.
A selector or absent edge can only activate with nonnegative scalar part.
This sign assertion is used only at points where \(h^0>0\).
A finite or \(\Gamma\) value of \(a^0>1\) therefore has the wrong sign.

For absent or selector \(a\), examine the remaining \(b,t\) types.
A finite \(b\) and finite or absent \(t\) would make \(\kappa\) a multiple of \(F\), although \(0<\kappa<F\).
A nonfinite \(t\) instead gives a positive nonfinite coefficient, which cannot match the absent or negative selector coefficient on the left.
For nonfinite \(b\), the only scalar product available on the right is a balanced \(bt=1/q\), giving scalar \(F\kappa/q\) instead of \(F\kappa\).
All other products have nonzero degree, and \(b=0\) would require \(a^0=1\).
Thus \(h^0>0\) is impossible and \(h^0=0\) everywhere.

At the all-ones winner \(h=wK_1\), with \(K_1\geq2\), and \(u^0=0\).
If \(b\) is present and \(g\) is active, inverse equality gives \(g^0=0\); its direct query sum is then \(Q\), and
\[
g^{(1)}=K_2+t^0K_1,\qquad K_2\geq2r.
\]
The constant-degree residue of \((h-u)/w\) is
\[
K_1-I K_1/q,\qquad I\in\{0,1\},
\]
where \(I=1\) only for a balanced \(bt\) pair.
It is at least \(2(1-1/q)>1\).
If \(b\) is absent or \(g\) inactive, that residue is \(K_1\).
Reversing gain signs negates it, and no gain-free remainder below one can repair either orientation.

Next put the gains on \(g,u\).
The inverse equation is
\[
(1-b^0)g^0=a^0h^0.
\]
If \(a=0\), then \(g^0=0\) everywhere.
At a winner \(g\) is active.
Its direct sum has normalized value \(Q\), and a present \(t\) forces \(h^0=0\) by finite magnitude or nonfinite degree separation.
Its derivative is \(K_2+t^0K_1\), with \(K_2\geq2r\) and \(0\leq K_1\leq6r+1\).
The gain residue is
\[
K_2-I K_1/q\geq2r-(6r+1)/q>1.
\]
A missing \(t\) gives the same expression with \(I=0\).
This excludes \(a=0\).

If \(a\) is present and \(h\) has a leading-positive coalition, choose a minimal one, so \(h^0=F\).
The inverse equation makes \(g\) active and gives
\[
(1-b^0)(v-Q+t^0)=a^0,
\]
where \(v\) is the direct normalized sum in \(V\).
Finite or \(\Gamma\) values \(b^0>1\) have the wrong sign by the positive-\(h^0\) argument above.
For \(b=s_0\), a finite or absent \(t\) makes \(X=v-Q+t^0\) a scalar, so the left side \(X-Xs_0\) has scalar coefficient \(X\) and \(s_0\)-coefficient \(-X\); it is zero or has two monomials, and in neither case equals the present positive primitive \(a^0\).
For \(t=\Gamma\), the scalar coefficient contains the nonintegral term \((v-Q)-1/q\), while the degree-one coefficient forces \(a=\Gamma\).
For \(t=s_0\), a negative \(s_0^2\) coefficient is unmatched.
Thus \(b=0\).
Then \(v-Q=a^0-t^0\); since \(|v-Q|<F\), finite differences force zero, and nonfinite differences must cancel identically.
Hence \(v=Q\) and \(a=t\) in full.
At this coalition \(g-u=V-P_2=wK_2\), \(K_2\geq2r\), whose signed gain value cannot be either Boolean label.
Other ordinary terms have only an \(F\)-multiple scalar coefficient here.

If \(h\) has no leading-positive coalition, both \(h^0,g^0\) vanish everywhere.
At the all-ones winner the preceding derivative calculation again gives \(K_2-I K_1/q>1\).
Thus both downstream gain pairs are excluded.

\emph{A between wire.}\par\nopagebreak

Write
\[
h=\rho(U-P_1),\quad u=V+ah,\quad g=\rho(W+bu+ch-P_2),
\]
with direct query sums \(U,W\) and positive affine \(V\).
For gain pair \(h,u\), the origin and the unit vectors of the restricted cube give \(V=0\).
The inverse equation \((1-a^0)h^0=0\) makes \(h^0=0\) everywhere, and its all-ones gain residue is \(\pm K_1\), \(K_1\geq2\).
Both biased hinges and the wire are zero at origin and units, so the output there is its bias plus at most one skip term.
These ordinary output parameters are positive or absent and the target is zero at these tests, so the output bias and every skip edge from a clean query coordinate are absent; the other skip edges are fixed to zero on the restriction.
Thus the ordinary output contribution, here the output bias and the skip edges, vanishes on the clean restriction.
The remaining gain-free terms cannot repair the residue.

For gain pair \(u,g\), first remove \(V\).
A positive bias of \(V\) would give, at origin,
\[
(b^0-1)V^0=P.
\]
A nonzero \(b\) would make its product with one primitive equal a sum of two primitives, contrary to \Cref{lem:boolean-short-products}; absent \(b\) has the wrong sign.
Hence \(V\) has no bias.
At a unit vector with a present query coefficient of \(V\), inverse equality requires
\[
W_j^0+(b^0-1)V_j^0=P.
\]
A finite \(b\) has an oversized \(F^2\) contribution, a nonfinite \(b\) has unmatched degree, and absent \(b\) gives \(W_j^0-V_j^0<P\), since \(W_j^0\leq F(Q+1)/2\).
Thus \(V=0\) identically and \(u=ah\).
If \(a=0\), the wire gain source is zero at a winner, already impossible.

For present \(a\), a minimal leading-positive \(h\) coalition gives \(h^0=F\), \(g^0=a^0F\), and
\[
v_W-Q+(b^0-1)a^0+c^0=0.
\]
Here \(v_W\) is the direct normalized sum in \(W\), with \(|v_W-Q|<F\).
Let \(a^0=F\nu\) be finite and suppose \(b^0=F\mu\) is finite, where \(\nu,\mu\geq1\) because positive finite primitives have positive normalized leading values.
A nonfinite \(c\) would be the only term at its \(\Gamma\)-degree, so \(c\) is finite or absent, with \(c^0=Fn_c\) and \(n_c\geq0\).
The equation becomes \(v_W-Q+F(F\mu\nu-\nu+n_c)=0\), and reduction modulo \(F\) gives \(v_W=Q\).
Then \(F\mu\nu-\nu+n_c=0\), although \(F\mu\nu-\nu+n_c\geq(F-1)\nu>0\), which is impossible.
A nonfinite \(b\) has an unmatched positive degree.
For nonfinite \(a\), finite \(b\) gives positive coefficient \((F\mu-1)a^0\), and a same-type \(b\) gives a square degree, neither cancellable by positive \(c\).
An opposite-type \(b\) leaves \(v_W-Q+1/q=0\) after the nonfinite terms cancel, impossible for integer \(v_W-Q\).
The only remaining choice is \(b=0\), \(c=a\) in full and \(v_W=Q\).
Then \(g-u=W-P_2=wK_2\), \(K_2\geq2r\), again incompatible with either Boolean label.

Finally, if \(h\) has no leading-positive coalition, \(h^0=u^0=g^0=0\) everywhere.
At all-ones both gain sources are active, so \(h=wK_1\), the direct \(W\)-sum is \(Q\), and
\[
g^{(1)}=K_2+(b^0a^0+c^0)K_1,
\qquad K_2\geq2r.
\]
The constant residue of \((g-u)/w\) is \(K_2+I K_1/q>1\), with \(I\in\{0,1\}\).
Reversing gain signs again negates it.
This excludes both between-wire pairs and completes all wire positions.
\end{proof}

\subsection{Removing Coupling Between the Biased Hinges}

The output gains now originate at the two negatively biased vertices themselves.
Both are active at every clean winner by \Cref{lem:boolean-sole-negative,lem:boolean-sole-positive-output}.
This lets us compare their affine preactivations on a winning stencil, without first assuming a shared score.
The resulting restrictions concern actual edges and will therefore hold on the full input cube.

\begin{lemma}[Parallelization]
\label{lem:boolean-parallelization}

Consider any permitted candidate representing \(f\) in which both gains are hidden-to-output edges, both finite negatives are biases at two distinct hidden vertices, and the two gains originate at these two biased vertices.
Suppose that its remaining hidden vertex, if present, is a positive wire.
Write the two biased vertices as \(h,g\), in their relative topological order, and the optional positive wire as \(u\).
Each possible position of \(u\) gives one of the forms below, and each form satisfies the stated conclusion.
In the following forms, \(A,B\) are direct input sums, \(V\) is positive affine, \(\{P_1,P_2\}=\{P-w,P\}\), and all lettered coefficients are positive nongain primitives or zero.
If the wire is upstream, so that
\[
u=V,\quad h=\rho(A+au-P_1),\quad g=\rho(B+bu+th-P_2),
\]
then the edge \(t\) is absent.
If the wire is downstream, so that
\[
h=\rho(A-P_1),\quad g=\rho(B+th-P_2),\quad u=V+ah+bg,
\]
then the edge \(t\) is absent.
If the wire lies between \(h\) and \(g\), so that
\[
h=\rho(A-P_1),\quad u=V+ah,\quad g=\rho(B+bu+ch-P_2),
\]
then the edge \(c\) is absent and at least one of \(a,b\) is absent.
The conclusions hold for actual edges globally, not just their contributions on the clean block.
\end{lemma}

\begin{proof}

Restrict to the clean query block.
The two biased vertices \(h,g\), from which the output gains originate, are active at the all-ones query vertex and every single deletion.
Their inverse output equation equates their zeroth coefficients there.
These points affinely span the query cube, so their affine query coefficients agree as Laurent polynomials.
For each query coordinate \(j\), write \(A_j,B_j,V_j\) for the corresponding zeroth direct coefficients.
Each is zero or \(F\) times a power of two with exponent in \(\{0,\ldots,r-1\}\).

For the upstream form the equality is
\[
(1-t^0)(A_j+a^0V_j)=B_j+b^0V_j.
\]
Some \(A_j+a^0V_j\) is nonzero; otherwise \(h\) is constant on the clean cube and only \(g\) remains an effective hinge.
For that coordinate, finite or \(\Gamma\) values \(t^0>1\) make the left side negative and the right side nonnegative at the prescribed specialization.
If \(t=s_0\) and \(a\) is finite or absent, its left side has a negative selector coefficient, whereas the right has only nonnegative coefficients.
If \(a=\Gamma\), the negative selector term forces \(A_j=0\).
Then \(V_j>0\); the degree-one term forces \(b=\Gamma\), while the remaining scalar equation requires \(B_j=-V_j/q<0\).
If \(a=s_0\), the negative square-selector term first forces \(V_j=0\), after which the negative selector multiple of \(A_j\) is unmatched.
All positive types of \(t\) are excluded, so its edge is absent.

For the downstream form the query equation is
\[
(1-t^0)A_j=B_j.
\]
Some \(A_j>0\), since otherwise the first biased gate is identically zero and cannot be a gain source at a winner.
Finite or \(\Gamma\) values of \(t\) give a negative left side.
A selector gives an unmatched negative selector coefficient.
Thus this \(t\) is absent too.

For the between form,
\[
(1-b^0a^0-c^0)A_j=B_j+b^0V_j.
\]
Choose \(A_j>0\), put \(R_j=B_j/A_j\), \(T_j=V_j/A_j\), and rearrange as
\[
1=c^0+b^0a^0+R_j+b^0T_j.
\]
Each nonzero ratio is a power of two at least \(2^{-(r-1)}\).
Every term on the right is a nonnegative Laurent monomial.
A nonfinite \(c\) has an exponent which cannot cancel, and a finite \(c\) is at least \(F>1\).
Hence \(c\) is absent.
If \(b^0T_j\ne0\), a finite \(b\) gives a value at least \(F/2^{r-1}>1\), and a nonfinite \(b\) gives an uncancellable exponent.
Thus \(b^0T_j=0\); this does not assert \(T_j=0\) when \(b\) is absent.

The product \(b^0a^0\) is either zero, larger than one if both factors are finite, of nonzero degree, or the balanced constant \(1/q\).
The last possibility would require \(R_j=1-1/q\).
This positive number is neither zero nor a power of two: in lowest terms its denominator is the odd prime \(q\), whereas a rational power of two has a power-of-two denominator.
Therefore \(b^0a^0=0\), so at least one of the actual positive edges \(a,b\) is absent.
Every excluded primitive has a nonzero zeroth coefficient, making these genuine edge absences.

When \(a=0\), the between wire is independent of \(h\) and can be reordered upstream.
When \(b=0\), it can be reordered downstream.
The graphs and parameter occurrences are unchanged by this topological reordering.
Thus the full network has two parallel biased hinges with a positive wire upstream or downstream, or with that wire absent.
\end{proof}

\subsection{The Full-Cube Common Ramp}

The graph restrictions have so far used one query block.
Their actual edge absences hold globally, but equality of coefficient vectors on that block alone is insufficient for witness extraction.
We now use every origin/unit test and the affinely spanning nonminimal winners of the full target.
All positive source, anchor, query, control, \(\Gamma\), and selector primitives are allowed in arbitrary input positions in this part of the argument.

\begin{lemma}[A local sharing comparison]
\label{lem:boolean-local-sharing}

Let \(U_0\ne0\) be a sum of at most two positive nongain leading primitives.
Let \(a_0,b_0,d_0,e_0\) be positive nongain leading primitives or zero.
Then
\[
(a_0-b_0)U_0=e_0-d_0
\]
forces \(a_0=b_0\) and \(d_0=e_0\).
\end{lemma}

\begin{proof}

Write \(U_0=F\nu+\gamma\Gamma+\sigma s_0\), with \(\nu\geq0\), \(\gamma,\sigma\in\{0,1,2\}\), and at most two summands in all.
A finite coefficient on the right is an \(F\)-multiple of magnitude at most \(P\), while its coefficient relative to \(\Gamma\) or \(s_0\) has absolute value at most one.
If \(a_0,b_0\) are finite or zero and unequal, their difference is a nonzero \(F\)-multiple.
A finite summand in \(U_0\) gives magnitude at least \(F^2>P\), and a nonfinite summand gives coefficient magnitude at least \(F>1\).
Thus this case requires equality.

For \(a_0=\Gamma,b_0=s_0\), the degree-two and degree-minus-two equations give \(\gamma=\sigma=0\).
The remaining finite summand would have coefficient at least \(F\) at both nonfinite degrees, impossible on the right.
For \(a_0=\Gamma\) and finite or zero \(b_0\), the degree-two equation first gives \(\gamma=0\).
The degree-one equation then gives \(\nu=0\), since its coefficient \(F\nu\) is either zero or at least \(F\).
The remaining constant coefficient \(\sigma/q\), with \(1\leq\sigma\leq2<q\), cannot be an \(F\)-multiple.
For \(a_0=s_0\) and finite or zero \(b_0\), use degree minus two, then degree minus one, then the same constant comparison \(\gamma/q\).
Reversing \(a_0,b_0\) changes only the sign of both sides.
Equal nonfinite types are already identical.
These alternatives exhaust the types and prove \(a_0=b_0\); the displayed equation then gives \(d_0=e_0\).
\end{proof}

This comparison is applied at an origin or unit point, where a positive affine wire has at most its bias and one input contribution.
It also applies to a single wire input when global coefficient equality is established.

\begin{lemma}[Global common-ramp form]
\label{lem:boolean-global-ramp}

Suppose the finite negatives are the biases of two parallel hidden hinges, and their output edges are \(+D,-D\).
Suppose the optional third vertex is a positive wire upstream or downstream, with no path coupling the hinges.
If the network equals \(f\) on the full Boolean cube, then it computes the function
\[
D\left(\rho(S(x)-(P-w))-\rho(S(x)-P)\right),
\qquad S(x)=a u(x)+\sum_{j=1}^{d}b_jx_j.
\]
Here \(u\) is an optional positive affine wire with one optional actual primitive per input and one optional actual primitive bias.
The coefficient \(a\) is the weight of the sharing edge from \(u\) to each hinge, and \(b_j\) is the weight of the direct input edge from \(x_j\) to each hinge, as in \Cref{lem:boolean-placement}.
A nonzero \(a\) uses two identical-value actual occurrences, one for each sharing edge, and each nonzero \(b_j\) likewise uses two identical-value occurrences.
All other output contributions vanish.
If the wire was downstream, it can be omitted and \(a=0\).
\end{lemma}

\begin{proof}

Write \(h_A\) for the positive-gain hinge, \(h_B\) for the negative-gain hinge, and \(\{P_A,P_B\}=\{P-w,P\}\) for their bias magnitudes.
The full target has all the properties of \Cref{lem:boolean-full-target}.
We treat the two wire positions separately.

\emph{An upstream wire.}\par\nopagebreak

Let \(d_A,d_B\in\mathbb R_{\geq0}^d\) be the vectors of direct input coefficients, with absent edges represented by zeros.
Write
\[
u=v_{\mathrm b}+\sum_{j=1}^{d}v_jx_j,
\quad h_A=\rho(a u+d_A\cdot x-P_A),
\quad h_B=\rho(a' u+d_B\cdot x-P_B),
\]
where \(a\) and \(a'\) are the sharing weights into \(h_A\) and \(h_B\), and \(N=D(h_A-h_B)+E(x)\), where \(E\) is the nonnegative affine sum of ordinary output bias, skips and a possible wire output.

At an origin or unit point, the positive score of a hinge, its preactivation without its negative bias, is a direct primitive plus its sharing primitive times a sum of at most two primitives.
We show that a hinge whose zeroth score, the \(w^0\)-coefficient of this positive score, equals \(P\) at such a point is inactive.
In that case every finite-times-finite product is too large, every unbalanced nonfinite degree cannot cancel, and any balanced contributions total \(k/q\) with \(0\leq k\leq2<q\).
The latter cannot complete an integer \(F\)-multiple to \(P\).
The shared contribution must therefore be zero in full, and the only possible direct primitive with leading value \(P\) is the control primitive \(C\).
Its actual value \(P-3w/2\) is below both thresholds.
Such a hinge is inactive.

At each of these target-zero points, a positive-only active hinge would make \(N>0\).
A negative-only active hinge would require zeroth preactivation zero by the inverse equation, which the preceding comparison has just excluded.
If both are active, inverse equality and \Cref{lem:boolean-local-sharing} identify their full scores at that point; if the wire value is zero, direct primitive equality suffices.
Their full preactivation difference is then \(w\) or \(-w\), so \(N=1+E\) or \(N=-1+E\).
The first cannot be zero.
The second requires \(E=1\).
At an origin or unit point every term of \(E\) is a positive primitive or a product of two positive primitives.
Nonzero nonfinite degrees cannot cancel.
A finite scalar term is at least \(F\), while balanced products total at most \(2/q<1\).
Thus \(E\) cannot be one.
Both hinges are consequently off at the origin and at every unit vector.
The nonnegative affine function \(E\) vanishes there, so it is identically zero.

Every coordinate has a strictly positive effective coefficient in the positive-gain score.
Otherwise \(N=D(h_A-h_B)\) is nonincreasing in that coordinate, and monotonicity of \(f\) would make it inessential, contrary to \Cref{lem:boolean-full-target}.
Let \(y=x+e_j\) be a nonminimal winner with \(x\) also winning.
The positive hinge is active at \(x\) and its score strictly increases at \(y\).
If the negative hinge were off at \(y\), it would also be off at \(x\), making \(N(y)>N(x)=1\).
Therefore both are active at every nonminimal winner.

Let \(A,B\) denote their affine preactivations, including biases.
At those winners the formal output identity gives \(A-B=w\).
For each Laurent monomial the difference has an affine coefficient in \(x\).
The nonminimal winners affinely span the full input space, so \(A-B=w\) identically as a formal affine expression.
If the wire bias is present, its leading constant identity forces \(a^0=(a')^0\), hence \(a=a'\) in full.
If the wire bias is absent but some wire input \(v_j\) is present, use the leading coefficient identity
\[
(a^0-(a')^0)v_j^0=(d_B)_j^0-(d_A)_j^0
\]
and \Cref{lem:boolean-local-sharing} to obtain the same full sharing equality.
Once sharing agrees, every leading direct pair agrees, and \Cref{lem:boolean-short-products} identifies its full value.
If the wire is absent or zero, direct equality alone suffices and it is omitted rather than retained as an illegal empty vertex.

The bias orientation is correct when the positive-gain hinge \(h_A\) carries the bias magnitude \(P-w\), so that \(P_A=P-w\) and \(P_B=P\).
It is reversed when \(P_A=P\) and \(P_B=P-w\).
Reversed bias orientation would require \((a-a')v_{\mathrm b}=2w\).
For a present bias the sharing values are equal, and for an absent bias its product is zero.
Either way this is impossible.
Thus \(P_A=P-w\), \(P_B=P\), with a common full score \(S\).
The two distinct physical edges in every nonzero sharing or direct pair use distinct inventory occurrences, as claimed.

\emph{A downstream wire.}\par\nopagebreak
Write
\[
h_A=\rho(d_A\cdot x-P_A),\quad h_B=\rho(d_B\cdot x-P_B),\quad u=V+\theta_A h_A+\theta_B h_B.
\]
Let \(c_u\geq0\) be the wire's ordinary output coefficient, and put \(\alpha=c_u\theta_A\geq0\), \(\beta=c_u\theta_B\geq0\).
Let \(c_x\in\mathbb R_{\geq0}^d\) list the remaining direct output coefficients, and let \(c_{\mathrm b}\geq0\) be the output bias.
The remaining affine output is \(E_0=c_uV+c_x\cdot x+c_{\mathrm b}\geq0\).
Then
\[
N=(D+\alpha)h_A-(D-\beta)h_B+E_0.
\]
If \(D-\beta\leq0\), this is a nonnegative combination of positive-input hinges and an affine function, hence supermodular by \Cref{lem:boolean-supermodular}.
The target's OR face excludes that case.
Thus \(D-\beta>0\).

At origin both hinges are off, so \(E_0(0)=0\).
At a unit vector the only primitive large enough to activate either direct hinge is \(\Gamma\): every finite primitive and \(s_0\) lies below both thresholds.
If one is active, the inverse equation forces both active and their direct leading primitives equal.
Both are \(\Gamma\), consuming the two available occurrences.
Their gain difference is \(1\) or \(-1\).
The positive orientation cannot give zero; the negative orientation would require the total ordinary output at that unit to equal one.

No \(\Gamma\) remains for \(c_u,\theta_A,\theta_B,V,c_x\).
The ordinary output there is
\[
c_uV_j+(c_x)_j+c_u\theta_A(\Gamma-P_A)+c_u\theta_B(\Gamma-P_B).
\]
Its \(\Gamma^0w^0\)-coefficient lies in \((F/q)\mathbb Z\).
Direct finite values and finite products have factor \(F\).
A finite-selector product multiplied by the displayed \(\Gamma\) contributes \(Fn/q\).
Two selectors have nonzero degree after that multiplication, and the negative threshold parts still have factor \(F\) whenever their degree is zero.
The origin test has removed all bias contributions.
Since \(F>q\), this coefficient cannot be one.
Therefore both hinges are off at all units, and \(E_0\), a nonnegative affine function vanishing at origin and all units, is identically zero.

Every coefficient of \(d_A\) is strictly positive by essentiality: with it zero, the positive hinge is independent of that input while the negative hinge is nondecreasing, making \(N\) nonincreasing.
At any winner \(h_A>0\) because \(D-\beta>0\).
For adjacent winning vertices \(y=x+e_j\), an inactive \(h_B(y)\) would also make \(h_B(x)=0\) and give
\[
N(y)-N(x)=(D+\alpha)(d_A)_j>0,
\]
contrary to the equal labels one.
Thus both hinges are active on the spanning nonminimal winners.
Their inverse coefficients give \((d_A^0-d_B^0)\cdot y=0\) there.
Affine spanning and full positive primitive identification yield \(d_A=d_B\).
Let \(S=d_A\cdot x\), with every coefficient positive.

If the bias orientation is reversed, so that \(P_A=P\) and \(P_B=P-w\), the signed gain contribution on these winners is minus one.
The ordinary term must equal two there, so the affine function
\[
(\alpha+\beta)S-\alpha P_A-\beta P_B
\]
equals two on an affinely spanning set.
It is therefore the constant two everywhere.
Its slope coefficients force \(\alpha+\beta=0\), since every coefficient of \(S\) is positive.
Nonnegativity gives \(\alpha=\beta=0\), making that function zero, a contradiction.
Hence the bias orientation is correct.

At any nonminimal winner the signed gain contribution is exactly one, and both hinges are positive.
The ordinary term \(\alpha h_A+\beta h_B\) must be zero, forcing \(\alpha=\beta=0\).
Together with \(E_0=0\), this removes every ordinary contribution and gives the claimed common ramp with no sharing wire.
A present downstream wire would have a positive output edge by reachability; all its incoming contributions then vanish, making it parameterless.
It is consequently omitted in the legal common-ramp representation.
\end{proof}

\subsection{Completion of the Placement Argument}

We can now combine the local restrictions with the full-cube rigidity statement.
The logical order matters: coefficient comparisons on one query block force actual internal edges absent, and only then are full-coordinate scores identified.

\textbf{Restatement of \Cref{lem:boolean-placement}.}\par\nopagebreak
Let \(W,f\) be the calibrated inventory and target above, with computational-node cap four and affine output.
If any permitted network represents \(f\) on the Boolean cube, then some permitted network using a submultiset of the same occurrences has the form
\[
D\left(\rho(S(x)-(P-w))-\rho(S(x)-P)\right),
\qquad S(x)=a u(x)+\sum_{j=1}^{d}b_jx_j.
\]
The coefficients \(a,b_1,\ldots,b_d,v_{\mathrm b},v_1,\ldots,v_d\) are nonnegative reals; a zero denotes an absent parameter, and each nonzero coefficient is a single positive actual inventory primitive.
The optional wire is \(u(x)=v_{\mathrm b}+\sum_{j=1}^{d}v_jx_j\), also called the sharing wire.
It enters each hinge through one \emph{sharing edge} of weight \(a\), and each input \(x_j\) enters each hinge through one direct input edge of weight \(b_j\).
Every nonzero \(a\) uses two identical-value occurrences for its two sharing edges, and every nonzero \(b_j\) uses two identical-value occurrences for its two direct input edges.
The two displayed biases and gains use their specified supplied occurrences.
The hinges carry no other bias, so the \emph{score bias}, the constant term of \(S\), is \(a v_{\mathrm b}\).
The output has no \emph{ordinary output contribution}, meaning no output bias, no skip edge and no edge from the wire to the output.
A wire without incoming parameters, or with no sharing edge, is omitted.

\begin{proof}[Proof of \Cref{lem:boolean-placement}]

A permitted candidate has at most three hidden vertices.
Choose its clean block from \Cref{lem:boolean-clean-block} and keep the actual activation patterns.
The formal identities follow from \Cref{lem:boolean-specialization}.
If neither gain is used, \Cref{lem:boolean-coefficient-obstruction} excludes a winner.
\Cref{lem:boolean-gain-bias} excludes both output-bias positions.
\Cref{lem:boolean-input-gains} excludes gain input edges and an unused gain, using legal pruning only in the case where \(+D\) is internal and \(-D\) is unavailable on the clean cube, which it reduces to \Cref{lem:boolean-sole-positive-internal} or, when pruning discards \(+D\), to \Cref{lem:boolean-coefficient-obstruction}.
Thus both gains have hidden sources or are hidden biases.

If both are internal edges, \Cref{lem:boolean-internal-edge-gains} excludes them.
If both are internal and one or both are biases, \Cref{lem:boolean-internal-bias-gains} excludes them.
If exactly one is an output edge, \Cref{lem:boolean-positive-mixed,lem:boolean-negative-mixed} exclude its two sign orientations with every internal position of the other gain.
The two gains must therefore be distinct hidden-to-output edges.

\Cref{lem:boolean-finite-biases} puts both finite negatives at distinct hidden biases.
Both gains and both finite negatives are now placed, so the remaining hidden vertex, if present, receives only positive parameters and is a positive wire.
\Cref{lem:boolean-gain-sources} puts the gains at those same biased vertices, rather than at the optional positive wire.
The positive wire can be ordered upstream, downstream, or between the two biased vertices; incomparable and omitted cases are included by missing edges.
\Cref{lem:boolean-parallelization} removes every direct or wire-mediated coupling between the hinges.
These are actual absent edges, so the conclusion holds on all coordinates, not merely on the chosen block.
A between wire can be reordered upstream or downstream without changing an edge or replacing a numerical value.

The full-target properties of \Cref{lem:boolean-full-target} now apply to \Cref{lem:boolean-global-ramp}.
They force the correct bias/gain orientation, identical full sharing and direct coefficients, and zero ordinary output contribution.
This is precisely the stated common-ramp form with actual paired occurrences.
Any omitted zero contribution only discards parameters and cannot increase the node count.
The wire is retained only when it has an incoming parameter and a sharing path to output, so the resulting representation is legal.
No step replaces a primitive by an effective numerical coefficient that would require a new inventory occurrence.
\end{proof}
\end{document}